\documentclass{nature}

\usepackage{pdflscape}
\usepackage{graphicx}
\usepackage{multirow}
\usepackage{float}
\usepackage{media9}
\usepackage{lineno}

\usepackage{titlesec}
\titleformat{\section}{\normalfont\large\bfseries}{\thesection.}{3pt}{\space}[]
\titlespacing*{\section}{0em}{1ex}{1em}[0em]
\titleformat{\subsection}{\bfseries}{\thesubsection.}{3pt}{\space}[]
\titlespacing*{\subsection}{0em}{1ex}{1em}[0em]

\usepackage{xcolor}
\usepackage[colorlinks,breaklinks=true,plainpages=false,citecolor=blue,linkcolor=blue,urlcolor=blue,bookmarksopen=true,bookmarksnumbered=false,bookmarksdepth=5]{hyperref}
\usepackage[symbol]{footmisc}
\usepackage{xurl}
\urldef{\springervideo}\url{https://static-content.springer.com/esm/art%3A10.1038%2Fs41550-025-02697-2/MediaObjects/41550_2025_2697_MOESM2_ESM.mp4}

\usepackage{enumitem}
\usepackage{amsmath}
\usepackage{amssymb}
\usepackage{hyperref}
\usepackage{subcaption}
\usepackage{caption}
\usepackage{tikz}
\usepackage{float}
\usepackage{afterpage}

\def\farcs{\hbox{$.\!\!^{\prime\prime}$}}

\newcommand{\btheta}{\boldsymbol{\theta}}
\newcommand{\bmu}{\boldsymbol{\mu}}
\newcommand{\bSigma}{\boldsymbol{\Sigma}}
\newcommand{\boldm}{\boldsymbol{m}}
\newcommand{\by}{\boldsymbol{y}}

\usepackage{verbatim}

\usepackage{rotating}

\usepackage{amsmath,amssymb,amsfonts}
\usepackage{empheq}
\usepackage{amsthm}
\usepackage{mathrsfs}
\usepackage[title]{appendix}
\usepackage{textcomp}
\usepackage{manyfoot}
\usepackage{booktabs}
\usepackage{algorithm}
\usepackage{algorithmicx}
\usepackage{algpseudocode}
\usepackage{listings}
\usepackage{physics}
\usepackage{bm}

\definecolor{orcidcolor}{HTML}{A6CE39}
\newcommand{\orcidicon}{%
  \tikz[baseline=-0.5ex]{
    \draw[fill=orcidcolor, draw=none] (0,0.14) circle [radius=0.5ex];
    \node[white, scale=0.3] at (0,0.13) {\textbf{ID}};
  }%
}
\newcommand{\orcid}[1]{\href{https://orcid.org/#1}{\textcolor[HTML]{A6CE39}{\orcidicon}}}

\def\la{\mathrel{\mathchoice{\vcenter{\offinterlineskip\halign{\hfil$\displaystyle##$\hfil\cr<\cr\sim\cr}}}
{\vcenter{\offinterlineskip\halign{\hfil$\textstyle##$\hfil\cr<\cr\sim\cr}}}
{\vcenter{\offinterlineskip\halign{\hfil$\scriptstyle##$\hfil\cr
<\cr\sim\cr}}}
{\vcenter{\offinterlineskip\halign{\hfil$\scriptscriptstyle##$\hfil\cr><cr\sim\cr}}}}}

\def\ga{\mathrel{\mathchoice{\vcenter{\offinterlineskip\halign{\hfil$\displaystyle##$\hfil\cr>\cr\sim\cr}}}
{\vcenter{\offinterlineskip\halign{\hfil$\textstyle##$\hfil\cr>\cr\sim\cr}}}
{\vcenter{\offinterlineskip\halign{\hfil$\scriptstyle##$\hfil\cr
<\cr\sim\cr}}}
{\vcenter{\offinterlineskip\halign{\hfil$\scriptscriptstyle##$\hfil\cr><cr\sim\cr}}}}}

\newcount\longrefs
\def\aap{\ifnum\longrefs=1 {Astron.\ Astrophys.}\else 
                           {A\hbox{\rm \&}A}\fi}
\def\aapl{\ifnum\longrefs=1 {Astron.\ Astrophys.\ Lett.}\else 
                           {A\hbox{\rm \&}A}\fi}
\def\aapr{\ifnum\longrefs=1 {Astron.\ Astrophys.\ Rev.}\else 
                            {A\hbox{\rm \&}AR}\fi}
\def\aaps{\ifnum\longrefs=1 {Astron.\ Astrophys.\ Suppl.}\else 
                            {A\hbox{\rm \&}AS}\fi}
\def\aj{\ifnum\longrefs=1 {Astron.\ J.}\else 
                          {AJ}\fi} 
\def\ao{\ifnum\longrefs=1 {Applied Optics}\else 
                           {Appl.\ Opt.}\fi} 
\def\aspcs{\ifnum\longrefs=1 {Astron.\ Soc.\ Pacific Conf. Series}\else 
                           {ASP Conf.\ Ser.}\fi} 
\def\apj{\ifnum\longrefs=1 {Astrophys.\ J.}\else 
                           {ApJ}\fi} 
\def\apjl{\ifnum\longrefs=1 {Astrophys.\ J.\ Lett.}\else 
                            {ApJ}\fi} 
\def\aplett{\ifnum\longrefs=1 {Astrophys.\ J.\ Lett.}\else 
                            {ApJ}\fi} 
\def\apjs{\ifnum\longrefs=1 {Astrophys.\ J.\ Suppl.}\else 
                            {ApJS}\fi}
\def\apss{\ifnum\longrefs=1 {Astrophys.\ and Space Science}\else 
                            {Ap\hbox{\rm \&}SS}\fi}
\def\araa{\ifnum\longrefs=1 {Ann.\ Rev.\ Astron.\ Astrophys.}\else 
                            {ARA\hbox{\rm \&}A}\fi}
\def\azh{\ifnum\longrefs=1 {Astronomicheskii Zhurnal}\else 
                            {Astron.\ Zhur.}\fi}
\def\baas{\ifnum\longrefs=1 {Bull.\ Am.\ Astron.\ Soc.}\else 
                            {BAAS}\fi}
\def\bain{\ifnum\longrefs=1 {Bull.\ Astronom.\ Institutes Netherlands}\else
                            {Bull.\ Astr.\ Inst.\ Neth.}\fi}
\def\gca{\ifnum\longrefs=1 {Geochim.\ Cosmochim.\ Acta}\else 
                           {Geochim.\ Cosmochim.\ Acta}\fi}
\def\grl{\ifnum\longrefs=1 {Geophys.\ Res.\ Lett.}\else 
                           {Geoph.\ Res.\ Lett.}\fi}
\def\iaucirc{\ifnum\longrefs=1 {IAU Circulars}\else 
                          {IAU Circ.}\fi}
\def\ip{\ifnum\longrefs=1 {in press}\else 
                          {in press}\fi}
\def\jchemp{\ifnum\longrefs=1 {J.\ Chem.\ Phys.}\else 
                           {J.\ Chem.\ Phys.}\fi}  
\def\jcp{\ifnum\longrefs=1 {J.\ Chem.\ Phys.}\else 
                           {J.\ Chem.\ Phys.}\fi}  
\def\jgr{\ifnum\longrefs=1 {J.\ Geophys.\ Res.}\else 
                           {J.\ Geophys.\ Res.}\fi}  
\def\jmolspec{\ifnum\longrefs=1 {J.\ Mol.\ Spectrosc.}\else 
                           {J.\ Mol.\ Spectrosc.}\fi}  
\def\jqsrt{\ifnum\longrefs=1 {J.\ Quant.\ Spectrosc.\ Radiat.\ Transfer}\else 
                           {J.\ Quant.\ Spectrosc.\ Radiat.\ Transfer}\fi}  
\def\jrasc{\ifnum\longrefs=1 {J.\ Royal Astron.\ Soc.\ Canada}\else 
                           {JRAS Can.}\fi}  
\def\mnras{\ifnum\longrefs=1 {Mon.\ Not.\ Roy.\ Astron.\ Soc.}\else 
                             {MNRAS}\fi} 
\def\nat{\ifnum\longrefs=1 {Nature}\else 
                           {Nat}\fi}
\def\pasj{\ifnum\longrefs=1 {Pub.\ Astron.\ Soc.\ Japan}\else 
                            {PASJ}\fi} 
\def\pasp{\ifnum\longrefs=1 {Pub.\ Astron.\ Soc.\ Pacific}\else 
                            {PASP}\fi} 
\def\physscr{\ifnum\longrefs=1 {Physica Scripta}\else 
                            {Phys.\ Scrip.}\fi} 
\def\planss{\ifnum\longrefs=1 {Planetary \& Space Science}\else 
                            {Plan. \& Space Sci.}\fi} 
\def\procspie{\ifnum\longrefs=1 {Proc.\ SPIE}\else 
                            {Proc.\ SPIE}\fi} 
\def\qjras{\ifnum\longrefs=1 {Quarterly J.\ Royal Astron.\ Soc.}\else 
                            {QJRAS}\fi} 
\def\sa{\ifnum\longrefs=1 {Soviet Astron..}\else 
                               {Sov.\ Astron.}\fi}
\def\skytel{\ifnum\longrefs=1 {Sky \& Telescope}\else 
                            {Sky \& Tel.}\fi} 
\def\solphys{\ifnum\longrefs=1 {Solar Phys.}\else 
                               {Solar Phys.}\fi}
\def\ssr{\ifnum\longrefs=1 {Space Science Rev.}\else 
                               {Space\ Sci.\ Rev.}\fi}

\def\dutch{\def\refname{Referenties}\def\abstractname{Samenvatting}%
  \def\bibname{Bibliografie}\def\chaptername{Hoofdstuk}%
  \def\appendixname{Bijlage}\def\contentsname{Inhoudsopgave}%
  \def\listfigurename{Lijst van figuren}\def\listtablename{Lijst van tabellen}%
  \def\indexname{Index}\def\figurename{Figuur}\def\tablename{Tabel}%
  \def\partname{Deel}\def\enclname{Bijlage(n)}\def\ccname{Ter attentie van}%
  \def\headtoname{Aan}\def\headpagename{Pagina}%
  \def\today{\number\day\space\ifcase\month\or januari\or februari\or maart\or%
     april\or mei\or juni\or juli\or augustus\or september\or oktober\or%
     november\or december\fi \space\number\year}%
  \typeout{
              >>>>> use hlatex209 for Dutch hyphenation <<<<< 
         }}
\hyphenation{Schrij-ver Krij-ger Kuij-pers Bal-le-gooij-en}

\newcounter{onefig} \newcounter{fignumber}
\newcount\nocaptions \newcount\nofigures \newcount\figwidth
\newcount\viewgraphs
  \def\paper{}  \def\figlabel{} 
\long\def\nextfig#1{\setcounter{figure}{\value{fignumber}}
  \addtocounter{fignumber}{1}
  \ifnum \viewgraphs=1 \newpage \pagestyle{empty} \fi 
  \ifnum\value{onefig}=0 #1 \fi                 
  \ifnum\value{onefig}=\value{fignumber} #1 \fi}
\def\figwidths#1#2{\ifnum \nocaptions=1 #2mm \else #1mm \fi}  
\def\paper#1{}  
\long\def\plotfig#1#2{\ifnum \nofigures=1 \else #2 \fi}
\long\def\captiontext#1{\ifnum \nofigures=1 \raggedright \fi 
   \ifnum \nocaptions=1 \paper
     \ifnum \viewgraphs=0 
       \newline  \mbox{}\hrulefill\mbox{} \newline 
       \newline label:~\{\figlabel\} 
     \fi 
     \else \ifnum \nofigures=0 \fi 
   #1 \fi}

\newcount\panelwidth \newcount\panelheight 
\newcount\bxmin \newcount\bymin \newcount\bxmax \newcount\bymax
\newcount\tbxmin \newcount\tbymin
\newcount\tpanelwidth \newcount\tpanelheight \newcount\tpdif
\panelwidth=70 \panelheight=70  
\def\panelsize #1,#2;{\panelwidth=#1 \panelheight=#2}  
\def\setbb #1,#2;#3,#4;#5,#6;{
  \tbxmin=#1 \tbymin=#2    
  \bxmin=#3 \bymin=#4      
  \bxmax=#5 \bymax=#6}     
\def\barepanel #1{%
  \ifnum\panelheight=0 
    \tpdif=\bymax \advance\tpdif by -\bymin
    \multiply \tpdif by \panelwidth
    \tpanelheight=\tpdif
    \tpdif=\bxmax \advance\tpdif by -\bxmin
    \divide \tpanelheight by \tpdif
  \else \tpanelheight=\panelheight \fi
  \epsfig{file=#1,%
     bbllx=\bxmin bp,bblly=\bymin bp,bburx=\bxmax bp,bbury=\bymax bp,clip=,%
     width=\panelwidth mm,height=\tpanelheight mm}}
\def\labelypanel #1{
  \ifnum\panelheight=0 
    \tpdif=\bymax \advance\tpdif by -\bymin
    \multiply \tpdif by \panelwidth
    \tpanelheight=\tpdif
    \tpdif=\bxmax \advance\tpdif by -\bxmin
    \divide \tpanelheight by \tpdif
  \else \tpanelheight=\panelheight \fi
  \tpdif=\bxmax \advance\tpdif by -\tbxmin
  \tpanelwidth=\panelwidth \multiply \tpanelwidth by \tpdif
  \tpdif=\bxmax \advance\tpdif by -\bxmin
  \divide \tpanelwidth by \tpdif
  \epsfig{file=#1,%
    bbllx=\tbxmin bp,bblly=\bymin bp,bburx=\bxmax bp,bbury=\bymax bp,%
    clip=,width=\tpanelwidth mm,height=\tpanelheight mm}}
\def\labelxpanel #1{%
  \ifnum\panelheight=0 
    \tpdif=\bymax \advance\tpdif by -\bymin
    \multiply \tpdif by \panelwidth
    \tpanelheight=\tpdif
    \tpdif=\bxmax \advance\tpdif by -\bxmin
    \divide \tpanelheight by \tpdif
  \else \tpanelheight=\panelheight \fi
  \tpdif=\bymax \advance\tpdif by -\tbymin
  \multiply \tpanelheight by \tpdif
  \tpdif=\bymax \advance\tpdif by -\bymin
  \divide \tpanelheight by \tpdif
  \epsfig{file=#1,%
    bbllx=\bxmin bp,bblly=\tbymin bp,bburx=\bxmax bp,bbury=\bymax bp,%
    clip=,width=\panelwidth mm,height=\tpanelheight mm}}
\def\labelxypanel #1{%
  \ifnum\panelheight=0 
    \tpdif=\bymax \advance\tpdif by -\bymin
    \multiply \tpdif by \panelwidth
    \tpanelheight=\tpdif
    \tpdif=\bxmax \advance\tpdif by -\bxmin
    \divide \tpanelheight by \tpdif
  \else \tpanelheight=\panelheight \fi
  \tpdif=\bxmax \advance\tpdif by -\tbxmin
  \tpanelwidth=\panelwidth \multiply \tpanelwidth by \tpdif
  \tpdif=\bxmax \advance\tpdif by -\bxmin
  \divide \tpanelwidth by \tpdif 
  \tpdif=\bymax \advance\tpdif by -\tbymin 
  \multiply \tpanelheight by \tpdif
  \tpdif=\bymax \advance\tpdif by -\bymin
  \divide \tpanelheight by \tpdif
  \epsfig{file=#1,%
    bbllx=\tbxmin bp,bblly=\tbymin bp,bburx=\bxmax bp,bbury=\bymax bp,%
    clip=,width=\tpanelwidth mm,height=\tpanelheight mm}}

\def\CC{\par \vspace*{-2ex} \footnotesize \baselineskip=8pt \begin{verbatim}}

\long\def\startignore #1\stopignore{}   

\def\setlistparams{         
  \topsep=0.7ex                 
  \itemsep=0.7ex                
  \leftmargini=3ex}             
\setlistparams                  

\newcounter{alistindex}       

\def\spacing #1{\small \renewcommand{\baselinestretch}{#1}
                \normalsize \setlistparams}             

\newcounter{romenumnr}

\newlength{\minipagewidth}

\newsavebox{\boxcontent}
\newcommand{\ovalhead}[1]{
  \unitlength=1cm
  \sbox{\boxcontent}{\mbox{~~{#1}~~}}
  \begin{center}
    \ifdim\wd\boxcontent>6ex 
    \ifdim\wd\boxcontent<8cm 
    \begin{picture}(8,3) \thicklines     
      \put(4.0,0.8){\oval(8,1.6)} 
      \put(0.0,0.7){\parbox{8cm}{
         \begin{center} \usebox{\boxcontent} \end{center}}}
    \end{picture}
    \else \ifdim\wd\boxcontent<12cm 
    \begin{picture}(12,3) \thicklines     
        \put(6.0,0.8){\oval(12,1.6)} 
        \put(0.0,0.7){\parbox{12cm}{
           \begin{center} \usebox{\boxcontent} \end{center}}}
    \end{picture}
    \else
    \begin{picture}(16,3) \thicklines     
        \put(8.0,0.8){\oval(16,1.6)} 
        \put(0.0,0.7){\parbox{16cm}{
           \begin{center} \usebox{\boxcontent} \end{center}}}
    \end{picture}
    \fi \fi \fi
  \end{center}} 

\setcounter{secnumdepth}{3}
\setcounter{tocdepth}{3}

\newcounter{headnr}            
\newcounter{subheadnr}[headnr]
\newcounter{subsubheadnr}[subheadnr]
\def\head #1\par{
  \stepcounter{headnr}                          
  \vspace{2ex} \noindent                        
  {\bf \theheadnr~~~~#1}\\[1ex] \noindent}      
\def\subhead #1\par{  
  \stepcounter{subheadnr}
  \vspace{1.3ex} \noindent
  {\bf \theheadnr.\arabic{subheadnr}~~~#1}\\[0.3ex] \noindent}
\def\subsubhead #1\par{
  \stepcounter{subsubheadnr}
  \vspace{1.0ex} \noindent
  {\bf \theheadnr.\arabic{subheadnr}.\arabic{subsubheadnr}~~~#1}\\ \noindent}

\font\dropfont= cmr12 scaled \magstep5
\def\dropcap#1#2{{\noindent
    \setbox0\hbox{\dropfont #1}\setbox1\hbox{#2}\setbox2\hbox{(}%
    \count0=\ht0\advance\count0 by\dp0\count1\baselineskip
    \advance\count0 by-\ht1\advance\count0by\ht2
    \dimen1=.5ex\advance\count0by\dimen1\divide\count0 by\count1
    \advance\count0 by1\dimen0\wd0
    \advance\dimen0 by.25em\dimen1=\ht0\advance\dimen1 by-\ht1
    \global\hangindent\dimen0\global\hangafter-\count0
    \hskip-\dimen0\setbox0\hbox to\dimen0{\raise-\dimen1\box0\hss}%
    \dp0=0in\ht0=0in\box0}#2}

\def\level #1 #2#3#4{$#1 \: ^{#2} \mbox{#3} ^{#4}$}   

\def\deg{\hbox{$^\circ$}}       

\def\Msun{\hbox{M$_{\odot}$}}               
\def\Rsun{\hbox{R$_{\odot}$}}               
\def\Lsun{\hbox{L$_{\odot}$}}               

\def\Mdot{\hbox{$\dot{M}$}}                     
\def\mathstacksym#1#2#3#4#5{\def#1{\mathrel{\hbox to 0pt{\lower 
    #5\hbox{#3}\hss} \raise #4\hbox{#2}}}}

\mathstacksym\lta{$<$}{$\sim$}{1.5pt}{3.5pt} 
\mathstacksym\gta{$>$}{$\sim$}{1.5pt}{3.5pt} 
\mathstacksym\lrarrow{$\leftarrow$}{$\rightarrow$}{2pt}{1pt} 
\mathstacksym\lessgreat{$>$}{$<$}{3pt}{3pt} 

\raggedbottom{}

\title{Evidence for the Keplerian orbit of a close companion around a giant star}

\begin{document}

\renewcommand{\thefootnote}{\alph{footnote}}

\maketitle

\noindent\author{Mats Esseldeurs$^{1,\star}$\orcid{0000-0002-4650-6029},
    Leen Decin$^{1,\star}$\orcid{0000-0002-5342-8612},
    Joris De Ridder$^{1}$\orcid{0000-0001-6726-2863},
    Yoshiya Mori$^{2,3}$,
    Amanda I. Karakas$^{2,3,4}$\orcid{0000-0002-3625-6951},
    Jolien Malfait$^{1}$\orcid{0000-0002-8850-2763},
    Ta\"{\i}ssa Danilovich$^{2,3,1}$\orcid{0000-0002-1283-6038},
    St\'ephane Mathis$^{5}$\orcid{0000-0001-9491-8012},
    Anita M. S. Richards$^{6}$\orcid{0000-0002-3880-2450},
    Raghvendra Sahai$^{7}$\orcid{0000-0002-6858-5063},
    Jeremy Yates$^{8}$\orcid{0000-0003-1954-8749},
    Marie Van de Sande$^{9}$\orcid{0000-0001-9298-6265},
    Maarten Baes$^{10}$\orcid{0000-0002-3930-2757},
    Alain Baudry$^{11}$\orcid{0000-0003-1577-9427},
    Jan Bolte$^{12}$\orcid{0000-0002-7991-9663},
    Thomas Ceulemans$^{1}$\orcid{0000-0002-7808-9039},
    Frederik De Ceuster$^{1,13}$\orcid{0000-0001-5887-8498},
    Ileyk El Mellah$^{14,15,16}$\orcid{0000-0003-1075-0326},
    Sandra Etoka$^{6}$\orcid{0000-0003-3483-6212},
    Carl Gottlieb$^{17}$\orcid{0000-0003-2845-5317},
    Fabrice Herpin$^{11}$\orcid{0000-0003-2977-5072},
    Pierre Kervella$^{18}$\orcid{0000-0003-0626-1749},
    Camille Landri$^{1}$\orcid{0000-0001-8078-0905},
    Louise Marinho$^{19}$\orcid{0000-0003-3802-9389},
    Iain McDonald$^{6,20}$\orcid{0000-0003-0356-0655},
    Karl Menten$^{21,\dagger}$\orcid{0000-0001-6459-0669},
    Tom Millar$^{22}$\orcid{0000-0001-5178-3656},
    Zara Osborn$^{2,3}$\orcid{0000-0001-5546-6869},
    Bannawit Pimpanuwat$^{23}$\orcid{0000-0001-8782-0754},
    John Plane$^{24}$\orcid{0000-0003-3648-6893},
    Daniel J. Price$^{2}$\orcid{0000-0002-4716-4235},
    Lionel Siess$^{25}$\orcid{0000-0001-6008-1103},
    Owen Vermeulen$^{1}$\orcid{0009-0008-5582-357X},
    Ka Tat Wong$^{26,27}$\orcid{0000-0002-4579-6546}
}

\noindent$^\star$~Corresponding authors. Emails: \href{mailto:mats.esseldeurs@kuleuven.be}{mats.esseldeurs@kuleuven.be}; \href{mailto:leen.decin@kuleuven.be}{leen.decin@kuleuven.be}.\\
$\dagger$~Passed away on 30/12/2024.

\begin{affiliations}
    \item
    Institute of Astronomy, KU Leuven, Celestijnenlaan 200D, 3001 Leuven, Belgium
    \item
    School of Physics \& Astronomy, Monash University, Clayton VIC 3800, Australia
    \item
    ARC Centre of Excellence for All Sky Astrophysics in 3 Dimensions (ASTRO 3D), Clayton 3800, Australia
    \item
    Kavli IPMU (WPI), UTIAS, The University of Tokyo, Kashiwa, Chiba 277-8583, Japan
    \item
    Universit\'e Paris-Saclay, Universit\'e Paris Cit\'e, CEA, CNRS, AIM, 91191 Gif-sur-Yvette, France	
    \item
    JBCA, Department of Physics \& Astronomy, University of Manchester, Manchester, UK
    \item
    Jet Propulsion Laboratory, California Institute of Technology, Pasadena, CA 91109, USA
    \item
    University College London, Department of Computer Science, London, WC1E 6BT, UK
    \item
    Leiden Observatory, Leiden University, P.O. Box 9513, 2300 RA Leiden, The Netherlands
    \item
    Sterrenkundig Observatorium, Universiteit Gent, Krijgslaan 281 S9, 9000, Gent, Belgium
    \item
    Laboratoire d'Astrophysique de Bordeaux, Univ. Bordeaux, CNRS, B18N, all\'ee Geoffroy Saint-Hilaire, 33615 Pessac, France
    \item
    Department of Mathematics, Kiel University, Heinrich-Hecht-Platz 6, 24118 Kiel, Germany
    \item
    Leuven Gravity Institute, KU Leuven, Celestijnenlaan 200D, 3001 Leuven, Belgium
    \item
    Departamento de F\'{\i}sica, Universidad de Santiago de Chile, Av.\ Victor Jara 3659, Santiago, Chile
    \item
    Center for Interdisciplinary Research in Astrophysics and Space Exploration (CIRAS), USACH, Santiago, Chile
    \item
    Departament de Física, EEBE, Universitat Politècnica de Catalunya, Av. Eduard Maristany 16, 08019 Barcelona, Spain
    \item
    Harvard-Smithsonian Center for Astrophysics, 60 Garden Street, Cambridge MA 02138 USA
    \item
    LIRA, Observatoire de Paris, Université PSL, Sorbonne Université, Université Paris Cité, CY Cergy Paris Université, CNRS, 92190 Meudon, France.
    \item
    Instituto de Fisica Fundamental, CSIC, C/ Serrano 123, 28006 Madrid, Spain
    \item
    Open University, Walton Hall, Kents Hill, Milton Keynes MK7 6AA, UK
    \item
    Max-Planck-Institut f\"ur Radioastronomie, Auf dem Hu\"ugel 69,5 3121 Bonn, Germany
    \item
    School of Mathematics and Physics, Queen's Universtity Belfast, Belfast BT7 1NN, UK
    \item
    National Astronomical Research Institute of Thailand, 260 Moo 4, Donkaew, Mae Rim, Chiang Mai 50180, Thailand
    \item
    School of Chemistry, University of Leeds, Woodhouse Lane, Leeds LS2 9JT, UK
    \item
    Institut d'Astronomie et d'Astrophysique and Brussels Laboratory of the Universe (BLU-ULB), Universit\'e libre de Bruxelles, CP 226, Boulevard du Triomphe, 1050 Brussels, Belgium
    \item
    Theoretical Astrophysics, Department of Physics and Astronomy, Uppsala University, Box 516, 75120 Uppsala, Sweden
    \item
    Institut de Radioastronomie Millim\'etrique, 38406 Saint-Martin-d'H\`eres, France

\end{affiliations}

\setlength{\parskip}{10pt}

\bigskip

\begin{abstract}
Close companions influence stellar evolution through tidal interactions, mass transfer, and mass loss effects. While such companions are detected around young stellar objects, main-sequence stars, red giants, and compact objects, direct observational evidence of close-in companions around asymptotic giant branch (AGB) stars has remained elusive. Here, we present (sub)millimeter time-domain imaging spectroscopy revealing the Keplerian motion of a close-in companion around the AGB star $\pi^1$~Gruis. The companion, slightly more massive than the AGB star, is likely a main-sequence star. Unlike more evolved stars with companions at comparable distances, $\pi^1$~Gru’s companion follows a circular orbit, suggesting an eccentricity-generating mechanism late- or post-AGB. Our analysis suggests that model-predicted circularization rates may be underestimated. Our results highlight the potential of multi-epoch (sub)millimeter interferometry in detecting the Keplerian motion of close companions to giant stars and open avenues for our understanding of tidal interaction physics and binary evolution.
\end{abstract}

\bigskip

\section*{Main}

Most stars with initial mass greater than 0.8 solar masses (\(\text{M}_\odot\)) probably host at least one planetary or stellar companion~\cite{Moe2017ApJS..230...15M, Decin2020Sci...369.1497D}. Of these stars, $\sim$95\% will evolve through the Asymptotic Giant Branch (AGB) phase assuming a Salpeter initial mass function with exponent of 2.3~\cite{Kroupa2001MNRAS.322..231K}. Companions with close orbits influence stellar evolution at all stages, shaping mass loss, altering envelope dynamics, and driving interactions that affect planetary nebula formation, extrinsic carbon star creation, and the recycling of enriched material into the interstellar medium~\cite{Veras2016RSOS....350571V, Saladino2019A&A...626A..68S, Aydi2022MNRAS.513.4405A, VandeSande2022MNRAS.510.1204V, Esseldeurs2024A&A...690A.266E}.

Although close companions have been detected around a range of stars, from  young stellar objects to main-sequence (MS) stars, red giants (RGB), white dwarfs (WD), and neutron stars~\cite{Mayor1995Natur.378..355M, Maund2004Natur.427..129M, Sana2012Sci...337..444S, David2016Natur.534..658D, Grunblatt2016AJ....152..185G}, detecting companions in close orbits ($\la$~5 stellar radii) around AGB stars has proven difficult. AGB stars can undergo strong pulsations causing photometric variations of several magnitudes and shock velocities up to $\sim$10~km~s$^{-1}$, extreme luminosities (up to $10^5$ solar luminosities), and powerful winds with mass-loss rates (\Mdot) from $10^{-8}$ to several times $10^{-5}$ \Msun~yr$^{-1}$~\cite{Decin2021ARA&A..59..337D}. These conditions make transit and radial-velocity methods unsuitable for detecting close companions. While high-energy (UV/X-ray) emission suggests the presence of companions around AGB stars, indicating associated accretion flows and disks~\cite{Sahai2018ApJ...860..105S}, it does not reveal any orbital characteristics.

Long-term photometric monitoring of pulsating AGB stars reveals pulsation periods of hundreds of days, but in about a third of cases, a longer secondary period (LSP; 5--10 times the pulsation period) is detected, possibly indicating a companion's orbital period~\cite{Soszynski2021ApJ...911L..22S}. The LSP radial velocity amplitude can provide binary system properties only if it exceeds that of the stellar pulsation.
In systems where the presence of a companion is independently confirmed, such as the red symbiotic binary CH~Cyg~\cite{Hinkle2009ApJ...692.1360H}, orbital parameters can be derived from the radial velocity curve. However, estimates of orbital eccentricity can be inaccurate owing to perturbations of the radial velocity by gas streams from the AGB star's extended atmosphere caught in the companion's gravitational potential. Moreover, only the mass function can be derived, and astrometric data is required to derive inclination and system masses. A combined radial velocity and astrometric analysis has been successfully applied only to the carbon-rich AGB star V~Hya~\cite{Planquart2024A&A...682A.143P}. However, such analysis relies on narrow priors for primary mass and distance, and precise initial conditions for convergence. Moreover, this method cannot determine the barycentre’s proper motion. In addition, since the companion is undetected, the nature of the LSP remains debated, with alternatives including plasma ejection, triple system dynamics, episodic dust formation, or convective oscillations~\cite{Kiss2006MNRAS.372.1721K, Salas2019MNRAS.487.3029S}. Another approach has been explored for the red symbiotic R~Aqr~\cite{Alcolea2023hsa..conf..190A}, where the companion's motion is inferred over time by assuming its position coincides with the centre of the H30$\alpha$ jet or the edge of the continuum emission, although only a single direct detection exists. Combining this assumed relative position with radial velocity measurements allows estimates of the system’s orbital parameters.

The interaction between the AGB wind and potential companion(s) offers an alternative method for characterizing giant binary systems. Observations with the Atacama Large Millimeter/submillimeter Array (ALMA) have indirectly indicated the presence of companions to AGB stars through the detection of arcs, bipolar outflows, tori, rotating disks, and spirals in the otherwise smooth, radially outflowing wind~\cite{Decin2020Sci...369.1497D}. However, unlike wide companions (e.g., Mira AB~\cite{Karovska1997ApJ...482L.175K}, W Aql~\cite{Danilovich2024NatAs...8..308D}), no close companions have been directly detected in Keplerian motion through multi-epoch tracking of both components.

For two AGB stars, L$_2$ Pup and $\pi^1$~Gru, ALMA continuum data around 331.6\,GHz and 241\,GHz, respectively, reveal a maximum at the AGB star’s position, plus an offset peak, potentially indicating a close companion~\cite{Kervella2016A&A...596A..92K, Homan2020A&A...644A..61H}. Alternatively, the secondary peak may represent a dense aggregate of dust and gas formed by episodic mass loss, proposed as the cause of $\pi^1$~Gru’s fast bipolar outflow~\cite{Doan2020A&A...633A..13D}. Assuming the secondary peak in the ALMA $\pi^1$~Gru image traces a stellar companion orbiting an AGB star of mass $\sim$1.5\,\Msun, and approximating the tangential velocity anomaly by the orbital tangential velocity, it was inferred that this companion could be an accreting dwarf star of $\sim$0.86\,\Msun\ on an eccentric ($e \sim 0.35$), $\sim$11-yr orbit~\cite{Montarges2025A&A...699A..22M}. The longevity (weeks to about a year)~\cite{Kervella2016A&A...596A..92K} and motion -- Keplerian for a companion or radial for an outflowing aggregate -- can help distinguish between these scenarios. To test this conjecture, we observed the AGB star $\pi^1$~Gru at high angular resolution ($\sim$0\farcs020) with ALMA during Cycle 6 (C6, June–July 2019, band 6, 241.0~GHz) and Cycle 10 (C10, October 2023, band 7, 334.7~GHz) (see `ALMA observations' in Methods).

\section*{Results}
\subsection*{Direct evidence of the Keplerian motion of a close-in companion}
In both ALMA C6 and C10 epochs, continuum data reveal two emission maxima (see Fig.~\ref{fig:pi1_gru}, Extended Data Fig.~\ref{Fig:ALMA_data} and Extended Data Table.~\ref{table:fit}). The relative position of the secondary peak (referred to as $M_2$) with respect to the AGB star ($M_1$) was measured in right ascension ($\alpha_\star$) and declination ($\delta$) as ($-12.0 \pm 1.2$, $-34.7 \pm 1.2$)~mas in the C6 data, and ($-24.1 \pm 0.7$, $32.9 \pm 1.0$)~mas in the C10 data (see `ALMA observations' in Methods). This positional shift is inconsistent with pure radial motion, but can be explained by an elliptical orbital projection, as expected from Newtonian gravitational interaction between two masses (see `Orbital parameters' in Methods). This marks the direct detection of the Keplerian motion of a close-in companion around an AGB star, confirming the earlier hypothesis of the presence of a close-in companion around $\pi^1$~Gru~\cite{Homan2020A&A...644A..61H, Chiu2006ApJ...645..605C, Mayer2014A&A...570A.113M, Montarges2025A&A...699A..22M}. $\pi^1$~Gru also has a distant G0V companion, $\pi^1$~Gru~B, at a projected separation of 2\farcs71, known since 1953~\cite{Feast1953MNRAS.113..510F}. This confirms $\pi^1$~Gru as a hierarchical triple system, consisting of a close-in companion ($\pi^1$~Gru~C) and a widely separated tertiary companion, which is unlikely to significantly influence the inner binary’s dynamics (see `Orbital parameters' in Methods).

\subsection*{System’s orbital motion and barycentre’s proper motion}
Combining the multi-epoch ALMA proper motion images with \textit{Hipparcos} and \textit{Gaia} position-velocity vectors from epochs 1991.25 and 2016.0 (see Fig.~\ref{fig:pi1_gru}), we assembled data spanning nearly 25 years. Using 18 observational constraints, including astrometric positions, velocities, and parallax, we solved the two-body problem under Newtonian gravity and determine both the system’s orbital motion and the barycentre’s proper motion.

The proper motion vector is defined by six orbital elements ($\Omega$, $i$, $\omega$, $a$, $e$, $T_0$), along with $\pi^1$~Gru~A’s mass ($m_1$), the mass ratio ($q = m_2/m_1$, with $m_2$ the mass of $M_2$), the barycentre’s proper motion ($\mu^G_\alpha$, $\mu^G_\delta$), and parallax ($\varpi$, or distance $D$). Here, $\Omega$ is the longitude of the ascending node, $i$ inclination, $\omega$ argument of periastron, $a$ semi-major axis, $e$ eccentricity, and $T_0$ time of periastron passage. Using Bayesian inference, we derived posterior distributions for these 11 parameters. Broad, agnostic priors were adopted except for distance, which followed a Gamma prior. Geometrical degeneracies were present, with $i$ degenerate with $360^\circ - i$ and $(\omega, \Omega)$ degenerate with their antipodal values $(\omega + 180^\circ, \Omega + 180^\circ)$, but resolved using ALMA spectral line data (see `Orbital parameters' in Methods).

Specifically, we derive that $m_1 = 1.12 \pm 0.25 \, \Msun$, $q = 1.05 \pm 0.05$, $a = 6.81 \pm 0.49 \, \text{au}$, $\Omega = 101 \pm 36^\circ$, $i = 11 \pm 7^\circ$, $D = 179.74 \pm 10.09 \, \text{pc}$, $\mu_\alpha^G = 45.203 \pm 0.144 \, \text{mas yr}^{-1}$, and $\mu_\delta^G = -18.76 \pm 0.061\, \text{mas yr}^{-1}$ where the analysis strongly favors a circular orbit; see corner plot in Extended Data Fig.~\ref{fig:pi1_gru_corner_ultranest}. The fit to the proper motion data is shown in Fig.~\ref{fig:pi1_gru} and Suppl. Video~\ref{Video:pi1_gru}; the orbital system is visualized in Extended Data Fig.~\ref{fig:schematic}. With a more informed Gaussian prior on $m_1$ with mean 1.5$\pm$0.5\,\Msun, the retrieved $m_1$ increases to 1.27$\pm$0.22\,\Msun\ (see `Orbital parameters' in Supplementary Information), remaining within the uncertainty interval of our initial determination.

We derive the barycentre's radial velocity as $-14.8 \pm 1.0 \, \mathrm{km\,s}^{-1}$ (see `Orbital parameters' in Methods). Using the celestial coordinates, and derived distance, proper motion, and radial velocity, we determined the system's 6D Galactocentric phase-space position. By back-integrating its orbit over 2.8~Gyr, we find that the $\pi^1$ Gru system likely formed near the Galactic plane, close to or slightly within the Solar orbit (see `Galactic orbit' in Supplementary Information).

\subsection*{Nature of $\boldsymbol{\pi^1}$~Gru~C and its accretion disk}
We deduce that the companion orbits the AGB star in an anti-clockwise direction and is surrounded by an accretion disk. An SED analysis indicates that $\pi^1$Gru~C is either an F6–F8V MS star or a massive WD with a temperature up to 40,000~K, with additional (sub)millimeter emission from an accretion disk interacting with the companion. The UV emission is either intrinsic, from the AGB star’s chromosphere, or extrinsic, suggesting ongoing but weak accretion (see `Accretion disk' in Supplementary Information).

Wind Roche lobe overflow (RLOF)~\cite{Mohamed2007ASPC..372..397M} shapes the AGB circumstellar envelope’s density structure and the accretion disk around the companion (see `Hydrodynamical modelling' in Methods). To estimate the mass accretion rate and investigate the accretion disk’s characteristics, we perform high-resolution hydrodynamic simulations~\cite{Siess2022A&A...667A..75S, Malfait2024A&A...691A..84M} (see `Hydrodynamical modelling' in Methods). The resulting density distribution (Extended Data Fig.~\ref{Fig:disk_simulation}) shows a bow shock spiral in front of a dense, circular accretion disk with an outer radius of $\sim$0.83~au. The disk material orbits with tangential velocities of 80--100\% of the Keplerian velocity and is flared in the edge-on view, with a density scale height of $\sim$0.15~au at the outer radius. The mass accretion is $\sim$15\% of the mass loss rate ($8 \times 10^{-7}\, {\rm M_\odot \, yr}^{-1}$~\cite{Doan2020A&A...633A..13D}), corresponding to a mass accretion rate of $1.2 \times 10^{-7}$~\Msun~yr$^{-1}$, and the disk’s total mass is $2 \times 10^{-6}$~\Msun. ALMA band 6 and 7 at $\pi^1$~Gru~C’s position gives us a spectral index of $\alpha^s_{M_2} = 2.3 \pm 0.3$ ($F_\nu \propto \nu^{\alpha^s}$), indicating dust dominance~\cite{oGorman2015A&A...573L...1O}. For the accretion disk to account for this emission, the estimated disk dust mass is $\sim 8.5 \times 10^{-8}$~\Msun\ (see `Hydrodynamical modelling' in Methods).

\subsection*{System's initial configuration and future evolution}
Using stellar evolution calculations~\cite{Karakas2014MNRAS.445..347K}, we estimate the initial mass of $\pi^1$~Gru~A. The core-mass luminosity diagram and the observed C/O ratio indicate a best-fit initial mass of 1.7\,\Msun, with bounds of 1.25\,\Msun\ and 2\,\Msun\ (see `Stellar evolution' in Methods). As an additional constraint, we analyzed $\pi^1$~Gru~A's pulsation characteristics. Classified as an SRb long-period variable, $\pi^1$~Gru~A has a pulsation period of $\sim$195~days~\cite{Tabur2009MNRAS.400.1945T}, which we attribute to the radial first overtone mode (see `Stellar evolution' in Methods; Extended Data Fig.~\ref{Fig:pl-ogle-asassn}). Stellar evolution models~\cite{Karakas2014MNRAS.445..347K} combined with linear pulsation calculations~\cite{Trabucchi2019MNRAS.482..929T} indicate that the observed period and luminosity are consistent with an initial mass of 1.5\,--\,2\,\Msun. For an initial mass of 1.7\,\Msun, the model predicts a current mass of $\sim$1.4\,\Msun, near the end of the thermally pulsating (TP)-AGB evolution. Growth rates further confirm that the first overtone mode can dominate even during late thermal pulses (see Fig.~\ref{Fig:m1_7_nov2_25}). Thus, despite its advanced TP-AGB age, $\pi^1$~Gru~A remains a semi-regular variable rather than transitioning to a Mira variable pulsating in the fundamental mode. The TP-AGB mass estimate of $\sim$1.4\,\Msun\ lies at the upper limit of the $m_1$ agnostic prior retrieval, being more compatible with the Gaussian prior.

The current system parameters allow us to reconstruct the past and predict the future of the $\pi^1$~Gru inner binary through orbital evolution calculations. The survival of close companions depends on the balance between tidal forces, mass loss, and mass transfer. We account for non-conservative mass transfer by incorporating it into the orbital angular momentum balance equation. Additionally, we include both equilibrium and dynamical tide dissipation, with the latter incorporating the excitation and dissipation of progressive internal gravity waves in evolved stars~\cite{Esseldeurs2024A&A...690A.266E} (see `Orbital evolution' in Methods).

Our orbital evolution calculations reveal that the $\pi^1$~Gru system parameters remained relatively unchanged during the main-sequence and horizontal branch phases (Fig.~\ref{fig:OrbitalEvolution_circ}). The orbit began to expand modestly during the RGB phase, driven by mass loss from the primary star. However, during the TP-AGB phase, the interplay of non-conservative mass transfer and enhanced
tidal dissipation caused significant orbital contraction, which will eventually lead to a common envelope phase. Mass loss from $\pi^1$~Gru~A and accretion onto $\pi^1$~Gru~C shifted the mass ratio from $q < 1$ to $q > 1$ (Fig.~\ref{fig:OrbitalEvolution_circ}d). In the WD scenario, there remains a small probability that the system could explode as a Type Ia supernova, potentially allowing the system to avoid a common-envelope fate due to a WD kick.

\section*{Discussion}
Close giant binaries with orbital periods $T_{\rm orb} \lesssim$ 4,000 days are expected to circularize through tidal dissipation~\cite{Izzard2010A&A...523A..10I}. This is supported by our refined tidal models tailored to the specific configurations of $\pi^1$~Gru. Specifically, we derive that higher eccentricities result in faster circularization rates and that the circularization mainly occurs at the very end of the TP-AGB phase (see `Orbital evolution' in Methods and Extended Data Fig.~\ref{fig:OrbitalEvolutionecc}). Thus, the detection of an AGB binary with a circular orbit and $T_{\rm orb} \sim 11.76$ years ($\sim$4,295 days) might appear unsurprising. However, this contrasts with post-AGB binaries, where systems with $T_{\rm orb} \ga$ 1,000 days are observed to be exclusively eccentric, creating a discrepancy between theory and observations~\cite{VanWinckel2003ARA&A..41..391V}.

Two main hypotheses address this discrepancy. The first suggests incomplete knowledge of tidal dissipation: tidal circularization rates may be underestimated for the (progenitor) solar-type main-sequence binaries~\cite{Meibom2005ApJ...620..970M}, while overestimated during the giant phase for ellipsoidal red giant binaries~\cite{Nie2017ApJ...835..209N}. We add a single, critical data point to this discussion: confirmation that a TP-AGB binary system is circularized at $T_{\rm orb} >$ 1,000 days. It tentatively favors the second hypothesis: a mechanism generating eccentricity at the end of the TP-AGB or during the post-AGB phase. Proposed mechanisms include mass transfer at periastron, asymmetric mass loss, interactions with a circumbinary disk, white-dwarf kicks, hybrid wind-RLOF mass transfer, or the eccentric Kozai-Lidov mechanism in triple systems~\cite{Izzard2010A&A...523A..10I, VanWinckel2003ARA&A..41..391V}. Future studies of AGB binary systems may allow a sampling of the $e - \log T_{\rm orb}$ distribution, providing critical evidence to this discussion.

Our results also constitute an important test for tidal interaction physics. A key inference is that our orbital evolution models predict minimal change in the binary’s eccentricity, implying an initially near-circular orbit since the system is currently circular (see Extended Data Fig.~\ref{fig:OrbitalEvolutionecc}). Combined with the observed large eccentricities and deficit of systems with $e < 0.15$ among main-sequence solar-type binaries with orbital periods above 10 days~\cite{Moe2017ApJS..230...15M}, this may suggest that our circularisation rates are underestimated.
This tension between theory and observations is reinforced by a recent study of red giant binary eccentricities~\cite{Dewberry2025ApJ...984..137D}. One potential avenue for investigation is resonance locking, a phenomenon where the tidal forcing frequency and the frequency of a stellar pulsation mode vary in concert, enabling sustained resonant interactions over extended timescales compared to scenarios where either of these frequencies were constant~\cite{Witte1999A&A...350..129W}.

Alternatively, $\pi^1$~Gru’s inner binary system may have already undergone circularization, which would lend support to the hypothesis that $\pi^1$~Gru~C evolved already through the AGB phase and is now a WD. However, post-AGB binary systems with WD companions -- such as extrinsic S-type stars, barium stars, red symbiotics and WD-MS systems -- are all observed to have non-circular orbits across orbital periods ranging from $\sim$1,000\,--\,10,000 days~\cite{VanWinckel2003ARA&A..41..391V}. If the eccentricity-generating mechanism proposed for these systems applied here, it must have been unusually weak or inactive to result in the observed low eccentricity of the WD-AGB phase.

Moreover, using the WD mass distribution function derived from \textit{Gaia} data~\cite{Bergeron2019ApJ...876...67B}, the probability of $\pi^1$~Gru~C being a massive WD, with mass $1.18\pm0.27$\,\Msun, is only $\sim$3\%. Such massive WDs are thought to form through the merger of two average-mass WDs in close binaries or from the evolution of massive intermediate-mass single stars~\cite{Bergeron2019ApJ...876...67B}.
Massive WDs accreting at the rate inferred here ($\sim 1.2 \times 10^{-7}$~\Msun~yr$^{-1}$) are present in symbiotic recurrent novae, whose optical and UV spectra show strong H\,I, He\,I, and He\,II emission lines~\cite{Mondal2018MNRAS.474.4211M}, none of which are observed toward $\pi^1$~Gru. Furthermore, the accreting companion was not detected with SPHERE-ZIMPOL~\cite{Montarges2023A&A...671A..96M}, whereas the accreting WD in R~Aqr was detected by the same instrument despite an order of magnitude lower accretion rate~\cite{Schmid2017A&A...602A..53S}.
These statistical and observational constraints make it more likely that $\pi^1$~Gru~C is a MS star in a primordial circular orbit.

Hence, neither evolutionary scenario for $\pi^1$~Gru~C is without tension. Future spectrally resolved UV observations with the \textit{Hubble Space Telescope} may reveal the nature of the companion, with narrower line profiles pointing to a MS star, while broader profiles would be more indicative of a WD~\cite{Sahai2018ApJ...860..105S}.

This study presents the direct detection of the Keplerian motion of a close-in companion around an AGB star. By combining multi-epoch (sub)millimeter imaging with optical astrometric data, we achieve precise proper motion fitting, disentangling the barycentre’s motion from the binary’s orbital dynamics. The retrieved orbital parameters provide a crucial benchmark for stellar and binary evolution models, revealing tensions in tidal interaction physics. This work shows the potential of future multi-epoch (sub)millimeter imaging, particularly when combined with \textit{Gaia} DR4’s optical astrometry, to derive orbital constraints for giant binaries and improve our understanding of the tidal interactions and orbital transformations governing the evolution of stars, which commonly reside in binary systems.

\begin{methods}

\section{ALMA observations}\label{Methods:ALMA_data_reduction}

\subsection{Observations}

$\pi^1$~Gru was observed at high resolution as part of ALMA project codes 2018.1.00659.L and 2023.1.00091.S, during Cycle 6 (June 23 and July 6, 2019, Band 6, central frequency 241 GHz) and Cycle 10 (October 12 and October 26, 2023, Band 7, central frequency 334 GHz), respectively. In 2019, $\pi^1$~Gru was also observed in two other lower-resolution ALMA configurations, as described in ref.~\cite{Gottlieb2022A&A...660A..94G}. The observations were made in multiple, non-contiguous spectral windows (spw) at an initial spectral resolution $\sim1$ MHz or finer. The data from each epoch were combined and are referred to as C6 and C10 data throughout this work. The 2019 observations and data reduction process are detailed in Ref.~\cite{Gottlieb2022A&A...660A..94G}. Only the extended configuration data are used for continuum analysis here. For the 2023 data, a similar method was employed, with the main difference being the use of a higher frequency and smaller frequency span $\Delta\nu$. J2230-4416 and J2235-4835 served as phase reference sources, while J2242-4204 and J2230-4416 were used as check sources in 2019 and 2023, respectively. The check source, a compact source at a comparable angular separation from the phase reference as the target, was observed occasionally as if it were a target. The angular separations (in right ascension $\alpha_\star$ and declination $\delta$) of the check source and target from the phase reference are denoted by $\psi_{\mathrm{check}}$ and $\psi_{\mathrm{targ}}$, respectively. In both cases, the maximum recoverable scale (MRS) is larger than the region containing the $\pi^1$~Gru system. Details are summarized in Supplementary Table~\ref{table:obs}. The principal data products are continuum images of the system, analysed in this paper, and spectral line cubes, of which only the CO, SiO, and SiS data are used here.

\subsection{Calibration and imaging}

In brief, we began with the target data after applying the ALMA pipeline calibration (excluding the pipeline self-calibration available in 2023). We identified line-free channels from the pipeline images for the 2019 data, and from the calibrated visibilities in 2023. Here, $\Delta\nu$ represents the cumulative bandwidth of line-free channels distributed across $\nu_{\mathrm{tot}}$, and $\nu_{\mathrm{mean}}$ denotes the mean frequency of the line-free continuum channels. For each epoch, we made channel-averaged copies of the phase-referenced target data. These were used to create continuum images, excluding line emission, (from all data in 2019 and from the first execution with the best signal-to-noise ratio (S/N) in 2023), as starting models for phase and amplitude self-calibration. The system's proper motion is $\sim$1 mas during the 2-week interval between observations at each epoch; the self-calibration aligned the image peak with the positions given on the dates provided in Supplementary Table~\ref{table:obs} for C6 and C10, respectively, i.e. the mean date for C6 and the date of the first execution for C10.

The array configuration used in 2019 provided slightly longer baselines relative to observing wavelength than in 2023. In order to provide images with similar synthesized beam sizes ($\theta_{\mathrm b}$) we weighted the data during imaging, using robust parameter values of +0.5 in 2019 and $-$0.5 in 2023; the beams are within 5\% of circular, hence the geometric mean sizes are quoted here. Supplementary Table~\ref{table:obs} summarizes the main continuum imaging parameters, including the off-source noise $\sigma_{\mathrm{rms}}$. The ALMA C6 2019 and ALMA C10 2023 continuum images are displayed in Extended Data Fig.~\ref{Fig:ALMA_data}.

The target data at full spectral resolution were adjusted to constant velocity ($v_{\mathrm{LSRK}}$) in the direction of $\pi^1$~Gru. We applied the continuum self-calibration solutions to these data and subtracted the continuum before making spectral image cubes for each spw. Here, we use measurements from 2019 extended-configuration cubes only, with an angular resolution of $\sim0\farcs023$ and a noise rms in quiet channels of 0.9--1.4 mJy, increasing with frequency; full details are tabulated in ref.~\cite{Gottlieb2022A&A...660A..94G}. The details of the 2023 spectral cubes are given in Supplementary Table~\ref{table:lineobs}, including the synthesised beam sizes achieved with robust +0.5 in order to optimise sensitivity to extended molecular emission. In some cases, cubes were made at two spectral resolutions; at the highest resolution for masers and with channel averaging for thermal lines, and at higher angular resolution, robust $-$0.5 for SiO and CO v=1 J=3-2 shown in Supplementary Fig.~\ref{fig:pv_CO_2023}. The positions and flux densities of maser spots at both epochs were measured by fitting 2-D Gaussian components in the {\sc AIPS} package (task {\tt SAD}); the position errors were estimated as described in Section~\ref{sec:measurement_error}, in the range 1 -- 10 mas depending on S/N. The transitions used in this paper
are listed in Supplementary Table~\ref{Table:lines}.

\subsection{Measurements of ALMA continuum sources}
We measured the positions of the primary ($\pi^1$~Gru~A, hereafter referred to as $M_1$) and the nearby companion ($\pi^1$~Gru~C, hereafter referred to as $M_2$) in two ways: using the {\sc casa} task {\texttt{imfit}} to fit two 2-D Gaussian components to $M_1$ and $M_2$
and using the {\texttt{uvmultifit}} add-on package~\cite{2014A&A...563A.136M} to fit a uniform disc (UD) plus a delta (point-like) component to the calibrated visibilities.
For both epochs and methods we allowed the fit to $M_1$ to be elliptical with the ratio of major and minor axes $e_{\mathrm r} \ge 0.9$, (where $e_{\mathrm r}=1$ is a perfect circle).
We assumed that $M_2$ is unresolved as its relative faintness and proximity to $M_1$ rendered attempts to measure a size based on formal S/N errors (Section~\ref{sec:measurement_error}) unreliable, and therefore specified a point source.

\subsubsection{Stochastic and other relative measurement errors}\label{sec:measurement_error}
The errors reported by the packages are based on the assumption that the source can be described exactly by the model and on ideal phase noise, and tend to underestimate realistic errors. In fact these errors are `formal errors' reported by the packages; they are not `absolute errors' which may include systematic effects not fully calibrated. The relative position uncertainty due to phase noise in a Gaussian fit to an interferometry image from an array with gaps in the baseline coverage (as for ALMA in the extended configuration) is given by $\sigma_{\mathrm{pos},\phi} = \theta_{\mathrm b} \times \sigma_{\mathrm{rms}}/P$ where $P$ is the 
fitted peak flux density. The two methods gave results within 1 mas for the relative positions of the components (although as expected, the FWHM of the Gaussian component for component $M_1$ underestimates the stellar size). It is likely that a UD is a better fit to the stellar continuum at (sub)millimeter wavelengths as visibility-plane fitting is not affected by deconvolution errors --- although still potentially biased by blending of the components at the interferometer resolution and emission which is not included in the UD plus delta model. Any deviations of the apparent stellar shape from a circle are in reality not likely to be elliptical but random (due to stellar activity, localised mass loss, dust clumps etc.) on scales large enough not to average into noise. We therefore adopt the UD plus delta model, utilizing the phase noise estimate (where $P$ is calculated as the UD flux density multiplied by the ratio of the beam area to the UD area) and the ellipticity to estimate the errors.

The provisional relative position uncertainties $\sigma_{\mathrm{prov}}$ are taken as the sum in quadrature of $\sigma_{\mathrm{pos},\phi}$ and, for component $M_1$, the error due to the assumption of ellipticity. We estimated the latter as $\sigma_{\mathrm {pos, e_r}}$ = (fitted diameter)$\times (1-e_{\mathrm r}) \times \sqrt{2}$. $\sigma_{\mathrm{prov}}$ thus represents the errors in UD plus delta visibility plane fitting, including phase noise. We then compared the positions obtained by visibility plane fitting with those from Gaussian fitting in the image plane, giving their differences in each direction, $\sigma_{\mathrm{pos,fit}}$. $\sigma_{\mathrm{prov}}$ and $\sigma_{\mathrm{pos,fit}}$ are decomposed into equal components in $\alpha_\star$ and $\delta$, considering that the synthesized beams are nearly circular and the other uncertainties have unknown orientations.

We then took the larger of $\sigma_{\mathrm{prov}}$ or $\sigma_{\mathrm{pos,fit}}$ as the actual relative position uncertainty $\sigma_{\mathrm{pos,rel}}$. $\sigma_{\mathrm{pos,fit}}$ was larger only for component $M_2$ in 2023, probably due to slightly lower resolution and brighter dust. These values are listed in Extended Data Table~\ref{table:fit} and can be used to determine the uncertainty in the separation of the two components at each epoch.

The astrometric positions of $M_1$ and $M_2$ at both ALMA epochs are listed in Extended Data Table~\ref{table:fit} and visualised in Extended Data Fig.~\ref{Fig:ALMA_data}. The relative position of $M_2$ with respect to $M_1$ was measured as ($\alpha_\star$, $\delta$) = ($-12.025\pm1.212$, $-34.700\pm1.212$) mas in the ALMA 2019 C6 data and as ($-24.103\pm0.721$, $32.910\pm1.000$) mas in the ALMA 2023 C10 data. Consequently, the angular separation between $M_1$ and $M_2$ changed from $-37.702\pm1.898$ mas in 2019 to $40.775\pm1.233$ mas in 2023. This positional shift is inconsistent with pure radial motion.

The measured UD diameters $D_{\mathrm{UD}}$ of $M_1$ and the flux densities of both components, $S_{\mathrm{UD}}$ and $S_{\mathrm{delta}}$, are given in Supplementary Table~\ref{table:UD}. We estimate the error $\sigma_{\mathrm{UD}}$ from the same phenomena as those causing $\sigma_{\mathrm{pos,rel}}$, given in Extended Data Table~\ref{table:fit}, using the combined $\alpha_{\star}$ and $\delta$ errors, with an additional factor of $\sqrt{2}$ as the diameter is the difference between two position measurements.
The relative continuum flux density errors in $S_{\mathrm{UD}}$ and $S_{\mathrm{delta}}$ are similar at each epoch, comprising the noise $\sigma_{\mathrm{rms}}$ and fitting uncertainties, shown as $\sigma_{\mathrm S}$. When comparing the two epochs, the ALMA flux scale error, typically 7\% at these frequencies~\cite{ALMA-TH}, should also be included. Using $\nu_{\mathrm{mean}}$ from Supplementary Table~\ref{table:obs}, the spectral index, $\alpha^s$ of each component (ignoring any potential variability) is $\alpha^s_{M_1}= 1.9\pm0.3$ and $\alpha^s_{M_2}= 2.3\pm0.3$ (using the convention $S\propto\nu^{\alpha^s}$).

\subsubsection{Astrometric uncertainty}\label{sec:astrometricerror}
The absolute astrometric accuracy is determined by use of the phase reference source. The main contributions to astrometric errors are the phase reference position errors $\sigma_{\mathrm{phref}}$, short-term phase jitter within each few-minute scan $\sigma_{\mathrm{short}}$, and errors in transferring phase solutions from the phase reference source to the target $\sigma_{\mathrm{trans}}$. We took $\sigma_{\mathrm{phref}}$ from the ALMA calibrator catalogue. $\sigma_{\mathrm{short}} = \theta_{\mathrm b} \times\phi_{\mathrm{rms}}/360$, where the phase rms $\phi_{\mathrm{rms}}$ in degrees was taken from the QA0 reports (for the best observation at each epoch)~\cite{ALMA-TH}. $\sigma_{\mathrm{trans}}$ is dominated by antenna position errors and by atmospheric differences between the directions of the sources, which we assume to be equivalent to a phase screen with a linear gradient. We estimated the effect on the target by comparing the apparent and catalogue positions of the check source to derive its error ($\sigma_{\mathrm{pos,check}}$) and scaling these to the phase calibrator - target separation, $\sigma_{\mathrm{pos,scaled}} = \sqrt{(\sigma_{\mathrm{pos,check}}^2 + \sigma_{\mathrm{short}}^2)}\times (\psi_{\mathrm{targ}}/\psi_{\mathrm{check}})$ (from Supplementary Table~\ref{table:obs}). The target astrometric (absolute) uncertainty is then given by $\sigma_{\mathrm{pos,abs}} =\sqrt{\sigma_{\mathrm{pos},\phi}^2 + \sigma_{\mathrm{pos,scaled}}^2 + \sigma_{\mathrm{short}}^2 } $. The values are given in Supplementary Table~\ref{table:astromerr}. We use more significant figures than are strictly warranted in order to avoid rounding errors in later modelling.

The check source catalogue position uncertainties are $\sim$1--2 mas. However, without a comprehensive directional atmospheric model (i.e., multiple calibrators), it is not realistic to include these uncertainties, as their contribution would be minor.

\section{Orbital parameters}\label{Methods:fitting}

In order to determine the orbital parameters of the close-in companion, $\pi^1$~Gru~C, we will first introduce the orbital model, followed by a description of the observational constraints obtained from the ALMA, \textit{Gaia}, and \textit{Hipparcos} data. We will then describe the Bayesian analysis used to derive the orbital parameters. Finally, a sensitivity analysis is performed to assess the robustness of the derived orbital parameters.

\subsection{Orbital model}\label{Methods:orbital_model}

The 2019 ALMA Cycle 6 continuum data revealed two maxima separated by 37.702~mas, with the secondary peak interpreted as a potential close companion~\cite{Homan2020A&A...644A..61H}. However, this feature could also be a dense clump of gas and dust resulting from a mass ejection from the AGB star, which would follow a more radial outflow. The 2023 ALMA Cycle 10 data provide a second epoch to confirm the proper motion, definitively ruling out the possibility of a radially outflowing aggregate. Instead, the positional shift observed in the 2023 C10 data is consistent with an elliptical projection on the sky, thereby establishing that the secondary continuum peak is a close-in companion on a Keplerian orbit around the AGB star. This confirms the hypothesis of Ref.~\cite{Homan2020A&A...644A..61H, Chiu2006ApJ...645..605C, Mayer2014A&A...570A.113M, Montarges2025A&A...699A..22M} that a close-in companion interacts with the wind of $\pi^1$~Gru~A. However, their study lacked the observational constraints necessary to determine the orbital parameters accurately.

Since 1953, it is already known that $\pi^1$~Gru has another far distant G0V companion, $\pi^1$~Gru~B, at a projected separation of 2\farcs71~\cite{Feast1953MNRAS.113..510F}. This establishes $\pi^1$~Gru as a triple system. The dynamical evolution of triple systems with a close-in binary and a wider tertiary may be influenced by the `Eccentric Kozai-Lidov' (EKL) mechanism~\cite{Kozai1962AJ.....67..591K,Lidov1962P&SS....9..719L}, which can cause the inner binary to experience large-amplitude oscillations in eccentricity and inclination, driving it to small pericentre distances and potential merger, while the tertiary may move outward or become unbound over evolutionary timescales. On one hand, it has been demonstrated that for a circular inner orbit, a large mutual inclination (40$^\circ$\,--\,140$^\circ$) can lead to long-timescale modulations that drive the eccentricity to very high values and can even result in orbital flips~\cite{Li2014ApJ...785..116L}. Although we will derive the inclination of the inner orbit below, the inclination of the outer orbit (of $\pi^1$~Gru~B) is currently unconstrained, preventing an accurate assessment of the EKL effect for a potentially high mutual inclination.

On the other hand, it has been shown that even starting with an almost coplanar configuration, for eccentric inner and outer orbits, the inner orbit's eccentricity can still be excited to high values, and the orbit can flip by approximately 180$^\circ$, rolling over its major axis~\cite{Li2014ApJ...785..116L}. However, as shown below, we derive a circular orbit for the inner orbit (of $\pi^1$~Gru~C), suggesting that the EKL mechanism is not active here.

Even without considering the EKL mechanism, it is necessary to evaluate the effect of the distant companion on the orbital motion of the two inner bodies due to Newtonian gravitational interactions. This distant companion, with a mass of $\sim$1\,\Msun, exerts a negligible influence on the orbital motion of the two inner bodies. Over a span of 24.25 years, the maximum induced change in the astrometric position of the inner bodies due to $\pi^1$~Gru~B is only about $0.65 \times 10^{-5}$ mas. Consequently, we can model the astrometric motion of the primary AGB star, $\pi^1$~Gru~A ($M_1$), and its inner companion, $\pi^1$~Gru~C ($M_2$), as a two-body system, effectively neglecting the influence of $\pi^1$~Gru~B.

We therefore model the orbital motion of the two celestial bodies, $M_1$ and $M_2$, in a binary system using Newton's laws, aiming to determine their positions over time, $t$, while accounting for parallax, proper motion, and orbital parameters. This involves analyzing the motion of $M_1$ and $M_2$ interacting solely through their mutual gravitational attraction.

The remainder of the derivation of the orbital equations, including the explicit transformation from the orbital plane to the plane of the sky, the calculation of the true anomaly as a function of time, and the projection of the orbital motion onto the observed astrometric coordinates, is detailed in Supplementary Sect.~\ref{sec:suppinforbeq}. This section provides the full mathematical framework required to compute the predicted positions of both components at any given epoch, accounting for the effects of parallax, proper motion, and the orientation of the orbit in three-dimensional space. The derivation also discusses the conventions adopted for the orbital elements and clarifies the treatment of degeneracies in inclination and node, ensuring that the model predictions can be robustly compared to the multi-epoch astrometric data from ALMA, \textit{Gaia}, and \textit{Hipparcos}.

\subsection{Observational input}\label{Sec:observational_input}

The orbital parameters for $\pi^1$~Gru, as defined in Supplementary Sect.~\ref{Methods:orbital_model}, can be constrained using data from ALMA C6, ALMA C10, \textit{Gaia} DR3, and {\textit{Hipparcos}}, along with their respective uncertainties.

The ICRS astrometric positions of component $M_1$ and $M_2$ at the epoch of the ALMA C6 and C10 observations are listed in Extended Data Table~\ref{table:fit}. The positional accuracy is given by $\sqrt{\sigma_{\mathrm{pos, rel}}^2 + \sigma_{\mathrm{pos, abs}}^2}$. These ALMA positions are not yet corrected for the parallactic shift since the parallax $\varpi$ is one of the retrieval parameters.

To summarize, for the ALMA C6 data:
\begin{align*}
    \alpha_{\rm{C6}}(M_1) &= 22^h22^m44.269589^s \pm 0.000240^s \tag{OC1}\label{OC1}\\
    \delta_{\rm{C6}}(M_1) &= -45^\circ56^\prime53.00641^{\prime\prime} \pm 0.001697^{\prime\prime} \tag{OC2}\label{OC2}\\
    \alpha_{\rm{C6}}(M_2) &= 22^h22^m44.268393^s \pm 0.000212^s \tag{OC3}\label{OC3}\\
    \delta_{\rm{C6}}(M_2) &= -45^\circ56^\prime53.04198^{\prime\prime} \pm 0.001217^{\prime\prime} \tag{OC4}\label{OC4}
\end{align*}

For the ALMA C10 data:
\begin{align*}
    \alpha_{\rm{C10}}(M_1) &= 22^h22^m44.282263^s \pm 0.000200^s \tag{OC5}\label{OC5}\\
    \delta_{\rm{C10}}(M_1) &= -45^\circ56^\prime53.12381^{\prime\prime} \pm 0.001777^{\prime\prime}\tag{OC6}\label{OC6}\\
    \alpha_{\rm{C10}}(M_2) &= 22^h22^m44.279952^s \pm 0.000196^s \tag{OC7}\label{OC7}\\
    \delta_{\rm{C10}}(M_2) &= -45^\circ56^\prime53.09090^{\prime\prime} \pm 0.001612^{\prime\prime} \tag{OC8}\label{OC8}
\end{align*}

Although the components of the binary system are not resolved in the \textit{Gaia} and \textit{Hipparcos} data, the fact that the companion remains undetected in the VLT/SPHERE data~\cite{Montarges2023A&A...671A..96M} taken contemporaneously with the ALMA C6 data (see Supplementary Sect.~\ref{Sec:SEDs}), implies that the mass-to-light ratios of $\pi^1$~Gru~A and $\pi^1$~Gru~C are significantly different, and that the \textit{Hipparcos} and \textit{Gaia} photocenters therefore track the photocentre of the AGB star $M_1$. The {\textit{Hipparcos}} catalogue~\footnote{\url{https://www.cosmos.esa.int/documents/532822/552851/vol9_all.pdf/50682119-3f37-4048-8f3a-e10961614b44}} lists the following five-dimensional (5D)\footnote{The radial velocity could not be constrained from the \textit{Hipparcos} and \textit{Gaia} measurements.} ICRS position-velocity vector for $\pi^1$~Gru (= HIP 110478) at epoch 1991.25:
\begin{equation}\label{astrometric_solution_hip}
    \by_{{\rm Hip}, M_1} =
    \begin{pmatrix}
    \alpha^* \\
    \delta \\
    \varpi \\
    \mu_{\alpha^*} \\
    \mu_{\delta}
    \end{pmatrix}_{M_1}
    =
    \begin{pmatrix}
    233.40516824561706 \\
    -45.94791727 \\
    6.54 \\
    27.89 \\
    -10.92
    \end{pmatrix}_{\rm Hip}
    \tag{OC9)--(OC13}
\end{equation}
Both $\alpha^*$ and $\delta$ are expressed in degrees, the \textit{Hipparcos} parallax $\varpi$ is expressed in milli-arcseconds (mas),
and the proper motions $\mu_{\alpha^*}$ and $\mu_{\delta}$ are expressed in mas~yr$^{-1}$.
The corresponding covariance matrix $\Sigma_{\rm Hip}$ is
\begin{equation}\label{covmatrix_hip}
\Sigma_{\rm Hip} =
\begin{pmatrix}
0.64      & 0.0496    & -0.0808   & -0.17952  & -0.08528     \\
0.0496    & 0.3844    & -0.15655  & -0.12648  & -0.172856    \\
-0.0808   & -0.15655  &  1.0201   & 0.236946  & 0.132512     \\
-0.17952  & -0.12648  &  0.236946 & 1.0404    & 0.25092      \\
-0.08528  & -0.172856 &  0.132512 & 0.25092   & 0.6724       \\
\end{pmatrix}
\end{equation}
where the uncertainties of $\alpha^*$, $\delta$, and $\varpi$ are expressed in mas.

The \textit{Gaia} DR3 archive~\cite{Vallenari2023A&A...674A...1G} lists the following astrometric solution for $\pi^1$~Gru (\,=\,\textit{Gaia} DR3 6518817665843312000) at epoch 2016.0:
\begin{equation}\label{astrometric_solution_gaia}
    \tag{OC14)--(OC18}
    \by_{{\rm Gaia}, M_1} =
    \begin{pmatrix}
    \alpha            \\
    \delta            \\
    \varpi            \\
    \mu_{\alpha^*}    \\
    \mu_{\delta}
    \end{pmatrix}_{M_1}
    =
    \begin{pmatrix}
    335.6844004512271  \\
    -45.94804349249061 \\
    6.185845810192854  \\
    31.105558619539813 \\
    -10.338212763867
    \end{pmatrix}_{\rm Gaia}
\end{equation}
where the units are the same as for the \textit{Hipparcos} solution. The corresponding covariance matrix $\Sigma_{\rm Gaia }$ is:
\begin{equation}\label{covmatrix_gaia}
\Sigma_{\rm Gaia} =
\begin{pmatrix}
0.05428154  & 0.01727978  & 0.00399398   & 0.01529596  & -0.01802598 \\
0.01727978  & 0.0856448   & -0.02478438  & -0.0093199  & -0.03776861 \\
0.00399398  & -0.02478438 & 0.19944953   & -0.02224772 & -0.05414651 \\
0.01529596  & -0.0093199  & -0.02224772  & 0.06329501  & 0.02410661  \\
-0.01802598 & -0.03776861 & -0.05414651  & 0.02410661  & 0.12071868
\end{pmatrix}
\end{equation}
where the units are the same as for $\Sigma_{\rm Hip}$.

These 18 observational constraints, (OC1)--(OC18), will be used to infer the 6 orbital parameters ($a$, $e$, $\Omega$, $i$, $\omega$ and $T_0$), the primary mass $m_1$, the mass ratio $q = m_2/m_1$, the parallax $\varpi$ toward the $\pi^1$~Gru system (or, equivalently, its distance $D$), and the proper motion vector of the barycentre $\bmu^G$.

\subsection{Predictables}\label{Sec:predictables}
The 18 observational constraints (OC1)\,--\,(OC18) are compared with 18 predicted values to derive the orbital parameters of the $\pi^1$~Gru system. Specifically, we fit the astrometric positions and parallax shift of $M_1$ and $M_2$ at the ALMA C6 and C10 epochs, and fit the astrometric positions, parallax and proper motion of $M_1$ from the \textit{Hipparcos} and \textit{Gaia} epochs.

We have intentionally left the ALMA observations uncorrected for the parallactic shift, as this shift depends on parallax, one of the parameters in the Bayesian retrieval fitting. To correctly apply Bayes' theorem, a clear distinction between observables and predictables must be maintained, avoiding any mixing between them. Once the parallax has been determined in a way that is consistent with the ALMA, \textit{Hipparcos}, and \textit{Gaia} data, the ALMA observations can be corrected for the parallactic shift and only the motion fits including the proper motion and orbital motion can be displayed. Accordingly, all figures in this paper displaying proper motion fits have been constructed after applying this approach.

It is important to note that the observed proper motion derived from \textit{Gaia} and \textit{Hipparcos} represents an average value over the duration of each mission. The \textit{Hipparcos} mission lasted from 1989.8 to 1993.2, while the \textit{Gaia} mission spanned from 2014.5 to 2017.4. The timestamps for the \textit{Gaia} observations of $\pi^1$~Gru are readily available and were used to derive the mean proper motion of $M_1$ for the \textit{Gaia} epoch. However, the exact timestamps for the \textit{Hipparcos} mission could not be retrieved. To address this, we sampled \textit{Hipparcos}' full observational time span with 300 linearly spaced timestamps, which were then used to calculate the right ascension and declination of $M_1$ at each timestamp, and hence to derive the average proper motion for the \textit{Hipparcos} epoch. Sampling with more data points did not impact the results. Additionally, given that the \textit{Hipparcos} data have a limited impact on the overall results, as demonstrated in the sensitivity analysis presented in Supplementary Sect.~\ref{Sec:sensitivity}, this assumption is considered well justified.

\subsection{Baysesian inference}\label{bayesian_modelling}
Using the Bayesian framework, we compute the posterior distribution $P(\btheta\mid\by)$ which is proportional to the product of the likelihood function and the prior distribution:
\begin{equation}
P(\btheta \mid \by) \propto P(\by \mid \btheta) \ P(\btheta).
\end{equation}
where $\btheta = (m_1, q, a, e, T_0, \omega, \Omega, i, D, \mu^G_\alpha, \mu^G_\delta)$ are the 11 model parameters, and $\by$ are the observations
outlined in Section~\ref{Sec:observational_input}. Since the \textit{Gaia}, \textit{Hipparcos}, and ALMA observations are statistically
independent, we can split the likelihood into
\begin{equation}
\begin{aligned}
    P(\by\mid\btheta) = & P(\by_{\rm C6, M_1} | \btheta) \cdot
                          P(\by_{\rm C6, M_2} | \btheta) \cdot
                          P(\by_{\rm C10, M_1} | \btheta) \cdot
                          P(\by_{\rm C10, M_2}|\btheta) \ \cdot \\[2mm]
                        &
                          P(\by_{\rm\textit{Gaia}, M_1}|\btheta) \cdot
                          P(\by_{\rm Hip, M_1}|\btheta)
\end{aligned}
\end{equation}
where we abbreviated $\by_{{\rm C}n, M_i}\hspace{-1pt}=\hspace{-1pt}(\alpha_{{\rm C}n, M_i}, \delta_{{\rm C}n, M_i})$
and $\by_{\rm\textit{Gaia}/Hip, M_1}\hspace{-1pt}=\hspace{-1pt}{(\alpha_{\rm M_1}, \delta_{\rm M_1}, \varpi, \mu_{\alpha^*_1}, \mu_{\delta_1})}_{\rm\textit{Gaia}/Hip}$.
$\by_{{\rm C}n, M_i}$ and $\by_{\rm\textit{Gaia}/Hip, M_1}$ are given in Section~\ref{Sec:observational_input}.
Each of the likelihoods was chosen to be a Gaussian:
\begin{equation}
\begin{aligned}
    P(\by_{X}|\btheta) & = \mathcal{N}(\boldm(\btheta), \bSigma_{X}) \\
       & = \frac{1}{\sqrt{{(2\pi)}^k |\bSigma_X|}} \exp\left(-\frac{1}{2} \ {(y_X - \boldm(\btheta))}^t \ \bSigma_X^{-1} \ (y_X - \boldm(\btheta))\right)\label{gaussian_likelihood}
\end{aligned}
\end{equation}
where $\boldm(\btheta) = {\rm E}[\by_X]$. In the case of $P(\by_{\rm\textit{Gaia}, M_1}|\btheta)$ and $P(\by_{\rm Hip, M_1}|\btheta)$, the non-diagonal covariance matrices
$\Sigma_{\rm Hip}$ and $\Sigma_{\rm\textit{Gaia}}$ are given by Expr.~(\ref{covmatrix_hip}) and (\ref{covmatrix_gaia}). In all other cases, the covariance matrices are diagonal with the square of the standard errors given in Section~\ref{Sec:observational_input} on the diagonal.

In each case the components of the expected value $\boldm = E[\by_X]$ related to the sky coordinates $(\alpha, \delta)$ of $M_1$ or $M_2$ are computed using the following model:
\begin{equation}
    \binom{E[\alpha]}{E[\delta]} = \binom{\alpha_G^0}{\delta_G^0} + \binom{\mu_{\alpha}^G}{\mu_{\delta}^G} (t - t_{C6}) + \binom{\Delta\alpha}{\Delta\delta}.\label{eq:alpha_delta_model}
\end{equation}
Here, $(\alpha_G^0, \delta_G^0)$ denote the ICRS coordinates of the binary system's centre of mass at $t=t_{C6}$:
\begin{align}
    \alpha_G^0 \equiv \frac{m_1 \ \alpha_{C6, M_1} + m_2 \ \alpha_{C6, M_2}}{m_1 + m_2} - \Delta^p_\alpha \\[3mm]
    \delta_G^0 \equiv \frac{m_1 \ \delta_{C6, M_1} + m_2 \ \delta_{C6, M_2}}{m_1 + m_2} -\Delta^p_\delta\,.\label{Eq:def_alpha0_delta0}
\end{align}
where the terms $\Delta^p_\alpha$ and $\Delta^p_\delta$ represent corrections for the parallax motion, accounting for the fact that $(\alpha_{C6, M_1}, \delta_{C6, M_1})$ and $(\alpha_{C6, M_2}, \delta_{C6, M_2})$ are the observed ALMA astrometric positions.
The second term in Eq.~(\ref{eq:alpha_delta_model}) reflects the proper motion of the barycentre and the third term the change in coordinates with respect to the barycentre due to the orbital motion of the stars, and is computed as outlined in Supplementary Sect.~\ref{Methods:pos}.

We assumed no a priori correlations between the model parameters, and decomposed the prior $P(\btheta)$ into:
\begin{equation}
    P(\btheta) = \prod_{n=1}^9 P(\theta_n)
\end{equation}
where we implemented the following priors for each of the 11 model parameters:
\begin{align}
P(m_1)    & = \mathcal{U}(0.6, 3.5)          & [M_{\odot}] \tag{P1} \\[2mm]
P(q)      & = \mathcal{U}(0.2, 5.0)        & [\ ]        \tag{P2} \\[2mm]
P(a)      & = \mathcal{U}(2, 15)           & [{\rm au}]  \tag{P3} \\[2mm]
P(e)      & = \mathcal{U}(0, 0.95)          & [\ ]      \tag{P4}   \\[2mm]
P(T_0)    & = \mathcal{U}(2009, 2030)      & [{\rm yr}] \tag{P5}  \\[2mm]
P(\omega) & = \mathcal{U}(0, 360)        & [{\deg}] \tag{P6} \\[2mm]
P(\Omega) & = \mathcal{U}(0, 360)        & [{\deg}] \tag{P7} \\[2mm]
P(i)      & = \mathcal{U}(0, 360)           & [{\deg}] \tag{P8} \\[2mm]
P(D)      & = {\rm Gamma}(k=3,L=500)  & [{\rm pc}] \tag{P9}\\[2mm]
P(\mu^G_\alpha) & = \mathcal{U}(42, 46)          & [{\rm mas\ yr^{-1}}] \tag{P10} \\[2mm]
P(\mu^G_\delta) & = \mathcal{U}(-20, -17)          & [{\rm mas\ yr^{-1}}] \tag{P11}
\end{align}
where $\mathcal{U}(u_1, u_2)$ indicates a uniform prior between ranges $u_1$ and $u_2$. We initially explored a broader prior range for $T_0$ (1900--2200~yr), but all runs consistently converged to orbital periods around 11~yr. Based on this, we refined the prior to the range 2009--2030~yr.

The Gamma prior for the distance $D$ of the system, with shape parameter $k=3$ and scale parameters $L=500$~pc, reflects an exponentially decreasing stellar volume density in the nearby Milky Way~\cite{Bailer-Jones2015PASP..127..994B}. I.e.\ the prior for the distance is determined as
\begin{equation}
    F^{-1}(p, k=3, L=500)
\end{equation}
where $p$ is a uniform random variable ($p \in [0,1]$) and
\begin{equation}
    F(x, k, L) = \frac{1}{\Gamma(k)} \int_0^x \frac{t^{k-1} e^{-t/L}}{L^3}\ dt\,,
\end{equation}
with $\Gamma(k) = (k-1)!$\,.

To sample the posterior distribution, we used different sampling methods, for which nested sampling works the most efficient. More details on the sampling methods and the \texttt{ultranest} package can be found in Supplementary Sect.~\ref{sec:sampling}.

\subsection[Inferred orbital parameters for the Pi1~Gru system]{Inferred orbital parameters for the $\boldsymbol{\pi^1}$~Gru system}\label{Methods:orbital_results}
Using the observational data outlined in Sect.~\ref{Sec:observational_input} and the \texttt{ultranest}~\cite{Ultranest2021JOSS....6.3001B} Bayesian modeling framework, we inferred the six orbital elements, the primary mass ($m_1$), the mass ratio ($q$), the distance ($D$), and the proper motion of the barycentre ($\bmu^G$). To assess robustness, \texttt{ultranest} was executed 10 times, as detailed in Supplementary Sect.~\ref{Sec:sensitivity}, with a summary of results presented in Supplementary Table~\ref{table:robustness}. In this section the results of the best run are presented.

The outcomes of Bayesian inference are characterized by two key metrics. First, we report the mean and standard deviation of the retrieved parameters, derived from the posterior samples (upper part of the table). The mean is weighted by the likelihood of each sample, while the standard deviation reflects the posterior spread. Second, we present the best-fit values, corresponding to the maximum posterior probability for each parameter (lower part of the table).

The run that achieved the highest maximum marginalized likelihood, and we refer to this as the \texttt{eccentric} model. The Bayesian retrieval yields $m_1 = 1.02 \pm 0.20 \, \Msun$, $q = 1.04 \pm 0.05$, $a = 6.60 \pm 0.41 \, \text{au}$, $e = 0.023 \pm 0.017$, $T_0 = 2026.75 \pm 3.16 \, \text{yr}$, $\omega = 101 \pm 98^\circ$, $\Omega = 94 \pm 23^\circ$, $i = 14 \pm 8^\circ$, $D = 174 \pm 9 \, \text{pc}$, $\mu_\alpha^G = 45.212 \pm 0.168 \, \text{mas yr}^{-1}$, and $\mu_\delta^G = -18.773 \pm 0.068 \, \text{mas yr}^{-1}$. The small eccentricity implies $\omega$ and $T_0$ are poorly constrained (see also Supplementary Sect.~\ref{Sec:sensitivity}).

However, the Bayesian marginalized likelihood ($\log z$) -- factoring in both fit quality and model complexity -- indicates a preference for a circular orbit ($e = 0$) over the above eccentric model (see Supplementary Sect.~\ref{Sec:sensitivity}). For a circular orbit, $\omega$ and $T_0$ are undefined; instead, we define $T_0$ as the time when the body crosses the ascending node. Similar to the eccentric case, \texttt{ultranest} was executed 10 times (see Supplementary Table~\ref{table:robustness_circular}). The highest maximum likelihood run, and we refer to this as the \texttt{circular} model. The inferred parameters are $m_1 = 1.12 \pm 0.25 \, \Msun$, $q = 1.05 \pm 0.05$, $a = 6.81 \pm 0.49 \, \text{au}$, $T_0 = 2016.39 \pm 1.18 \, \text{yr}$, $\Omega = 101 \pm 36^\circ$, $i = 11 \pm 7^\circ$, $D = 180 \pm 10 \, \text{pc}$ (or equivalently, $\varpi = 5.555 \pm 0.309 \, \text{mas}$), $\mu_\alpha^G = 45.203 \pm 0.144 \, \text{mas yr}^{-1}$, and $\mu_\delta^G = -18.776 \pm 0.061 \, \text{mas yr}^{-1}$.
These results are shown in a corner plot (Extended Data Fig.~\ref{fig:pi1_gru_corner_ultranest}) and summarized in Supplementary Table~\ref{table:fit_sensitivity}. The best-fit values are given in Supplementary Table~\ref{table:best_fit}.

The fit of the astrometric positions and tangential velocities to the observational data, based on these best-fit parameters, is depicted in Fig.~\ref{fig:pi1_gru}. The orbital system of $\pi^1$~Gru, using the retrieved best-fit parameters, is visualized in Extended Data Fig.~\ref{fig:schematic} for two different viewing angles. We deduce that the companion moves in a circular, anti-clockwise orbit relative to the AGB star, a conclusion further reinforced by ALMA and SPHERE-ZIMPOL data presented in ref.~\cite{Homan2020A&A...644A..61H, Doan2020A&A...633A..13D, Montarges2023A&A...671A..96M, Doan2017A&A...605A..28D}. The arcs observed in these datasets, interpreted as segments of an anti-clockwise spiral, exhibit shapes consistent with this orbital motion. The ALMA data~\cite{Homan2020A&A...644A..61H, Doan2020A&A...633A..13D, Doan2017A&A...605A..28D} trace the gas motion, while the SPHERE-ZIMPOL data~\cite{Montarges2023A&A...671A..96M}, taken almost at the same time as the ALMA C6 observations, reveal that the dust tail, formed in the companion's wake, also exhibits a shape consistent with this anti-clockwise orbit.

The corner plot (Extended Data Fig.~\ref{fig:pi1_gru_corner_ultranest}) demonstrates high-quality sampling with well-defined boundaries and Gaussian-like posterior distributions for $m_1, q, a, D$, and $\bmu^G$, underscoring the effectiveness of the \texttt{ultranest} sampling method. It also reveals the expected strong correlations between $m_1$, $a$, and $D$ and $\Omega$ and $T_0$. In particular, there is a perfect positive correlation between $\Omega$ and $T_0$. Additionally, a strong positive degeneracy exists between $a$ and $m_1$ (with linear correlation coefficient, $\rho$, of 0.99), between $D$ and $a$ ($\rho = 0.94$) and $D$ and $m_1$ ($\rho=0.90$).

For the derived best-fit circular system parameters, the predicted tangential velocity of the barycentre, $v_{\tan}^G$, is 29.86 km s$^{-1}$ with a position angle of $120.84^\circ$. The tangential velocity anomaly, $\Delta v_{\tan}$, and its corresponding position angle, $\theta$, at the \textit{Gaia} and \textit{Hipparcos} epochs are 7.03~km~s$^{-1}$ and $359.04^\circ$, and 6.80~km~s$^{-1}$ and $336.39^\circ$, respectively.

The corresponding orbital period is $11.76 \pm 1.85$\,years. Whether this period can be linked to a long secondary period (LSP) remains unclear. The periodogram derived from the ASAS light curve of $\pi^1$~Gru reproduces the AGB pulsation period of 195.5\,days well (see Sect.~\ref{Sec:stellar_evolution}), and there are tentative indications of an LSP of approximately 10\,years. However, the data span of $\sim9$\,years is insufficient for a detection at the $\geq 3\sigma$ level. Furthermore, given the low inclination we derive for the system, the probability of detecting an LSP is inherently low.

The relative motion fit of $M_2$ with respect to $M_1$ for the ALMA 2019 C6 and ALMA 2023 C10 data is presented in Supplementary Fig.~\ref{Fig:relative_motion}. Accounting for both $\sigma_{\rm pos,abs}$ and $\sigma_{\rm pos, rel}$, the uncertainties in the relative position of $M_2$ with respect to $M_1$ in ($\alpha_\star$, $\delta$) are (3.34, 2.09)~mas for the ALMA C6 epoch and (2.92, 2.40)~mas for the ALMA C10 epoch. The difference between the observed and predicted astrometric positions is (0.48, 1.60)~mas at ALMA C6 and ($-0.95$, $-1.77$)~mas at ALMA C10, indicating consistency within the measurement uncertainties.

The astrometric position of $M_2$ at the ALMA C6 epoch corresponds to an orbital contribution to the radial velocity of 0.06~km~s$^{-1}$ for the ALMA C6 epoch and 3.37~km~s$^{-1}$ for the ALMA C10 epoch (see Supplementary Sect.~\ref{Methods:vrad}). This orbital motion represents only one component of the overall velocity vector field. As discussed in Sect.~\ref{Sec:3D_hydro}, the velocity amplitude and direction exhibit significant variation at and around the position of $M_2$, due to the formation of a bow shock spiral and an accretion disk. This complexity in the velocity vector field is also evident in the ALMA C6 and C10 data, particularly in the analysis of the SiO maser lines; see Sect.~\ref{Sec:vrad}.

A comprehensive sensitivity analysis of the orbital parameter retrieval is presented in Supplementary Sect.~\ref{Sec:sensitivity}. There, we systematically explore the robustness of the inferred orbital parameters by varying input assumptions, priors, and data subsets. This includes repeated runs of the Bayesian inference, alternative prior choices (e.g., for the primary mass), and exclusion of specific observational constraints. The resulting posterior distributions and best-fit values are compared to assess the stability of the solution and to identify any potential degeneracies or biases.

\subsubsection[Distinguishing between (omega, Omega) and its antipodal nodes using radial velocity measurements]{Distinguishing between ($\omega, \Omega$) and its antipodal nodes using radial velocity measurements}\label{Sec:vrad}

Although the radial velocity of $\pi^1$~Gru's barycentre remains unknown, we can use the difference in the orbital contribution to the radial velocity of $M_1$ at the ALMA C6 and ALMA C10 epochs to initially resolve the ($\omega, \Omega$) versus ($\omega+180^\circ, \Omega+180^\circ$) ambiguity.

For the specific \texttt{circular} configuration, the predicted orbital contribution to the radial velocity of $M_1$ is approximately $0.21$~km~s$^{-1}$ at the ALMA C6 epoch, and about $-1.67$~km~s$^{-1}$ at the ALMA C10 epoch (see Supplementary Fig.~\ref{Fig:vrad_vtan}). This leads to a predicted change in radial velocity between both epochs of $\Delta v_{\rm rad}^{\rm orb} \approx -1.9$~km~s$^{-1}$.

This preliminary prediction can be compared to observational constraints. Specifically, we use the central velocities of various molecular lines to estimate
$v_{\rm rad}^\prime$ of $M_1$ in the kinematic local standard of rest (LSRK)
frame at both the ALMA C6 and ALMA C10 epochs (see Eq.~\eqref{Eq:vrad_prime}).
To achieve this, we extracted data from a small aperture (of diameter $0\farcs04$) centered on the AGB continuum peak and select high-excitation lines probing the inner few stellar radii around the AGB star (see Supplementary Fig.~\ref{Fig:central_velocities}). We used the extended configuration ALMA C6 2019 data with an angular resolution of $\sim0\farcs019$~\cite{Gottlieb2022A&A...660A..94G}.
The central velocities are derived by fitting a Gaussian to the line profiles.

From this analysis, we obtain an LSRK velocity for $M_1$ of approximately $-16.8 \pm 1.0$~km~s$^{-1}$ at the ALMA C6 epoch and $-18.3 \pm 0.5$~km~s$^{-1}$ at the ALMA C10 epoch, with corresponding spectral resolution being 1.3~km~s$^{-1}$ and 0.17~km~s$^{-1}$, respectively.
We note that the emission from behind the star is obscured and that within the
inner few stellar radii, the wind could be in infall, causing the apparent velocity
to be slightly blue-shifted. However, since the line profiles are fairly symmetric
with aligned peaks at each epoch, this suggests that the blue-shifted bias is of
similar strength at both epochs, effectively canceling out when calculating the
velocity difference. The observed velocity difference is $\Delta v_{\rm rad}^{\rm orb} \approx -1.5 \pm 1.1$. The good agreement between the predicted value and the observational constraint, both of which fall within the same ballpark, is reassuring and provides an initial indication that the current combination of ($\omega, \Omega$) is likely the correct one.

This analysis also offers a first estimate of the radial velocity of $\pi^1$~Gru's barycentre, which we determine to be $-14.8 \pm 1.0$\,km\,s$^{-1}$.
This yields predicted radial velocities for $M_1$ in the ICRS frame of $-14.59 \pm 1.0$\,km\,s$^{-1}$ and $-16.48 \pm 1.0$\,km\,s$^{-1}$ at the ALMA C6 and ALMA C10 epochs, respectively. When transformed into the LSRK frame, these values become $-16.63 \pm 1.0$~km~s$^{-1}$ and $-18.52\pm1.0$~km~s$^{-1}$, which are very close to the observed LSRK velocities of $-16.8 \pm 1.0$~km~s$^{-1}$ at the ALMA C6 epoch and $-18.3 \pm 0.5$~km~s$^{-1}$ at the ALMA C10 epoch, listed above. This predicted radial velocity for the barycentre is also confirmed when comparing to the
ALMA 2023 C10 moment-1 (intensity-weighted mean velocity) map of the $^{29}$SiO v=0 J=8-7 emission (panel~(b) of Supplementary Fig.~\ref{fig:pv_CO_2023}).
Future ALMA line observations would provide additional velocity information, thus introducing even more observational constraints in the Bayesian fitting process.

Another validation for the current combination of ($\omega, \Omega$) comes from the 2019 ALMA C6 SiO maser data presented by Ref.~\cite{Homan2020A&A...644A..61H}. We complement this with a similar analysis of the $^{28}$SiO v=0, 1, 2 J=8-7 lines (panel~a of Supplementary Fig.~\ref{fig:pv_CO_2023}), and the $^{12}$CO v=1 J=3-2 and $^{29}$SiO v=0 J=8-7 position-velocity (PV) diagrams from the 2023 ALMA C10 data (Supplementary Fig.~\ref{fig:pv_CO_2023}, panels~d-e). Panel~(c) of Supplementary Fig.~\ref{fig:pv_CO_2023} depicts a $^{28}$SiO v=1 J=6-5 PV diagram from the 2019 ALMA C6 data. The ALMA C10 data have a higher spectral resolution and spectral line signal-to-noise ratio (SNR) than the C6 data, allowing for a more detailed view of the velocity structure around the primary and companion stars.

In particular, high resolution SiO data reveal the velocity structure close to $M_2$ at both epochs. This is seen in Supplementary Fig.~\ref{fig:pv_CO_2023} and also in Fig.~11--12 of Ref.~\cite{Homan2020A&A...644A..61H}. SiO $v>0$ emission, mostly masers, comes from high-energy level states and thus is confined close to the star. The $v=0$ emission shows some signs of masing and, by selecting observations at the highest angular resolutions, the region around $M_1$ and $M_2$ can be resolved whilst large-scale thermal emission is resolved-out. However, when including short baselines sensitive to extended flux, such as the 2019 combined data, $v=0$ thermal emission on larger scales dominates, showing the circumbinary disc but obscuring the $M_1-M_2$ interaction. Consequently, the 2019 SiO moment-1 map (intensity-weighted mean velocity) presented by Ref.~\cite{Homan2020A&A...644A..61H} appears near zero at the position of $M_2$. However, as we discuss below, this should not -- and does not -- imply that the radial velocity is actually zero at that location. Instead, this is likely due to weaker SiO emission, possibly caused by partial obscuration by the companion star and its accretion disk.

In 2019, the SiO maser flow from $M_1$ towards $M_2$ showed an increasingly blue-shifted trend closer to $M_2$. In contrast, by 2023, this maser flow had become red-shifted; see panel~(a) of Supplementary Fig.~\ref{fig:pv_CO_2023}. Accounting for the ALMA C10 barycentric radial velocity of approximately $-18.3$~km~s$^{-1}$, the relative radial velocity at and around the location of $M_2$ is non-zero, reflecting the radial projection of the complex velocity structure, also seen in the hydrodynamical simulations presented in Sect.~\ref{Sec:3D_hydro}, where the radially outflowing wind and the spiral bow shock play key roles.

The same velocity gradient is observed in the 2023 C10 $^{12}$CO v=1 J=3-2 data (see panel~(e) of Supplementary Fig.~\ref{fig:pv_CO_2023}), where the radial velocity ranges from approximately $-19$~km~s$^{-1}$ to $-12$~km~s$^{-1}$ midway between $M_1$ and $M_2$; however, the emission is too weak to measure the velocity structure directly at $M_2$. Additionally, the CO v=1 J=3-2 emission near $M_1$ appears weaker, likely due to partial obscuration by the star. In this respect, the 2023 $^{29}$SiO v=0 J=8-7 strong emission provides better diagnostics, with panel~(d) of Supplementary Fig.~\ref{fig:pv_CO_2023} clearly showing the increasingly red-shifted stream towards $M_2$.

This streamer was progressively more blue-shifted in the 2019 data, as confirmed by the SiO maser analysis presented by Ref.~\cite{Homan2020A&A...644A..61H}. The blue-shift is also evident in, for example, the $^{28}$SiO v=1 J=6-5 PV diagram presented in panel~(c) of Supplementary Fig.~\ref{fig:pv_CO_2023}. This shift from a blue-shifted (2019) to a red-shifted (2023) streamer between $M_1$ and $M_2$ is consistent with the retrieved orbital configuration, including ($\omega, \Omega$), as listed in Supplementary Table~\ref{table:fit_sensitivity} and shown in Extended Data Fig.~\ref{fig:schematic}.

\section{Hydrodynamical modelling}\label{Sec:3D_hydro}
For a circular orbit with synchronous stars, the Roche lobe radius of a star in a binary system depends solely on the system's semi-major axis and mass ratio. It is expressed as~\cite{Eggleton2006epbm.book.....E}:
\begin{equation}
    R_{L,1} = \frac{0.49 q^{2/3}}{0.6 q^{2/3} + \log(1 + q^{1/3})} a\,,
\end{equation}
where $R_{L,1}$ represents the radius of a sphere with a volume equivalent to that of the Roche lobe. Using the derived values for $a$ and $q$, we calculate a Roche lobe radius of approximately $\sim 2.61~\text{au}$. For $\pi^1~\text{Gru~A}$, which has a radius of $\sim 1.65~\text{au}$ (see Sect.~\ref{Sec:Luminosity_A}), this indicates that the star does not fill its Roche lobe.

Mass-loss still occurs for the AGB star due to their dust-driven winds. This material can be attracted by the companion and will be accreted. In the most simple case where the wind velocity is much larger then the orbital velocity of the companion this mechanism can be described by the classical Bondi-Hoyle-Lyttleton (BHL)~\cite{Hoyle1939PCPS...35..405H,Bondi1944MNRAS.104..273B} formalism. In the case of the $\pi^1$~Gru system, the orbital velocity of the companion of $\sim$16\,km~s$^{-1}$ almost equals the radial expanding wind velocity of $\sim$14\,km~s$^{-1}$ (as derived by Ref.~\cite{Doan2017A&A...605A..28D}). This implies that a wind accretion scenario relying on this BHL formalism is not valid.

This suggests that the wind-RLOF regime, first studied in the context of symbiotic binaries by Ref.~\cite{Mohamed2007ASPC..372..397M}, offers a more promising framework to understand the structure in the circumstellar envelope of $\pi^1$~Gru~A and the wind-companion interaction. For this mechanism to be effective, the dust condensation radius must extend beyond the Roche lobe, which is a reasonable assumption given that dust condensation in such environments typically occurs at a few stellar radii.

The gravitational influence of a binary companion affects the wind morphology of an AGB star in two distinct ways~\cite{Kim2012ApJ...759...59K}. On the one hand, the companion's gravity focuses a fraction of the wind material towards the equatorial plane into a detached bow shock or accretion wake flowing behind the companion~\cite{Maes2021A&A...653A..25M}. On the other hand, it induces an orbital motion of both stars around the barycentre. This generates a spiral shock with a stand-off radius defined by the orbital and wind velocity. The arc pattern due to the reflex motion of the mass-losing AGB star nearly reaches the orbital axis and introduces an oblate-shaped flattening of the circumstellar envelope density.

To predict the wind morphology, estimate the mass accretion rate onto the companion, and explore the nature and properties of the accretion disk expected to form around the companion star, we perform high-resolution, three-dimensional smoothed particle hydrodynamics (SPH) simulations using the \textsc{Phantom} code~\cite{Price2018PASA...35...31P}. The numerical setup of the simulations is the same as described in Ref.~\cite{Malfait2024A&A...691A..84M}. The stars are modelled as sink particles that can accrete wind particles and gain their mass and momentum~\cite{Malfait2024A&A...691A..84M, Price2018PASA...35...31P}. The stellar wind is modelled using the free-wind approximation, where the gravitational force of the AGB star is artificially balanced by the radiation force. This allows for a computationally simple way to launch a wind, but it does not adequately take into account pulsations, dust formation and the impact of radiative transfer.
Better wind prescriptions that incorporate pulsations~\cite{Aydi2022MNRAS.513.4405A}, dust nucleation~\cite{Siess2022A&A...667A..75S} or radiation transport~\cite{Esseldeurs2023A&A...674A.122E} are currently under development, but a fully integrated model that combines all these effects remains to be achieved.

To model the accretion around the companion, we include H\,I cooling and assume the wind is atomic, with a mean molecular weight of $1.26$ u~\cite{Malfait2024A&A...691A..84M}. The model parameters are as follows: the companion's accretion radius is $R_{\rm2, accr} = 2.15 \, R_\odot$, optimized for resolving the accretion disk in the model; the stellar masses are $m_1 = 1.12 \, M_\odot$ and $m_2 = 1.17 \, M_\odot$; the semi-major axis is $a=6.81 \, {\rm au}$; and the orbital eccentricity is $e=0.0$. The AGB star’s mass-loss rate is $8 \times 10^{-7}\, {\rm M_\odot \, yr}^{-1}$~\cite{Doan2017A&A...605A..28D}, and the initial wind velocity is $8.85 \, {\rm km\,s}^{-1}$ at the stellar surface, corresponding to a terminal velocity of $\sim18 \, {\rm km\,s}^{-1}$. To optimize computational efficiency, particles beyond a radius of $30 \, {\rm au}$ are removed from the simulation. The simulation spans approximately $14.5$ orbital periods, at which point the accretion disk’s mass stabilizes, indicating equilibrium between the wind accretion rate onto the disk and the mass accretion rate from the disk onto the companion. The simulation employs roughly 500,000 particles with a particle mass of $3.05 \times 10^{-11}$\Msun.

The resulting density distribution around the stars is presented in Extended Data Fig.~\ref{Fig:disk_simulation}. A bow shock spiral originates in front of a dense accretion disk that surrounds the companion. The zoomed-in figure (panel b) reveals the 3D shape of the accretion disk, that is roughly circular in the orbital plane and flaring in the edge-on view, and reveals the tangential motion of matter in the accretion disk around the companion. Material is orbiting with tangential velocities in the range
[$0.8,1.0$] times the Keplerian velocity, i.e.\ $(0.8-1.0)\times \sqrt{G m_2/r}$.

Adopting the method described in Ref.~\cite{Malfait2024A&A...691A..84M}, we estimate that the disk has a radius of $r_{\rm disk} = 0.83 \, {\rm au}$, a mass of $M_{\rm disk} = 2.0 \times 10^{-6} \, {\rm M_\odot}$, a flared edge-on profile with maximum scale height $H(r_{\rm disk}) = 0.144 \, {\rm au}$, and a maximum midplane density at $r=0.09$~au of $\rho_{\max} = 8\times 10^{-11} \, {\rm g\,cm^{-3}}$.
The estimated mass accretion efficiency onto the companion sink particle is $14.5 \%$, corresponding to a mass accretion rate of $1.16\times 10^{-7} \,{\rm M_\odot \, yr^{-1}}$.
Given that the outer disk radius is less than 70\% of the companion's Roche lobe radius (of $\sim$1.6~au), tidal interactions that could truncate the disk can be safely neglected.

\section{Stellar evolution}\label{Sec:stellar_evolution}
In order to access the evolutionary state of $\pi^1$~Gru~A, and estimate its initial mass, stellar properties are derived like the luminosity, temperature, and radius as well as its pulsation period. These properties are then compared to stellar evolution models to infer the star's evolutionary state and initial mass.

\subsection[Stellar properties of Pi1~Gru~A]{Stellar properties of $\boldsymbol{\pi^1}$~Gru~A}\label{Sec:Luminosity_A}

To estimate the luminosity and temperature of $\pi^1$~Gru~A, we rely on the ($V-K$) colour index, which is a well-established temperature indicator for late-type stars. The mean Johnson $V$-band magnitude is reported in the ASAS-SN catalogue of variable stars as $V=7.06$ with amplitude of variation of 1.48\cite{Jayasinghe2020MNRAS.491...13J}. The $K$-band amplitude of variation is far less, $\sim$0.1, and we here use $V-K = 8.61$ as reported by Ref.~\cite{Ducati2002yCat.2237....0D}.

To estimate both interstellar and circumstellar extinction along the line of sight, we use the $E(BP-RP)$ colour excess from the \textit{Gaia} DR3 archive~\cite{Vallenari2023A&A...674A...1G}. The \textit{Gaia} $BP$ band covers wavelengths from approximately 400 to 500\,nm, while the $RP$ band spans 600 to 750\,nm. According to \textit{Gaia} DR3, the colour excess is listed as $E(BP-RP) = 1.6056$. Using established photometric transformations between \textit{Gaia} and other systems\footnote{\url{https://gea.esac.esa.int/archive/documentation/GEDR3/Data_processing/chap_cu5pho/cu5pho_sec_photSystem/cu5pho_ssec_photRelations.html}}, we derive a corresponding \textit{Hipparcos} $E(B-V)$ value of 1.40.

The total reddening caused by dust depends on the ratio of total to selective extinction, $R_V = A_V / E(B-V)$. For the diffuse interstellar medium, $R_V \sim 3.1$, but in dense or dusty environments, such as circumstellar regions of AGB stars, grain growth leads to higher $R_V$ values. Typical values range between 3.5 and 4.5~\cite{Cardelli1989ApJ...345..245C}. For oxygen-rich red supergiant stars, which share their dust composition with oxygen-rich AGB stars, $R_V$ has been determined to be $\sim 4.2$~\cite{Massey2005ApJ...634.1286M}. Applying this value, we calculate $A_V = 5.87\pm0.70$. Using the extinction relations from Ref.~\cite{Gordon2023ApJ...950...86G}, we determine that ${(V-K)}_0 = 4.55 \pm 1.65$.

To compute the bolometric correction, $BC_K$, we use the (${(V-K)}_0$, $BC_K$) relation from Ref.~\cite{Bessell1984PASP...96..247B}, which yields $BC_K = 2.78 \pm 0.27$. Adopting the solar bolometric magnitude of $M_{\rm bol, \odot}=4.74$ and a best-fit distance of 169.38~pc from the eccentric model, we derive a luminosity of 7,300$\pm$2,100\,\Lsun. At a distance of 180~pc, the luminosity increases by $\sim$1,000\,\Lsun.

The angular diameter of $\pi^1$~Gru~A, measured at 1.65\,$\mu$m, is $18.37 \pm 0.18$~mas~\cite{Paladini2018Natur.553..310P}. At a distance of $169.38$~pc, this translates to a stellar radius of $334 \pm 20\,R_\odot$. Using the derived luminosity, radius and mass (of $1.12\pm0.25$\,\Msun), the effective temperature is determined to be 2,900$\pm$200~K with surface gravity of $\log_{10} g$ [cgs]\,=\,$-0.56\pm0.11$, or $\log_{10} g$ [cgs]\,=\,$-0.51\pm0.10$ in the case of the higher current mass $m_1$ derived from the Gaussian prior sensitivity study.

\subsection[Initial mass of Pi1~Gru~A]{Initial mass of $\boldsymbol{\pi^1}$~Gru~A}\label{Sec:initial_mass}

We provide an independent estimate of the initial mass from analysing $\pi^1$~Gru~A's variability, when coupled with stellar evolution models. The stellar evolution models help to constrain the nature and evolution of $\pi^1$~Gru~A along the AGB phase.

We construct a core-mass luminosity diagram using AGB models between 1\,\Msun\ and 2\,\Msun\ with solar metallicity, using models from Ref.~\cite{Karakas2014MNRAS.445..347K}, which is shown in Supplementary Fig.~\ref{Fig:core-mass-lum}. The luminosity of $\pi^1$~Gru~A is indicated by the dashed lines, along with upper and lower bounds from the literature. This diagram illustrates that models with initial masses of around 1.7\,\Msun\ are the best fit to the luminosity as explained below, but the uncertainties are large.
This figure demonstrates the degeneracy in the core-mass luminosity relationship for low-mass AGB stars. However, models below 1.25\,\Msun\ can be ruled out owing to the fact that these do not experience any third dredge-up mixing~\cite{Karakas2014MNRAS.445..347K} which is needed to explain the fact that $\pi^1$~Gru~A is an intrinsic S-type AGB star with a clear detection of the radioactive isotope of s-process element technetium ($^{99}$Tc)~\cite{VanEck1999A&A...345..127V}. Third dredge-up causes the star's surface C/O ratio to increase from its RGB value of $\approx 0.3$ to $\ge 0.5$, which is the minimum C/O ratio in S-type stars like $\pi^1$~Gru~A. It is also possible to rule out higher-mass models such as the 2\,\Msun\, where the predicted luminosity is
higher than $\pi^1$~Gru~A, especially near the end of its TP-AGB phase. Furthermore, this model becomes C-rich where the surface C/O ratio $\ge 1$, higher than observed in $\pi^1$~Gru~A ($\sim$ 0.97 \cite{Murty1983Ap&SS..94..295M}).

The Monash stellar evolution code was used to calculate the AGB models shown in Supplementary Fig.~\ref{Fig:core-mass-lum} and we refer to
Ref.~\cite{Karakas2014MNRAS.445..347K}
for details of the input physics.
We use the Mixing-length Theory of convection, with a mixing-length parameter $\alpha_{\rm MLT} = 1.86$, and assume that mixing is instantaneous in convective regions. No overshoot or convective boundary mixing is included in the calculations prior to the AGB.
We set the initial metallicity to be solar, here defined to be $Z=Z_{\rm sun} = 0.014$ given that $\pi^1$~Gru~A is a close AGB star found in the thin disk of the Milky Way. The evolution of the 1.7\,\Msun\ model provides the closest match to $\pi^1$~Gru~A's luminosity, although it does so only during the last few thermal pulses. This is consistent with the fact that $\pi^1$~Gru~A has undergone mass loss.

A major uncertainty in AGB stellar models is calculating the third dredge-up, which depends on the numerical model for convection and the treatment of convective borders~\cite{Karakas2014PASA...31...30K}.
In the Monash models we include a simple prescription for convective overshoot discussed in Ref.~\cite{Kamath2012ApJ...746...20K}, where we extend the base of the envelope by $N_{\rm ov}$ pressure scale heights during dredge-up. In order for the masses considered here to become S-type, we use
$N_{\rm ov} \le 3$. In the 1.7\,\Msun\ model shown in Supplementary Fig.~\ref{Fig:core-mass-lum} we use $N_{\rm ov} = 2.5$, which results in a final C/O = 1.16 after 18 thermal pulses noting that the star only becomes C-rich at the very last thermal pulse. The total TP-AGB lifetime is $\sim$1.65 million years.
Mass loss on the AGB is another major uncertainty in AGB models; see for example Ref.~\cite{Karakas2014PASA...31...30K}.
Using the mass-loss rate from Ref.~\cite{Schroder2005ApJ...630L..73S} on the AGB in the 1.7\,\Msun\ model with all the other input parameters the same results in a final C/O = 0.95, after 18 thermal pulses and a total TP-AGB lifetime of $\sim$1.7 million years.

As an additional constraint for the initial mass of $\pi^1$~Gru~A, we analyse the characteristics of its pulsation. As mentioned previously, $\pi^1$~Gru~A is a long period variable (LPV) star of type SRb, with a pulsation period of roughly 195.5 days. This period is listed on The International Variable Star Index (VSX)~\cite{Watson2006SASS...25...47W} and is based on combined V-band light curves from the All Sky Automated Survey~\cite{Pojmanski2002AcA....52..397P} and Ref.~\cite{Tabur2009MNRAS.400.1945T}, as well as being further substantiated by long term light curves from the American Association of Variable Star Observers (AAVSO) (\url{https://www.aavso.org}) and the All Sky Automated Survey for Supernovae (ASAS-SN)~\cite{Kochanek2017PASP..129j4502K}. The additional period listed in VSX of 128 d may be a result of convolution with the observing window and the true peak in the periodogram ((1/195.5 d + 1/365.25 d)$^{-1}$ $\sim$ 128 d), so we omit this from our analysis. We then assign a pulsation mode for this period by $\pi^1$~Gru~A's position on the period-luminosity (PL) diagram. LPVs form multiple parallel sequences on the PL diagram, each being associated to low order radial pulsation modes with mode order descending from left to right~\cite{Trabucchi2017ApJ...847..139T}, excluding the lower right sequence D which is associated with long secondary periods. In Extended Data Fig.~\ref{Fig:pl-ogle-asassn} we compare $\pi^1$~Gru~A with the period-luminosity diagram for the Optical Gravitational Lensing Experiment (OGLE) catalogue of LPVs for the Large Magellanic Cloud (LMC)~\cite{Soszynski2009AcA....59..239S}, as well as the ASAS-SN catalogue of LPVs~\cite{Christy2023MNRAS.519.5271C}. We construct the PL diagrams with J- and K$_s$-band photometry from 2MASS, using the NIR Wesenheit function $W_{K_s,J-K_s} = K_s - 0.686 \cdot (J-K_s)$ to calculate a reddening free magnitude. We use a distance modulus of $\mu = 18.49$ ($D$ = 49.97~kpc) for the LMC~\cite{deGrijs2017SSRv..212.1743D}, and the Bayesian geometric distances using \textit{Gaia} DR3 parallaxes from Ref.~\cite{Bailer-Jones2021AJ....161..147B} for the ASAS-SN catalogue to calculate absolute magnitudes. Additionally, we plot the PL sequence boundaries for the LMC from Ref.~\cite{Trabucchi2021A&A...656A..66T} to illustrate the positions of the sequences in the ASAS-SN PL diagram. We use the distance of $180 \pm 10$ pc for $\pi^1$~Gru~A, which places it on sequence C' on the PL diagram. In a similar fashion to Ref.~\cite{Trabucchi2021A&A...656A..66T}, we can assign the pulsation mode of the 195.5 d period to be the radial first overtone mode. However, the universality of the period-luminosity relation for LPVs remains somewhat uncertain for different metallicity environments~\cite{Whitelock2008MNRAS.386..313W,Trabucchi2025ApJ...978...30T}, though the clear offset in sequence C' for the ASAS-SN PL diagram may be attributed to uncertain parallax distances for the brightest, more evolved AGB stars~\cite{Andriantsaralaza2022A&A...667A..74A} or from saturation of the NIR photometry. Though the exact effects of metallicity on the LPV PL sequences are still under investigation, Ref.~\cite{Trabucchi2021A&A...656A..66T} compare LPVs the LMC and SMC to find that the variability amplitudes may be affected by metallicity, due to lower metallicities favouring the production of C-rich stars which have higher variability amplitudes and periods than O-rich stars on average. More recently, Ref.~\cite{Trabucchi2025ApJ...978...30T} find that stability against pulsation is likely sensitive to chemical composition, though further investigation is required to better understand the effects on LPVs in general. We have included the PL diagrams from OGLE (LMC) and ASAS-SN (all-sky) to represent different metallicity environments, and that the pulsation mode appears to be consistent. Ultimately, the present mode assignment is thus made under the assumption that the metallicity environment does not significantly shift the PL sequences, or their slopes, such that at the very least the period can be reasonably assigned to a sequence and thus a pulsation mode.

Once we have assigned a pulsation mode to the period, we make comparisons to results from stellar evolution codes. We estimate the pulsation characteristics for the TP-AGB with the results of the linear pulsation models from Ref.~\cite{Trabucchi2019MNRAS.482..929T}, which provides an interpolation code to find the pulsation periods and amplitude growth rates for the radial pulsation modes given a set of stellar parameters which we take from the Monash models. In Supplementary Fig.~\ref{Fig:tpagb-plhr-z014}, we begin by comparing $\pi^1$~Gru~A to the theoretical P-L diagram for models with initial masses of $1, 1.5$ and $2$\,\Msun. These models use the same initial metallicity, mass-loss $\eta$ and mixing-length $\alpha_\text{MLT}$ as above, with $N_{\rm ov} = 2.0$. The pulsation period and luminosity ($7,300\,\Lsun$) for $\pi^1$~Gru~A is consistent with models between initial mass $1.5$\,--\,$2$~\Msun. We also include the model tracks on the HR diagram with the luminosity and effective temperature derived above for $\pi^1$~Gru~A, and find them to be roughly consistent with the PL diagram.

As with the above discussion on the core-mass luminosity relationship, we adopt a likely mass of approximately $1.7$\,\Msun. Fig.~\ref{Fig:m1_7_nov2_25} presents the time evolution of the total stellar mass, luminosity, pulsation periods for the first and fundamental modes, the amplitude growth rates of these modes and the C/O for this model. The amplitude growth rate describes the fractional rate of change in the radial amplitude per cycle, and may be used as a general indicator of the dominant pulsation mode, though we do note that this quantity is still subject to the uncertain interaction between convection and pulsation. The total mass of this model at the onset of the TP-AGB is $1.62$\,\Msun\ due to mass loss on the RGB, and reaches the current mass of $\sim1.46$\,\Msun\ during the last two thermal pulses. The combined period and luminosity of $\pi^1$~Gru~A appears to be consistent with the last few thermal pulses for this model mass. Importantly, the growth rates suggest that even in the last few thermal pulses, the first overtone mode can still be dominant at the period and luminosity of $\pi^1$~Gru~A. That is, despite appearing to be late on the TP-AGB, $\pi^1$~Gru~A can still appear as a semi-regular variable, instead of a Mira variable pulsating in the fundamental mode.

We also demonstrate the effects of key model parameters on this estimate, namely the convective overshoot parameter $N_{\rm ov}$ and the mass-loss prescription on the TP-AGB. Fig.~\ref{Fig:m1_7_nov2_25} demonstrates the effect of different $N_{\rm ov}$, where a lower value of $N_{\rm ov} = 2.0$ will terminate the TP-AGB with greater mass-loss in the final interpulse phases, and will only reach a C/O ratio of $\sim 0.7$, which is somewhat lower than the estimated value for $\pi^1$~Gru~A. The previously described model with the mass-loss prescription from Ref.~\cite{Schroder2005ApJ...630L..73S} behaves somewhat differently compared to this model, due to continuous mass-loss over the whole TP-AGB instead of episodic mass-loss in the final thermal pulses. This model approaches a total mass of 1.46\Msun\ before the model luminosity, pulsation period and C/O reaches the measured values, and it also survives for roughly 0.1 Myr longer. This prescription does not however take into account the episodic nature of pulsation-enhanced, dust-driven winds that $\pi^1$~Gru~A is more likely experiencing. Furthermore, it must be mentioned that neither mass-loss prescription accounts for any effects a close companion may have on mass-loss rates and pulsation, a topic which requires further detailed investigation.

\section{Orbital evolution}\label{Sec:orbital_evolution}

Given the current properties of the binary system, as well as the estimated initial mass of $\pi^1$~Gru~C, the orbital properties are projected back in time to understand the past evolution of the system, as well as forward in time to understand the future evolution of the system. To do this, an orbital evolution model is established, which is then used to project the system properties. Finally, a sensitivity study is performed to understand the impact of the observed properties on the final outcome of the system.

Given that gravitational wave radiation and magnetic braking are negligible for the system under consideration, the change in orbital elements arises from two main components. First, mass loss -- in particular during the giant phases -- may lead to a widening of the companion's orbit. Second,
tidal dissipation may cause the orbit to shrink. To compute the changes in the orbital parameters, both effects must be taken into account\footnote{The change in eccentricity is computed as the change in the square of the eccentricity to prevent numerical issues when the eccentricity is close to 0.}:
\begin{equation}
    \frac{1}{a}\frac{\dd a}{\dd t} = \left.\frac{1}{a}\frac{\dd a}{\dd t}\right|_{\rm wind} + \left.\frac{1}{a}\frac{\dd a}{\dd t}\right|_{\rm tides}\ ,\ \ \ \frac{\dd e^2}{\dd t} = \left.\frac{\dd e^2}{\dd t}\right|_{\rm wind} + \left.\frac{\dd e^2}{\dd t}\right|_{\rm tides}\ .
\end{equation}
Here, the first term represents the contribution of wind mass loss, while the second term accounts for tidal dissipation.

\subsection{Impact of stellar winds on orbital dynamics}
If the donor star, here $M_1$, is losing mass and mass transfer is non-conservative, the companion star ($M_2$) will accrete a fraction of the material and the rest will be lost, thereby carrying away angular momentum from the system and impacting the orbital separation and eccentricity. We first consider the case of a circular orbit, and then work out the equations for an eccentric orbit.

    \subsubsection{The case of a circular orbit}\label{Sec:wind_circular}
        In the case of a circular orbit, the change in orbital separation due to wind mass loss is given by~\cite{Siess2013A&A...550A.100S}
        \begin{align}
            \left.\frac{1}{a}\frac{\dd a}{\dd t}\right|_{\rm wind} & = -2 \left.\frac{\dot L_\mathrm{orb}}{L_\mathrm{orb}}\right|_{\rm wind} - 2 \frac{\dot m_1}{m_1} - 2 \frac{\dot m_2}{m_2} + \frac{\dot m_1 + \dot m_2}{m_1+m_2}\\
            & = -2 \left.\frac{\dot L_\mathrm{orb}}{L_\mathrm{orb}}\right|_{\rm wind} - 2 \frac{\dot m_1}{m_1}\left(1-\frac{\beta}{q} - \frac{1-\beta}{2} \frac{1}{(1+q)}\right)\,,\label{Eq:dadt}
        \end{align}
        where $q = m_2/m_1$, $\dot m_1$ is the mass loss from the primary (here taken negative) and $\dot m_2$ the mass gained by the companion, i.e. $\dot m_2 = -\beta \dot m_1$, with $\beta$ the mass accretion efficiency. If $\beta < 1$, mass transfer is non-conservative. The change in angular momentum can be expressed as~\cite{Saladino2019A&A...626A..68S}
        \begin{equation}
            \left.\dot L_\mathrm{orb}\right|_{\rm wind} = \eta a^2 \Omega_o (\dot m_1 + \dot m_2)\ ,\label{Eq:Lorb_dot}
        \end{equation}
        where $\Omega_o$ is the orbital frequency and $\eta$ the specific orbital angular momentum of the material lost in units of the orbital angular momentum of the system per reduced mass. Combining Eqs.~\eqref{Eq:dadt}--\eqref{Eq:Lorb_dot}, the change in the orbital separation can be written as \cite{Saladino2019A&A...626A..68S}
        \begin{equation}
            \left.\frac{1}{a}\frac{\dd a}{\dd t}\right|_{\rm wind} = - 2 \frac{\dot m_1}{m_1}\left(1-\frac{\beta}{q} - \eta (1-\beta)\frac{1+q}{q} - \frac{1-\beta}{2} \frac{1}{1+q}\right)\ .
        \end{equation}

        Assuming a fast isotropic wind, where the velocity of the wind at the location of the companion is much larger than the orbital velocity, the mass accretion efficiency is given by the Bondi-Hoyle-Lyttleton (BHL) analytical approximation~\cite{Saladino2019A&A...626A..68S}
        \begin{equation}
            \beta_\mathrm{BHL} = \frac{q^2}{{(1+q)}^2}\frac{v_\mathrm{orb}^4}{v_\mathrm{w}{(v_\mathrm{w}^2+v_\mathrm{orb}^2)}^{3/2}}\ ,
        \end{equation}
        where $v_\mathrm{orb} = \sqrt{G(m_1+m_2)/a}$ is the orbital velocity and $v_\mathrm{w}$ is the wind velocity. In this specific case (in this context also called Jeans' mode), the specific angular momentum taken from the orbit and transferred to the outflowing gas is given by~\cite{Saladino2019A&A...626A..68S}
        \begin{equation}
            \eta_\mathrm{iso} = \frac{q^2}{{(1+q)}^2}\ .
        \end{equation}

        However, in the case of AGB binary stars, the fast-wind scenario may not hold. For example, in this particular case of $\pi^1$~Gru~A and C, the orbital velocity and the wind velocity at the location of the companion are both around 15~km~s$^{-1}$. In such a case, no simple analytical expressions exist for $\beta$ and $\eta$. Instead, these quantities must be computed numerically using detailed hydrodynamical simulations (see, e.g., Sect.~\ref{Sec:3D_hydro}). Performing such models throughout the evolution of the system is extremely computationally expensive, and infeasible with current computational resources. However, we can rely on the work of Ref.~\cite{Saladino2019A&A...626A..68S}, who calibrated $\beta$ and $\eta$ as a function of the mass ratio $q$ and the ratio of the terminal wind velocity to the orbital velocity, $v_\infty / v_{\rm orb}$, using their 3D hydrodynamical simulations. These authors obtained that:
        \begin{align}
            \beta\left(q, \frac{v_\infty}{v_\mathrm{orb}}\right) &= \min\left[\left(0.75 + \frac{1}{1.7 + 0.3/q + {\left((0.5 + 0.2/q)\frac{v_\infty}{v_\mathrm{orb}}\right)}^5}\right)\beta_\mathrm{BHL}, 0.3, 1.4q^{2}\right] \,,\label{eq:beta}\\
            \eta\left(q, \frac{v_\infty}{v_\mathrm{orb}}\right) &= \min\left[\frac{1}{\max\left(1/q, 0.6q^{-1.7}\right) + {\left((1.5 + 0.3/q)\frac{v_\infty}{v_\mathrm{orb}}\right)}^3} + \eta_\mathrm{iso}, 0.6\right] \,.\label{eq:eta}
        \end{align}
        As these equations depend on the terminal wind velocity, this property needs to be computed throughout the entire stellar evolution. Similar to Ref.~\cite{Hurley2002MNRAS.329..897H}, we assume the terminal wind velocity to be a fraction of the escape velocity, i.e.
        \begin{equation}
            v_\infty = \sqrt{2\alpha_W \frac{Gm_1}{R_1}}\ ,
        \end{equation}
        where $\alpha_W$ is a constant taken to be 1/8~\cite{Hurley2002MNRAS.329..897H}.

    \subsubsection{The case of an eccentric orbit}
    When the orbit is eccentric, the change in the orbital separation and eccentricity due to mass loss becomes more complicated as it becomes phase-dependent. The change in the orbital parameters is then given by~\cite{Dosopoulou2016ApJ...825...71D}
        \begin{align}
            \left.\frac{1}{a}\frac{\dd a}{\dd t}\right|_{\rm wind} &= - 2 \frac{\dot m_1}{m_1}\left(1-\frac{\beta}{q} - \eta (1-\beta)\frac{1+q}{q} - \frac{1-\beta}{2} \frac{1}{1+q}\right) \left[\frac{e^2+2 e \cos f+1}{1-e^2}\right]\\
            \left.\frac{\dd e^2}{\dd t}\right|_{\rm wind} &= - 4 \frac{\dot m_1}{m_1}\left(1-\frac{\beta}{q} - \eta (1-\beta)\frac{1+q}{q} - \frac{1-\beta}{2} \frac{1}{1+q}\right) \left[e^2+e\cos f\right]\ .
        \end{align}
        Integrating these equations over the true anomaly $f$ as a function of time is computationally expensive. Instead, the change in the orbital separation and eccentricity can be computed by averaging over the orbital period, allowing these averages to be used in the orbital evolution equations. The orbit-averaged quantities can be computed using~\cite{Dosopoulou2016ApJ...825...71D}
        \begin{equation}
            \langle(\ldots)\rangle=\frac{{\left(1-e^2\right)}^{3 / 2}}{2 \pi} \int_0^{2 \pi}(\ldots) \frac{\dd f}{{(1+e \cos f)}^2}\ .
        \end{equation}

        However, the dependence of $\beta$ and $\eta$ on the true anomaly $f$ remains an open question, not fully understood or calibrated, therefore requiring further investigation. It can be hypothesized that the mass accretion efficiency, $\beta$, is proportional to
        \[
        \frac{{(1 + e \cos f)}^2}{a^2 {(1 - e^2)}^2},
        \]
        as $\beta_\mathrm{BHL}$ primarily depends on the inverse square of the orbital distance (a consequence of the orbital velocity scaling as the fourth power in orbital velocity, in the limit where the orbital velocity is larger than the wind velocity). Conversely, the specific angular momentum carried away by the wind, $\eta$, can be assumed independent of the true anomaly, consistent with the assumption that $\eta_\mathrm{iso}$ does not vary with $f$. Under these assumptions, the changes in orbital separation and eccentricity due to mass loss can be expressed as:
        \begin{align}
            \left.\left\langle\frac{1}{a}\frac{\dd a}{\dd t}\right\rangle\right|_{\rm wind} &= - 2 \frac{\dot m_1}{m_1}\left(1-\frac{\beta_\mathrm{enh}}{q} - \eta (1-\beta_\mathrm{enh})\frac{1+q}{q} - \frac{1-\beta_\mathrm{enh}}{2} \frac{1}{1+q}\right)\\
            \left.\left\langle\frac{\dd e^2}{\dd t}\right\rangle\right|_{\rm wind} &= - 4 e^2 \frac{\dot m_1}{m_1}\langle\beta\rangle\left( \frac{1}{q} - \eta\frac{1+q}{q} - \frac{1}{2}\frac{1}{1+q}\right)\ .
        \end{align}
        where
        \[
        \beta_\mathrm{enh} = \langle\beta\rangle \frac{1+e^2}{1-e^2}
        \]
        is the enhanced mass accretion efficiency effect for the change in the orbital separation,
        \[
        \langle\beta\rangle = \beta_\mathrm{c} \frac{1}{\sqrt{1-e^2}}
        \]
        is the average mass accretion efficiency over an orbit, and $\beta_\mathrm{c}$ is the mass accretion efficiency for a circular orbit (with the same $a$), taken from Eq.~\eqref{eq:beta}. These equations have been tested against a limited sample of hydrodynamic models of eccentric systems from Ref.~\cite{Saladino2019A&A...629A.103S}, giving consistent results.

\subsection{Impact of tidal dissipation on orbital dynamics}\label{Sec:tides}
To calculate the rate of change of the orbital separation and eccentricity due to tidal interactions, the tidal potential must be calculated. Using the tidal potential, the tidal strength can be computed and the orbital evolution equations can be solved. Tidal dissipation of an evolved star is dominant compared to the dissipation during the MS for the same orbital distance~\cite{Esseldeurs2024A&A...690A.266E}. Changes in the orbital evolution will therefore mainly occur during these giant phases (see Sect.~\ref{sec:orbital_evolution_system}). As evolved stars are known to be slow rotators~\cite{Fuller2019MNRAS.485.3661F}, the rotation of the star can be neglected and only the tidal dissipation of the evolved star will be taken into account.

    \subsubsection{Tidal potential}
        For two bodies, where only the deformation of one object is considered, the tidal potential is given by the difference between the gravitational potential induced by the secondary object at each point of the extended primary body and its value at its barycentre. The tidal potential $\Psi$ can be expressed as~\cite{Ogilvie2014ARA&A..52..171O}:
        \begin{equation}\setlength{\jot}{-4pt}
            \begin{aligned}
                \Psi(r, \theta, \varphi, t)=\Re \left\{\sum_{l=2}^{\infty} \sum_{m=0}^l \sum_{n=-\infty}^{\infty}\frac{G m_2}{a} A_{l, m, n} (e, i){\left(\frac{r}{a}\right)}^l \times Y_l^m(\theta, \varphi) \exp({-\mathrm{i} n \Omega_o t})\right\}\ ,
            \end{aligned}
        \end{equation}
        where $r,\theta,\varphi$ are the spherical coordinates centered at the origin of the reference frame attached to the primary centre of mass, $Y_l^m$ are the spherical harmonics, and $A_{l, m, n}$ are the tidal coefficients~\cite{Witte1999A&A...350..129W, Ogilvie2014ARA&A..52..171O}
        \begin{equation}
            A_{l, m, n} (e, i=0) = \frac{4\pi}{2l+1} \epsilon_m \epsilon_{m,n} Y_l^m(\pi/2, 0) h_n^{l, m}(e)\ ,
        \end{equation}
        with $h_n^{l, m}(e)$ the Hansen coefficients, $\epsilon_m\,=\,1$ for $m=0$ and $\epsilon_m\,=\,2$ for $m>0$, where $\epsilon_{m,n}=1$ for $m>0$ or $m=n=0$, $\epsilon_{m,n}=2$ for $m=0$, $n>0$ and $\epsilon_{m,n}=0$ for $m=0$, $n<0$ to ensure orthogonality. For companions not too close to their host star, the quadrupolar approximation~\cite{Mathis2009A&A...497..889M} is valid and only $l=2$ needs to be taken into account, which results in $A_{2, 0, n}(e) = \sqrt{\pi/5}\ h_n^{2, 0}(e)$ and $A_{2, 2, n}(e) = \sqrt{6\pi/5}\ h_n^{2, 2}(e)$. The contribution from $m=1$ vanishes for $i=0$.

        The Hansen coefficients are given by~\cite{Witte1999A&A...350..129W}
        \begin{equation}
            {\left(\frac{a}{r^\prime}\right)}^{l+1} e^{i m f} = \sum_{n=-\infty}^{\infty} h_n^{l, m}(e) \exp({-i n M})\ ,
        \end{equation}
        which leads to
        \begin{equation}
            h_n^{l, m}(e) = \frac{1}{2\pi} \int_0^{2\pi} {\left(\frac{a}{r}\right)}^{l+1} \exp\left[i (m f - n M)\right]\, \dd M\ ,
        \end{equation}
        with $M$ the mean anomaly, $f$ the true anomaly, and $r$ the instantaneous orbital distance (see Sect.~\ref{Methods:orbital_model}). By using Kepler's equations, the Hansen coefficients can be written as~\cite{Smeyers1991A&A...248...94S}
        \begin{equation}\label{eq:hansen}
            h_n^{l, m}(e) = \frac{{(1-e^2)}^{1/2-l}}{2\pi} \int_0^{2\pi} {(1+e\cos f)}^{l-1} \cos(m f - n M)\, \dd f\ ,
        \end{equation}
        which is integrated numerically. In the special case of a circular orbit ($e=0$), the Hansen coefficients simplify to $h_n^{l, m}(e) = \delta_n^m$. Consequently, $A_{2, 0, 0}(0) = \sqrt{\pi/5}$ and $A_{2, 2, 2}(0) = \sqrt{6\pi/5}$.

    \subsubsection{Tidal orbital evolution equations}\label{sec:tidalequations}
        The tidal orbital evolution is determined by the rate of change of the orbital energy and angular momentum, which are given by~\cite{Witte1999A&A...350..129W, Ogilvie2014ARA&A..52..171O}
        \begin{equation}
            \left.\dot{E}_\mathrm{orb}\right|_\mathrm{tides} = -\sum_{l=2}^{\infty} \sum_{m=0}^{\infty} \sum_{n=-\infty}^{\infty} E^{l, m}_n\ ,\ \ \
            \left.\dot{L}_\mathrm{orb}\right|_\mathrm{tides} = -\sum_{l=2}^{\infty} \sum_{m=0}^{\infty} \sum_{n=-\infty}^{\infty} L^{l, m}_n
        \end{equation}
        where
        \begin{align}
            E^{l, m}_n &= n\Omega_o \mathcal{T}^{l, m}_n\ ,\ \ \
            L^{l, m}_n = m \mathcal{T}^{l, m}_n\ , \\
            \mathcal{T}^{l, m}_n &= \frac{(2l+1) R_\star |\varphi_T (R_\star)|^2}{8\pi G} \Im (k^{l, m}_n)\ .
        \end{align}
        Here, $R_\star$ is the stellar radius, $\varphi_T(R_\star)$ the tidal potential at the stellar radius, and $k^{l, m}_n$ the Love number (see Sect.~\ref{sec:LoveNumber}). $\mathcal{T}^{l, m}_n$ can be rewritten by plugging in the tidal potential as
        \begin{equation}
            \mathcal{T}^{l, m}_n = \frac{(2l+1)}{8\pi} \frac{G m_2^2}{a} |A_{l, m, n}(e)|^2{\left(\frac{R_\star}{a}\right)}^{2l+1} \Im (k^{l, m}_n)\ .
        \end{equation}

        Having the tidal power and torque, the tidal orbital evolution can be calculated. The change in semi-major axis is given by ref.~\cite{Witte1999A&A...350..129W}
        \begin{align}
            \left.\frac{1}{a}\frac{\dd a}{\dd t}\right|_\mathrm{tides} &= \frac{2a}{G m_2 m_1} \left.\dot{E}_\mathrm{orb}\right|_\mathrm{tides}\\
            &= - \frac{2l+1}{4\pi} \frac{m_2}{m_1} {\left(\frac{R_\star}{a}\right)}^{2l+1} \sum_{l=2}^{\infty} \sum_{m=0}^{\infty} \sum_{n=-\infty}^{\infty} n\Omega_o |A_{l, m, n}(e)|^2 \Im (k^{l, m}_n)\\
            &= - \frac{5}{4\pi} \frac{m_2}{m_1} {\left(\frac{R_\star}{a}\right)}^{5} \sum_{m=0}^{\infty} \sum_{n=-\infty}^{\infty} n\Omega_o |A_{2, m, n}(e)|^2 \Im (k^{2, m}_n)\,,
        \end{align}
        when assuming the quadrupolar approximation. The change in eccentricity is given by ref.~\cite{Witte1999A&A...350..129W}
        \begin{equation}
            \left.\frac{\dd e^2}{\dd t}\right|_\mathrm{tides} = \frac{2a}{G m_2 m_1} \left((1-e^2)\left.\dot{E}_\mathrm{orb}\right|_\mathrm{tides} - \Omega_o\sqrt{1-e^2}\left.\dot{L}_\mathrm{orb}\right|_\mathrm{tides}\right)\ ,
        \end{equation}
        which can be rewritten in terms of the Love numbers as:
        \begin{align}
            \left.\frac{\dd e^2}{\dd t}\right|_\mathrm{tides} = & -\frac{2a}{G m_2 m_1} \sqrt{1-e^2}\Omega_o\sum_{l, m, n}\left(n\sqrt{1-e^2} - m\right) \mathcal{T}_n^{l, m}\\
            = &-\frac{2l+1}{4\pi} \frac{m_2}{m_1} {\left(\frac{R_\star}{a}\right)}^5 \sqrt{1-e^2}\Omega_o \sum_{l, m, n}\left(n\sqrt{1-e^2} - m\right) |A_{l, m, n}|^2 \Im (k^{l, m}_n)\\
            = &-\frac{5}{4\pi} \frac{m_2}{m_1} {\left(\frac{R_\star}{a}\right)}^5 \sqrt{1-e^2}\Omega_o \sum_{m, n}\left(n\sqrt{1-e^2} - m\right) |A_{2, m, n}|^2 \Im (k^{2, m}_n)\ .\label{Eq:dedt}
         \end{align}
        again using the quadrupolar approximation. Once the tidal Love numbers are known, the tidal orbital evolution can be calculated.

    \subsubsection{Tidal Love Numbers}\label{sec:LoveNumber}
        The imaginary part of the Love number has two components: the equilibrium tide dissipation and the dynamical tide dissipation. The equilibrium tide dissipation originates from the turbulent friction applied by the turbulent convection on the displacement induced by the hydrostatic deformation of the body triggered by the gravitational companion of a companion~\cite{Zahn1966AnAp...29..313Z}. The tidal Love number of the equilibrium tide for evolved stars can be calculated following Sect. 2.3.1 of Ref.~\cite{Esseldeurs2024A&A...690A.266E}. The dynamical tide dissipation arises from the excitation of waves due to the gravitational potential of the companion. For evolved stars, the dynamical tide consists of the excitation and dissipation of progressive internal gravity waves~\cite{Esseldeurs2024A&A...690A.266E} and can be calculated following Sect. 2.3.2 of Ref.~\cite{Esseldeurs2024A&A...690A.266E}. The total tidal Love number is calculated by summing the equilibrium and dynamical tide contributions.
        The equations for the tidal Love numbers provided by Ref.~\cite{Esseldeurs2024A&A...690A.266E} are specifically designed to calculate the tidal Love number for $l = m = n = 2$. The tidal Love number is an intrinsic property of the perturbed body, defined as the ratio of the primary's gravitational potential perturbation — induced by the companion's presence — to the tidal potential, evaluated at the surface. Since the gravitational potential perturbation follows a linear response, the dependence on the tidal potential cancels out within the equations, eliminating any dependence on $m$ and $n$. As a result, only the tidal frequency ($\omega_t = n\Omega_o - m\Omega_s$, where $\Omega_s$ is the spin frequency if rotation were to be considered) varies when evaluating different combinations of $m$ and $n$.

    \subsection{Choice of stellar structure and evolutionary models}
        The stellar evolutionary models from the Monash stellar evolutionary code presented in Sect.~\ref{Sec:stellar_evolution} do not provide the necessary output parameters needed to calculate the tidal Love numbers. Therefore, different stellar evolutionary models will be used in this section. These models are computed using the Modular Experiments in Stellar Astrophysics (MESA) code~\cite{Jermyn2023ApJS..265...15J} with the inlist of Ref.~\cite{Esseldeurs2024A&A...690A.266E} that is based on the inlist of Ref.~\cite{Cinquegrana2022ApJ...939...50C} to be consistent with the Monash code. These models have initial masses between 1 and 2~\Msun\ at solar metallicity ($Z = 0.0134$) and are evolved from the pre-main sequence to the white dwarf phase.

    \subsection[Orbital evolution of the Pi1~Gru system]{Orbital evolution of the $\boldsymbol{\pi^1}$~Gru system}\label{sec:orbital_evolution_system}
   
        To compute the orbital evolution of the $\pi^1$~Gru system, we need to estimate the current stellar age of the system. To evaluate the current stellar age, we use the stellar evolutionary model to determine the time at which the mass of the primary star equals the current mass. At this moment in time, we impose the current system parameters. The orbital evolution is then integrated (using a first-order Euler method\footnote{A Runge–Kutta 4th order scheme was also tested and results in exactly the same solution. To reduce computational time, the Euler method was chosen.}) until the primary star reaches the white dwarf phase forward in time, as well as backward in time until the primary star reaches the PMS phase. As discussed in Supplementary Sect.~\ref{Sec:sensitivity}, which addresses the sensitivity of the prior on $m_1$, there are two potential scenarios for the current mass of $\pi^1$~Gru~A. The first scenario where $\pi^1$~Gru~A starts with an initial mass of 1.7 \Msun\ and has a current mass of 1.23 \Msun\ ($q_\text{current} = 1.03$~\Msun, $a_\text{current} = 7.10$~au, $e_\text{current} = 0$), and the second scenario where $\pi^1$~Gru~A starts with an initial mass of 1.5 \Msun\ and has a current mass of 1.12 \Msun\ ($q_\text{current} = 1.05$~\Msun, $a_\text{current} = 6.81$~au, $e_\text{current} = 0$).The orbital evolution is calculated for both scenarios. The results of this evolution are shown together with a zoom-in to the TP-AGB phase in Supplementary Fig.~\ref{fig:OrbitalEvolution_prior} for the first scenario and Fig.~\ref{fig:OrbitalEvolution_circ} for the second. For clarity on the timescale, we plot ``stellar age – stellar age WD [yr]'' on the $x$-axis, where ``stellar age WD'' indicates the age at which the white dwarf (WD) has cooled to a luminosity of $10^{-1}\,\Lsun$. During these evolutionary calculations, we follow the evolution of the primary star, while the secondary star is assumed to be a non evolving companion (where the mass is increased by accreting material lost from the primary). This is a adequate assumption if the companion is a lower mass main-sequence star, but it is also possible that the companion is a white dwarf with a complicated history (see Supplementary Sect.~\ref{Sec:pi1gruc}). In this second scenario our computations are thus only valid from the moment $\pi^1$~Gru~C became a white dwarf, from which point in time the mass will remain constant.

        Tracing the system's past, the orbital evolution calculations indicate that the system parameters remained relatively unchanged during the MS and HB phases. During the RGB phase, the mass loss from $\pi^1$~Gru~A causes the orbital separation to increase. The gradual change in orbital distance starts to change more dramatically during the AGB phase, especially during the TP-AGB phase. During this phase, the non-conservative mass loss from the system is able to lose a lot of angular momentum as it is significantly enhanced due to the comparable wind and orbital velocity (see Sect.~\ref{Sec:wind_circular}). This causes the orbit to shrink instead of expand. Additionally, tidal dissipation became more significant during the AGB phase compared to the RGB phase, as the star's radius increased substantially. This leads to a continuous decrease in orbital separation throughout the TP-AGB phase, ultimately resulting in the current configuration of the system. Projecting into the future, this trend of decreasing orbital separation persists, eventually resulting in a common envelope phase. However, this assumes that stellar rotation is neglected. When rotation is taken into account, the tidal frequency decreases, leading to reduced tidal dissipation.

        Throughout the evolution of this system, the equilibrium tide dissipation dominates over the dynamical tide dissipation, where the dynamical tide dissipation only contributes with a factor $10^{-5}$ in strength compared to the equilibrium tide dissipation. This is in line with the results of Ref.~\cite{Esseldeurs2024A&A...690A.266E}.

        A sensitivity analysis of the orbital evolution is performed in Supplementary Sect.~\ref{sec:orbevolsensitivity}, where the impact of uncertainties in the system parameters on the orbital evolution is discussed, including the current semi-major axis, the current mass of $\pi^1$~Gru~A, the current mass of $\pi^1$~Gru~C, and the initial mass of $\pi^1$~Gru~A. Within these uncertainties, the orbital evolution remains qualitatively similar, with the system evolving towards a common envelope phase.

        \subsection{The effect of a non-circular orbit}

            The Bayesian retrieval of the multi-epoch proper motion ALMA, \textit{Gaia} and \textit{Hipparcos} observations indicates that the current eccentricity of the system is zero. Therefore, the sensitivity study of the orbital dynamics presented in the previous section considered only circular orbits. An important question to address is how this system transitioned to a circular orbit.

            To examine how rapidly the system circularizes, the current eccentricity is increased while keeping the other orbital parameters constant. The results of this analysis are shown in Extended Data Fig.~\ref{fig:OrbitalEvolutionecc}. The figure illustrates that, in the case of a non-zero initial eccentricity, the system begins to circularize during the TP-AGB phase, but the circularization process is not rapid at the start of this phase. At the current estimated age of the system, the eccentricity is expected to retain about 80\% of its initial value. For higher eccentricity cases, tidal circularization involves more wave numbers (see Sect.~\ref{sec:tidalequations}), causing the circularization process to accelerate, but it must also remove a greater amount of eccentricity. If the system had started with an eccentric orbit, it would still exhibit significant eccentricity at the current age. This indicates that either the system must have started with a very low eccentricity before entering the AGB phase, or tidal circularisation is even stronger then currently predicted during the evolved phases (see main text).
\end{methods}

\clearpage
\newpage

\begin{addendum}

\item[Data availability]
Standard ALMA pipeline products and the enhanced products and scripts for proposal 2018.1.00659.L are available via the ALMA Science Archive. The data from the proposal 2023.1.00091.S can be retrieved from the ALMA data archive at \url{http://almascience.eso.org/aq/}. Other ALMA products, including molecular line maps, will be made available on reasonable request by sending a request to A.M.S. Richards \\(\href{mailto:a.m.s.richards@manchester.ac.uk}{a.m.s.richards@manchester.ac.uk}).

\item[Code availability]
The scripts used to generate all figures in the main paper, Methods section, and Supplementary Information are available upon request. For specific components, please contact:
\begin{itemize}
        \item\ Mats Esseldeurs (\href{mailto:mats.esseldeurs@kuleuven.be}{mats.esseldeurs@kuleuven.be}) for 3D hydrodynamical simulations and orbital evolution calculations,
        \item\ Leen Decin (\href{mailto:leen.decin@kuleuven.be}{leen.decin@kuleuven.be}) for Bayesian retrieval of the orbital parameters,
        \item\ A.M.S. Richards (\href{mailto:a.m.s.richards@manchester.ac.uk}{a.m.s.richards@manchester.ac.uk}) for ALMA data reduction, and
        \item\ Amanda Karakas (\href{mailto:amanda.karakas@monash.edu}{amanda.karakas@monash.edu}) for stellar evolution calculations.
\end{itemize}
{\sc phantom} is publicly available at \url{https://github.com/danieljprice/phantom}. {\sc plons} is publicly available at \url{https://github.com/Ensor-code/plons}.

\item [Acknowledgments]
This papers uses the ALMA data ADS/JAO.ALMA2018.1.00659.L, `ATOMIUM: ALMA tracing the origins of molecules forming dust in oxygen-rich M-type stars’ and ADS/JAO.ALMA.\allowbreak2023\allowbreak.1.\allowbreak00091.S, `Putting to test the first hypothesized ALMA detections of a close stellar/planetary companion orbiting an AGB star'. ALMA is a partnership of ESO (representing its member states), NSF (USA) and NINS (Japan), together with NRC (Canada) and NSC and ASIAA (Taiwan), in cooperation with the Republic of Chile. The Joint ALMA Observatory is operated by ESO, auI/NRAO and NAOJ. This paper makes use of the CASA data reduction package: \url{https://casa.nrao.edu/} - Credit: International consortium of scientists based at the National Radio Astronomical Observatory (NRAO), the European Southern Observatory (ESO), the National Astronomical Observatory of Japan (NAOJ), the CSIRO Australia Telescope National Facility (CSIRO/ATNF), and the Netherlands Institute for Radio Astronomy (ASTRON) under the guidance of NRAO. The authors thanks the Data Reduction team at ESO for customizing the imaging pipeline. This paper makes use of the Cologne Database for Molecular Spectroscopy (CDMS; \url{https://cdms.astro.uni-koeln.de/}) and the spectral line catalogs of the Jet Propulsion Laboratory (JPL; \url{https://spec.jpl.nasa.gov/});
This work has made use of data from the European Space Agency (ESA) mission \textit{Gaia} (\url{https://www.cosmos.esa.int/gaia}), processed by the \textit{Gaia} Data Processing and Analysis Consortium (DPAC,
\url{https://www.cosmos.esa.int/web/gaia/dpac/consortium}). Funding for the DPAC has been provided by national institutions, in particular the institutions participating in the \textit{Gaia} Multilateral Agreement. This work made use of the latest \textit{Hipparcos} reduction whose validation is described in Ref.~\cite{vanLeeuwen2007A&A...474..653V}. Computer resources have been provided by the KU Leuven C1 Excellence Grant BRAVE KAC/16/23/008, the VSC Flemish supercomputer, and STFC IRIS.
The consortium thanks Hans Van Winckel, Graham Harper and Pat Wallace for useful discussions.
M.E.\ acknowledges funding from the FWO research grants G099720N and G0B3823N.
L.D.\ acknowledges support from the KU Leuven C1 excellence grant BRAVE C16/23/009, KU Leuven Methusalem grant SOUL METH/24/012, and the FWO research grants G099720N and G0B3823N.
Y.M, A.I.K, T.D., Z.O. acknowledge this research is supported in part by the Australian Research Council Centre of Excellence for All Sky Astrophysics in 3 Dimensions (ASTRO 3D), through project number CE170100013.
J.M.\ acknowledges support from the FWO research grant G099720N.
T.D.\ is supported in part by the Australian Research Council through a Discovery Early Career Researcher Award (DE230100183).
S.M.\ acknowledges support from the European Research Council (ERC) under the Horizon Europe programme (Synergy Grant agreement 101071505: 4D-STAR), from the CNES SOHO-GOLF and PLATO grants at CEA-DAp, and from PNPS (CNRS/INSU). While partially funded by the European Union, views and opinions expressed are however those of the author only and do not necessarily reflect those of the European Union or the European Research Council. Neither the European Union nor the granting authority can be held responsible for them.
R.S.’s contribution to the research described here was carried out at the Jet Propulsion Laboratory, California Institute of Technology, under a contract with NASA, and funded in part by NASA via ADAP award number 80NM0018F0610.
M.V.d.S.\ acknowledges support from the Oort Fellowship at Leiden Observatory.
T.C.\ acknowledges funding from the Research Foundation - Flanders (FWO), grant 1166724N.
F.D.C.\ is a Postdoctoral Research Fellow of the Research Foundation - Flanders (FWO), grant 1253223N.
I.E.M.\ acknowledges support from ANID/FONDECYT, grant 11240206.
This work is funded by the French National Research Agency (ANR) project PEPPER (ANR-20-CE31-0002).
P.K.\ acknowledges funding from the European Research Council (ERC) under the European Union’s Horizon 2020 research and innovation program (project UniverScale, grant agreement 951549).
C.L.\ acknowledges support from the KU Leuven C1 excellence grant BRAVE C16/23/009.
T.J.M.\ acknowledges support from STFC grant no. ST/T000198/1.
J.M.C.P.\ acknowledges support for the UKRI STFC grant ST/T000287/1.
Z.O. acknowledges this research was supported by an Australian Government Research Training Program (RTP) Scholarship.
D.J.P.\ acknowledges Australian Research Council funding via DP220103767. We also thank the Australia French Association for Research and Innovation (AFRAN) for financially supporting the 5th Phantom users workshop.
L.S.\ is FNRS senior researcher.
O.V.\ acknowledges funding from the Research Foundation - Flanders (FWO), grant 1173025N.
K.T.W.\ acknowledges support from the European Research Council (ERC) under the European Union's Horizon 2020 Research and Innovation programme (grant agreement number 883867, project EXWINGS).

\item[Author Contributions]
L.D.\ is PI of the ALMA Large Program ATOMIUM 2018.1.000659.L and PI of the ALMA Program 2023.1.00091.S, developed the orbital model, developed and executed the Bayesian methodology to retrieve the orbital parameters, and derived the properties of $\pi^1$~Gru~C and its accretion disk.
M.E.\ and S.M. computed the tidal effects and the orbital evolution of the system.
J.M.\ and M.E.\ computed the 3D hydrodynamical simulations of the binary system.
A.M.S.R.\ led the ALMA data reduction and wrote software used in the data reduction.
J.D.R.\ advised in the development of the orbital model and Bayesian retrieval framework.
A.I.K.\ and Y.M.\ computed the initial mass of $\pi^1$~Gru~A and its stellar evolution.
R.S.\ advised in the derivation of the properties of the accretion disk.
L.S.\ advised in the calculations of the orbital evolution.
T.D.\ and I.M.D\ advised in the inclusion of the ALMA parallax shift.
M.V.d.S and T.D.\ advised on the discussion of the chemical signature of the companion.
P.K.\ has contributed to the astrometry and quantification of the proper motion anomaly.
All authors discussed the interpretation of the data, contributed to the scientific results, and helped prepare the paper.

\item[Author Information]
$^{\star}$ Correspondence and requests for materials should be addressed to \\
\href{mailto:mats.esseldeurs@kuleuven.be}{mats.esseldeurs@kuleuven.be} and/or
\href{mailto:leen.decin@kuleuven.be}{leen.decin@kuleuven.be}.

\item[Competing Interests] The authors declare that they have no competing interests.

\end{addendum}

\begin{figure}[!htp]
    \centering
    \includegraphics[width=0.83\textwidth]{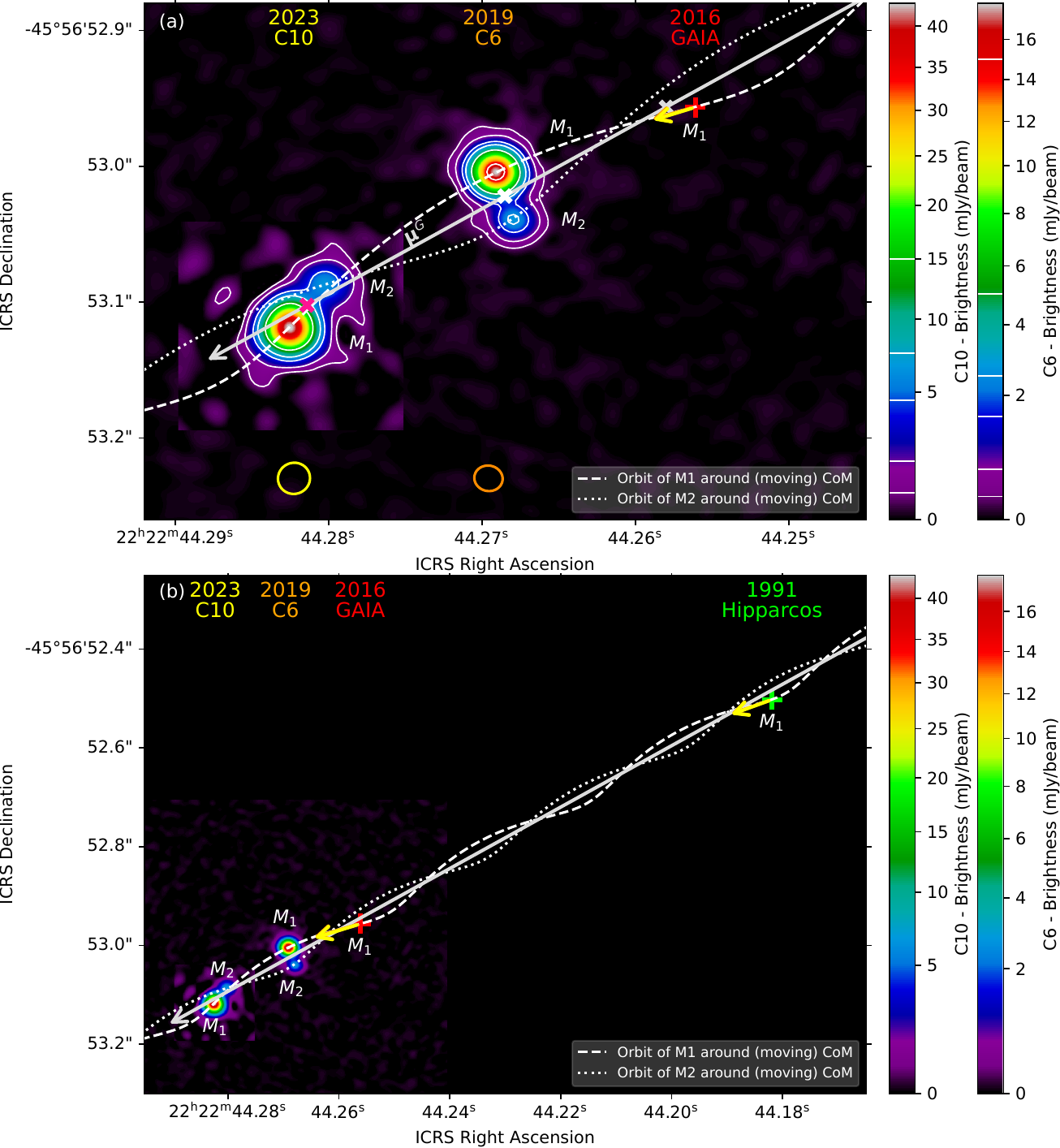}\\
    \caption{\textbf{Proper motion of the $\boldsymbol{\pi^1}$~Gru system.}
Panel (a): 2019 ALMA C6 and 2023 ALMA C10 data with white contours at (3, 10, 30, 50, 100, 300)$\times$ the continuum rms value. The ALMA beam sizes are shown in orange (2019 C6) and yellow (2023 C10) at the bottom. Data are corrected for the parallactic shift. The red cross marks the \textit{Gaia} 2016.0 position of $\pi^1$~Gru~A. The grey arrow ($\mu^{G}$) indicates the proper motion of the binary system's center of mass (CoM, $G_\star$). The white, pink, and grey crosses mark the barycentre's position at the 2019, 2023, and \textit{Gaia} 2016.0 epochs, respectively. The dashed white line shows the orbit of $\pi^1$~Gru~A ($M_1$), while the dotted white line shows the orbit of $\pi^1$~Gru~C ($M_2$), both in the ICRS frame based on Bayesian best-fit parameters.
Panel (b): Similar to (a), but also including the \textit{Hipparcos} 1991.25 position of $\pi^1$~Gru~A (green cross). ALMA contour levels are omitted for clarity. In both panels, observed proper motion is in yellow, predicted motion in orange, with vectors representing 1-year (panel a) and 3-year (panel b) intervals. The orange vector is nearly indistinguishable due to the almost perfect fit with the observed motion. An accompanying video is in Suppl. Video~\ref{Video:pi1_gru}.}\label{fig:pi1_gru}
\end{figure}

\begin{figure}[!htp]
\centering
\includegraphics[width=0.69\textwidth]{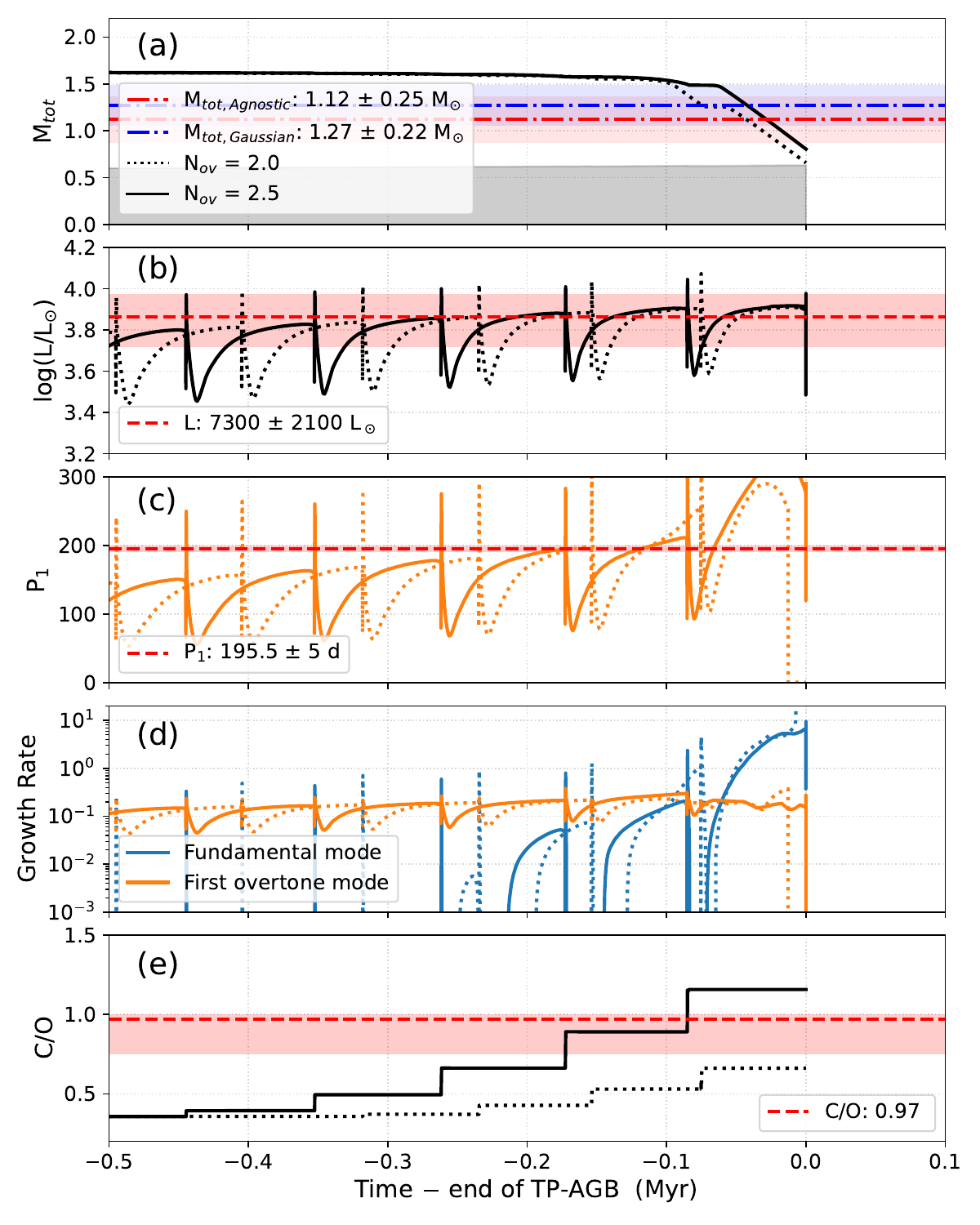}
\vspace{-1pt}
\caption{\textbf{Time evolution of stellar parameters for a 1.7\,\Msun\ model near the end of the thermally-pulsing AGB, compared to observed values for $\boldsymbol{\pi^1}$~Gru~A.} We include models with two different values for the overshoot parameter in pressure scale heights, $N_{\rm ov}$, of $2.0$ (dotted lines) and $2.5$ (solid lines). The panels show from top to bottom: (a): total (lines) and core masses (grey shaded region), (b): luminosity, (c): first overtone mode period, (d): amplitude growth rates for the first overtone mode (orange) and fundamental mode (blue) and (e): C/O ratio; all with respect to time since the onset of the TP-AGB phase. Also included are the measurements for $\pi^1$~Gru~A (red dashed lines and shaded regions): luminosity of 7,300 \Lsun, period of $195.5$ d and C/O ratio of $0.97$ (between $0.75$ and $1$). In the top panel, we plot the derived masses from both agnostic and Gaussian priors on $m_1$, which are $1.12$ and $1.27$ \Msun\ respectively. The period, luminosity and total mass are consistent with the final few thermal pulses of this model, though the higher mass using the Gaussian prior is favoured for this model mass. The first overtone mode is also dominant at these model times (i.e. the growth rate is higher), which is consistent with the observed pulsation mode of $\pi^1$~Gru~A.}\label{Fig:m1_7_nov2_25}
\end{figure}

\begin{figure}[!htp]
    \centering
    \includegraphics[width=\linewidth]{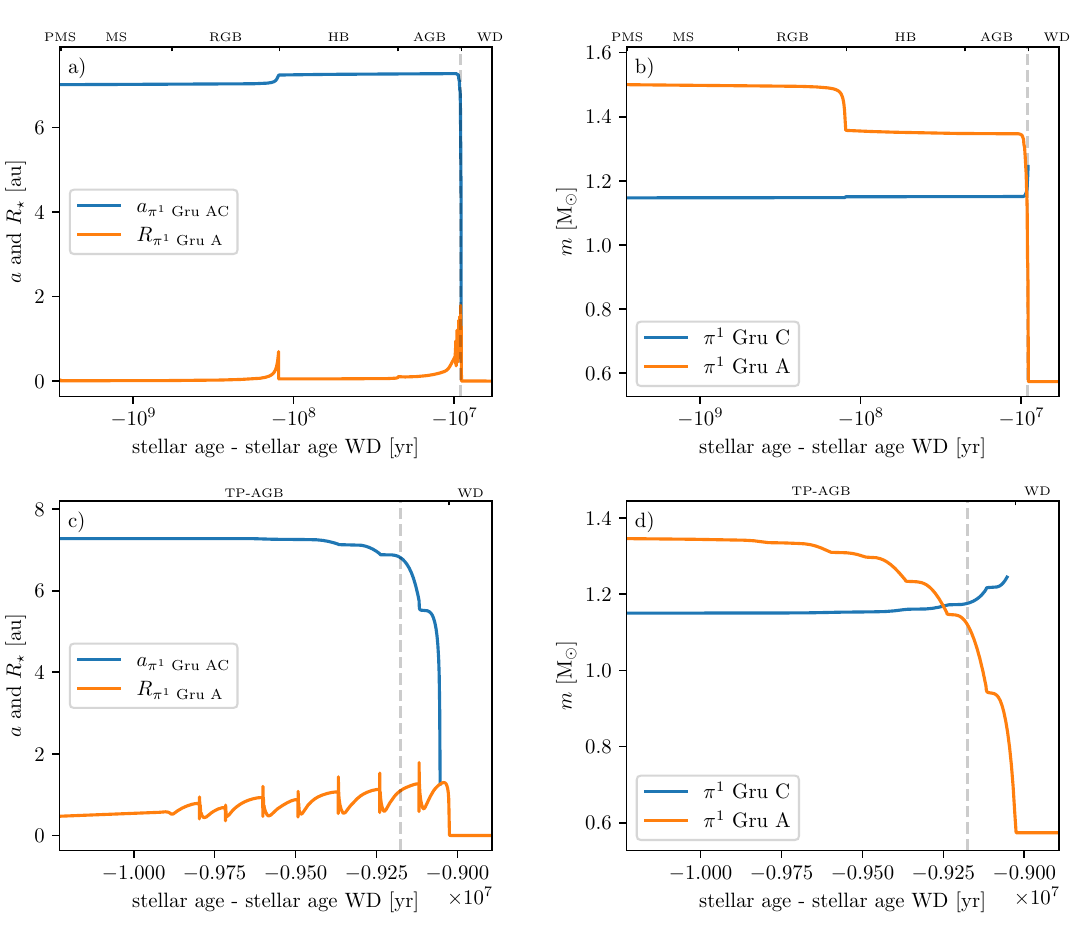}
    \caption{\textbf{Orbital evolution of the $\boldsymbol{\pi^1}$~Gru system during the TP-AGB phase.} Panel~(a) illustrates the variation in orbital separation $a$ over time (blue line), along with the evolution of the radius of $\pi^1$~Gru~A (orange line). Panel~(b) shows the mass evolution of $\pi^1$~Gru~A (orange line) and $\pi^1$~Gru~C (blue line). Panels (c) and (d) show a zoom-in to the TP-AGB phase. These panels show the evolution of the system parameters when the initial mass of $\pi^1$~Gru~A is set to 1.5\,\Msun, and the current mass of $\pi^1$~Gru~C is set to 1.12\,\Msun. Initially, the system has a mass ratio $q<1$, but mass loss from $\pi^1$~Gru~A and mass accretion onto $\pi^1$~Gru~C lead to an increase, resulting in $q>1$. For clarity on the timescale, the $x$-axis represents ``stellar age – stellar age WD [yr]'', where ``stellar age WD'' indicates the age at which the white dwarf (WD) has cooled to a luminosity of $10^{-1}\,\Lsun$. The vertical dashed grey line in each panel marks the current age of the $\pi^1$~Gru system.}\label{fig:OrbitalEvolution_circ}
\end{figure}

\afterpage{\clearpage}
\newpage


\afterpage{\clearpage}
\newpage

\setcounter{table}{0}
\renewcommand{\tablename}{Extended Data -- Table}
\setcounter{figure}{0}
\renewcommand{\figurename}{Extended Data -- Figure}

\section*{Extended Data}

\begin{figure}[!htp]
    \centering
    \includegraphics[width=12truecm]{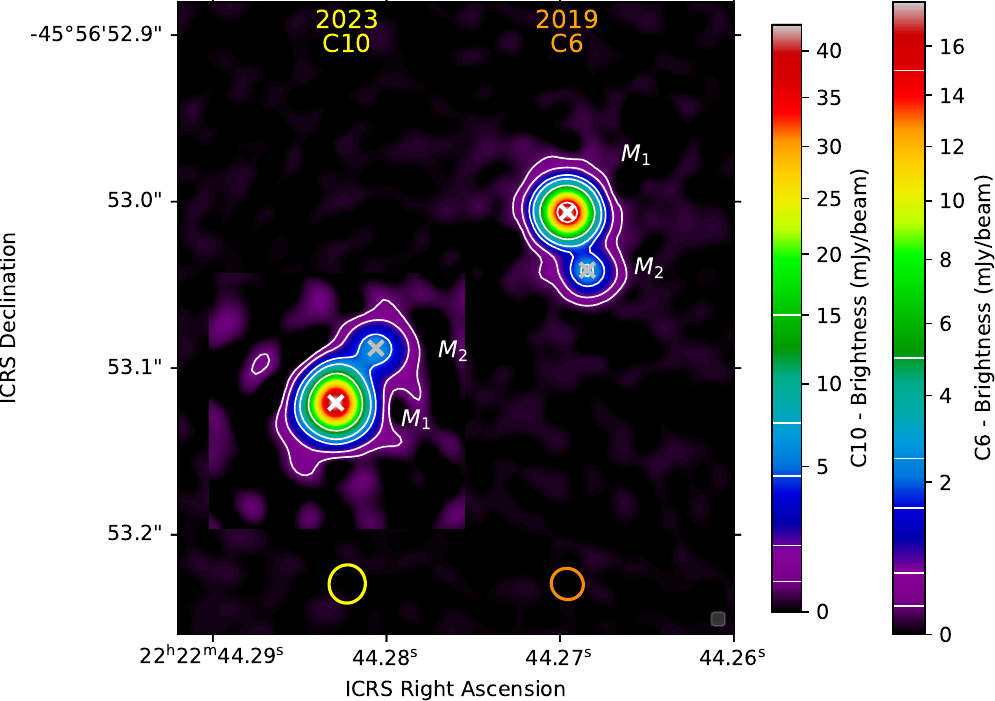}
    \caption{\textbf{ALMA continuum observations of the $\boldsymbol{\pi^1}$~Gru system.}
    2019 ALMA C6 and 2023 ALMA C10 data are shown with white contours drawn at (3, 10, 30, 50, 100, 300)$\times$ the continuum rms value of 0.05~mJy and 0.15~mJy, respectively. The ALMA beam sizes for the 2019 C6 and 2023 C10 data are depicted at the bottom of the figure in orange and yellow, respectively, positioned below their corresponding images. White crosses indicate the astrometric position of the primary ($\pi^1$~Gru~A, or `$M_1$'), while silver crosses mark the position of the companion ($\pi^1$~Gru~C, or `$M_2$') at both epochs.}\label{Fig:ALMA_data}
\end{figure}

\begin{table*}[!htp]
\caption{\textbf{ICRS astrometric positions of the $\boldsymbol{\pi^1}$~Gru components $\boldsymbol{M_1}$ and $\boldsymbol{M_2}$ and relative position errors.}
$\pi^1$~Gru positions fitted using a uniform disc and delta component. Where two values are given, the errors have been decomposed into the $\alpha_\star$ and $\delta$ directions.
}\label{table:fit}
\centering
\begin{tabular} {l|lcccc}
\toprule
       & Fitted centre  & $\sigma_{\mathrm{pos},\phi}$&$\sigma_{\mathrm{pos, e_r}}$&$\sigma_{\mathrm{pos,fit}}$& $\sigma_{\mathrm{pos,rel}}$ \\
       & $\alpha, \delta$ & [mas]                      & [mas]                     & [mas] & [mas]\\
\midrule
C6 $M_1$ & 22:22:44.269589, --45:56:53.00641&0.03       &1.7                       & $-$0.01, $\hphantom{-}$0.01 & 1.2, 1.2\\
C6 $M_2$ & 22:22:44.268393, --45:56:53.04198&0.3     &-- &  $\hphantom{-}$0.14, $-$0.16& 0.2, 0.2\\
C10 $M_1$& 22:22:44.282263, --45:56:53.12381&0.06       &0.9                       & $\hphantom{-}$0.01, $-$0.03& 0.6, 0.6\\
C10 $M_2$& 22:22:44.279952, --45:56:53.09090&0.5        &--         &$\hphantom{-}$0.05,  $\hphantom{-}$0.84& 0.4, 0.8\\
\bottomrule
\end{tabular}
\end{table*}

\begin{figure*}[!htp]
\centering
\includegraphics[width=.98\textwidth]{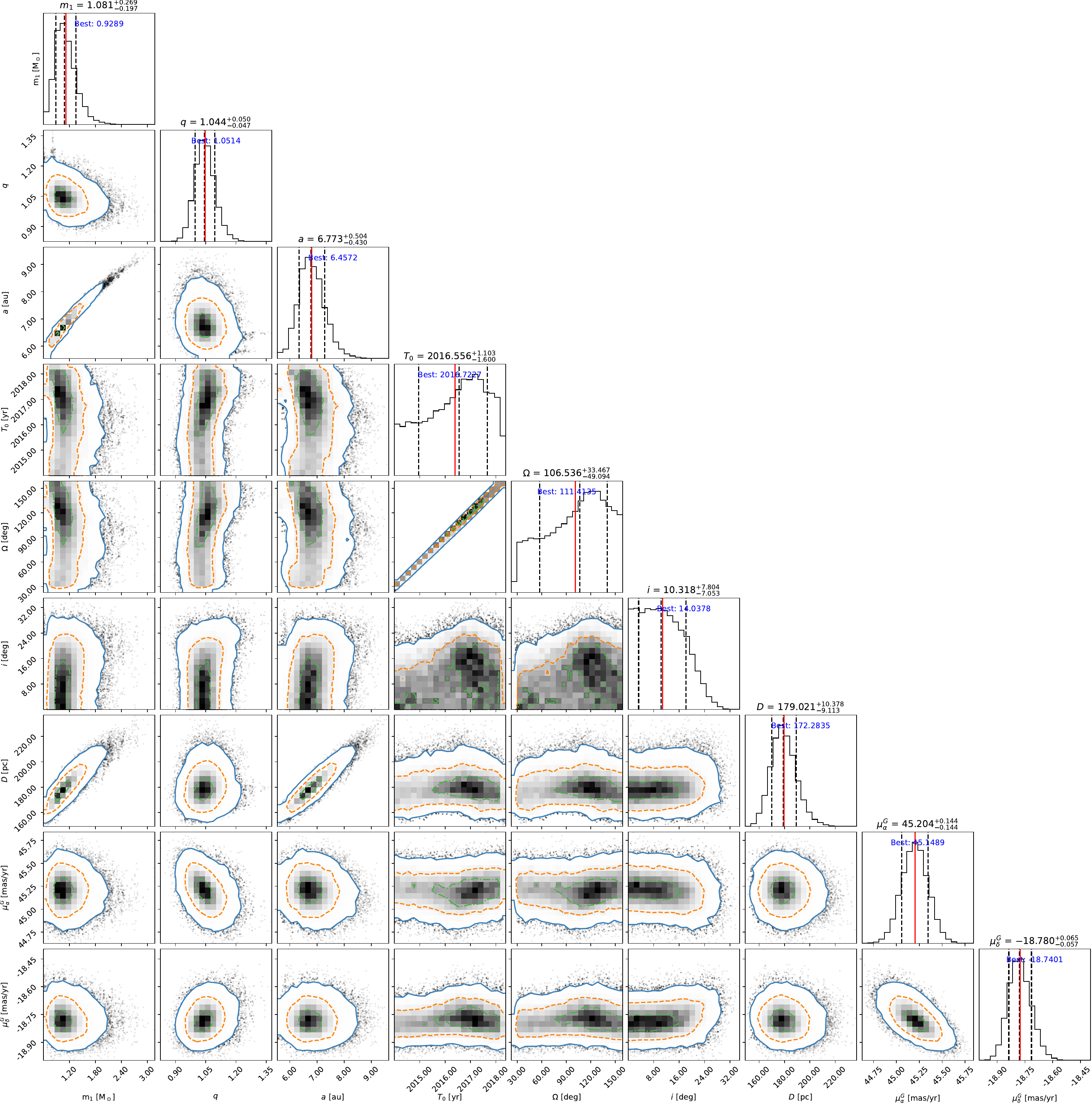}
\caption{\textbf{Corner plot for the retrieval of the $\boldsymbol{\pi^1}$~Gru orbital parameters using \texttt{ultranest} sampling for a circular orbit $(e=0)$.} The diagonal panels display the posterior distributions of each parameter. These panels also include annotations for the median, 16th percentile, and 84th percentile (title of the diagonal panel and black dashed vertical lines), as well as the best-fit value in blue and the mean value indicated with a red dashed vertical line. The off-diagonal panels illustrate the pairwise relationships between parameters, using contour plots to show joint distributions and correlations. Contours corresponding to the 1, 2, and 3-sigma confidence levels are shown in blue, orange, and green, respectively.}\label{fig:pi1_gru_corner_ultranest}
\end{figure*}

\begin{figure}[!htp]
    \centering
    \begin{subfigure}[t]{0.59\textwidth}
        \centering
        \begin{tikzpicture}[baseline=(current bounding box.north)]
            \node[anchor=north west] at (0,0) {
                \includegraphics[width=0.9\textwidth]{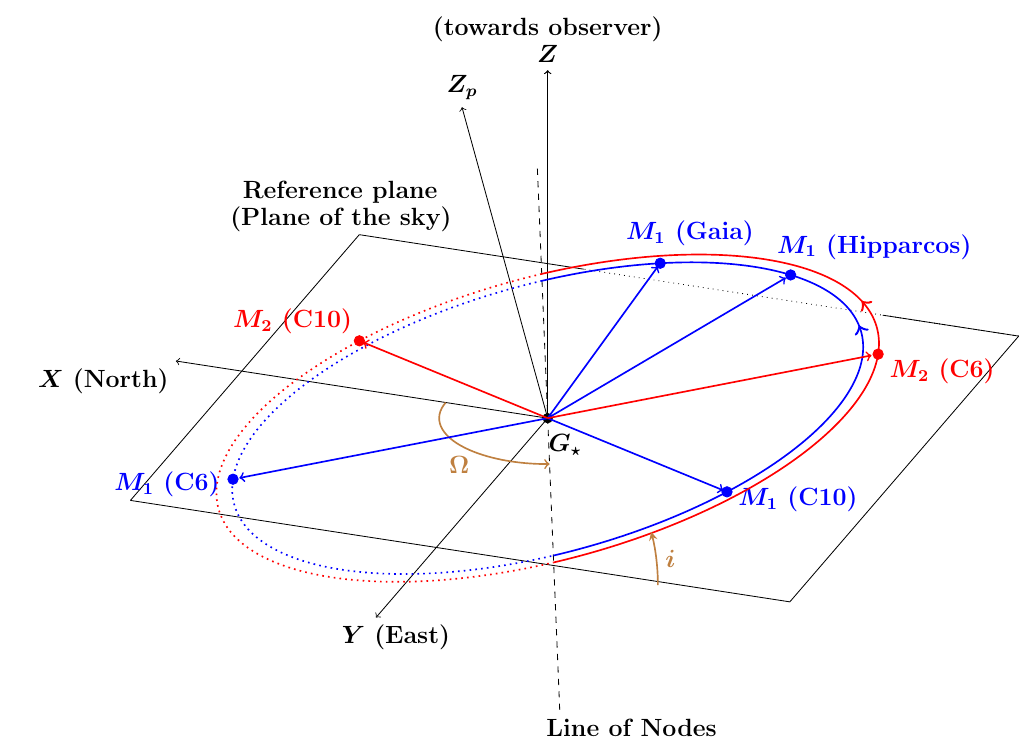}
            };
            \node[anchor=north west] at (0,0) {(a)};
        \end{tikzpicture}
    \end{subfigure}
    \begin{subfigure}[t]{0.59\textwidth}
        \centering
        \begin{tikzpicture}[baseline=(current bounding box.north)]
            \node[anchor=north west] at (0,0) {
                \includegraphics[width=0.9\textwidth]{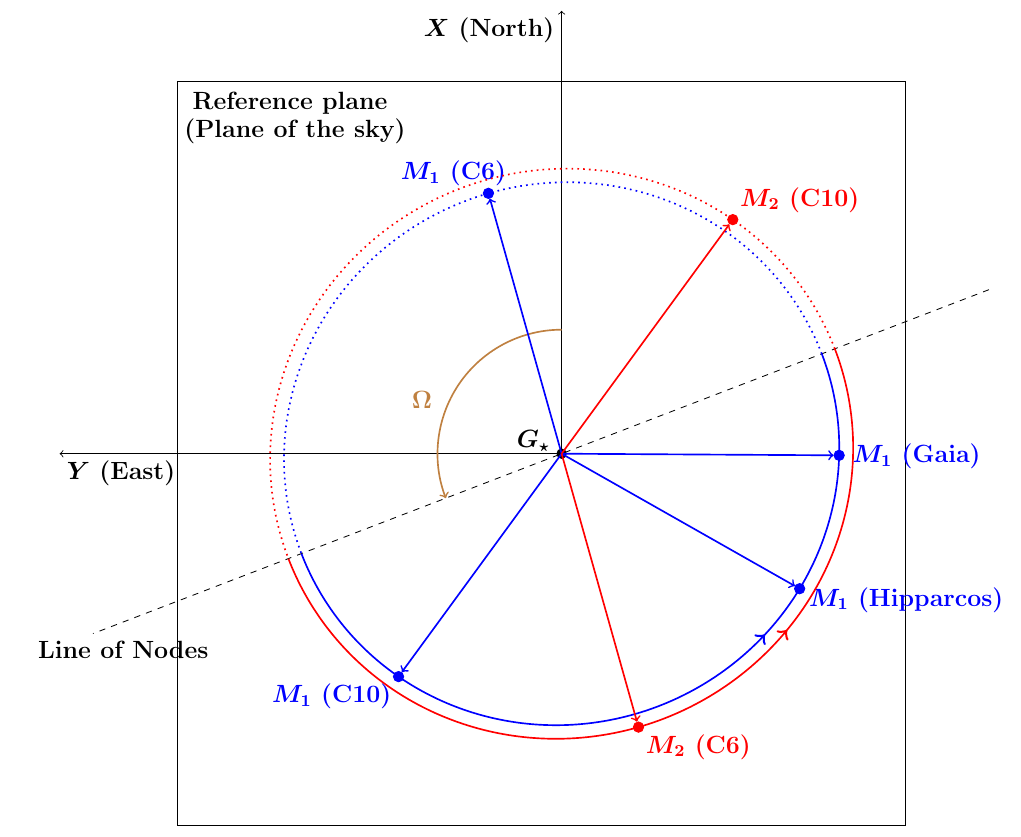}
            };
            \node[anchor=north west] at (0,0) {(b)};
        \end{tikzpicture}
    \end{subfigure}
        \caption{\textbf{Visualization of the $\boldsymbol{\pi^1}$~Gru System}. The orbital elements are derived from the best-fit parameters obtained through \texttt{ultranest} sampling for a circular orbit, as detailed in Supplementary Table~\ref{table:fit_sensitivity}. The orbital motion of $M_1$ is shown in blue, with a semi-major axis $a_1 = \frac{1}{q+1}a$, and the motion of $M_2$ in red, with a semi-major axis $a_2 = \frac{q}{q+1}a$. Coloured circles indicate the epochs of the \textit{Hipparcos} (1991.25), \textit{Gaia} (2016.0), ALMA C6, and C10 observations, with the blue and red vectors representing the position vectors of $M_1$ and $M_2$ relative to the barycentre $G_\star$, respectively. The circles correspond to specific epochs, but the data retrieval considers the full observing periods of \textit{Hipparcos} and \textit{Gaia}. The right-handed Cartesian reference frame (\(X, Y, Z\)) has its origin at the binary system's barycenter, with the $Z$-axis pointing towards the observer, and the $X$-axis pointing towards the equatorial North pole. The $Z_p$-axis is directed along the angular momentum vector of the orbit.
        The orbital elements $\Omega$ and $i$ are highlighted in brown. Panel~(a) shows a viewing angle of (65$^\circ$, 200$^\circ$), with the first coordinate being the azimuthal angle and the second the polar angle. Panel~(b) represents the observer's view, corresponding to a viewing angle of (0$^\circ$, 270$^\circ$).}\label{fig:schematic}
\end{figure}

\begin{figure}[htp]
\centering
\includegraphics[width=\textwidth]{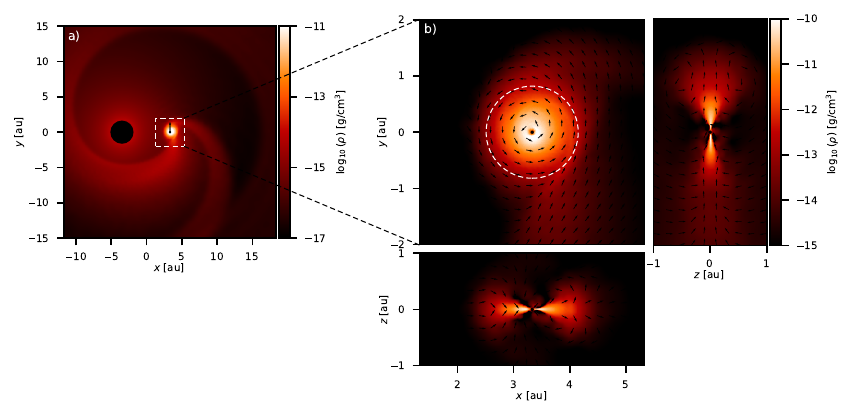}
\caption{\textbf{Hydrodynamic simulation of the accretion disk around companion $\boldsymbol{\pi^1}$~Gru~C.}
Density distribution in 2D slices through the 3D hydrodynamic simulation, with the AGB star and companion located on the $y=0, z=0$ axis at $x<0$ and $x>0$, respectively. Panel (a) shows the density distribution in the orbital plane slice, showing the high-density bow shock spiral originating in front of the companion-surrounded disk. The direction of the orbital motion of the companion star is indicated with a black arrow. Panel (b) shows the zoomed-in density distribution of the accretion disk in 3 perpendicular slices, over-plotted with arrows presenting the velocity distribution. The estimated disk radius ($r_{\rm disk}=0.83\,{\rm au}$) is annotated as a dashed white circle. Figure constructed with visualisation tool \textsc{Plons}.
}\label{Fig:disk_simulation}
\end{figure}

\begin{figure}[htp]
\centering
\includegraphics[width=\textwidth]{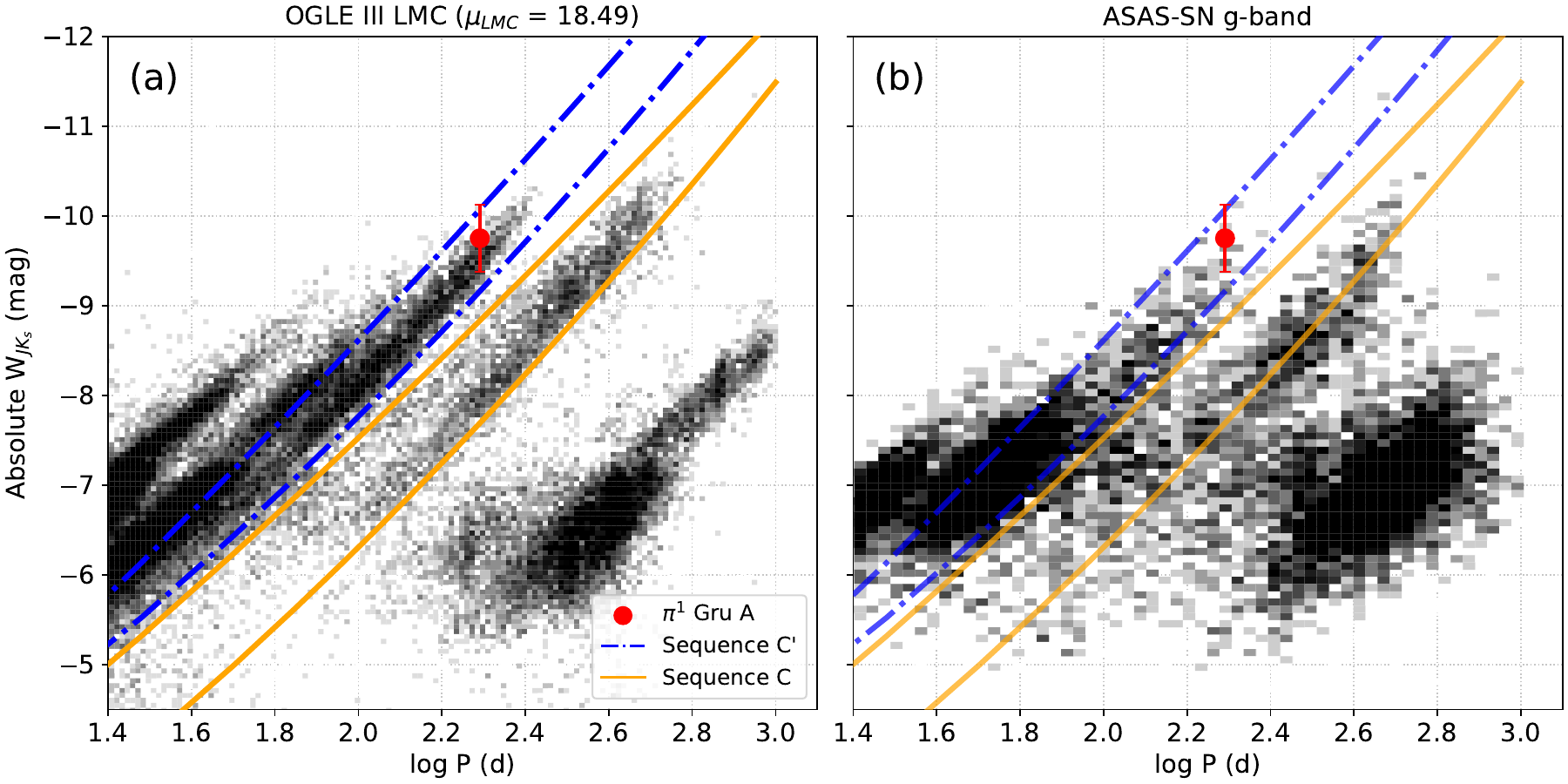}
\caption{\textbf{Observational period-luminosity diagrams for long period variables, with $\boldsymbol{\pi^1}$~Gru.} Panel (a): OGLE III catalogue of LPVs for the Large Magellanic Cloud. Panel (b): ASAS-SN g-band catalogue of LPVs. Also included are the boundaries for the period-luminosity sequences C' and C for the LMC by Ref.~\protect\cite{Trabucchi2021A&A...656A..66T}. Both period-luminosity diagrams show sequences associated with the radial order pulsation modes of red giants, with decreasing radial order from left to right. $\pi^1$~Gru (with $P=195.5$ d, $W_{JK_s}=-9.750\pm0.3767$ mag) appears to most likely be on sequence C', which is associated with pulsation in the radial first overtone mode. The effects of metallicity on the period-luminosity sequences is assumed to be small enough that the period can be reasonably assigned to a pulsation mode.
}\label{Fig:pl-ogle-asassn}
\end{figure}

\begin{figure}[!htp]
    \centering
    \includegraphics[width=\linewidth]{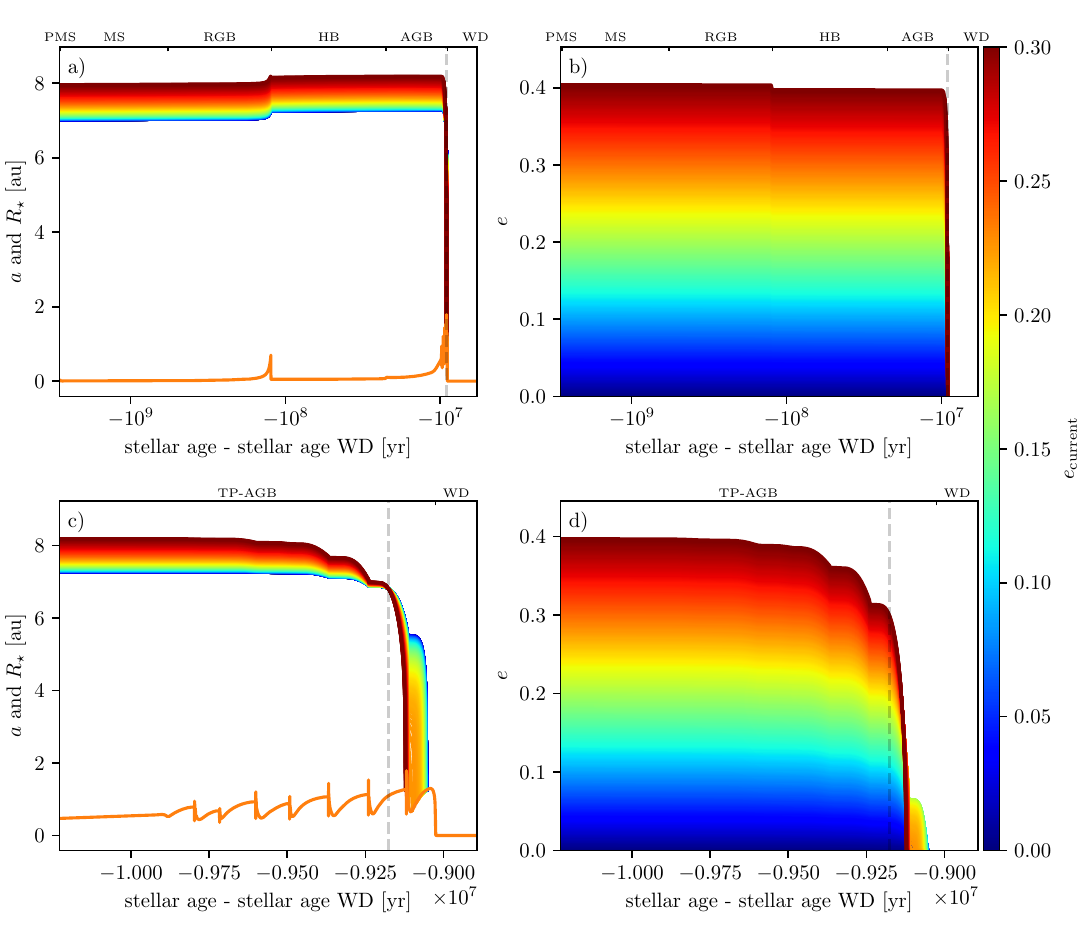}
    \caption{\textbf{Orbital evolution of the $\boldsymbol{\pi^1}$~Gru system with variations in the current eccentricity.} This figure illustrates how changes in the current eccentricity (represented by the colour bar) affect the orbital evolution of the system over its lifetime. Panel~(a) shows the evolution of the orbital separation for each current eccentricity value, using corresponding colours, alongside the changing radius of $\pi^1$~Gru~A (orange line). Panel~(b) presents the evolution of the eccentricity. Panels~(c) and (d) show a zoom-in to the TP-AGB phase. The vertical dashed grey line in each panel indicates the current age of the $\pi^1$~Gru system.}\label{fig:OrbitalEvolutionecc}
\end{figure}

\afterpage{\clearpage}
\newpage

\noindent{\Large{\textbf{Supplementary Video}}\label{sec:supp_video}}
\setcounter{figure}{0}
\renewcommand{\figurename}{Extended Data -- Video}
\begin{figure}[htp]
\centering
\includemedia[width=.95\textwidth,activate=onclick,passcontext,transparent,addresource=Figs/pi1gru60fps.mp4,flashvars={source=Figs/pi1gru60fps.mp4}]{\includegraphics[width=.95\textwidth]{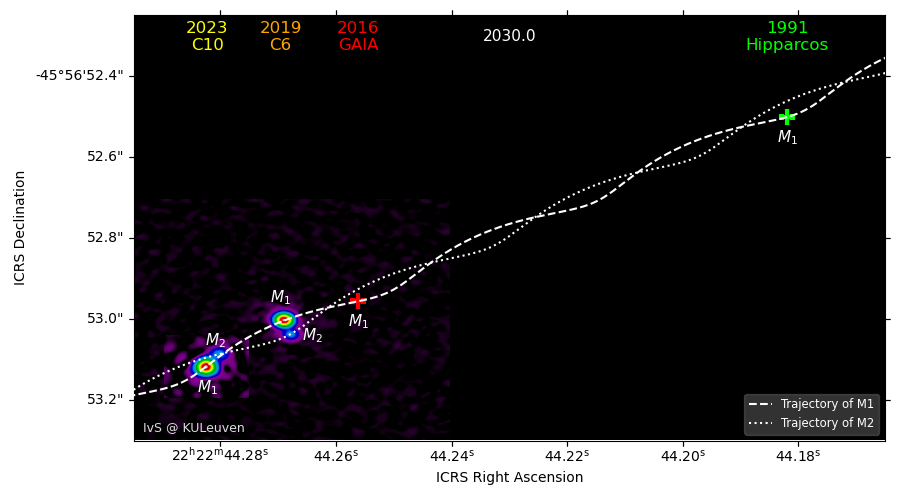}}{VPlayer.swf}
\caption[Video showing the proper motion of pi1 Gru A and pi1 Gru C over time, spanning from 1984.5 to 2030.]{\textbf{Video showing the proper motion of $\boldsymbol{\pi^1}$~Gru~A and $\boldsymbol{\pi^1}$~Gru~C over time, spanning from 1984.5 to 2030.} The video can be found on: \springervideo.}\label{Video:pi1_gru}
\end{figure}

\clearpage
\noindent{\Large{\textbf{Supplementary Information}}\label{sec:supp_information}}

\setcounter{section}{0}

\setcounter{figure}{0}
\renewcommand{\figurename}{Supplementary Figure}
\setcounter{table}{0}
\renewcommand{\tablename}{Supplementary Table}

\section{Supplementary information ALMA observations}

\begin{table*}[!htp]
\caption{\textbf{$\boldsymbol{\pi^1}$~Gru ALMA observational parameters.}
For each of the 2019 C6 (extended configuration) and 2023 C10 data we give the reference date, the overall frequency span $\nu_{\rm{tot}}$, the weighted mean frequency $\nu_{\mathrm{mean}}$ and the cumulative (non-contiguous) frequency range of the line-free continuum channels $\Delta\nu$, the synthesised beam FWHM $\theta_{\mathrm b}$, the maximum recoverable scale MRS, the off-source noise in the continuum image $\sigma_{\mathrm{rms}}$, the separation from the phase reference (in $\alpha_\star$ and $\delta$) of the check source $\psi_{\mathrm {check}}$ and of the target $\psi_{\mathrm{targ}}$.}\label{table:obs}
\centering
\begin{tabular} {ll|cccccccc}
\toprule
&Date &$\nu_{\mathrm{tot}}$ &$\nu_{\mathrm{mean}}$ &$\Delta\nu$ &$\theta_{\mathrm b}$ &MRS& $\sigma_{\mathrm{rms}}$ &$\psi_{\mathrm {check}}$ & $\psi_{\mathrm{targ}}$ \\
&yymmdd& [GHz]                & [GHz]                & [GHz]      &[mas] &[$''$] & [mJy] & [mas] & [mas]\\
\midrule
C6& 190630 & 213.9--269.6&241.03&24.1& 19 & 0.4 & 0.04&2.1, 2.2 & 1.4, 1.7 \\
C10& 231012 & 330.5--345.9&334.70&2.88&22 &0.7 &0.13&3.8, 4.3 &1.6, 2.6\\
\bottomrule
\end{tabular}
\end{table*}

\begin{table*}[!htp]
\caption{\textbf{Spectral line cubes of 2023 observations.} For each spectral window, we give the central frequency $\nu_{\mathrm {cen}}$, the full spw frequency range $\Delta{\nu}$, the angular resolution $\theta_{\rm b}$, the spectral resolution $\delta{\mathrm V}$ and rms noise of the channel $\sigma_{\mathrm{chan}}$. `Wide' channels refer to the channel averaging for thermal lines, and `narrow' channels for maser lines.
}\label{table:lineobs}
\centering
\begin{tabular} {l|ccccccc}
\toprule
spw &$\nu_{\mathrm {cen}}$& $\Delta{\nu}$ & $\theta_{\mathrm b}$&\multicolumn{2}{c}{wide channels} &\multicolumn{2}{c}{narrow channels} \\
    &                     &               &         & $\delta{\mathrm V}$& $\sigma_{\mathrm{chan}}$ & $\delta{\mathrm V}$& $\sigma_{\mathrm{chan}}$ \\
    & [GHz]              & [GHz]          & [mas]  & [km s$^{-1}$] & [mJy] &  [km s$^{-1}$] & [mJy]\\
\midrule
0  &330.600             & 0.117          &36      &0.89      &2.5      &      &\\
1  &331.593             & 0.058          &36      &0.88      &2.2      &      &\\
2  &331.128             & 0.058          &36      &0.88      &1.6      &      &\\
3  &332.813             & 1.875          &36      &0.88      &1.6      &      &\\
4  &342.661             & 0.058          &35      &0.85      &1.7      &      &\\
5  &342.518             & 0.058          &35      &0.85      &1.8      &0.107 &4.1   \\
6  &342.995             & 0.117          &35      &0.85      &1.8      &0.107 &4.0   \\
7  &344.930             & 0.117          &34      &0.85      &2.2      &0.106 &4.8  \\
8  &345.810             & 0.117          &34      &0.85      &2.2      &      &\\
\bottomrule
\end{tabular}
\end{table*}

\begin{table*}[!htp]
\caption{\textbf{Molecular transitions shown in this paper}. The molecule, vibrational state, rotational transition and rest frequency are listed. All oxygen is $^{16}$O.}\label{Table:lines}
\centering
\begin{tabular} {l|cccccccccccccc}
\toprule
2019  &\\
Molecule&$^{12}$CO	 & $^{28}$Si$^{32}$S& $^{28}$Si$^{32}$S& $^{28}$Si$^{32}$S& $^{28}$Si$^{32}$S& $^{29}$Si$^{32}$S& $^{28}$Si$^{32}$S\\
$v$ state& 1  & 2  & 0  & 1  & 0  & 1  & 2  \\
$J$ transition&2--1   &12--11 &13--12 &14--13 &14--13 &15--14 &15--14 \\
Rest $\nu$ (GHz)&230.538& 215.697& 235.961& 252.866& 254.103& 265.954& 269.593\\
\midrule
2023&\\
Molecule& $^{13}$CO	 & $^{12}$CO	 & $^{30}$Si$^{32}$S& $^{28}$Si$^{34}$S& $^{28}$SiO	 & $^{29}$SiO	 & $^{28}$SiO	 \\
$v$ state& 0  & 1  & 0  & 2  & 2  & 0  & 1 \\
$J$ transition& 3--2   &3--2   &19--18 &19--18 &8--7   &8--7   &8--7  \\
Rest $\nu$ (GHz)& 330.5880& 342.6476& 332.5503& 332.1219& 342.504& 342.980& 344.917\\
\bottomrule
\end{tabular}
\end{table*}

\begin{table*}[!htp]
\caption{\textbf{$\boldsymbol{\pi^1}$~Gru measurements} of the diameters and flux densities of $M_1$ (UD) and the flux densities of $M_2$ (unresolved delta). The mean observation frequency is also given.
}\label{table:UD}
\centering
\begin{tabular} {l|cccccc}
\toprule
 ALMA &$D_{\mathrm{UD}}(M_1)$ & $\sigma_{\mathrm{UD}}(M_1)$ & $S_{\mathrm{UD}}(M_1)$ & $S_{\mathrm{delta}}(M_2)$ & $\sigma_{\mathrm S}$ & $\nu_{\mathrm{mean}}$ \\
band & [mas]     & [mas]          &  [mJy]     & [mJy]         & [mJy] & [GHz] \\
\midrule
6       &  24.0            &  2.5                  &  26.49            & 2.78                 & 0.05  & 241.03 \\
7       &  25.0            &  4.3                  &  49.00            & 5.94                 &  0.15 & 334.70 \\
\bottomrule
\end{tabular}
\end{table*}

\begin{table*}[!htp]
\caption{\textbf{$\boldsymbol{\pi^1}$~Gru astrometric errors.} The columns list the various uncertainties in $\alpha_\star$ and $\delta$ that contribute to the total astrometric uncertainty, given in the last column; see Section~\ref{sec:astrometricerror}.}\label{table:astromerr}
\centering
\begin{tabular} {l|cccccc}
\toprule
       &$\sigma_{\mathrm{phref}}$&$\phi_{\mathrm{rms}}$&$\sigma_{\mathrm{short}}$&$\sigma_{\mathrm{pos,check}}$&$\sigma_{\mathrm{pos,scaled}}$&$\sigma_{\mathrm{pos,abs}}$\\
       & [mas,mas]& & [mas,mas]& [mas,mas]& [mas,mas]& [mas,mas]\\
\midrule
C6    &0.3, 0.2&25$^{\circ}$&1.1, 1.1 &3.2, 1.3 & 1.9, 0.5 &2.2, 1.2\\
C10   &0.2, 0.1&30$^{\circ}$&1.3, 1.3 &3.7, 1.6 & 2.0, 1.4 &2.0, 1.4\\
\bottomrule
\end{tabular}
\end{table*}

\clearpage
\section{Supplementary information orbital parameters}
\subsection{Deriving the orbital equations}\label{sec:suppinforbeq}
Since this binary system is isolated (i.e., no external forces act on the bodies), conservation of linear momentum implies that the common centre of mass ($G_\star$, the system's barycentre) moves with constant velocity. This makes the reference frame in which $G_\star$ is at rest an inertial frame, with the motion of $G_\star$ determined by initial conditions. In this inertial frame, it remains to find the motion of each body relative to $G_\star$. Furthermore, in this frame, the total linear momentum of the two bodies is zero, which is why it is referred to as the zero-momentum (ZM) frame.

Determining the motion of the two bodies is a classical two-body problem. The two-body problem can be simplified by reducing it to an equivalent one-body problem, where we describe the motion of one body relative to the other using the system's reduced mass $\mu_r$
\begin{equation}
    \mu_r = \frac{m_1 m_2}{m_1 + m_2}\,.
\end{equation}
with $m_1$ and $m_2$ the mass of $M_1$ and $M_2$, respectively.
It is well known that the path equation solution for a central attractive inverse-square force field is the polar equation of a conic section with eccentricity $e$ and a focus at $O$, the centre of the force field~\cite{gregory2006classical}.
Representing the unit vectors in polar coordinates as ($\hat{r}$, $\hat{\Theta}$), the radial distance to the focal point, $r$, is given by~\cite{gregory2006classical}:
\begin{equation}
    r(\Theta) = \frac{L^2}{G m}\frac{1}{1 + e \cos(\Theta - \omega)}
\end{equation}
where $\Theta$ is the polar angle, $L$ is the angular momentum per unit mass ($L = r^2 d\Theta/dt$), $G$ is the gravitational constant, and $\omega$ is the angle between the major axis of the conic and the initial line $\Theta = 0$. While in the classical two-body problem, the barycentre occupies one of the focal points, in the equivalent one-body approach, one of the celestial bodies is at the focal point of the conic section. In the latter case, the radial coordinate $r$ is representing the motion of one body relative to the other body, and $m$ is the sum of the bodies' masses ($m = m_1 + m_2$).

For eccentricity values $e < 1$, the orbits are bound, taking the form of an ellipse as described by Kepler's first law.  For $a$ representing the semi-major axis of the ellipse, and $b$ the system's semi-minor axis $ b = a \sqrt{1 - e^2} $, the angular momentum per unit mass $L$ is related to the conic parameters $a$ and $b$ as
\begin{equation}
    L^2 = \frac {G M b^2}{a}\,,
\end{equation}
implying that the radial distance is given by
\begin{equation}
    r(\Theta) = \frac{a (1-e^2)}{1 + e \cos(\Theta - \omega)} \,.\label{Eq:r_ellipse}
\end{equation}

Let $\vec{r}_1^{\,G}$ and $\vec{r}_2^{\,G}$ represent the position vectors of $M_1$ and $M_2$ relative to the barycentre, $G_\star$, in the classical two-body problem, and let $\vec{r}$ denote the relative position of $M_2$ with respect to $M_1$. These vectors are related by
\begin{align}
    \vec{r}_1^{\,G} &= -\left(\frac{m_2}{m_1 + m_2}\right) \vec{r}, \label{Eq:vec_r1}\\
    \vec{r}_2^{\,G} &=  \left(\frac{m_1}{m_1 + m_2}\right) \vec{r}.
    \label{Eq:vec_r2}
\end{align}
As a result, the orbits described by $\vec{r}_1^{\,G}$, $\vec{r}_2^{\,G}$, and $\vec{r}$ are geometrically similar ellipses, though with different semi-major and semi-minor axes. Let $a$ be the semi-major axis of the relative orbit. Then $a = a_1 + a_2$, where $ a_1 $ and $ a_2 $ represent the semi-major axes of the elliptical orbits of $ M_1 $ and $ M_2 $ around the barycentre, respectively. Similarly, the system's semi-minor axis $ b = a \sqrt{1 - e^2} $ is the sum of the individual semi-minor axes, $b = b_1 + b_2$. All three orbits share the same eccentricity, $ e $, and orbital period, $ T_{\rm{orb}} $, given by Kepler's third law
\begin{equation}
    T_{\rm{orb}} = \sqrt{\frac{4 \pi^2 a^3}{G (m_1 + m_2)}} \,.
\end{equation}

The equivalent one-body approach is frequently used in binary star astrophysics, as the barycentre's motion, determined by initial conditions at the formation of the binary (or multiple) star system, is typically unconstrained. However, with two epochs of ALMA observations from which we can derive the positions of both $ M_1 $ and $ M_2 $, we have the unique opportunity to determine both the orbital motion of the system and the proper motion of its barycentre (see below). Thus, we will describe the proper motion of the system's barycentre, $G_\star$, and the orbital motion of both $ M_1 $ and $ M_2 $ relative to the barycentre, i.e., their total proper motion in the ZM frame, rather than describing the motion of $ M_2 $ relative to $ M_1 $.

\subsubsection{Position in an elliptic orbit}\label{Methods:pos}

Six elements define the orbit and the position of a body within an elliptic orbit: ($\Omega$, $i$, $\omega$, $a$, $e$, $T_0$); see Supplementary Fig.~\ref{fig:orbital_elements}. Here, $\Omega$ is the longitude of the ascending node, $i$ is the inclination between the orbital plane and the reference plane, $\omega$ is the argument of periapsis (the closest approach of a body to the system’s focal point), $a$ is the semi-major axis of the ellipse, $e$ is the eccentricity of the ellipse, and $T_0$ is the time of periapsis passage. For the equivalent one-body problem, $\omega$ is the argument of periastron (the closest approach in the orbit of one star around another in a binary system), and $T_0$ represents the time of periastron passage. The first three elements define the orbit’s orientation, the next two define its size and shape, and the final element specifies the body’s position within the orbit at a given time.

\paragraph{Orbital plane reference frame:}

\begin{figure}
    \centering
    \includegraphics[width=\textwidth]{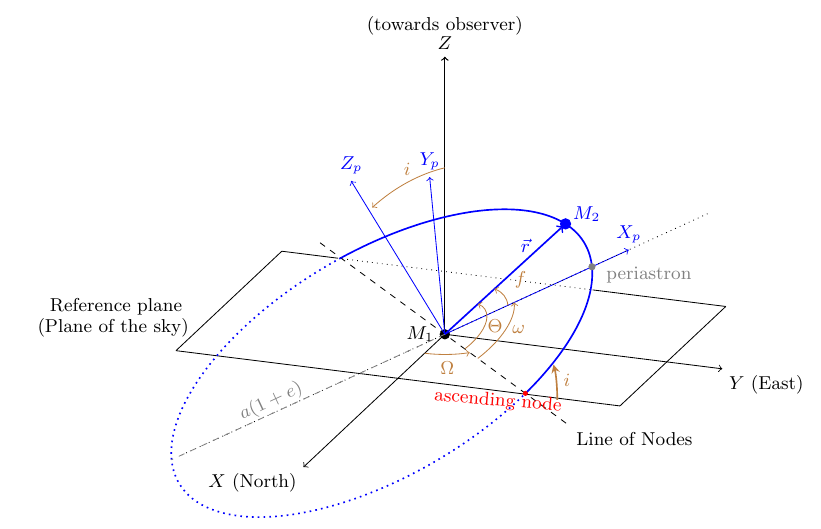}
\caption{\textbf{Orbital plane and focal reference frames illustrated for the equivalent one-body problem.}
The Cartesian orbital plane frame, $(X_p, Y_p, Z_p)$, is shown in blue, while the focal reference frame, $(X, Y, Z)$, is depicted in black. The position vector $\vec{r}$, representing the location of object $M_2$ relative to $M_1$, is defined by the orbital elements: $\Omega$ (longitude of the ascending node), $i$ (inclination), $\omega$ (argument of periapsis/periastron), $a$ (semi-major axis), $e$ (eccentricity), and $T_0$ (time of periapsis/periastron passage). The focal frame is related to the orbital plane through the angles $i$, $\Omega$, and $\omega$, which are indicated in brown, alongside the true anomaly $f$. The section of the orbit from the ascending to the descending node is depicted with a solid line, while the segment from the descending to the ascending node is shown with dotted lines. This figure presents a general illustration of orbital elements, while the specific visualization for the $\pi^1$~Gru system, with its retrieved orbital parameters, is provided in Extended Data Fig.~\ref{fig:schematic}.}\label{fig:orbital_elements}
\end{figure}

For a body on a Keplerian orbit, the position relative to the focal point can be described in the right-handed Cartesian orbital plane reference frame $(X_p, Y_p, Z_p)$. The origin of this reference frame is at a focal point\footnote{\label{footnote:focal}the focal point is either $M_1$ or $M_2$ for the equivalent one-body problem or the system's barycenter for the two-body problem}, and the $(X_p, Y_p)$ axes lay in the orbital plane with the $X_p$-axis pointing towards the periapsis/periastron. The $Z_p$-axis is directed along the angular momentum vector of the orbit, perpendicular to the orbital plane. This reference frame is illustrated in Supplementary Fig.~\ref{fig:orbital_elements} for the equivalent one-body approach. We will use $(x_p, y_p, z_p)$ to denote the Cartesian coordinates in this reference frame.

\noindent The position vector $\vec{r}$ in the orbital plane frame is given by
\begin{equation}
\vec{r} =
\begin{bmatrix}
x_p\\
y_p \\
z_p
\end{bmatrix}
=
\begin{bmatrix}
r \cos f \\
r \sin f \\
0
\end{bmatrix}\label{Eq:vec_r}
\end{equation}
where $r$ is the radial distance to the focal point derived from Eq.~\eqref{Eq:r_ellipse}--Eq.~\eqref{Eq:vec_r2}
and $f$ is the true anomaly (the angle between the direction of periapsis and the current position of the body, $f = \Theta - \omega$; see Supplementary Fig.~\ref{fig:orbital_elements}).

In practice, the time evolution of the orbit is computed by numerically solving Kepler's equation:
\begin{equation}
    E - e \sin E = 2\pi \, \frac{t-T_0}{T_{\rm orb}}\,.
\end{equation}
where the eccentric anomaly $E$ is defined by
\begin{equation}
    r \cos f = a \ (\cos E - e)\,.
\end{equation}
Once the eccentric anomaly $E$ is computed, the true anomaly $f$ can be derived with
\begin{equation}
    \tan\left( \frac{f}{2} \right) = {\left(\frac{1+e}{1-e} \right)}^{1/2} \tan\left(\frac{E}{2} \right) \,,
\end{equation}
and the radial coordinate $r$ and position vector $\vec{r}$ can be obtained from solving Eq.~\eqref{Eq:r_ellipse} and Eq.~\eqref{Eq:vec_r}.

\paragraph{Focal reference frame:}
The location of the orbiting body can also be described in the focal reference frame $(X, Y, Z)$; see Supplementary Fig.~\ref{fig:orbital_elements}. This right-handed Cartesian reference frame has its origin at a focal point$^{\ref{footnote:focal}}$ of the binary system, with the $Z$ axis pointing towards the observer, and the $X$-axis pointing towards the celestial North pole. We will use $(x, y, z)$ to denote the Cartesian coordinates in this reference frame.

\noindent
The conversion between the $(x_p, y_p, z_p)$ and $(x, y, z)$ coordinates is given by
\begin{equation}
\begin{bmatrix}
x\\
y \\
z
\end{bmatrix}
=
R(\Omega) \, R(i) \, R(\omega)
\begin{bmatrix}
x_p \\
y_p \\
z_p
\end{bmatrix}
\end{equation}
with $R$ denoting a passive rotation matrix.
The first rotation, around the $Z_p$ axis is described by the rotation matrix $R(\omega)$:
\begin{equation}
R(\omega) =
\begin{bmatrix}
\cos\omega & -\sin\omega & 0 \\
\sin\omega & \cos\omega & 0 \\
0 & 0 & 1
\end{bmatrix}
\end{equation}
where $\omega$ is the argument of the periapsis.
The second rotation, around the resulting X-axis, is given by the rotation matrix $R(i)$:
\begin{equation}
R(i) =
\begin{bmatrix}
1 & 0 & 0 \\
0 & \cos i & -\sin i \\
0 & \sin i & \cos i
\end{bmatrix}
\end{equation}
where the inclination angle $i$ is the angle between the angular momentum vector of the orbit and observer's line of sight,
such that $i=0^\circ$ corresponds to a face-on orbit and $i=90^\circ$ corresponds to an edge-on orbit.
The final rotation, around the resulting Z-axis (i.e.\ the vector pointing towards the observer), is given by the rotation matrix $R(\Omega)$ where $\Omega$ is the longitude of the ascending node:
\begin{equation}
R(\Omega) =
\begin{bmatrix}
\cos\Omega & -\sin\Omega & 0 \\
\sin\Omega & \cos\Omega & 0 \\
0 & 0 & 1
\end{bmatrix} \, .
\end{equation}

\noindent Substituting $(x_p, y_p, z_p) = (r \cos f, r \sin f, 0)$, we get:\begin{align}
\left.\begin{aligned}
x &= r \left[ \cos\Omega \cos(\omega + f) - \sin\Omega \sin(\omega + f) \cos i \right] \\
y &= r \left[ \sin\Omega \cos(\omega + f) + \cos\Omega \sin(\omega + f) \cos i \right] \\
z &= r \sin(\omega + f) \sin i
\end{aligned}\right\}\label{Eq:pos_xyz}
\end{align}

\paragraph{International Celestial Reference System (ICRS):}
The final reference frame we use is the right-handed Cartesian ICRS reference frame with its origin in the barycentre of the solar system $G_\odot$, the $Z_{\rm ICRS}$-axis pointing towards the celestial North pole, and the $X_{\rm ICRS}$-axis pointing towards the vernal equinox (see Supplementary Fig.~\ref{fig:ICRS}). We will use $(x', y', z')$ to denote coordinates in this reference frame.

\begin{figure}
    \centering
    \includegraphics[width=12cm]{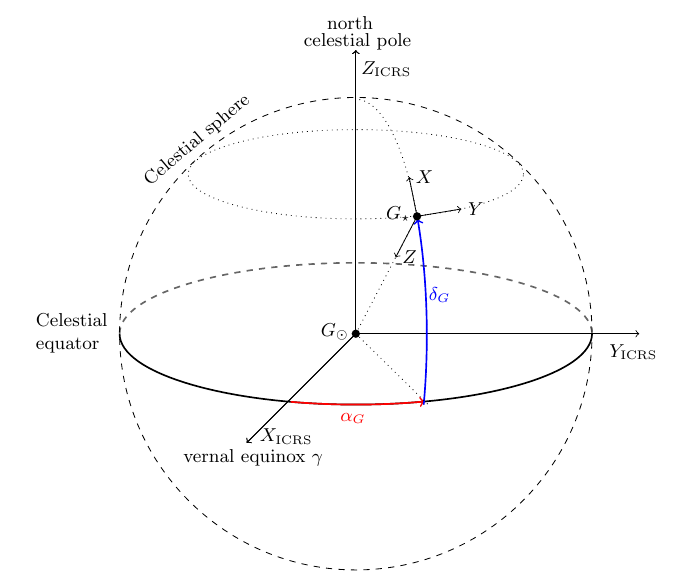}
    \caption{\textbf{Stellar system barycentric and ICRS reference frame.} Illustration of the stellar system barycentric reference frame ($X, Y, Z$) and the ICRS frame ($X_{\rm{ICRS}}, Y_{\rm{ICRS}}, Z_{\rm{ICRS}}$), with $\alpha_G$ (in red) and $\delta_G$ (in blue) denoted the right ascension and declination of the barycentre of the binary system $G_\star$.}\label{fig:ICRS}
\end{figure}

\noindent The conversion between the $(x, y, z)$ and $(x', y', z')$ coordinates is given by
\begin{equation}
\begin{bmatrix}
x' \\
y' \\
z'
\end{bmatrix}_{\rm ICRS}
=
 R(\alpha_G) \, R(\delta_G)
\begin{bmatrix}
x \\
y \\
z
\end{bmatrix}
+
\begin{bmatrix}
x'_G \\
y'_G \\
z'_G
\end{bmatrix}\label{Eq:to_ICRS}
\end{equation}
where $(\alpha_G, \delta_G)$ are the equatorial sky coordinates of the centre of mass of the binary system. The translation term denotes the shift between the barycentres of the binary and the solar system:
\begin{align}
    x^\prime_G &= D \cos\delta_G \cos\alpha_G \\
    y^\prime_G &= D \cos\delta_G \sin\alpha_G \\
    z^\prime_G &= D \sin\delta_G
\end{align}
with $D$ (\,=\,$1/\varpi$) the distance to the binary system and $\varpi$ the parallax. The rotation matrices $R(\alpha_G)$ and $R(\delta_G)$ are defined by:
\begin{equation}
R(\alpha_G) =
\begin{bmatrix}
\cos\alpha_G & -\sin\alpha_G & 0 \\
\sin\alpha_G & \cos\alpha_G & 0 \\
0 & 0 & 1
\end{bmatrix} \, .
\end{equation}
and
\begin{equation}
R(\delta_G) =
\begin{bmatrix}
-\sin\delta_G  & 0 & -\cos\delta_G \\
0              & 1 &        0      \\
\cos\delta_G   & 0 & -\sin\delta_G \\
\end{bmatrix} \, .
\end{equation}

\noindent The right ascension and declination of $M_1$ (or $M_2$) in the ICRS frame are:
\begin{align}
    \alpha_{M_1} &= \arctan\left(\frac{y^\prime_{\text{ICRS}}}{x^\prime_{\text{ICRS}}}\right)\label{Eq:RA_ICRS}\\
    \delta_{M_1} &= \arcsin\left(\frac{z^\prime_{\text{ICRS}}}{r_{\text{ICRS}}}\right)\label{Eq:DEC_ICRS}
\end{align}
where $ r_{\text{ICRS}} = \sqrt{{(x^\prime_{\text{ICRS}})}^2 + {(y^\prime_{\text{ICRS}})}^2 + {(z^\prime_{\text{ICRS}})}^2} \simeq D$.
The change in right ascension and declination due to the orbital motion of $M_1$ (or $M_2$) around the system's barycentre is then:
\begin{align}
    \Delta \alpha_{M_1} &= \alpha_{M_1} - \alpha_G\label{Delta_RA} \\
    \Delta \delta_{M_1} &= \delta_{M_1} - \delta_G\label{Delta_DEC}
\end{align}

\paragraph{Parallax correction:}

As a final step, it is necessary to align the ALMA detections with the inertial, barycentric International Celestial Reference Frame (ICRF), which involves correcting for the apparent motion of the star(s) due to the parallactic shift. For the unit vector $\hat{z} = \vec{e}_{\rm{los}}$ along the line of sight being
\begin{equation}
    \vec{e}_{\rm los} = (\cos\alpha_G \cos\delta_G, \sin\alpha_G \cos\delta_G, \sin\delta_G)\,,\label{Eq:elos}
\end{equation}
and denoting the Earth's position as $\vec{r}_\oplus$, the parallactic shift is $\vec{r}_\oplus \cdot \hat{z}$. Let ($X_\oplus$, $Y_\oplus$, $Z_\oplus$) in au represent the location of the Earth in barycentric coordinates. The corrections in right ascension ($\Delta^p_\alpha$) and declination ($\Delta^p_\delta$) due to parallax are then given by~\cite{Nielsen2020AJ....159...71N}:
\begin{eqnarray}
    \Delta^p_{\alpha^{\star}} & = & \varpi (X_\oplus \sin\alpha_G - Y_\oplus \cos\alpha_G) \\
    \Delta^p_\delta & = & \varpi (X_\oplus \cos\alpha_G \sin\delta_G + Y_\oplus \sin\alpha_G \sin\delta_G - Z_\oplus \cos\delta_G)
\end{eqnarray}
with the values of ($X_\oplus, Y_\oplus, Z_\oplus$) extracted from the JPL Horizons app\footnote{\url{https://ssd.jpl.nasa.gov/horizons/app.html\#/}}.
For the ALMA C6 and C10 orbital model predictions, this parallactic shift is incorporated in the calculations for both the $M_1$ and $M_2$ components. The difference between the proper motion of the binary system, including the parallactic shift or not, is illustrated in Supp. Inf. Fig.~\ref{fig:pi1_gru_motion_parallax_shift}. Specifically, at the ALMA C6 epoch, that correction is $(\Delta^p_{\alpha}, \Delta^p_\delta) = (7.0631, -2.2003)$~mas, and at ALMA C10 $(-5.5076, -3.0128)$~mas.

\begin{figure}[htp]
\centering
	\includegraphics[width=15cm]{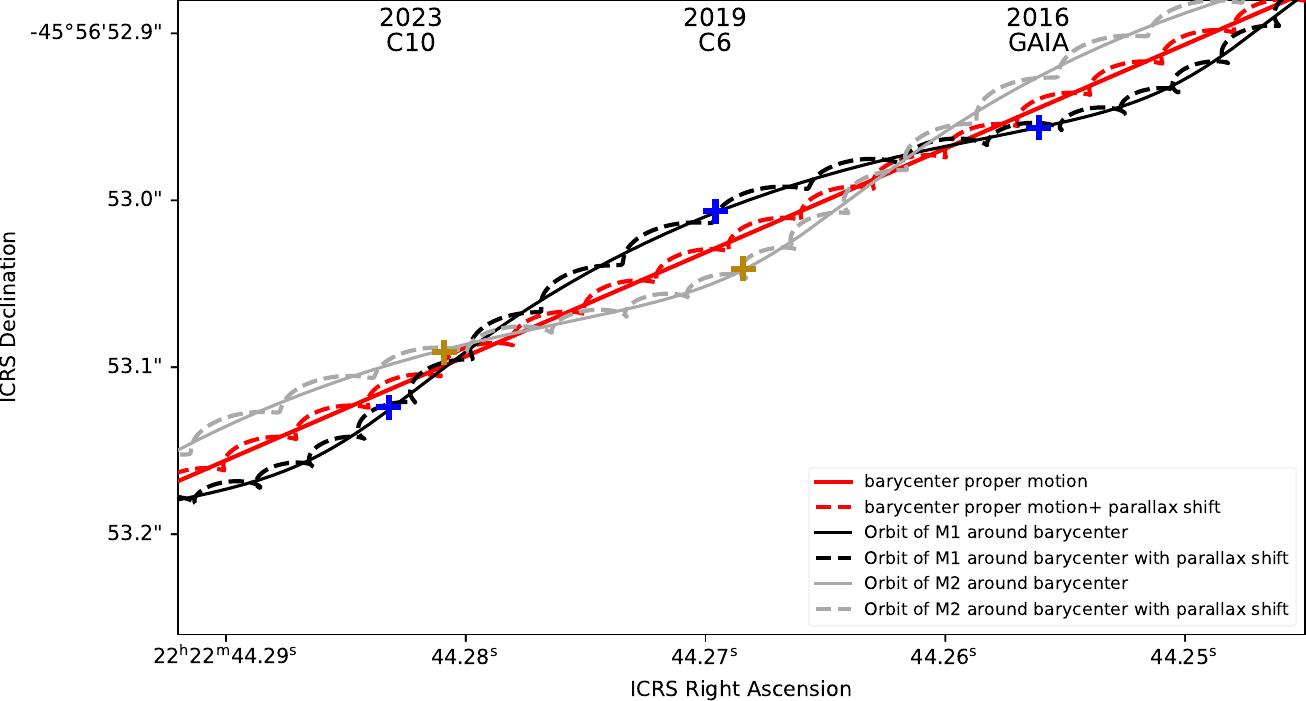}
\caption{\textbf{Parallactic shift for the $\boldsymbol{\pi^1}$~Gru system.} A schematic illustration of the parallax, proper motion, and orbital motion of the $\pi^1$~Gru system. Black lines represent the motion of $M_1$, grey lines represent $M_2$, and red lines represent the barycentre. Solid lines show the proper motion and orbital motion only, while dashed lines also include the parallactic shift. Blue crosses mark observations of $M_1$ at the \textit{Gaia}, ALMA C6, and ALMA C10 epochs, and brown crosses mark $M_2$ at the ALMA C6 and ALMA C10 epochs. In this figure, the \textit{Gaia} astrometric position has been corrected for parallax motion, while the ALMA C6 and C10 positions have not.}\label{fig:pi1_gru_motion_parallax_shift}
\end{figure}

\subsubsection{Proper motion (anomaly)}\label{Sec:theory:pma}
The proper motion vector, denoted by $\bmu$, describes the change in a star's position on the celestial sphere over time. It can be expressed as
\begin{align}
\mu_{\alpha^*} &= \frac{d\alpha}{dt} \cos\delta \\
\mu_{\delta} & = \frac{d\delta}{dt}
\end{align}
where $\mu_{\alpha^*} = \mu_\alpha \cos\delta$. For an isolated star with no intrinsic changes, the proper motion vector remains constant in both magnitude and direction. However, inaa binary or higher-order multiple system, the proper motion of a component star, $M_1$ (or $M_2$), is the sum of the motion of the system's barycentre, $\bmu^G$, and the star's orbital motion around the barycentre:
\begin{equation}
    \bmu^{\rm{obs}}_{M_1} = \bmu^G + \bmu_G^{M_1}.
\end{equation}
As a result, $\bmu^{\rm{obs}}_{M_1}$ varies in both magnitude and direction over time, leading to the concept of the proper motion anomaly.

The proper motion anomaly of a celestial body is defined as the difference between the `instantaneous' or short-term proper motion vector, $\bmu^{\rm{inst}}$, which is calculated from a short-term observational period (e.g., using \textit{Gaia} or \textit{Hipparcos} data), and the long-term mean proper motion, $\bmu^{\rm{long}}$, derived from the positional differences between two catalogs spanning a much longer time interval (see Supplementary Fig.~\ref{fig:mu_HG}):
\begin{equation}
\Delta \bmu = \bmu^{\rm{inst}} - \bmu^{\rm{long}}.
\end{equation}
In binary systems with sufficiently long orbital periods, the primary star's reflex motion is reflected in the observed short-term proper motion. Conversely, the long-term proper motion more accurately represents the movement of the system's barycentre. Due to the averaging effect in long-term proper motions, variations in $\Delta \bmu$ are predominantly detectable for long-period binaries.

\begin{figure}
    \centering
    \includegraphics[width=\textwidth]{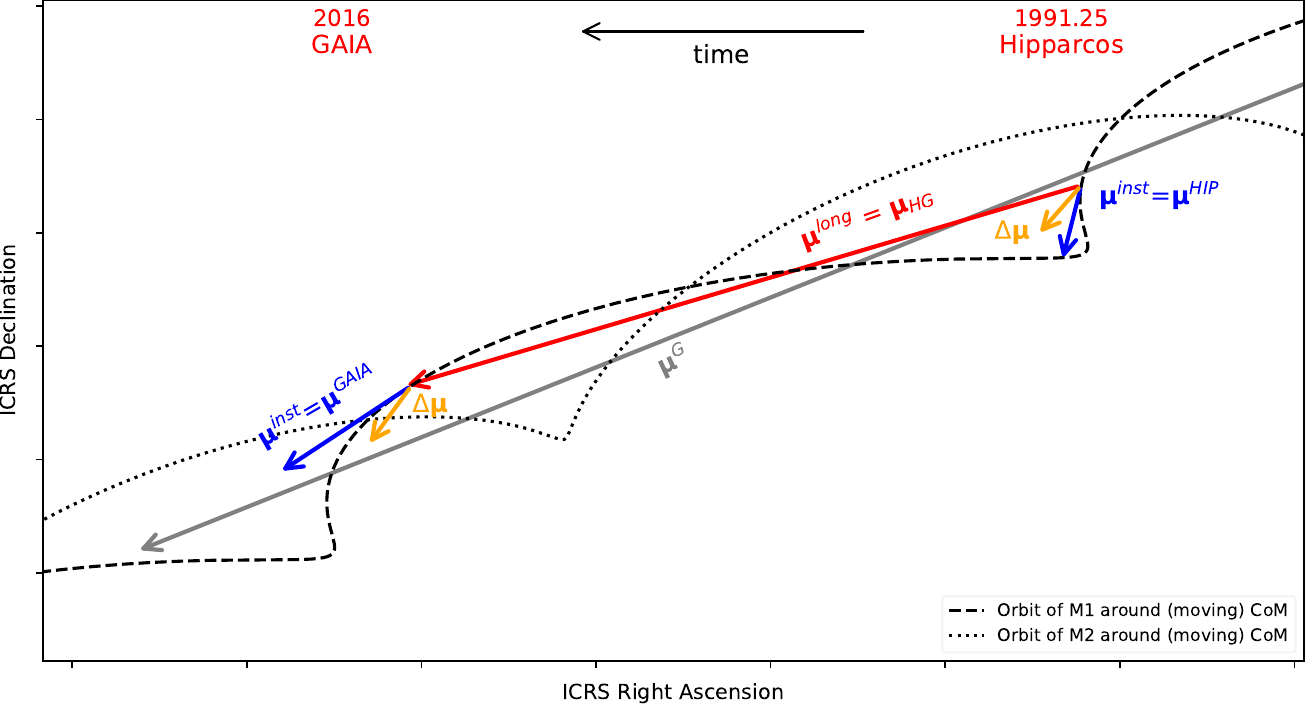}
    \caption{\textbf{Illustration of the proper motion anomaly in a binary system.} This schematic shows the orbital motion of two bodies, $M_1$ and $M_2$, around the system's barycentre. The proper motion of the barycentre over time, $\bmu^G$, is indicated by the gray arrow. The red arrow represents the long-term proper motion of $M_1$, $\bmu^{\text{long}}$, here illustrated over a timespan of $\sim$25 years. The instantaneous proper motion of $M_1$, derived from \textit{Hipparcos} or \textit{Gaia} data over a three-year span, is shown with the blue arrow. The resulting proper motion anomaly, $\Delta \bmu$, for this three-year period is depicted by the orange arrow.}\label{fig:mu_HG}
\end{figure}

As shown in Supplementary Fig.~\ref{fig:mu_HG}, the vector $\bmu^G$ does not coincide with $\bmu^{\mathrm{long}}$. Only when $\bmu^{\mathrm{long}}$ accurately represents the long-term motion of the barycentre, i.e., $\bmu^{\mathrm{long}} = \bmu^G$, does the vector $\Delta \bmu$ correspond to the projected velocity of the photocenter's orbit around the barycentre. This assumption may hold if one body is significantly more massive than the other, or coincidentally, if the epoch of the here \textit{Hipparcos} and \textit{Gaia} observations happens to satisfy $\bmu^{\mathrm{long}} \parallel \bmu^G$. In scenarios where either $\bmu^{\mathrm{long}} = \bmu^G$ or $\bmu^{\mathrm{long}} \parallel \bmu^G$ and $M_1$ is notably brighter than $M_2$ at the observed frequencies, $\Delta \bmu$ primarily reflects the motion of $M_1$'s photocentre relative to the system's centre of mass.

Given the proper motion $\bmu$, the tangential velocity in units of km s$^{-1}$ is
\begin{equation}
    v_{\tan} = \sqrt{\mu_{\alpha^\star}^2 + \mu_\delta^2} \times \frac{4.74047}{\varpi} \,,
\end{equation}
for the parallax $\varpi$ given in units of mas and the proper motion in units of mas yr$^{-1}$. Similar to the proper motion anomaly, the tangential velocity anomaly vector, $\Delta v_{\tan}$, is defined as
\begin{equation}
\Delta v_{\tan} = \sqrt{\Delta \mu_{\alpha^\star}^2 + \Delta \mu_\delta^2} \times \frac{4.74047}{\varpi}.\label{Eq:vtan_anomaly}
\end{equation}
with corresponding position angle (PA) $\theta$ being
\begin{equation}
    \tan\theta = \frac{\Delta \mu_{\alpha^\star}}{\Delta \mu_\delta}\label{Eq:PA_vtan}
\end{equation}
so that a PA of zero is for a velocity vector pointing to the North, and of 90$^\circ$ to the East ($\Delta \mu_{\alpha^\star} > 0$ and $\Delta \mu_\delta = 0$).

\subsection{Observational input}
\subsubsection{Proper motion (anomaly)}
Instead of using the proper motions from \textit{Hipparcos} and \textit{Gaia}, one can opt to use the proper motion anomaly as alternative observational constraint.
Precise astrometric measurements of the proper motion of $\pi^1$~Gru~A have been obtained from both the \textit{Hipparcos} and \textit{Gaia} missions. Each mission derived an `instantaneous' proper motion $\bmu^{\rm{inst}}$, denoted to as $\bmu_{\rm Hip}$ and $\bmu_{\rm\textit{Gaia}}$, respectively, and listed above. These vectors represent the combined motion of the system's barycentre and the orbital velocity around the barycentre (see Supplementary Sect.~\ref{Methods:pos}). The \textit{Hipparcos} and \textit{Gaia} missions span more than 30 years, allowing for the calculation of a mean long-term proper motion vector $\bmu^{\rm{long}}$ from the difference in sky coordinates $(\Delta\alpha, \Delta\delta)$ of $\pi^1$~Gru~A in both catalogs, referred to as $\bmu_{HG}$, with components~\cite{Kervella2019A&A...623A..72K}
\begin{align}
\mu_{HG,\alpha^\star_1} &= 31.306 \pm 0.028 \ {\rm{mas\ yr}^{-1}}\label{eq:mu_alphastar_hg} \\
\mu_{HG,\delta_1} & = -18.233 \pm 0.023 \ {\rm{mas\ yr}^{-1}} \,.\label{eq:mu_delta_hg}
\end{align}
and hence
\begin{equation}
 \mu_{HG,\alpha_1} = 45.024 \pm 0.040 \ {\rm{mas\ yr}^{-1}}\label{eq:mu_alpha_hg} \nonumber
\end{equation}
This long-term proper motion is primarily dominated by the barycentre's motion through the Milky Way, $\bmu^G$, but it is not exactly identical to it. The reason is that its measurement is solely based on the photocentre movement, in the case of the $\pi^1$~Gru system this is dominated by $M_1$. Therefore, $\bmu_{HG}$ represents the mean long-term motion of $M_1$'s photocentre over a period of $\sim$25 years (see Supplementary Fig.~\ref{fig:mu_HG}).

Subtracting $\bmu_{HG}$ from the instantaneous proper motion yields the
proper motion anomaly~\cite{Kervella2019A&A...623A..72K}
\begin{equation}
    \Delta\bmu_{\rm Hip} = \bmu_{\rm Hip} - \bmu_{HG}
\end{equation}
with components~\cite{Kervella2022A&A...657A...7K}
\begin{align*}
    \Delta \mu_{\alpha^*_1} &= -2.826 \pm 0.940\ {\rm{mas\ yr^{-1}}} \tag{OC12b} \\
    \Delta \mu_{\delta_1} &= 6.093 \pm 0.600\ {\rm{mas\ yr^{-1}}} \tag{OC13b}
\end{align*}
at epoch 1991.25, and
\begin{equation}
    \Delta\bmu_{\rm\textit{Gaia}} = \bmu_{\rm\textit{Gaia}} - \bmu_{HG}
\end{equation}
with components~\cite{Kervella2022A&A...657A...7K}
\begin{align*}
    \Delta \mu_{\alpha^*_1} &= -0.194 \pm 0.253\ {\rm{mas\ yr^{-1}}} \tag{OC17b} \\
    \Delta \mu_{\delta_1} &= 7.933 \pm 0.348\ {\rm{mas\ yr^{-1}}} \tag{OC18b}
\end{align*}
at epoch 2016.0. The corresponding tangential velocity anomaly at the \textit{Hipparcos} epoch is
\begin{equation}
    \Delta v_{\rm{tan,Hip,1}} = 4.87 \pm 0.96\ {\rm{km\ s^{-1}}} \tag{OC12c}
\end{equation}
with position angle being
\begin{equation}
    \theta_{\rm{Hip,1}} = 335.12 \pm 7.65^\circ\,, \tag{OC13c}
\end{equation}
and at the \textit{Gaia} epoch is
\begin{equation}
    \Delta v_{\rm{tan,Gaia,1}} = 6.08 \pm 0.53\ {\rm{km\ s^{-1}}} \tag{OC17c}
\end{equation}
with position angle being
\begin{equation}
    \theta_{\rm{Gaia,1}} = 358.60 \pm 1.34^\circ\,. \tag{OC18c}
\end{equation}
One can choose to either use the pair (OC12, OC13), (OC12b, OC13b) or (OC12c, OC13c) as observational constraint, since the three sets are equivalent. Similarly, one can choose pairs (OC17, OC18), (OC17b, OC18b) or (OC17c, OC18c). Note that $\bmu_{HG}$ (Expr.~\eqref{eq:mu_alphastar_hg} -- {\eqref{eq:mu_delta_hg}}) is not used as an independent observational constraint in the fitting routine, as its value is derived from the astrometric positions at the \textit{Hipparcos} and \textit{Gaia} epochs, which are already accounted for. Alternatively, one can opt to not use either the \textit{Hipparcos} or \textit{Gaia} astrometric position, and use $\bmu_{HG}$ instead. The subscript 1 indicates that (OC12b)--(OC18c) are observational constraints for the $M_1$ component.

In our analysis, we choose (OC12, OC13, OC17, OC18) as observational constraints because the alternative sets introduce correlations that are not quantified.

\subsection[Inferred orbital parameters for the Pi1~Gru system]{Inferred orbital parameters for the $\boldsymbol{\pi^1}$~Gru system}\label{suppinfResults}

\subsubsection{MCMC sampling approaches}\label{sec:sampling}
Two different approaches were explored: Hamiltonian Monte Carlo (HMC) sampling~\cite{Duane1987PhLB..195..216D, HoffmanNUTS} and nested sampling~\cite{Skilling2004AIPC..735..395S, Buchner2023StSur..17..169B}. Both methods are used for sampling complex, high-dimensional probability distributions. The HMC method starts from an initial point and explores the parameter space using a random walk that exploits gradient information. The initial point was taken to be the best-fit outcome from a frequentist hybrid global and local Levenberg-Marquardt search.

Nested sampling begins with a set of live points sampled from the prior distribution. The algorithm iteratively replaces the lowest-likelihood live point with a new sample having a higher likelihood, effectively shrinking the prior volume. This process continues until a stopping criterion is met, typically when the remaining prior volume is sufficiently small. Nested sampling provides both the Bayesian evidence (also known as the marginal likelihood) for model comparison and posterior samples for parameter estimation.

It turned out that for our applications the HMC method was sensitive to its tuning parameters, and that convergence was difficult to achieve as it struggled with the high-dimensional distribution so that no stable reliable solution could be found. Even narrowing the input parameter range did not yield a stable solution. In this context, nested sampling can be more efficient than HMC for complex multi-modal distributions and distributions with pronounced degeneracies. We opted to use the Python library \texttt{ultranest}~\cite{Ultranest2021JOSS....6.3001B}, which implements the nested sampling algorithm. \texttt{ultranest} can handle high-dimensional parameter spaces, multi-modal distributions, and narrow and wide peaks in posterior distributions. Moreover, it incorporates sophisticated stopping criteria to ensure that the sampling process is both thorough and efficient, avoiding unnecessary computations once the algorithm has sufficiently explored the parameter space. By default, we used a step sampler with 4,000 live points ($n_{\rm{live}}$\,=\,4,000) and a number of steps equal to twice the number of input parameters ($n_{\rm{steps}}$\,=\,2$\times n_{\dim}$\,=\,22). We ran \texttt{ultranest} 10 times, with each run yielding a similar mean within the standard deviation of the other runs, demonstrating the stability of the \texttt{ultranest} outcomes (see below). In its default configuration and utilizing \texttt{OpenMPI}, a single \texttt{ultranest} run takes approximately 16\,--\,48 hours on a 96-core computer cluster with 250GB memory of CPU type {\sc{`AMD EPYC 7643 @ 3.64GHz'}}.

\begin{figure}[!htp]
    \centering
    \hspace*{-1.4cm}
    \includegraphics[width=.7\textwidth]{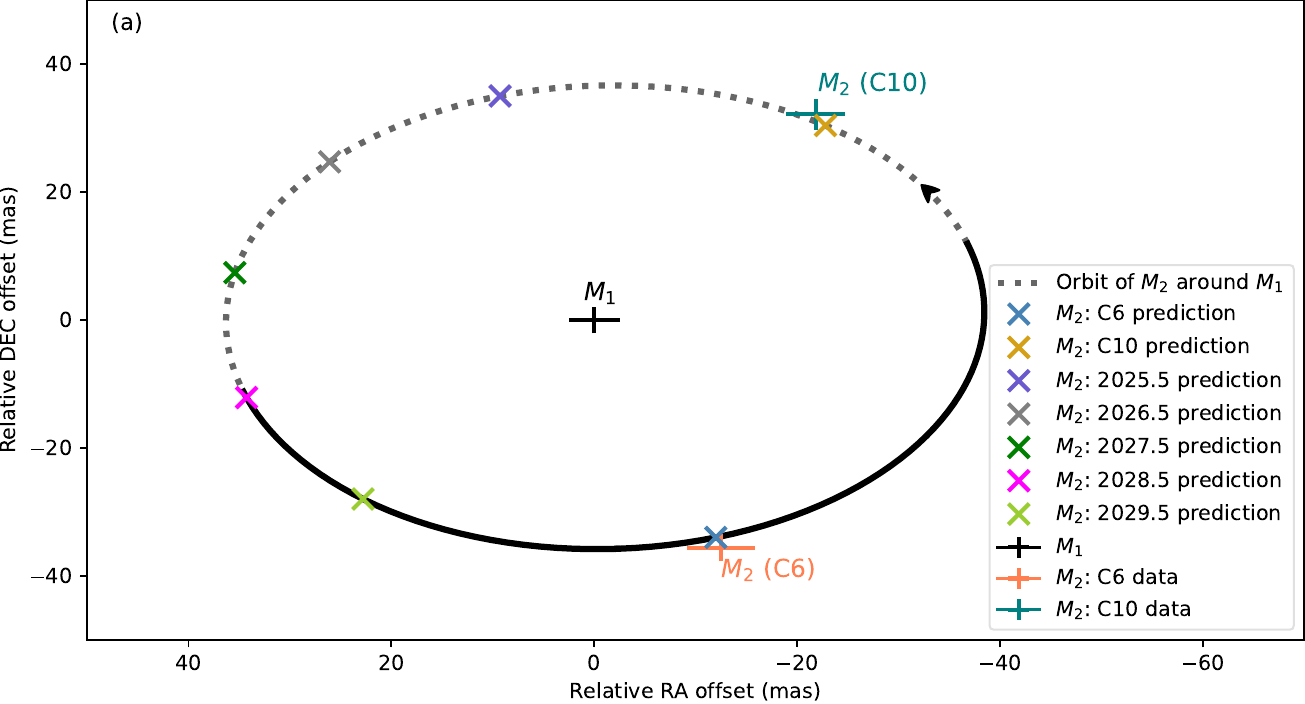}
    \includegraphics[width=.788\textwidth]{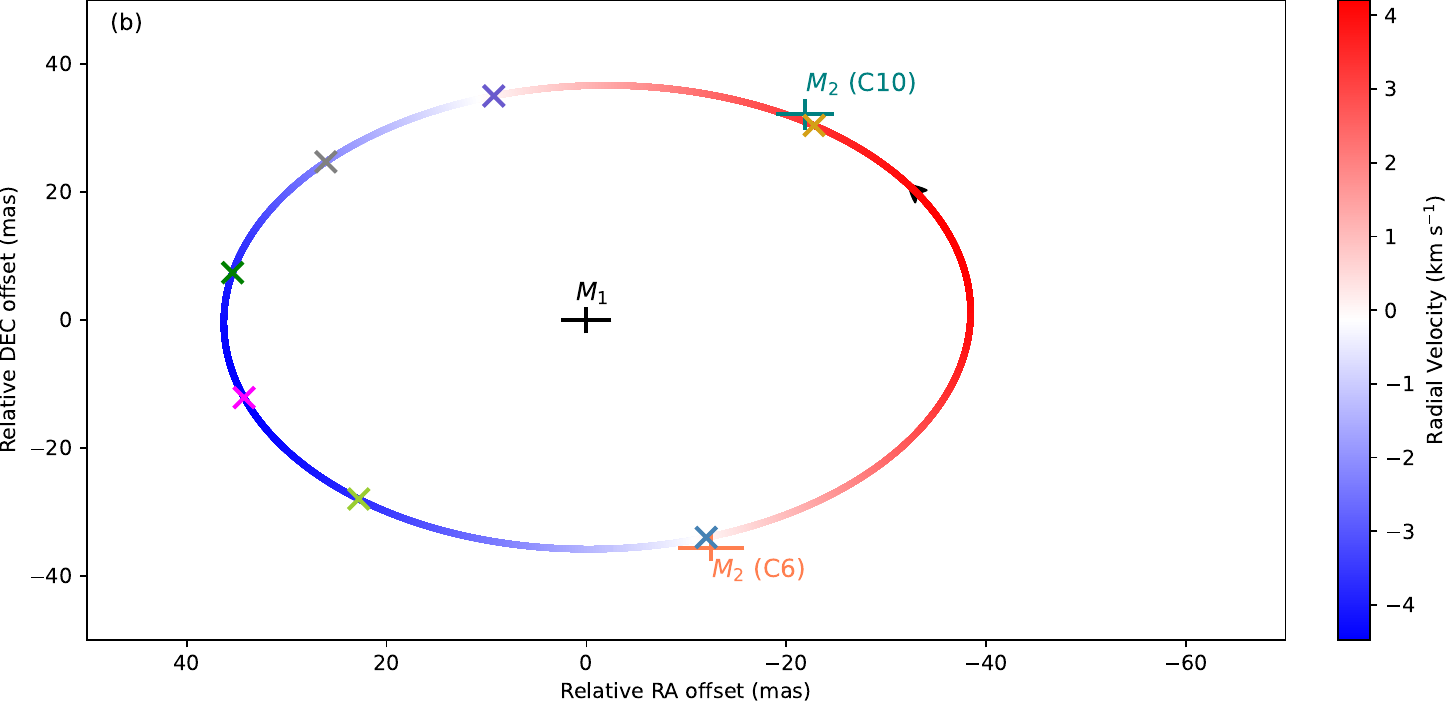}
\caption{\textbf{Relative motion fit of the $\boldsymbol{\pi^1}$~Gru system in the plane of the sky.} The data show the relative position of $M_2$ with respect to $M_1$, with 1-sigma uncertainties represented by error bars. The orange and teal crosses indicate the positions from the ALMA 2019 C6 data and ALMA 2023 C10 data, respectively. The elliptic curve represents the predicted orbital motion based on the parameters from our \texttt{eccentric} model. The blue and gold crosses mark the predicted positions for the C6 and C10 epochs, respectively.
The five crosses indicate the position at example epochs $\pm$ one or more orbital period and serve to guide potential future observations.
Panel~(a) shows the orbital trajectory with a solid line from the ascending to descending node, and a dotted line from the descending to ascending node. Panel~(b) colour-codes the trajectory to indicate the orbital contribution to the radial velocity, $v_{\rm rad}^{\rm orb}$, for the velocity of $M_2$ relative to $M_1$ (see Supplementary Sect.~\ref{Methods:vrad}, for details). That contribution is $\sim$0.06~km~s$^{-1}$ at the ALMA C6 epoch and $\sim$3.37~km~s$^{-1}$ at the ALMA C10 epoch.
}\label{Fig:relative_motion}
\end{figure}

\subsubsection{Sensitivity analysis of Bayesian retrievals}\label{Sec:sensitivity}

In this section, we perform an extensive sensitivity analysis, considering various aspects of the model setup and sampling approaches. First, we examine potential degeneracies in the orbital parameters, specifically the inclination $i$ and its complement $360^\circ - i$, as well as the relationships between $\omega$ and $\Omega$ with their antipodal angles. Next, we run the default settings 10 times to check the robustness of the retrieval method. We also explore the inclusion and exclusion of the \textit{Hipparcos} and \textit{Gaia} covariance matrices to assess the impact of correlated data versus independent sampling. Additionally, we evaluate the significance of detecting a small but non-circular eccentricity. Key priors, such as those on $m_1$ and distance, are tested for their influence on the outcomes. The likelihood is modified to exclude data from ALMA C10, \textit{Gaia}, or \textit{Hipparcos}, allowing us to assess the contributions of individual datasets. Moreover, we examine the impact of using approximate solutions for the astrometric positions and the tangential velocity anomaly on the retrieval outcomes in Supplementary Sect.~\ref{Sec:approximation}. Each test is run at least 4 times to check for the stability of the solution. The outcomes of this sensitivity analysis are summarized in Supplementary Table~\ref{table:fit_sensitivity} and Supplementary Table~\ref{table:best_fit}, where we report, for each test, the result of the run with the highest marginalized likelihood. The only exception is the case where ALMA C10 data is excluded. As discussed below, without the inclusion of ALMA C10 data, the \texttt{ultranest} retrieval becomes highly unstable, with each run converging on different solution spaces. This suggests that the fitting is finding local maxima or saddle points, but not the global posterior maximum.

We do not add an additional test for sensitivity to the precise observed position of the companion, for three main reasons: (1) The hydrodynamical simulations presented in Sect.~\ref{Sec:3D_hydro} indicate that the accretion disk is nearly circular, centered on the companion, and has a much higher density than the attached bow shock. Thus, the disk centre should closely correspond to the companion’s location. (2) The outer disk radius is only $\sim$0.83~au (or 4.85~mas); see Sect.~\ref{Sec:3D_hydro}. Even a fraction of this radius falls within the current uncertainty range for ALMA astrometric positions. (3) Introducing such a test would require an arbitrary set of parameters to adjust the observed positions in right ascension and declination, complicating the model unnecessarily.

For model comparison, we present both the maximum of the logarithm of the marginalized likelihood ($\max(\log z)$) and the normalized root mean square error ($L_{\rm{NRMSE}}$) in Supplementary Table~\ref{table:best_fit}. These metrics assess how well the model fits the observed data, with higher values of $\log z$ and lower values of $L_{\rm{NRMSE}}$ indicating better fits. The log-likelihood $\log z$ is derived from a Bayesian fitting procedure, while $L_{\rm{NRMSE}}$ follows a frequentist approach. The normalized root mean square error is defined as:
\begin{equation}
L_{\rm{NRMSE}} = \sqrt{\frac{1}{18} \sum_{j=1}^{18} {\left( \frac{y_{\rm{obs},j} - y_{\rm{pred},j}} {\sigma_j} \right)}^2}\label{Eq:ell}
\end{equation}
where $ j $ indexes each of the 18 observational constraints. In this equation, $ y_{\rm{obs},j} $ are the observed values, $ y_{\rm{pred},j} $ are the model predictions, and $ \sigma_j $ represents the uncertainties associated with each observation. In this formulation, we do not account for the covariance matrices of the \textit{Gaia} and \textit{Hipparcos} observables and we compare each model prediction to all of the 18 observational constraints. The frequentist approach minimizes $L_{\rm{NRMSE}}$, thereby reducing the discrepancy between the observed and predicted values. On the other hand, the Bayesian method seeks to maximize the posterior probability, which combines the likelihood of the data given the model parameters with prior information.

\paragraph{Geometrical degeneracies:}
The \texttt{ultranest} retrieval yields specific values for the orbital parameters ($\Omega$, $i$, $\omega$, $a$, $e$, $T_0$). However, it is important to recognize that these results exhibit degeneracies. Specifically, the inclination $i$ is degenerate with its complement $360^\circ - i$, and the angles $(\omega, \Omega)$ are degenerate with their antipodal values $(\omega + 180^\circ, \Omega + 180^\circ)$. As is clear from Eq.~\eqref{Eq:pos_xyz}, changing $i$ to $360^\circ - i$ or shifting both $(\omega, \Omega)$ by $180^\circ$ leaves the $x$ and $y$ components unchanged, while flipping the sign of $z$. As discussed in Supplementary Sect.~\ref{Methods:approx_pos}, $x$ and $y$ correspond to changes in declination and right ascension, respectively. This implies that our astrometric analysis cannot distinguish between these degenerate configurations. Despite the degeneracies, a particular solution emerges as the most likely outcome per run. However, the complementary configuration appears as the best-fit result in other test runs. Given the fact that our astrometric analysis can not distinguish between these degeneracies, we restrict the priors of both $i$ and $\Omega$ to $\mathcal{U}(0, 180)$ for all further runs presented in this section.

This degeneracy in $(\omega, \Omega)$ and their antipodal angles also reflects a broader issue in stellar physics, particularly when it comes to defining the ascending node. In general, there is no agreement on whether the ascending node marks the point where the orbiting body crosses the plane of the sky moving away or towards the observer. It differs from the descending node by $180^\circ$ and can only be determined through radial-velocity measurements.

Regarding the geometrical degeneracy of $i$ versus $360^\circ - i$, this specifically translates here to $i \sim 15^\circ$ or $i \sim 345^\circ$. Previous studies have estimated the inclination of the fragmented spiral arm and the expanding equatorial torus or disk-like structure surrounding the primary AGB star using ALMA CO $J=3\text{--}2$ and $J=2\text{--}1$ data~\cite{Homan2020A&A...644A..61H, Doan2020A&A...633A..13D, Doan2017A&A...605A..28D}. These studies suggest the spiral arm is inclined by $38 \pm 5^\circ$ with respect to the line of sight~\cite{Homan2020A&A...644A..61H} and that the angle between the line of sight and the equatorial plane of the torus is $i = 55^\circ$~\cite{Chiu2006ApJ...645..605C}\textsuperscript{,}\cite{Knapp1999A&A...346..175K} or $i = 40 \pm 10^\circ$~\cite{Doan2020A&A...633A..13D, Doan2017A&A...605A..28D}.

Several considerations are important here. First, different definitions for inclination have been used. Refs.~\cite{Homan2020A&A...644A..61H, Doan2020A&A...633A..13D, Chiu2006ApJ...645..605C, Doan2017A&A...605A..28D}\textsuperscript{,}\cite{Knapp1999A&A...346..175K} define inclination ($i_{\rm CO}$) as the angle between the line of sight and the equatorial plane of the torus, where $0^\circ < i_{\rm CO} < 90^\circ$ implies the southern side of the CO disk is closer to Earth. Assuming the orbital plane to be coplanar with the CO disc, this definition differs by $90^\circ$ from the convention adopted in this study, i.e.\ $i_{\rm CO} = 90^\circ - i$. Here, we follow the standard convention, defining $i$ as the angle between the orbital plane of the binary system and the plane of the sky, measured from the line of sight to the normal of the orbital plane ($Z_p$) (see Supplementary Fig.~\ref{fig:orbital_elements}).

Second, Ref.~\cite{Homan2020A&A...644A..61H} used the inner 5" of the CO emission to infer the inclination of a spiral structure by fitting an Archimedean spiral pattern. However, caution is warranted as the spiral arm's geometry is more complex, with its vertical extent and width challenging to constrain. Furthermore, the presence of an outer G0V companion at a projected separation of 2\farcs71~\cite{Feast1953MNRAS.113..510F} likely affects the CO density distribution in both the spiral arm and the expanding disk. Thus, the reported accuracy of $5^\circ$ may be optimistic.

Third, Refs.~\cite{Doan2020A&A...633A..13D, Doan2017A&A...605A..28D} determined inclination under the assumption that the CO density follows a radially expanding torus shaped as a flared disk. However, our 3D hydrodynamic simulations (Sect.~\ref{Sec:3D_hydro}) and those presented in other publications~\cite{Malfait2024A&A...691A..84M, Kim2012ApJ...759...59K, Maes2021A&A...653A..25M} indicate a flattened disk caused by the gravitational focusing of the AGB wind onto the companion's orbital plane. The inclination estimates of Refs.~\cite{Doan2020A&A...633A..13D, Doan2017A&A...605A..28D} are thus influenced by the assumed simplified velocity and density distribution of the CO gas. In contrast, our method directly fits the orbital model to the data, independent of such assumptions, resulting in a more accurate inclination determination.

Despite these differences in inclination definitions and their inherent limitations, the CO observations suggest the northern part of the CO emission is tilted away from the observer, supporting $i \sim 15^\circ$, under the assumption that the CO disk is (roughly) aligned with the orbital plane of the inner companion.

Resolving the remaining degeneracy in $(\omega, \Omega)$ and $(\omega + 180^\circ, \Omega + 180^\circ)$ will require precise (time-dependent) radial velocity measurements of each component, which could potentially be obtained through ALMA. As discussed in Supplementary Sect.~\ref{Methods:vrad}, our preliminary analysis, based on two epochs of ALMA observations, suggests that the current $(\omega, \Omega)$-set for the eccentric model (or $\Omega$ for the circular model) is the correct one.

\paragraph{Robustness of the \texttt{ultranest} retrieval:}

We test the robustness of the \texttt{ultranest} retrieval by running the default setting 10 times to observe consistency in the results; see Supplementary Table~\ref{table:robustness}. Across these runs, we find minimal variation in the best-fit parameters\footnote{with the exception being $\omega$ and $T_0$ owing to the nearly circular orbit; see below}, confirming the stability of the retrieval process. The posterior distributions remain consistent, indicating that the algorithm converges reliably on a global solution without being significantly affected by the random sampling. These findings suggest that the \texttt{ultranest} retrieval method is robust, providing confidence in the inferred parameters for the $\pi^1$~Gru system.

The parameter that shows the largest variation is $m_1$, with mean values ranging from 1.00 to 1.33\,\Msun. The parameter that is least constrained is $\omega$ with standard deviation of $\sim$100$^\circ$. This is unsurprising given the near-circular orbit of the system. The very low eccentricity ($e \sim 0.02$) implies that both $\omega$ and $T_0$ are poorly constrained. Consequently, we will not discuss the mean and best-fit values for these two parameters in the remainder of this paper, as they provide only limited diagnostic value.

The corner plots and the standard deviation on $e$ suggest that a circular model ($e=0$) could potentially be preferred over an eccentric model (see below). As part of the robustness check, we also executed an \texttt{ultranest} retrieval 10 times for a circular orbit (see Supplementary Table~\ref{table:robustness_circular}). Similar to the eccentric model, the posterior distributions remain highly consistent. Notably, the variation in the mean value of $m_1$ has improved, with values ranging between 1.05--1.18\,\Msun. Here, $T_0$ denotes the reference time at which the body crosses the ascending node, implying that $T_0$ and $\Omega$ are perfectly correlated.

As an additional test to verify that \texttt{ultranest} has identified the global maximum, rather than a local maximum or saddle point, we performed a frequentist bootstrapping analysis to assess the robustness of the results. Bootstrapping involves resampling the observational data within their uncertainties and refitting the model to each resampled dataset. This generates an empirical distribution of the parameter estimates, with confidence intervals derived from the percentiles of this distribution. Importantly, bootstrapped confidence intervals reflect the variability in the data itself and differ in interpretation from Bayesian credible intervals, which are derived from the posterior distribution.

Unlike Bayesian intervals, which incorporate prior information and assumptions about the parameter space, bootstrapping intervals are \textit{prior-independent} and are more directly influenced by the data's variability. These intervals can often be larger, as they account for all potential sources of variation in the data. Additionally, bootstrapping does not rely on normality assumptions or the local quadratic approximation of the likelihood surface.

Given our model's complexity, with 11 free parameters and 18 observational constraints, we used 5,000 bootstrap samples to strike a balance between computational cost and resultant stability. The local optimization was performed using the \texttt{least squares} method provided by \texttt{scipy}~\cite{Scipy2020SciPy-NMeth}, which produced confidence intervals that were significantly smaller than those from the Bayesian posterior distribution. This suggests a sharply peaked likelihood function.

To further ensure that the global maximum was found, we conducted a second round of bootstrapping, this time resampling not only the observational data within their uncertainties but also varying the initial guess parameters within three sigma of the values listed in Supplementary Table~\ref{table:robustness}. Again, the resulting parameter bounds were tighter than those from the Bayesian posterior distribution, confirming that \texttt{ultranest} has likely identified the global maximum of the posterior distribution.

\paragraph{Sensitivity analysis of the covariance matrices:}
The \textit{Hipparcos} and \textit{Gaia} covariance matrices are provided in Eq.~\eqref{covmatrix_hip} and Eq.~\eqref{covmatrix_gaia}, respectively. The correlation coefficient, $\rho_{ij}$, between the $i$-th and $j$-th parameters is defined as
\begin{equation}
    \rho_{ij} = \frac{\Sigma_{ij}}{\sqrt{\Sigma_{ii} \Sigma_{jj}}}
\end{equation}
where $\Sigma_{ij}$ represents the covariance between the $i$-th and $j$-th parameters, and $\Sigma_{ii}$ and $\Sigma_{jj}$ are the variances of the $i$-th and $j$-th parameters, respectively. For the \textit{Gaia} data, the absolute values of the correlation coefficients range from 3.8\% to 37\%, while for the \textit{Hipparcos} data, they vary between 10\% and 34\%. In this analysis, we examine the impact of omitting the covariance matrices to assess how data correlations affect the results compared to independent sampling. Similar to the robustness tests that included the \textit{Hipparcos} and \textit{Gaia} covariance matrices, the runs excluding these matrices proved to be very stable, with posterior distributions remaining consistent across all iterations. The corner plot for the run with the highest marginalized likelihood is shown in Supplementary Fig.~\ref{fig:pi1_gru_corner_indep}, and the corresponding proper motion fit is provided in Supplementary Fig.~\ref{fig:pi1_gru_motion_indep}. The retrieved orbital parameters are listed in Supplementary Table~\ref{table:fit_sensitivity} and Supplementary Table~\ref{table:best_fit}. The impact of omitting the covariance matrices is minor, with all runs showing a slight reduction in the standard deviations for $\mu_\alpha^G$ and $\mu_\delta^G$.

\afterpage{
\begin{landscape}
\begin{table*}[htp]
\caption{\textbf{Priors and retrieved parameters for the $\boldsymbol{\pi^1}$~Gru system across various tests in the sensitivity analysis.} The first part lists the Bayesian prior for each parameter, the second part the mean and standard deviation of the retrieved parameters. In the second part, each row displays the results from a different sensitivity test.
The first row presents the outcome from the \texttt{ultranest} sampling for an eccentric orbit and the second row for a circular orbit ($e = 0$).
The third row lists the results for the independent sampling.
The fourth and fifth row display the outcomes from the prior sensitivity analysis, where the fourth row shows results with an alternative prior on the distance and the fifth row with a Gaussian prior on $M_1$. The following three rows present the results from the likelihood sensitivity study: the sixth row excludes the 2023 ALMA C10 data, the seventh row excludes the \textit{Gaia} data, and the eighth row excludes the {\textit{Hipparcos}} data. The last row displays the results for the approximation of the orbital solution discussed in Supplementary Sect.~\ref{Sec:approximation}. All corner plots are available in the Supplementary Information.}\label{table:fit_sensitivity}
\begin{center}
\small
\setlength{\tabcolsep}{.3mm}
\begin{tabular} {l|ccccccccccc}
\toprule
model & $m_1$ & $q$ & $a$ & $e$ & $T_0$ & $\omega$ & $\Omega$ & $i$ & $D$ & $\mu^G_\alpha$ & $\mu^G_\delta$ \\
 & [$M_\odot$] & & [au] & & [yr] & [$^\circ$] & [$^\circ$] & [$^\circ$] & [pc] & [mas yr$^{-1}$] & [mas yr$^{-1}$] \\
\midrule
prior & $\mathcal{U}(0.6,3.5)$ & $\mathcal{U}(0.2,5)$ & $\mathcal{U}(2,15)$ & $\mathcal{U}(0,0.95)$ & $\mathcal{U}(2009,2030)$ & $\mathcal{U}(0,360)$ & $\mathcal{U}(0,360)$ & $\mathcal{U}(0,360)$ & Gamma & $\mathcal{U}(42,46)$ & $\mathcal{U}(-20,-17)$ \\
 & & & & & & & & & ($k$=3,$L$=500) & & \\
\midrule
eccentric & 1.02$\pm$0.20 & 1.04$\pm$0.05 & 6.60$\pm$0.41 & 0.023$\pm$0.017 & 2026.75$\pm$3.16 & 101$\pm$98 & 94$\pm$23 & 14$\pm$8 & 174$\pm$9 & 45.212$\pm$0.168 & $-$18.773$\pm$0.068 \\
circular & 1.12$\pm$0.25 & 1.05$\pm$0.05 & 6.81$\pm$0.49 & 0 & 2016.39$\pm$1.18$^\dagger$ & $-^\dagger$ & 101$\pm$36 & 11$\pm$7 & 180$\pm$10 & 45.203$\pm$0.144 & $-$18.776$\pm$0.061 \\
 & & & & & & & & & & & \\
independent & 1.02$\pm$0.24 & 1.05$\pm$0.05 & 6.62$\pm$0.49 & 0.016$\pm$0.012 & 2019.44$\pm$4.42 & 165$\pm$120 & 100$\pm$29 & 14$\pm$8 & 174$\pm$10 & 45.131$\pm$0.052 & $-$18.751$\pm$0.036 \\
 & & & & & & & & & & & \\
prior $D$ & 1.17$\pm$0.22 & 1.04$\pm$0.05 & 6.92$\pm$0.41 & 0.021$\pm$0.015 & 2022.93$\pm$5.58 & 84$\pm$95 & 100$\pm$33 & 14$\pm$7 & 180$\pm$9 & 45.244$\pm$0.156 & $-$18.792$\pm$0.068 8\\
prior $m_1$ & 1.27$\pm$0.22 & 1.03$\pm$0.05 & 7.10$\pm$0.40 & 0.025$\pm$0.015 & 2028.18$\pm$2.77 & 42$\pm$50 & 97$\pm$27 & 15$\pm$7 & 182$\pm$8 & 45.255$\pm$0.156& $-$18.802$\pm$0.063 \\
 & & & & & & & & & & & \\
no C10 & \multicolumn{10}{l}{no stable solution found} & \\
no \textit{Gaia} & 1.25$\pm$0.45 & 1.11$\pm$0.06 & 6.94$\pm$0.82 & 0.093$\pm$0.084 & 2019.77$\pm$4.61 & 268$\pm$96 & 118$\pm$56 & 25$\pm$11 & 176$\pm$16 & 45.315$\pm$0.166& $-$18.859$\pm$0.111 \\
no \textit{Hipparcos} & 1.17$\pm$0.31 & 1.01$\pm$0.07 & 6.88$\pm$0.59 & 0.045$\pm$0.025 & 2022.49$\pm$5.37 & 177$\pm$122 & 73$\pm$48 & 13$\pm$8 & 177$\pm$12 & 45.468$\pm$0.319 & $-$19.197$\pm$0.273 \\
 & & & & & & & & & & & \\
approximate & 0.68$\pm$0.06 & 1.19$\pm$0.07 & 6.13$\pm$0.18 & 0.03$\pm$0.018 & 2019.92$\pm$4.31 & 218$\pm$142 & 145$\pm$7 & 37$\pm$4 & 155$\pm$5 & 45.024$^\ddagger$ & $-$18.233$^\ddagger$\\
\bottomrule
\end{tabular}
\end{center}\leavevmode\\
$^\dagger$ For a circular orbit, $\omega$ and $T_0$ are undefined. Here, $T_0$ represents the reference time at which the body crosses the ascending node.\\
$^\ddagger$ fixed parameter, taking the value of the mean long-term proper motion vector, $\bmu^{\rm long}$, as derived from the \textit{Gaia} and \textit{Hipparcos} astrometric positions~\cite{Kervella2022A&A...657A...7K}.
\end{table*}
\end{landscape}

\begin{landscape}
\begin{table*}[htp]
\caption{\textbf{Retrieved best-fit parameters for the $\boldsymbol{\pi^1}$~Gru system for the various tests presented in the sensitivity analysis.}
Rows have the same meaning as for Supplementary Table~\ref{table:fit_sensitivity}, but we here list the retrieved best-fit parameters and add two extra columns listing the maximum of the logarithm of the marginalized likelihood $\log z$ and the normalized root mean square error $L_{\rm NRMSE}$.}\label{table:best_fit}
\begin{center}
\small
\setlength{\tabcolsep}{1.5mm}
\begin{tabular} {l|ccccccccccc|c|c}
\toprule
model & $m_1$ & $q$ & $a$ & $e$ & $T_0$ & $\omega$ & $\Omega$ & $i$ & $D$ & $\mu^G_\alpha$ & $\mu^G_\delta$ & $\max(\log z$) & $L_{\rm NRMSE}$\\
  & [$M_\odot$] & & [au] & & [yr] & [$^\circ$] & [$^\circ$] & [$^\circ$] & [pc] & [mas yr$^{-1}$] & [mas yr$^{-1}$] & & \\
\midrule
eccentric  & 0.89 & 1.06 & 6.35 & 0.031 & 2028.58 & 4.09 & 107.97 & 15.76 & 169.38 & 45.308  & $-$18.805 & 113.937 & 2.16 \\
circular & 0.93 & 1.05 & 6.46  & 0 & 2016.72$^\dagger$ & $-^\dagger$ & 111.41 & 14.04 & 172.28 & 45.149 & $-$18.740 & 112.739  & 2.04 \\
 &  &  &  &  &  &  &  &  &  &  &  &  & \\
independent & 0.91  & 1.04 & 6.39 & 0.019 & 2016.37 & 358.33 & 102.27 & 18.26 & 167.86 & 45.143 & $-$18.742 & 179.583 & 2.11\\
 &  &  &  &  &  &  &  &  &  & & & & \\
prior $D$   & 0.96 & 1.04 & 6.48 & 0.023 & 2028.45 & 6.24 & 102.41 & 16.71 & 171.44 & 45.288 &  $-$18.806 & 113.834 & 2.13 \\
prior $m_1$  & 1.02 & 1.05 & 6.63 & 0.032 & 2028.66 & 7.58 & 107.30 & 20.72 & 173.01 & 45.248& $-$18.798 & 113.734 & 2.09 \\
 &  &  &  &  &  &  &  &  &  & & &  & \\
no C10 & \multicolumn{10}{l}{no stable solution found} & & \\
no \textit{Gaia} & 1.15& 1.05 & 6.93 & 0.160 & 2028.92 & 113.55 & 0.65 & 48.08 & 164.67 & 45.606 & $-$19.025 & 115.181 & 10.72\\
no \textit{Hipparcos} & 1.02  & 0.94 & 6.54 & 0.068  & 2029.63 & 113.83 & 27.09& 14.79 & 166.21 & 45.751 & $-$19.410 & 120.613 & 7.20 \\
 &  &  &  &  &  &  &  &  &  & & &  & \\
approximate$^\ddagger$  & 0.61 & 1.24 & 5.97 & 0.039 & 2017.71 & 355.76 & 149.20 & 40.73 & 147.78 & $\mu_{HG,\alpha_1}$ & $\mu_{HG,\delta_1}$ & 145.551 &  68.14\\
\bottomrule
\end{tabular}
\end{center}\leavevmode\\
$^\dagger$ For a circular orbit, $\omega$ and $T_0$ are undefined. Here, $T_0$ represents the reference time at which the body crosses the ascending node.\\
$^\ddagger$ (OC12), (OC13), (OC17), and (OC18) have been replaced here by their alternative formulations (OC12c), (OC13c), (OC17c), and (OC18c), with the predictables calculated using the approximate orbital solution described in Supplementary Sect.~\ref{Sec:approximation}. The large value for $L_{\rm NRMSE}$ is caused by the large difference in observed and predicted position angle for the tangential velocity anomaly at the \textit{Hipparcos} epoch when using the approximate method.
\end{table*}
\end{landscape}}

\paragraph{Eccentric orbit:}

Our default \texttt{ultranest} run indicates a small eccentricity of $e = 0.023 \pm 0.017$. To evaluate the significance of this eccentricity, we perform a Bayesian comparison between our default model (Model $\mathcal{M}_e$), which allows for an eccentric orbit, and an alternative model (Model $\mathcal{M}_c$) that assumes a perfectly circular orbit ($e = 0$). The circular orbit model has two fewer free parameters ($e, \omega$), making it simpler. For a model with circular orbit, $T_0$ changes its meaning and represents the reference time at which the body crosses the ascending node. The principle behind this Bayesian comparison is to prefer simpler models unless a more complex model provides a significantly better fit to the data~\cite{Jeffreys1961}. Although a model with more free parameters will generally yield a better fit, the Bayesian evidence takes into account the trade-off between model complexity and goodness of fit~\cite{MacKay2003}. Thus, the more complex model is only favored if the improvement in fit quality is substantial enough to justify the increased complexity, overcoming the so-called Occam's razor penalty~\cite{Kass1995}.

To quantify the evidence in favor of one model over another, we compute the ratio of the posteriors:
\begin{eqnarray}
    \frac{P(\mathcal{M}_c \mid \mathcal{D})}{P(\mathcal{M}_e \mid \mathcal{D})} = \frac{P(\mathcal{D} \mid \mathcal{M}_c)}{P(\mathcal{D} \mid \mathcal{M}_e)} \cdot \frac{P(\mathcal{M}_c)}{P(\mathcal{M}_e)} \,.
\end{eqnarray}
The first factor
\begin{equation}
B = \frac{P(\mathcal{D} \mid \mathcal{M}_c)}{P(\mathcal{D} \mid \mathcal{M}_e)} = \frac{z(\mathcal{M}_c)}{z(\mathcal{M}_e)}
\end{equation}
is the Bayes factor, defined as the ratio of the marginalized likelihoods of the two models~\cite{Kass1995}. Here $ P(\mathcal{D} \mid \mathcal{M}_e) $ and $ P(\mathcal{D} \mid \mathcal{M}_c) $ are the marginal likelihoods of the data $ \mathcal{D} $ given models $\mathcal{M}_e$ and $\mathcal{M}_c$, respectively. The second factor $P(\mathcal{M}_e)/P(\mathcal{M}_c)$ is the prior odds ratio, which we assume to be 1 in the absence of prior information.

The strength of the evidence is interpreted using established benchmarks, such as those proposed by Jeffreys~\cite{Jeffreys1961}, where $1 < B < 3$ indicates weak evidence, $3 < B < 10$ indicates moderate evidence, and $B > 10$ signifies strong evidence in favor of a particular model. Similarly, Kass \& Raftery~\cite{Kass1995} provide slightly refined categories: $1 < B < 3.2$ is considered not worth more than a bare mention, $3.2 < B < 10$ indicates substantial evidence, $10 < B < 100$ signifies strong evidence, and $B > 100$ is classified as decisive.

In our case, the marginalized log-evidence values are $\log z(\mathcal{M}_e) = 75.192$ and $\log z(\mathcal{M}_c) = 78.805$, with the lower value for the eccentric model ($\mathcal{M}_e$) primarily driven by difficulties in properly sampling $\omega$ and $T_0$ for a very low eccentricity orbit. The Bayes factor, calculated as $z(\mathcal{M}_c) / z(\mathcal{M}_e) \approx 37$, provides evidence in favor of the circular model ($\mathcal{M}_c$).

A useful calibration between the Bayes factor and the frequentist measure of confidence is provided by Ref.~\cite{Benneke2013ApJ...778..153B}. Specifically, a Bayes factor of $B = 37$ corresponds to a significance level of $3.6\sigma$ for detecting a circular orbit instead of an eccentric one.

We still note that the maximum of the log-likelihood is higher for the eccentric than for the circular orbital model (see Supplementary Table~\ref{table:best_fit}). In that sense, the parameters of the eccentric model provide a slightly better fit to the observed data, but this comes at the cost of greater model complexity and poorly constrained parameters, such as $\omega$ and $T_0$, due to the very low eccentricity.

\paragraph{Sensitivity analysis of priors:}
For all input parameters, we have used an agnostic (flat) prior with the exception of the distance for which a Gamma prior was used. Changing the $L$ value from 500\,pc to 700\,pc does not bear any statistically different outcomes on the inferred parameters, see Supplementary Table~\ref{table:fit_sensitivity} and Supplementary Table~\ref{table:best_fit}. The corner plot is shown in Supplementary Fig.~\ref{fig:pi1_gru_corner_ultranest_prior_D} and the proper motion fit is displayed in Supplementary Fig.~\ref{fig:pi1_gru_motion_prior_D}. We also tested a broader range for the uniform prior on $T_0$ to explore the possibility of larger orbital periods. However, the results consistently returned an orbital period of $\sim$11.8 years. Hence, for all runs presented in this paper, we restricted the prior range to 2009\,--\,2030.

The only prior for which we can make another justified choice of prior is the mass of the primary, $m_1$, using our knowledge that $\pi^1$~Gru~A is an intrinsic S-type AGB star with a C/O ratio near unity~\cite{VanEck1998A&A...329..971V} of variability type SRb with a pulsation period of approximately 195 days (see Sect.~\ref{Sec:stellar_evolution}). Stellar evolution models coupled with linear pulsation models yield an initial mass estimate for $\pi^1$~Gru~A of 1.5\,--\,2\,\Msun; see Sect.~\ref{Sec:stellar_evolution}.

During the RGB and AGB phase, the mass of the star decreases due to mass loss from the stellar wind. For $\pi^1$~Gru~A the current AGB mass-loss rate is estimated around $\dot{M} = 7.7 \times 10^{-7}$ \Msun~yr$^{-1}$~\cite{Doan2017A&A...605A..28D}. For stars within this initial mass range, the TP-AGB phase lasts around 2 Myr~\cite{Karakas2014MNRAS.445..347K}. However, in order for the star to be an intrinsic S-type star, the originally O-rich AGB star should have experienced only a few third dredge-up (TDU) events to avoid becoming a carbon-rich star, with the period between TDU events being mass-dependent. Stellar evolution models predict a mass around 1.4\,\Msun\ close to the tip of the AGB (Sect.~\ref{Sec:stellar_evolution})

Therefore, we conducted another \texttt{ultranest} sampling with a Gaussian prior for $m_1$, using a mean value of 1.5\,\Msun\ and a standard deviation of 0.5\,\Msun. The retrieval results are summarized in Supplementary Table~\ref{table:fit_sensitivity} and Supplementary Table~\ref{table:best_fit}. The corner plot is shown in Supplementary Fig.~\ref{fig:pi1_gru_corner_ultranest_prior_M1} and the proper motion fit is displayed in Supplementary Fig.~\ref{fig:pi1_gru_motion_prior_M1}. Across the four executions, the posterior mean ranges from 1.2 to 1.27\,\Msun, with a best-fit value of 1.27\,\Msun\ and a standard deviation of 0.22\,\Msun.  Compared to the agnostic prior retrieval ($1.12 \pm 0.25\,\Msun$), the Gaussian prior results in a higher estimate for $m_1$ ($1.27 \pm 0.22\,\Msun$), though both values remain consistent within their respective uncertainty intervals. Notably, the TP-AGB mass estimate of $\sim 1.4\,\Msun$ lies near the upper limit of the agnostic prior retrieval and aligns more closely with the Gaussian prior estimate.

This result highlights two aspects. First, if greater trust is placed in the agnostic $m_1$ retrieval ($m_1 \sim 1.12\,\Msun$), the observational result can serve as a critical benchmark for stellar evolution and linear pulsation models by providing an opportunity to calibrate numerical parameters used in the simplified description of complex processes, such as convection and the treatment of convective boundaries. On the other hand, if the Gaussian $m_1$ retrieval ($m_1 \sim 1.27\,\Msun$) is deemed more reliable, it underscores the significance of prior knowledge in Bayesian retrievals, particularly for parameters like $m_1$. Given these considerations, we will analyze the orbital evolution for both $m_1$ values -- those derived from the agnostic prior and the Gaussian prior -- throughout the remainder of this study.

\paragraph{Sensitivity analysis of likelihood components:}

To determine which components of the likelihood function are the primary drivers of the solution, we performed three additional \texttt{ultranest} samplings that excluded specific datasets: (1)~the 2023 ALMA C10 data, (2)~the \textit{Gaia} data, and (3)~the {\textit{Hipparcos}} position. This analysis helps to elucidate the relative impact of each dataset on the overall fit and the resulting parameter estimates; the outcomes are tabulated in Supplementary Table~\ref{table:fit_sensitivity} and Supplementary Table~\ref{table:best_fit}. The corner plots are shown in Supplementary Fig.~\ref{fig:pi1_gru_corner_no_C10_run1}--\ref{fig:pi1_gru_corner_no_HIP} and the proper motion fits are displayed in Supplementary Figs.~\ref{fig:pi1_gru_motion_no_C10}--\ref{fig:pi1_gru_motion_no_HIP}.

\noindent $\bullet$ \underline{Exclusion of ALMA C10 data:} The dataset with the most significant impact on our inferred system parameters is the 2023 ALMA C10 data. Without this dataset, the \texttt{ultranest} retrieval becomes highly unstable. As shown in the corner plots in Supplementary Fig.~\ref{fig:pi1_gru_corner_no_C10_run1}--\ref{fig:pi1_gru_corner_no_C10_run4}, each run converges on a different solution space, with some displaying distinct solution islands. Occasionally, these islands overlap with those from other runs, suggesting that the fitting process is finding local maxima or saddle points rather than the global posterior maximum. A notable example is an island around $e=0.4$, which appears prominently in \texttt{run a2}, but is shown by \texttt{run a4} to be merely a local maximum or saddle point.

Throughout the four runs, various retrieved parameters show large variations. For example, the best-fit value of $m_1$ varies between 0.75\,--\,2.07\,\Msun, $q$ between 0.68\,--\,1.14, $a$ between 6.61\,--\,7.92\,au, and $e$ between 0.033\,--\,0.425. While the parameters retrieved from \texttt{run a1} and \texttt{run a3} are closer to the values from the default eccentric run, the maximum likelihood of \texttt{run a2} is higher than that of these two runs. The predicted proper motion for \texttt{run a2} is shown in Supplementary Fig.~\ref{fig:pi1_gru_motion_no_C10}. This figure shows that the ALMA C6, \textit{Gaia}, and \textit{Hipparcos} observables are well predicted, but, as expected, the ALMA C10 data are poorly predicted without being included in the fit. In conclusion, the inclusion of the new ALMA C10 data is critical for accurately deriving the orbital parameters of the $\pi^1$~Gru system.

The results discussed here are not unexpected. Excluding the ALMA 2023 C10 data (or similarly, the ALMA 2019 C6 data) leaves only two direct observational constraints on $M_2$ being the astrometric position in right ascension and declination at the remaining ALMA epoch and an indirect observational constraint from the difference in proper motion of $M_1$ derived from the \textit{Gaia} and \textit{Hipparcos} measurements. These limited constraints are insufficient to robustly estimate the key orbital parameters -- $q$, $a$, $e$, and consequently $T_0$ and $\omega$ -- which are directly tied to $m_2$ through the dynamics of the binary system. This lack of constraints leads to an underdetermined system, where the number of parameters exceeds the available observational constraints. Consequently, this increases the risk of overfitting, as the retrievals may adjust parameters to fit the limited data, resulting in posterior distributions that are not physically meaningful.

Inspired by the Gaussian prior for $m_1$ in runs that included all datasets (see above), we conducted an additional set of four retrievals using a Gaussian prior for $m_1$, with a mean value of 1.5\,\Msun\ and a standard deviation of 0.5\,\Msun. The aim was to assess whether this approach could stabilize the solution and provide confidence in reliably retrieving the four key parameters essential for understanding the system's stellar evolution: $m_1$, $q$, $a$, and $e$. The retrieved mean value of $m_1$ remained within the bounds of 1.06\,--\,1.27\,\Msun, $q$ between 1.04\,--\,1.06, $a$ between 6.68\,--\,7.10\,au, and $e$ between 0.017\,--\,0.021. Hence, for this particular retrieval using a Gaussian prior for $m_1$ alone seems sufficient to stabilize the retrieval process and provide more reliable estimates for these critical orbital parameters.

\noindent $\bullet$ \underline{Exclusion of \textit{Gaia} data:} The dataset with the second highest impact on the retrieval outcomes, after both ALMA datasets, is the \textit{Gaia} data. This is not surprising given the much higher accuracy of the more recent \textit{Gaia} measurements compared to the older \textit{Hipparcos} data. The four runs that exclude the \textit{Gaia} data show similar results, one of which is listed in Supplementary Table~\ref{table:fit_sensitivity} and Supplementary Table~\ref{table:best_fit}, with the corresponding corner plot shown in Supplementary Fig.~\ref{fig:pi1_gru_corner_no_GAIA}. The proper motion fit, depicted in Supplementary Fig.~\ref{fig:pi1_gru_motion_no_GAIA}, demonstrates how a `good' fit can still be achieved, adequately fitting the ALMA C6, ALMA C10, and \textit{Hipparcos} data, but with the trajectory of $M_1$ failing to pass through the \textit{Gaia} data points. Excluding the \textit{Gaia} dataset affects all orbital elements, although with overlapping confidence intervals. In general, the retrieved values for $m_1$, $q$, $a$ and $e$ are higher compared to the default eccentric model.

\noindent $\bullet$ \underline{Exclusion of \textit{Hipparcos} data:} The dataset with the least significant impact on the retrieval outcomes of the orbital parameters is the \textit{Hipparcos} data. In a similar manner to the sensitivity tests where the \textit{Gaia} data were excluded, the four runs that omit the \textit{Hipparcos} data produce consistent results. One of these is listed in Supplementary Table~\ref{table:fit_sensitivity} and Supplementary Table~\ref{table:best_fit}, with the corresponding corner plot shown in Supplementary Fig.~\ref{fig:pi1_gru_corner_no_HIP}. Notably, the retrieval indicates a small, non-zero, eccentricity of $e = 0.045\pm0.025$. The proper motion fit, shown in Supplementary Fig.~\ref{fig:pi1_gru_motion_no_HIP}, illustrates how a reasonable fit can still be achieved, adequately reproducing the ALMA C6, ALMA C10, and \textit{Gaia} astrometric positions and fitting the \textit{Gaia} proper motion and tangential velocity anomaly. However, it fails to fit the \textit{Hipparcos} proper motion and tangential velocity anomaly.

While the \textit{Hipparcos} data have minimal impact on the orbital parameters, their influence on the proper motion of the barycentre $\bmu^G$ is the most significant among all sensitivity tests. Excluding the \textit{Hipparcos} data results in a higher value for $\mu^G_\alpha$ and a lower value for$\mu^G_\delta$, although still within the uncertainty intervals of the nominal \texttt{eccentric} model. The strength of the \textit{Hipparcos} data lies in its ability to complement the ALMA observations, spanning a time interval of over 30 years. This extended baseline allows the proper motion of the barycentre to be determined with high precision.

\subsection{Velocity in an elliptic orbit}\label{Methods:vrad}

For the two-body system described in Supplementary Sect.~\ref{Methods:pos}, the corresponding velocity components ($\dot{x}, \dot{y}, \dot{z}$) = ($v_x, v_y, v_z$) can be computed. Defining
\begin{equation}
    \left.\begin{aligned}
        c_1 & = \cos\Omega \cos\omega - \sin\Omega \sin\omega \cos i \\
        d_1 & = \sin\Omega \cos\omega + \cos\Omega \sin\omega \cos i \\
        e_1 & = \sin\omega \sin i
    \end{aligned}\right\}
\end{equation}
and
\begin{equation}
    \left.\begin{aligned}
        c_2 & = - \cos\Omega \sin\omega - \sin\Omega \cos\omega \cos i \\
        d_2 & = - \sin\Omega \sin\omega + \cos\Omega \cos\omega \cos i \\
        e_2 & = \cos\omega \sin i
    \end{aligned}\right\}
\end{equation}
the orbital velocity components in the stellar system barycentric reference frame can be written as
\begin{equation}
    \left.\begin{aligned}
        v_x & = \frac{2 \pi}{T_{\rm orb}}\frac{a}{r} \left(b c_2 \cos E - a c_1 \sin E \right)\\
        v_y & = \frac{2 \pi}{T_{\rm orb}}\frac{a}{r} \left(b d_2 \cos E - a d_1 \sin E \right)\\
        v_z & = \frac{2 \pi}{T_{\rm orb}}\frac{a}{r} \left(b e_2 \cos E - a e_1 \sin E \right)
    \end{aligned}\right\}\ .\label{xyz_dot}
\end{equation}
The corresponding velocity vector $\vec{v}^{\,\prime}$\footnote{The time derivative of the parallactic shift is proportional to $d\varpi/dt = 1/D^2 dD/dt$ and is negligible in this context.} is given by
\begin{equation}
\begin{bmatrix}
v_x^\prime\\
v_y^\prime \\
v_z^\prime
\end{bmatrix}
=
\underbrace{
\left(\dot{R}(\alpha_G) \, R(\delta_G) + R(\alpha_G) \, \dot{R}(\delta_G)\right)
\begin{bmatrix}
x \\
y \\
z
\end{bmatrix}
}_{\text{TERM1}}
+
\underbrace{
R(\alpha_G) \, R(\delta_G)
\begin{bmatrix}
v_x \\
v_y \\
v_z
\end{bmatrix}
}_{\text{TERM2}}
+
\underbrace{
\begin{bmatrix}
v_{x,G}^\prime \\
v_{y,G}^\prime \\
v_{z,G}^\prime
\end{bmatrix}
}_{\text{TERM3}}\label{Eq:v_to_ICRS}
\end{equation}
with
\begin{equation}
\dot{R}(\alpha_G) =
\begin{bmatrix}
-\cos\alpha_G \, \mu_{\alpha,G} & -\cos\alpha_G \, \mu_{\alpha,G} & 0 \\
 \cos\alpha_G \, \mu_{\alpha,G} & -\sin\alpha_G \, \mu_{\alpha,G} & 0 \\
0 & 0 & 0
\end{bmatrix}
\end{equation}
and
\begin{equation}
\dot{R}(\delta_G) =
\begin{bmatrix}
-\cos\delta_G \, \mu_{\delta,G} & 0 & \sin\delta_G \, \mu_{\delta,G}\\
0              & 0 &        0      \\
-\sin\delta_G \, \mu_{\delta,G} & 0 & -\cos\delta_G \,\mu_{\delta,G}\\
\end{bmatrix} \, .
\end{equation}
and $\bmu^G$ being the proper motion vector of the barycentre. The second term in Eq.~\eqref{Eq:v_to_ICRS} represents the effect of orbital motion on the velocity vector, while the third term accounts for the velocity translation. For the $\pi^1$~Gru system under study, it can easily be calculated that the contribution of the first term is negligible. Hence,
\begin{equation}
\begin{bmatrix}
v_x^\prime\\
v_y^\prime \\
v_z^\prime
\end{bmatrix}
=
R(\alpha_G) \, R(\delta_G)
\begin{bmatrix}
v_x \\
v_y \\
v_z
\end{bmatrix}
+
\begin{bmatrix}
v_{x,G}^\prime \\
v_{y,G}^\prime \\
v_{z,G}^\prime
\end{bmatrix}\label{Eq:v_to_ICRS2}
\end{equation}
Writing the radial velocity of the barycentre as
\begin{equation}
    v_{{\rm rad},G}^\prime \equiv \frac{dD}{dt}
\end{equation}
and
\begin{align}
    v_{\alpha^\star,G}^\prime & = \mu_{\alpha_G^*} \times \frac{4.74047}{\varpi} \\
    v_{\delta,G}^\prime & = \mu_{\delta_G} \times \frac{4.74047}{\varpi}
\end{align}
the velocity translation term $\vec{v}^{\,\prime}_G$ is
\begin{equation}
\begin{bmatrix}
v_{x,G}^\prime\\
v_{y,G}^\prime \\
v_{z,G}^\prime
\end{bmatrix}
=
\begin{bmatrix}
\dot{x}_{G}^\prime\\
\dot{y}_G^\prime \\
\dot{z}_G^\prime
\end{bmatrix}
=
\begin{bmatrix}
v_{\rm rad,G}^\prime \cos \alpha_G \cos \delta_G - v_{\alpha^\star,G}^\prime \sin \alpha_G - v_{\delta,G}^\prime \cos \alpha_G \sin \delta_G \\
    v_{\rm rad,G}^\prime \sin \alpha_G \cos \delta_G + v_{\alpha^\star,G}^\prime \cos \alpha_G - v_{\delta,G}^\prime \sin \alpha_G \sin \delta_G \\
    v_{\rm rad,G}^\prime \sin \delta_G + v_{\delta,G}^\prime \cos \delta_G
\end{bmatrix}\,.
\end{equation}
The above equation can also be written as
\begin{equation}
\begin{bmatrix}
v_{x,G}^\prime\\
v_{y,G}^\prime \\
v_{z,G}^\prime
\end{bmatrix}
=
R(\alpha_G) \, R(\delta_G)
\begin{bmatrix}
v_{\delta,G}^\prime \\
v_{\alpha^\star,G}^\prime \\
-v_{{\rm rad},G}^\prime
\end{bmatrix} \,.
\end{equation}
Therefore, Eq.~\eqref{Eq:v_to_ICRS} can simply be written as
\begin{equation}
\begin{bmatrix}
v_x^\prime\\
v_y^\prime \\
v_z^\prime
\end{bmatrix}
=
R(\alpha_G) \, R(\delta_G)
\begin{bmatrix}
v_x + v_{\delta,G}^\prime\\
v_y + v_{\alpha^\star, G}^\prime\\
v_z - v_{{\rm rad},G}^\prime
\end{bmatrix}\label{Eq:v_to_ICRS3}
\end{equation}

Given the unit vector defined in Eq.~\eqref{Eq:elos}, the radial velocity can be computed using the dot product:
\begin{equation}
    v_{\rm rad}^\prime = \vec{v}^{\,\prime} \cdot \vec{e}_{\rm los} \,.
\end{equation}
Similarly, the unit vector along the right ascension is
\begin{equation}
    \vec{e}_{\alpha} = (-\sin\alpha_G, \cos\alpha_G, 0)\,
\end{equation}
and along the declination is
\begin{equation}
    \vec{e}_{\delta} = (-\cos\alpha_G \sin\delta_G, -\sin\alpha_G \sin\delta_G, \cos\delta_G)\,
\end{equation}
so that the velocity components along $\vec{e}_{\alpha}$ and $\vec{e}_{\delta}$ can be obtained as
\begin{align}
    v_{\alpha^\star}^\prime &= \vec{v}^{\,\prime} \cdot \vec{e}_{\alpha} \\
    v_{\delta}^\prime &= \vec{v}^{\,\prime} \cdot \vec{e}_{\delta} \,.
\end{align}
The corresponding tangential velocity $v_{\tan}^\prime$ and its position angle $\theta^\prime$ are then
\begin{align}
    v_{\tan}^\prime & = \sqrt{{v_{\alpha^\star}^\prime}^2 + {v_{\delta}^\prime}^2} \\
    \tan\theta^\prime & = \frac{v_{\alpha^\star}^\prime}{v_{\delta}^\prime} \,.
\end{align}

By representing $v_x$ as $v_\delta^{\rm orb}$, $v_y$ as $v_{\alpha^\star}^{\rm orb}$, and $v_z$ as $-v_{\rm rad}^{\rm orb}$, and subsequently applying the dot product with the corresponding unit vectors to Eq.~\eqref{Eq:v_to_ICRS3}, the following relations are obtained:
\begin{align}
    v_{\rm rad}^{\,\prime} &= v_{\rm rad}^{\rm orb} + v_{{\rm rad},G}^\prime\label{Eq:vrad_prime}\\
    v_{\alpha^\star}^{\,\prime} &= v_{\alpha^\star}^{\rm orb} + v_{{\alpha^\star},G}^\prime \\
    v_{\delta}^{\,\prime} &= v_{\delta}^{\rm orb} + v_{{\delta},G}^\prime
\end{align}
Therefore
\begin{align}
    v_{\tan}^\prime & = \sqrt{{(v_{\alpha^\star}^{\rm orb} + v_{\alpha^\star,G}^\prime)}^2 + {(v_{\delta}^{\rm orb} + v_{\delta,G}^\prime)}^2}\label{Eq:final_vtan}\\
    \tan\theta^\prime & = \frac{v_{\alpha^\star}^{\rm orb} + v_{\alpha^\star,G}^\prime}{v_{\delta}^{\rm orb} + v_{\delta,G}^\prime} \,.\label{Eq:final_theta}
\end{align}
Assuming that $\bmu^G=0$, such as when working in the orbital plane frame, and having access to only one epoch of data or to only relative orbital separations,
the tangential velocity and its position angle are simplified to the orbital contributions:
\begin{align}
    v_{\tan}^\prime & = \sqrt{{v_x}^2 + {v_y}^2}\label{Eq:approx_vtan} \\
    \tan\theta^\prime & = \frac{v_y}{v_x} \,.\label{Eq:approx_PA}
\end{align}

The orbital contributions to the radial velocity, $v_{\rm rad}^{\rm orb}$, and to the tangential velocity, $\sqrt{v_x^2 + v_y^2}$, are shown in Supplementary Fig.~\ref{Fig:vrad_vtan}. For an inclination of $360^\circ - i$ or a change of ($\omega, \Omega$) to their antipodal angles ($\omega+180^\circ, \Omega+180^\circ$) the sign of $v_{\rm rad}^{\rm orb}$ would be reversed (see the discussion on geometrical parameter degeneracies).

\subsubsection[Distinguishing between (omega, Omega) and its antipodal nodes using radial velocity measurements]{Distinguishing between ($\omega, \Omega$) and its antipodal nodes using radial velocity measurements}\label{suppinfSec:vrad}

\begin{figure}[!htp]
    \centering
    \includegraphics[width=\textwidth]{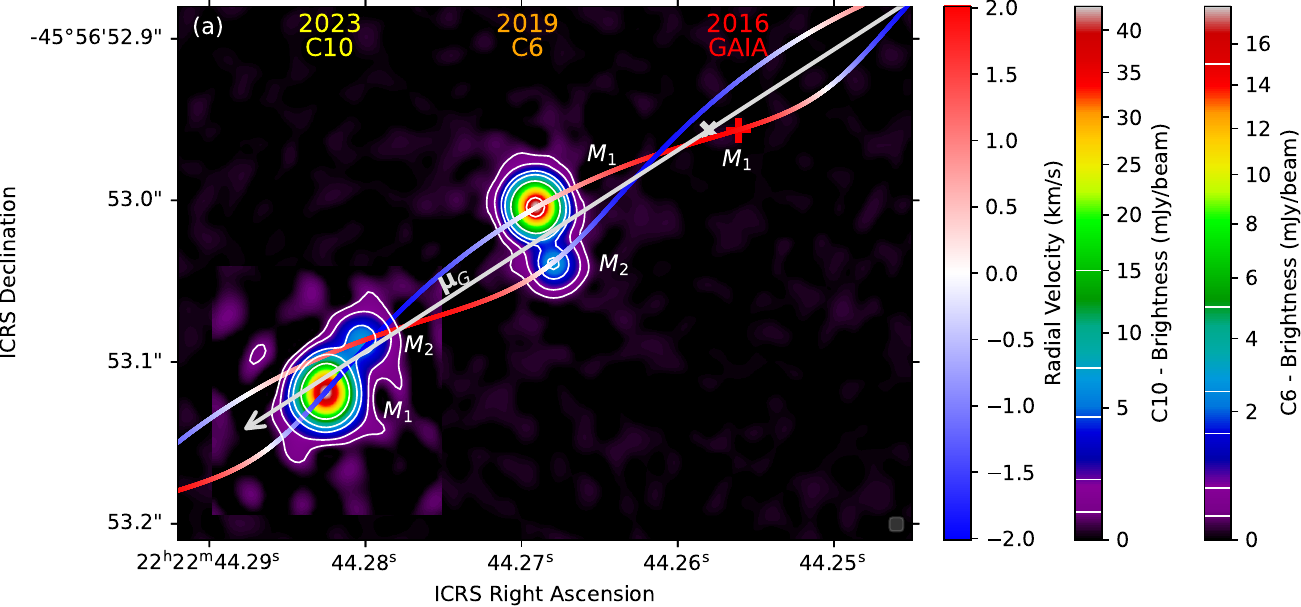}\\
    \includegraphics[width=\textwidth]{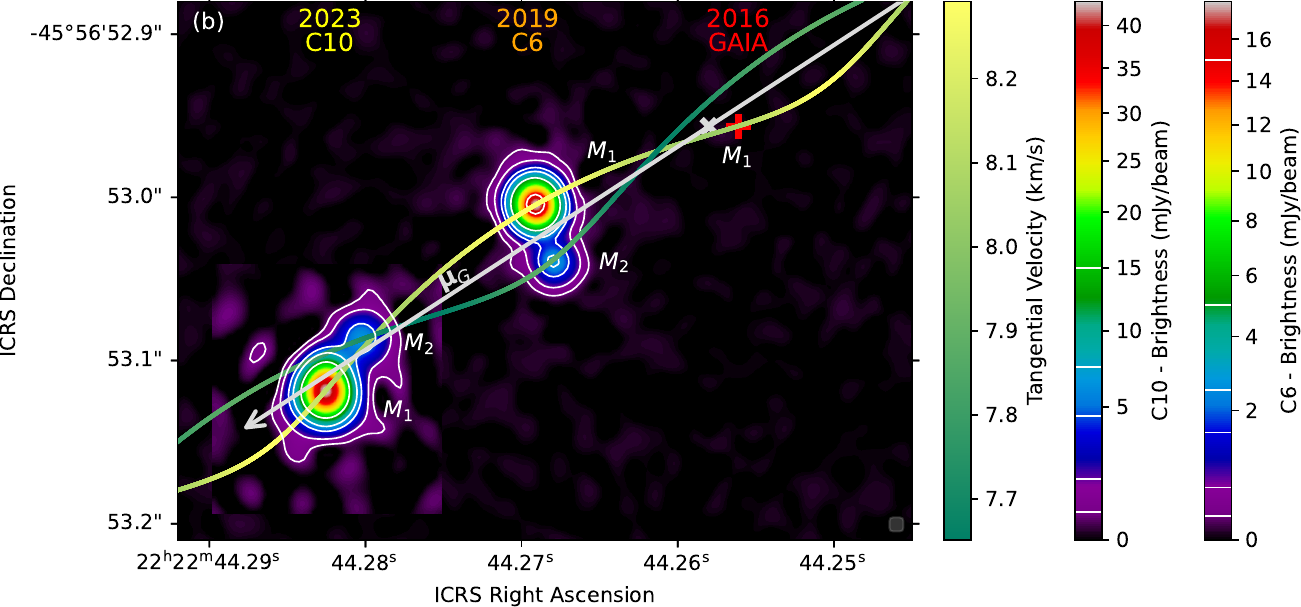}
    \caption{\textbf{Orbital contributions to the radial and tangential velocities of the $\boldsymbol{\pi^1}$~Gru system.} This figure shows the orbital trajectories of $M_1$ and $M_2$ with respect to the barycentre. In panel~(a), the trajectories are colour-coded to indicate the orbital contribution to the radial velocity, $v_{\rm rad}^{\rm orb}$, with red representing positive values and blue representing negative values.  In panel~(b), the trajectories are yellow-green colour-coded to show the orbital contribution to the tangential velocity, $\sqrt{v_x^2 + v_y^2}$, where the velocity magnitude is represented.}\label{Fig:vrad_vtan}
\end{figure}

\begin{figure}[!htp]
    \centering
    \begin{tikzpicture}
        \node at (0, 0) {\includegraphics[width=0.48\textwidth]{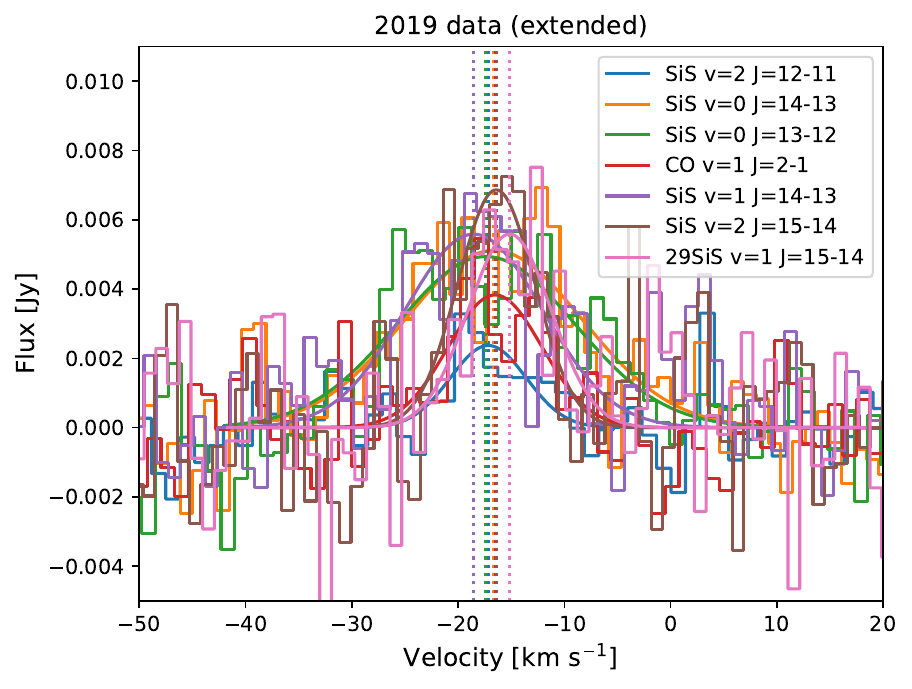}};
        \node at (-2.3, 2.2) {(a)};
        \hfill
        \node at (8.2, 0) {\includegraphics[width=0.48\textwidth]{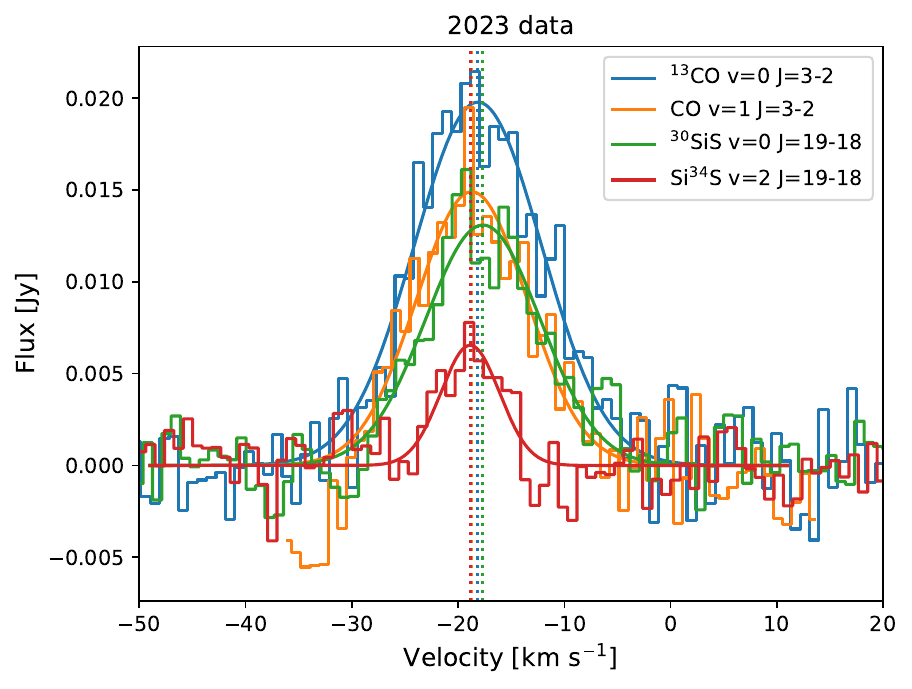}};
        \node at (6.0, 2.2) {(b)};
    \end{tikzpicture}
    \caption{\textbf{Determining the LSRK velocity of $M_1$.} The data are presented as histograms, with Gaussian fits to the line profiles shown in corresponding colours. Panel~(a) displays the central velocities at the ALMA C6 epoch, while panel~(b) shows the central velocities at the ALMA C10 epoch. For comparison, the low-excitation $^{13}$CO v=0 J=3-2 line is also included in panel~(b).}\label{Fig:central_velocities}
\end{figure}

\begin{figure}[!htp]
   \centering
    \begin{tikzpicture}
        \node at (-4.1, 0) {\includegraphics[width=0.48\textwidth]{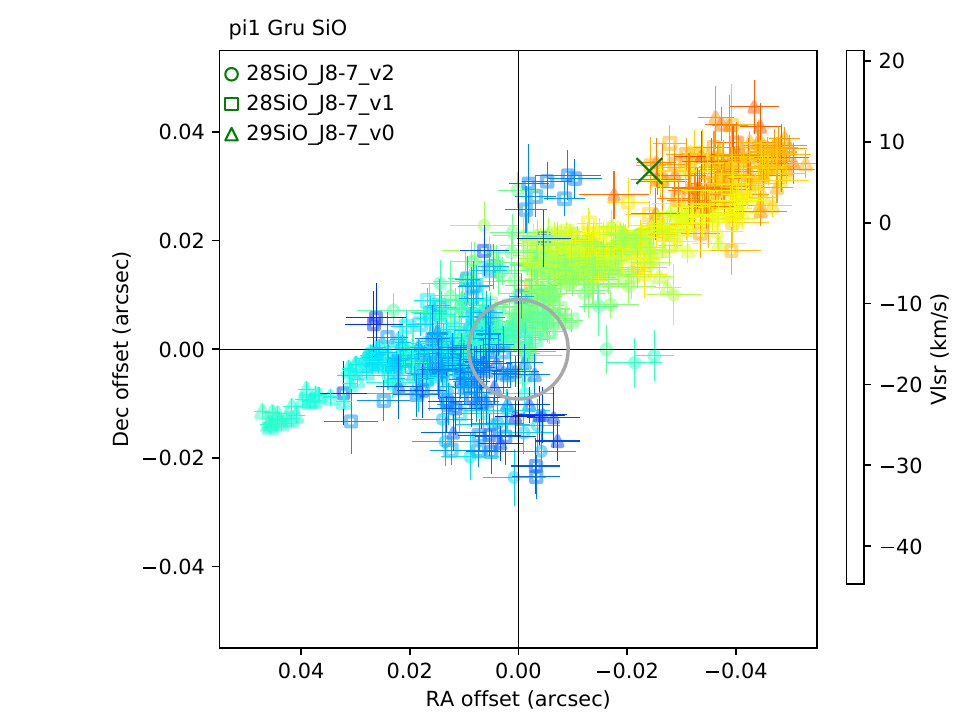}};
        \node at (-2., -1.8) {(a)};
        \hfill
        \node at (4.2,-0.1) {\includegraphics[width=0.49\textwidth]{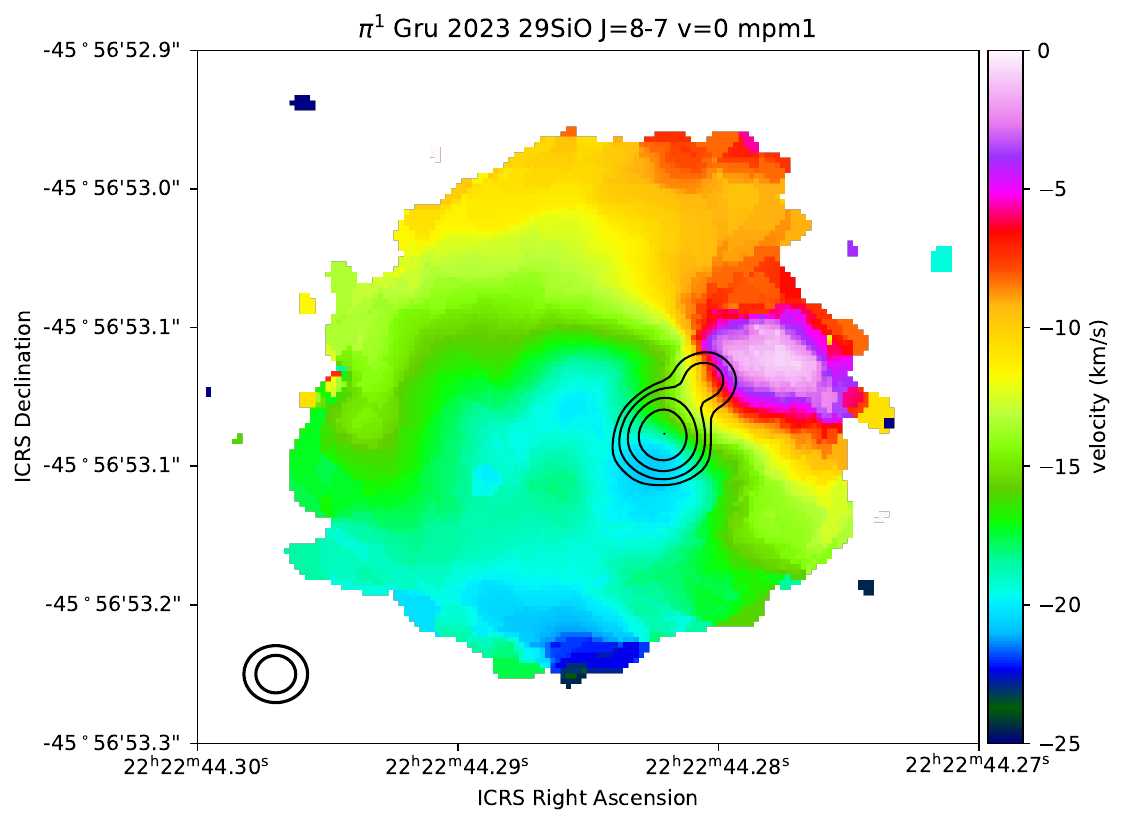}};
        \node at (6.5, -1.8) {{\color{black}{(b)}}};
        \node at (-4.5, -5.2) {\includegraphics[width=0.32\textwidth]{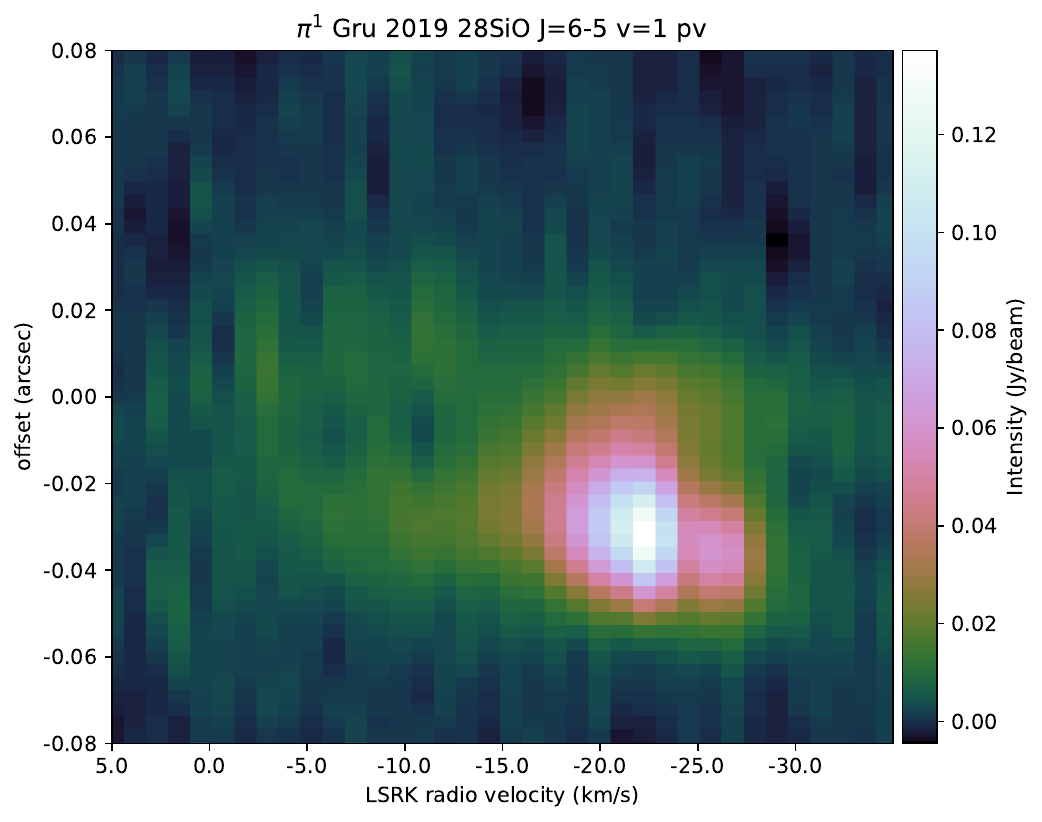}};
        \node at (-6.1, -3.7) {{\color{white}{(c)}}};
        \hfill
        \node at (0.3, -5.2) {\includegraphics[width=0.32\textwidth]{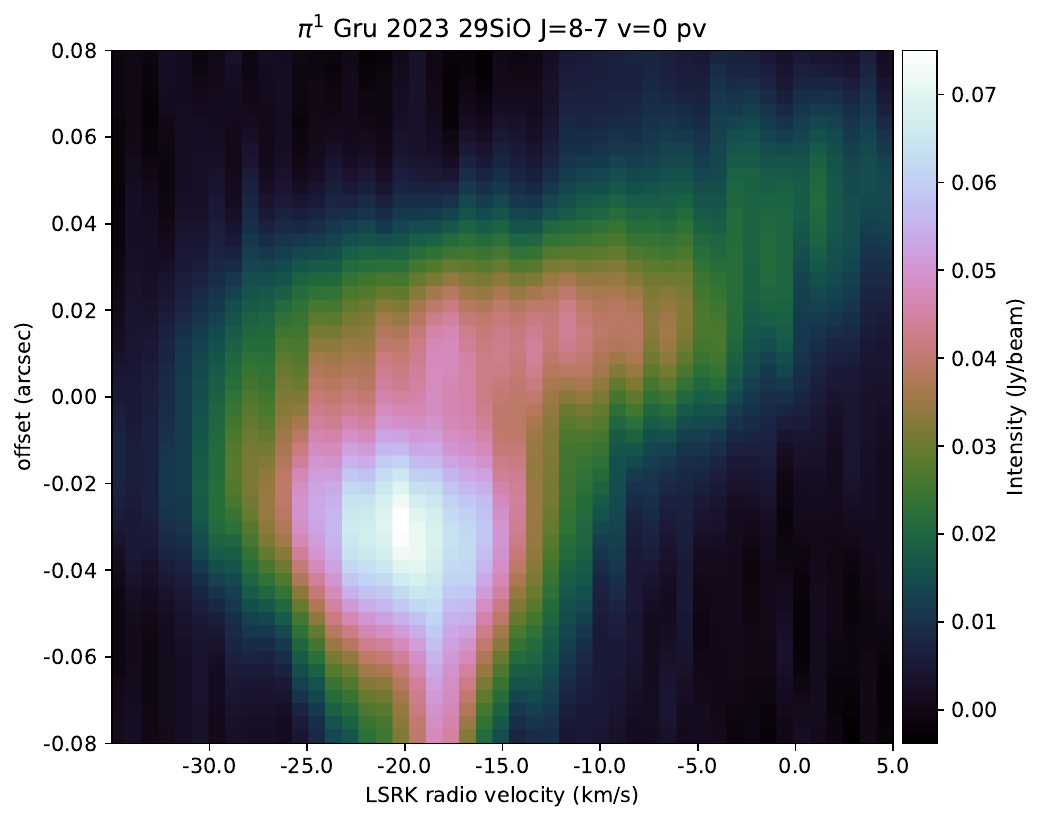}};
        \node at (-1.3, -3.7) {{\color{white}{(d)}}};
        \hfill
        \node at (5.4, -5.2) {\includegraphics[width=0.32\textwidth]{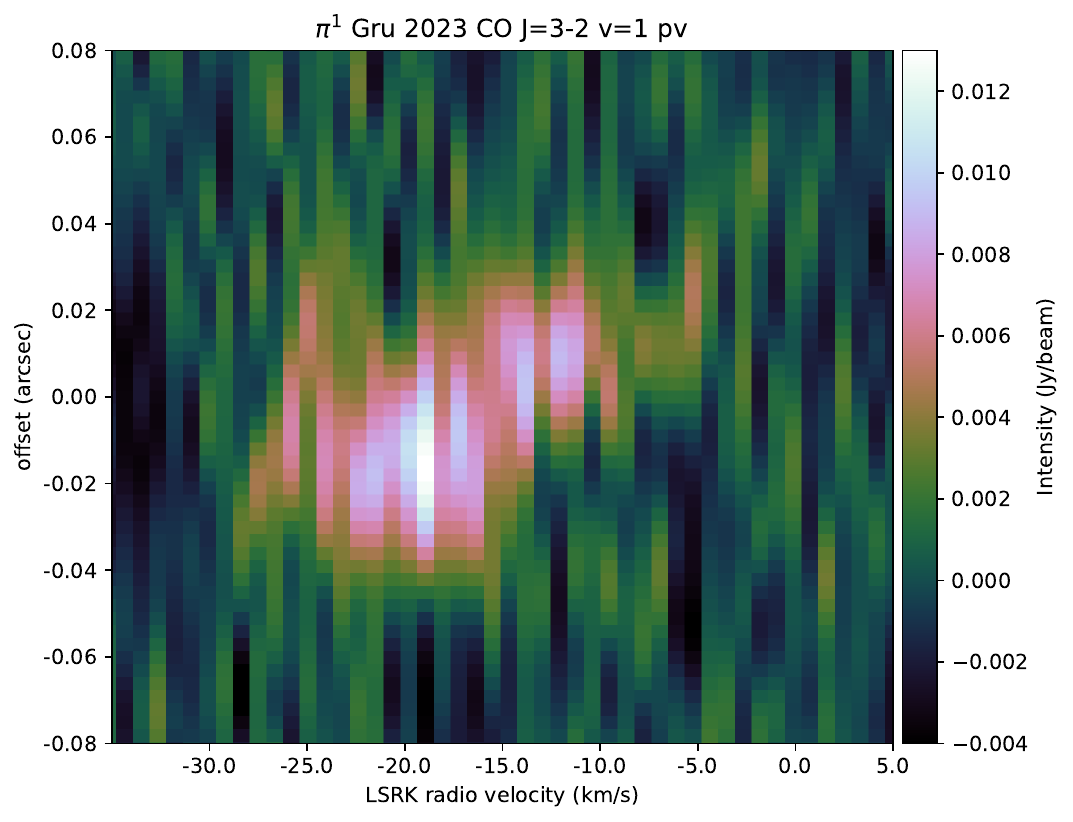}};
        \node at (3.8, -3.7) {{\color{white}{(e)}}};
    \end{tikzpicture}
    \caption{\textbf{Velocity structure around $\boldsymbol{\pi^1}$~Gru~A and $\boldsymbol{\pi^1}$~Gru~C in 2023 and 2019.} All velocities are in the LSRK reference frame.
Panel~(a): ALMA 2023 C10 position and relative radial velocity of the localized peak positions of the $^{28}$SiO v=0, 1, 2 J=8-7 emission lines. The gray circle denotes the size of the primary AGB star $\pi^1$~Gru~A, and the green cross marks the position of $\pi^1$~Gru~C.
Panel~(b): ALMA 2023 C10 moment-1 map of the $^{29}$SiO v=0 J=8-7 emission, with the white contours representing the continuum emission map.
Panel~(c): ALMA 2019 C6 PV diagram of the $^{28}$SiO v=1 J=6-5 line, using a 39~mas wide slit oriented through $M_1$ and $M_2$, with a position angle of 197.5$^\circ$. The primary is at an offset of 0\farcs0 with a velocity of approximately $-16.8$~km~s$^{-1}$, while the companion is at negative offsets (at $-$37.702~mas) with more blue-shifted velocities.
Panel~(d)~ALMA 2023 C10 position-velocity (PV) diagram of the $^{29}$SiO v=0 J=8-7 emission with a 39~mas wide slit oriented through $M_1$ and $M_2$, with a position angle of 140$^\circ$. The primary is located at an offset of 0\farcs0 with a velocity of approximately $-18.3$~km~s$^{-1}$, while the companion appears at positive offsets (toward the northwest, at 40.775~mas) with more red-shifted velocities.
Panel~(e): Similar to panel~(d), but showing the ALMA 2023 C10 $^{12}$CO v=1 J=3-2 PV diagram.
}\label{fig:pv_CO_2023}
\end{figure}
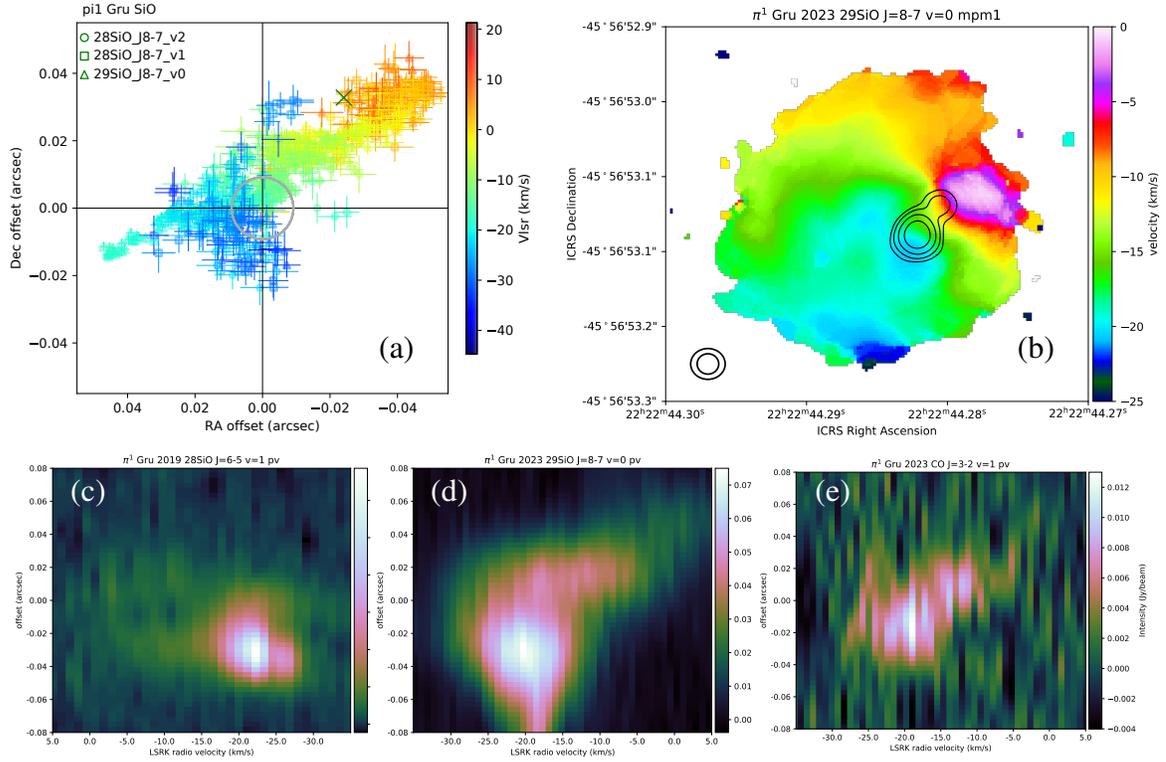

\subsection{Approximation of the orbital solution}\label{Sec:approximation}
\subsubsection{Approximation of the tangential velocity anomaly}\label{Methods:vtan_approx}

The expressions for the tangential velocity $v_{\tan}^\prime$ and its position angle $\theta^\prime$, as given in Eqs.~\eqref{Eq:final_vtan}--\eqref{Eq:final_theta}, differ from the tangential velocity anomaly $\Delta v_{\tan}$ and its position angle $\theta$, as defined in Eqs.~\eqref{Eq:vtan_anomaly}--\eqref{Eq:PA_vtan}. However, under the assumption that the observed $\bmu_{HG}$ represents the barycentre's proper motion $\bmu^G$ -- which implies that $m_1 \gg m_2$ and that the photocentre accurately follows the motion of $M_1$ -- we can express:
\begin{align}
    \Delta v_{\tan} &= \sqrt{{v_{\alpha^\star}^{\rm orb}}^2+{v_{\delta}^{\rm orb}}^2} = \sqrt{v_x^2 + v_y^2}, \\
    \tan\theta &= \frac{v_{\alpha^\star}^{\rm orb}}{v_{\delta}^{\rm orb}} = \frac{v_y}{v_x}.
\end{align}
Thus, $\Delta v_{\tan} = v_{\tan}^\prime$ and $\theta = \theta^\prime$ for the expressions of the tangential velocity and its position angle derived when $\bmu^G = 0$. Consequently, approximating
\begin{align}
    \Delta v_{\tan} &\simeq \sqrt{v_x^2 + v_y^2},\label{Eq:approx_vtan_anom}\\
    \tan\theta &\simeq \frac{v_y}{v_x}\label{Eq:approx_PA_vtan_anom}
\end{align}
could be a valid alternative for Bayesian retrieval fitting.

An additional note is in order here. Just as the observed proper motion derived from \textit{Gaia} and \textit{Hipparcos} data represents an average value over the missions' lifetimes, we model the tangential velocity anomaly and its position angle as averaged values over the approximately 3-year observing period. Given the orbital period of $\sim$12 years, the position angle of the orbital velocity changes significantly during this time. As a result, the averaged position angle $\theta$, and to a lesser extent the tangential velocity anomaly, differ from those at specific epochs, such as 2016.0 (\textit{Gaia}) or 1991.25 (\textit{Hipparcos}). For the eccentric \texttt{ultranest} best-fit results, the predicted position angle of the orbital motion of $M_1$ shifts from 315.3$^\circ$ at the start of the \textit{Gaia} observing period to 52.6$^\circ$ at the end, with an angle average value of 2.09$^\circ$ for an averaged tangential velocity of 8.03~km~s$^{-1}$. Although the instantaneous \textit{Gaia} values of $v^\prime_{\tan} = 7.63$~km~s$^{-1}$, $\theta^\prime = 1.88^\circ$ are closer to the predicted exact values of $\Delta v_{\tan} = 6.58$~km~s$^{-1}$ and $\theta = 1.86^\circ$, this does not imply that one should prefer the latter method, as the exact solutions are based on averaged proper motion values over the duration of the respective missions. The closer agreement between the exact and approximate instantaneous values in this case may simply be coincidental.

Using this approximation and the orbital parameters derived from the best-fit circular results of \texttt{ultranest}, the predicted tangential velocity anomaly during the \textit{Gaia} observing period is $\Delta v_{\rm tan} = 8.11$~km~s$^{-1}$ with a position angle of $\theta^\prime = 358.51^\circ$ (see Supplementary Fig.~\ref{Fig:approximate_vtan_xy}). At the \textit{Hipparcos} epoch, the predicted anomaly is $\Delta v_{\rm tan} = 8.166$~km~s$^{-1}$ with a position angle of $\theta^\prime = 337.03^\circ$.

The effect of this approximation on the retrieval outcomes is discussed in Supplementary Sect.~\ref{Methods:approx_retrieval}.

\subsubsection{Approximation of the astrometric position}\label{Methods:approx_pos}

In a manner similar to the approximation of tangential velocity (anomaly) and its position angle derived in Eqs.~\eqref{Eq:approx_vtan_anom}--\eqref{Eq:approx_PA_vtan_anom}, we can also approximate the changes in right ascension, $\Delta \alpha$, and declination, $\Delta \delta$ given in Eqs.~\eqref{Delta_RA}--\eqref{Delta_DEC}. By taking the dot product of the position vector $ \vec{r}^{\,\prime} = {(x^\prime, y^\prime, z^\prime)}_{\rm ICRS}$ with the unit vectors ($\vec{e}_{\alpha}$, $\vec{e}_{\delta}$, $\vec{e}_{\rm los}$), we derive the following expressions:
\begin{align}
    \vec{r}^{\,\prime} \cdot \vec{e}_{\delta}  & = x \\
    \vec{r}^{\,\prime} \cdot \vec{e}_{\alpha}  & = y \\
    \vec{r}^{\,\prime} \cdot \vec{e}_{\rm los} & = -z + D\label{Eq:contrib_xyz}
\end{align}
This indicates that the $x$-component contributes to changes in declination, the $y$-component to changes in right ascension, and the $z$-component to changes along the radial direction. This leads to
\begin{align}
    \Delta \alpha_\star &\simeq \arcsin\left(\frac{y}{D}\right)\label{Eq:approx_Delta_alpha} \\
    \Delta \delta &\simeq \arcsin\left(\frac{x}{D}\right)\label{Eq:approx_Delta_delta}
\end{align}

For the retrieved best-fit parameters of the \texttt{circular} model, the difference between exact and approximate astrometric position is negligible, i.e.\ lower than $7\times 10^{-6}$\,mas; see Supplementary Fig.~\ref{Fig:approximate_vtan_xy}.

The combined effect of this minor approximation, along with the approximation of the tangential velocity anomaly and its position angle, on the retrieval outcomes is discussed below.

\begin{figure}[htp]
    \centering
    \includegraphics[width=\textwidth]{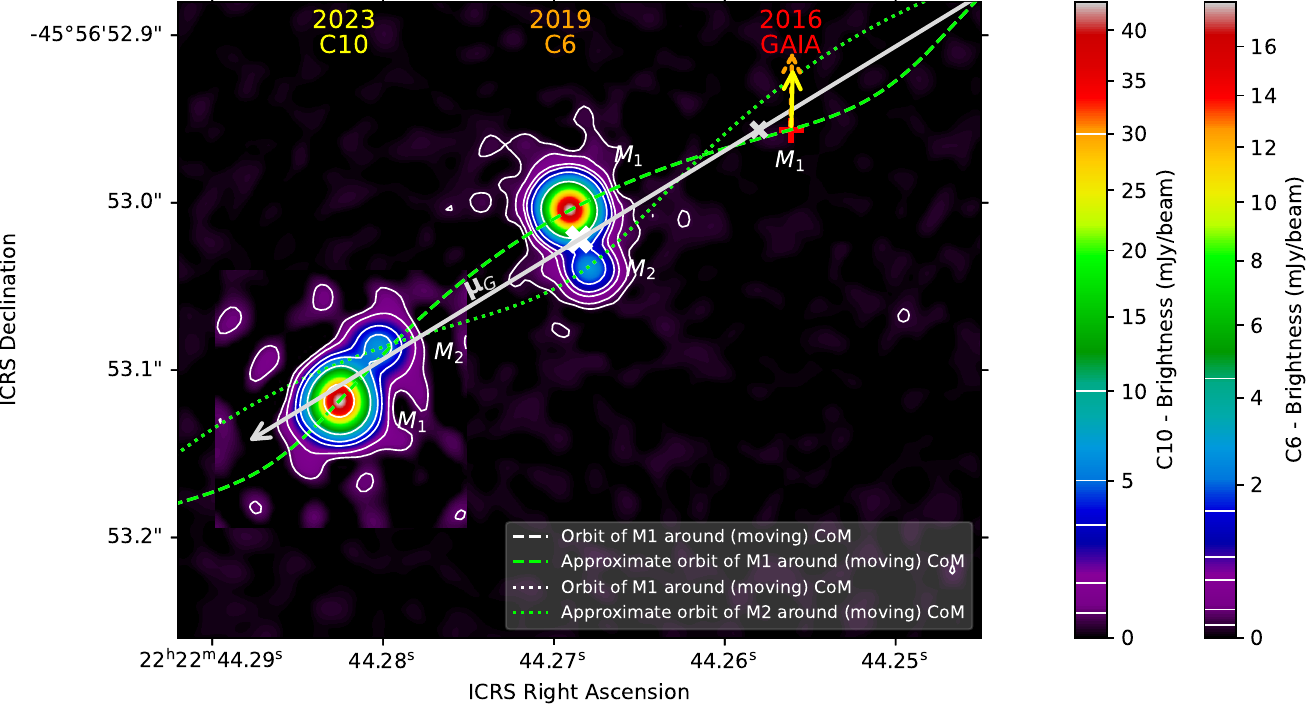}
    \caption{\textbf{Approximate proper motion of the $\boldsymbol{\pi^1}$~Gru system.} The predicted proper motion of the $\pi^1$~Gru system is based on the circular \texttt{ultranest} best-fit results. The changes in right ascension and declination are approximated using Eqs.~\eqref{Eq:approx_Delta_alpha}--\eqref{Eq:approx_Delta_delta} (shown in lime), while the exact solution, derived using Eqs.~\eqref{Eq:RA_ICRS}--\eqref{Eq:DEC_ICRS}, is represented in white. The tangential velocity anomaly and its position angle are approximated by Eqs.~\eqref{Eq:approx_vtan_anom}--\eqref{Eq:approx_PA_vtan_anom}. The dotted orange arrow represents the predicted averaged orbital tangential velocity over the \textit{Gaia} observing period ($v^\prime_{\tan} = 8.11$~km~s$^{-1}$, $\theta^\prime = 358.51^\circ$), which serves as an approximation for the observed tangential velocity anomaly (in yellow).
    For more detailed information, we refer to the caption of Fig.~\ref{fig:pi1_gru}.}\label{Fig:approximate_vtan_xy}
\end{figure}

\subsubsection{Impact of approximate position-velocity vector on retrieval outcome}\label{Methods:approx_retrieval}

Given the similarity between the exact and approximate solutions for the astrometric position, as illustrated in Supplementary Fig.~\ref{Fig:approximate_vtan_xy}, and the fact that the tangential velocity of $M_1$ and its position angle provide a reasonable approximation for the tangential velocity anomaly and its position angle as measured by the \textit{Gaia} and \textit{Hipparcos} missions, it becomes appealing to use these approximations in the Bayesian retrievals. Consequently, instead of using (OC12, OC13, OC17, OC18) as observational constraint, we use (OC12c, OC13c, OC17c, OC18c). The approximation reduces the number of parameters by two (namely $\mu_\alpha^G$ and $\mu_\delta^G$), which speeds up the convergence of the routine by approximately a factor of two. Although the inherent assumption of the approximate method is that $\bmu^G = \bmu_{HG}$ -- which implies that the $q \ll 1$ and that the photocentre accurately follows the motion of $M_1$ -- we allowed $q$ to vary between 0.2\,--\,5.0, similar to all other tests described in Supplementary Sect.~\ref{Sec:sensitivity}.

As with the exact solution, we first test the robustness of the retrieval; see Supplementary Table~\ref{table:approximate}. Similar to the exact case, the variation in retrieved parameters is minimal, and the posterior distributions remain consistent, confirming the stability of the process.

However, a comparison between the test runs for the exact and approximate solutions reveals systematic differences in the retrieved orbital parameters. Notably, the approximate solution yields a significantly lower primary mass \(m_1\) of $\sim$0.68\,\Msun\ as compared to the exact  \texttt{eccentric} solution ($\sim$1.1\,\Msun). Such a low mass for $\pi^1$~Gru~A is not compatible with its known pulsation characteristics; see Sect.~\ref{Sec:initial_mass}. Furthermore, the distance $D$ ($\sim 155$\,pc) is significantly lower compared to the exact solution ($\sim 180$\,pc), with standard deviations indicating no overlap between the two estimates. Similarly, the inclination $i$ is considerably higher in the approximate solution ($38^\circ$) than in the exact solution ($\sim 15^\circ$), again with non-overlapping uncertainties. The mass ratio $q$ is also higher in the approximate solution ($\sim 1.18$) compared to the exact case ($q \sim 1.05$). Additionally, the semi-major axis $a$ is smaller in the approximate solution ($\sim 6.14\,\text{au}$) compared to the exact solution ($a \sim 6.7\,\text{au}$). These differences highlight the impact of using an approximate solution on the inferred orbital parameters.

An additional issue becomes evident when examining Supplementary Fig.~\ref{fig:pi1_gru_motion_approximate}: the best-fit approximate solution fails to reproduce the \textit{Hipparcos} tangential velocity anomaly and its position angle.  The reason for this failure lies in the inherent assumption of the approximation, namely that $\bmu^G = \bmu_{HG}$, implying $m_1 \gg m_2$. However, the results from the exact \texttt{eccentric} run indicate that the assumption $\bmu^G = \bmu_{HG}$ overestimates the actual proper motion of the barycentre, and the assumption $m_1 \gg m_2$ is not valid (with $q > 1$). This discrepancy is most visible at the \textit{Hipparcos} epoch due to our model approach in Eq.~\eqref{eq:alpha_delta_model}, where ($\alpha_G^0, \delta_G^0$) are taken at $t = t_{C6}$, making the overestimation of $\bmu^G$ most pronounced at the epoch furthest from $t = t_{C6}$.

Altogether, while the approximation is attractive for its simplicity and elegance, it should be used with caution. Systematic differences may arise, and without proper oversight, they could affect the four key parameters that govern the binary's evolution: $m_1$, $q$, $a$, or $e$.

\subsection{Assessing alternative astrometric solutions}\label{Sec:comparison_astrometric}
Even though a multi-epoch detection of the motion of the inner companion $\pi^1$~Gru~C remained out of reach until our ALMA C6 and C10 observations, its existence was speculated based on irregularities in the CO and dust emission maps\cite{Homan2020A&A...644A..61H, Chiu2006ApJ...645..605C, Doan2017A&A...605A..28D} and the fact that $\pi^1$~Gru is showing a proper motion anomaly ($\Delta \bmu$)~\cite{Mayer2014A&A...570A.113M}\textsuperscript{,}\cite{Kervella2019A&A...623A..72K, Kervella2022A&A...657A...7K}. This latter characteristic prompted several authors to attempt deriving the astrometric orbit.

(1)~The first astrometric orbit for the inner $\pi^1$~Gru binary system was derived by Ref.~\cite{Mayer2014A&A...570A.113M}, who utilized the \textit{Tycho-2} long-term proper motion and the \textit{Hipparcos} Intermediate Astrometric Data (IAD) following the method outlined in Refs.~\cite{Pourbaix2000A&AS..145..161P, Pourbaix2004ASPC..318..132P}. This approach minimizes an objective function in a 12-parameter space, incorporating the \textit{Hipparcos} 5-parameter solution ($\alpha$, $\delta$, $\varpi$, $\mu_\alpha$, $\mu_\delta$) and seven additional orbital parameters (\(i, \omega, \Omega, e, a, T_0, T_{\rm orb}\)). By scanning a grid in the (\(e - \log T_{\rm orb}\)) plane, they obtained the best-fit solutions, which indicated eccentricities above 0.5 and orbital periods between 5 and 11 years, although the other orbital parameters were not explicitly reported.

Our derived orbital period is only slightly longer than the upper limit of the reported range. However, the high eccentricity found in that study differs from our results. This difference may arise from certain limitations inherent in this approach. The method is particularly well-suited for orbital periods between 1 and 3 years, ensuring good sampling of the orbit during the \textit{Hipparcos} mission's operational lifetime. For systems with parallax $\varpi \geq 5$~mas and orbital periods $\leq$ 10--11 years~\cite{Pourbaix2000A&AS..145..161P, Jorissen2004ASPC..318..141J}, astrometric orbits can often be derived by fitting the orbital arc. However, this methodology has known challenges in accurately constraining inclination~\cite{Pourbaix2000A&AS..145..161P}. A circular orbit observed at an inclined angle can appear elliptical in the plane of the sky, and a difference in inclination from our derived value could account for the discrepancy in the retrieved eccentricities.

Given their assumed primary mass of $m_1 = 1.5$~\Msun, a best-fit orbital period of 6.3~yr, and an inferred companion mass in the range of 0.5--1.5~\Msun, we determined for which value of \(m_2\) their partial orbital arc shown in in Fig.~7 of Ref.~\cite{Mayer2014A&A...570A.113M}, representing $\vec{r}_1^G$, can be scaled (using Eq.~\eqref{Eq:vec_r1}) such that the relative position of \(M_2\) with respect to \(M_1\) intersects the ALMA C6 or C10 position. For \(m_2 \sim 0.7\)~\Msun\ (corresponding to an orbital separation of \(a \sim 4.44\)~au), the  relative motion vector, $\vec{r}$, passes through the ALMA C10 relative position of \(M_2\). However, at the ALMA C6 epoch, the observed separation between \(M_1\) and \(M_2\) is larger than predicted by the scaled fit.

Admittedly, our method also has its limitations. Specifically, the proper motions listed in the \textit{Hipparcos} and \textit{Gaia} catalogues may be inaccurate for binary systems with orbital periods exceeding 3 years. This can occur if such systems were not correctly identified as binaries during the catalog reduction process. In such cases, the orbital motion of the binary components can contribute to the observed proper motion, altering both its magnitude and direction; see Supplementary Sect.~\ref{Sec:theory:pma}. However, for orbital periods exceeding 5~yr, the orbital motion becomes negligible over the duration of the \textit{Hipparcos} and \textit{Gaia} missions, and the proper motions and its modulus are rather well determined in those catalogues, except that the error bars may become larger~\cite{Pourbaix2000A&AS..145..161P}. While the \textit{Hipparcos} data set is the least constraining for our retrieval methodology, the \textit{Gaia} data set, with its higher precision, has a greater impact (see Supplementary Sect.~\ref{Sec:sensitivity}). After the \textit{Gaia} Data Release 4 (DR4), anticipated no earlier than mid-2026, we plan to revisit the modeling and refine the outcomes.

(2)~Assuming a circular orbit observed face-on ($i=0^\circ$), Ref.~\cite{Kervella2022A&A...657A...7K} derived the properties of the astrometric orbit using \textit{Gaia} DR3 and \textit{Hipparcos} proper motions, combined with the tangential velocity anomaly. Their method allows for an estimation of the companion’s mass, but this remains degenerate with the orbital separation. The mass and radius of the primary star were inferred from dereddened photometry following the approach described in Ref.~\cite{Kervella2019A&A...623A..72K}, yielding a primary mass of \(m_1 = 0.64 \pm 0.03\)\,\Msun. Based on this, the companion mass was estimated to be \(m_2 = 0.45\)\,\Msun, \(0.42\)\,\Msun, \(0.48\)\,\Msun, or \(2.82\)\,\Msun\ for orbital separations of 3~au, 5~au, 20~au, or 30~au, respectively.

Given their use of the \textit{Gaia} DR3 parallax of \(6.202 \pm 0.512\)~mas and the observed separation between \(M_2\) and \(M_1\) of 37.702~mas (40.775~mas) at the ALMA C6 (C10) epoch, the corresponding orbital separation is 6.08~au (6.57~au) for a circular orbit seen face-on. This would imply a companion mass of \(m_2 \sim 0.45\)\,\Msun, and hence an orbital period of 14.34~yr (16.12~yr).

The primary mass \(m_1\) was derived from observed visible and \(K\)-band magnitudes, along with \(E(B-V)\) values predicted by the {\sc Stilism}\footnote{\url{ https://stilism.obspm.fr}} 3D model of the local interstellar medium. However, the catalog estimates did not account for circumstellar reddening, which likely explains the lower mass value obtained by these authors.

Since their estimate for the primary mass \(m_1\) is nearly a factor of two lower than our derived value, and given their analytical relation \(m_2 \propto \sqrt{m_1}\), it is unsurprising that their companion mass estimate is significantly lower than our derived value of \(m_2 \sim 1.18\)\,\Msun. However, as discussed in Supplementary Sect.~\ref{Sec:pi1gruc}, a companion mass of \(\sim\)0.5\,\Msun\ cannot account for the observed elliptical morphology in both CO and dust emission.

Furthermore, the difference in relative position between \(M_2\) and \(M_1\) at the ALMA C6 and C10 epochs -- $37.702\pm1.898$~mas and $40.775\pm1.233$~mas, respectively -- is inconsistent with a circular orbit seen face-on.

(3)~Lastly, Ref.~\cite{Montarges2025A&A...699A..22M} estimated the astrometric orbit using the ALMA C6 data, the \textit{Gaia} proper motion anomaly, and an estimate of the position angle where the radial velocity is 0~km~s$^{-1}$ indicating the orbital node direction which is orthogonal to this direction. They derived \(m_1 = 1.54\,-\,2.19\)\,\Msun, \(q = 0.48^{+0.08}_{-0.08}\), \(a = 7.05^{+0.54}_{-0.57}\)~au, \(e = 0.35^{+0.18}_{-0.17}\), \(\Omega_M = 321.7^\circ{^{+5.91}_{-5.67}}\), \(\omega_M = 359.00^\circ{^{+7.87}_{-11.26}}\), \(i_M = 31.75 - 38.04\)°, and \(T_0 = 2015.06^{+1.09}_{-1.14}\)~yr, although they do not exclude a circular orbit. The subscript M refers to parameters where Ref.~\cite{Montarges2025A&A...699A..22M} used a different definition compared to this work. They suggest the companion to be of mass $\sim$0.86\,\Msun, either being a K1~V MS or a WD.

In their definition, the parameters $\omega_M$, $\Omega_M$, and $i_M$ are defined as follows:
\begin{equation}
    \left\{
    \begin{aligned}
        \omega_M &= 90^\circ - \omega_R \\
        \Omega_M &= 90^\circ - \Omega_R \\
        i_M &= i - 90^\circ
    \end{aligned}\ ,
    \right.
\end{equation}
where the subscript R now refers to the parameters used in the equations of Ref.~\cite{Roy2005ormo.book.....R}. They further define the $x$, $y$, and $z$ axes different from the ones used in this work. In our framework, the $x$ axis points North, the $y$ axis points East, and the $z$ axis points towards the observer, while in their framework, the $x$ axis points towards the East, the $y$ axis points North, and the $z$ axis points away from the observer (which still corresponds to a right-handed system). The transformation between the two systems is given by:
\begin{equation}
    \left\{
    \begin{aligned}
        x_R &= y \\
        y_R &= x \\
        z_R &= -z
    \end{aligned}
    \right. \iff
    \left\{
    \begin{aligned}
        \omega_R &= -\omega \\
        \Omega_R &= 90^\circ - \Omega \\
        i_R &= 180^\circ - i
    \end{aligned}
    \right.\ ,
\end{equation}
on top of which there is a difference in definition of the angles $\omega$ and $\Omega$ between the two systems. The difference lies in a different definition of the ascending node. In our framework, the ascending node is defined as the point where the orbiting body crosses the plane of the sky moving towards the observer (i.e., crossing from $Z < 0$ to $Z > 0$, such that the radial component of the orbital motion is at its minimum, corresponding to the maximally blue-shifted position). By contrast, in their framework, the ascending node is defined as the position where the radial velocity is at its maximum (corresponding to the maximally red-shifted position). This results in a $180^\circ$ difference in $\Omega$ and $\omega$ between both frameworks. The transformation between the two systems is given by:
\begin{equation}
    \left\{
    \begin{aligned}
        \omega &= -\omega_M - 90^\circ \\
        \Omega &= \Omega_M - 180^\circ \\
        i &= 90^\circ - i_M
    \end{aligned}
    \right.\ .
\end{equation}
Taking into account the transformation between the two systems, their best-fit parameters are become \(m_1 = 1.54\,-\,2.19\)\,\Msun, \(q = 0.48^{+0.08}_{-0.08}\), \(a = 7.05^{+0.54}_{-0.57}\)~au, \(e = 0.35^{+0.18}_{-0.17}\), \(\Omega = 141.7^\circ{^{+5.91}_{-5.67}}\), \(\omega = 271.00^\circ{^{+7.87}_{-11.26}}\), \(i = 58.25 \rightarrow 51.06\)°, and \(T_0 = 2015.06^{+1.09}_{-1.14}\)~yr.

The different orbital solution found in Ref.~\cite{Montarges2025A&A...699A..22M} could arise from a number of reasons.
First, they treat the tangential velocity as a proxy for the tangential velocity anomaly. As we discuss in Supplementary Sect.~\ref{Sec:approximation}, this approximation should be applied with caution, as it can introduce systematic errors.
Second, their estimate of the position angle where the radial velocity is zero simplifies the problem and does not fully account for the complexity of the velocity vector field, as discussed by Ref.~\cite{Homan2020A&A...644A..61H} (see their Fig.~11) and in Sect.~\ref{Sec:vrad}. Their model assumes that close to the location of the inner companion, the bulk gas motion predominantly reflects orbital motion due to the wobble of the primary star. The angle of the observed zero relative radial velocity ($52.5\pm5^\circ$) is then interpreted as being orthogonal to the orbital node direction (\(\Omega\)). However, this assumption overlooks the influence of radial outflows (with observed absolute values ranging from a few to 14~km~s$^{-1}$~\cite{Doan2017A&A...605A..28D}) as well as the formation of the spiral wake induced by the companion's gravity and bow shocks, both of which play a significant role in shaping the velocity vector field. As can be seen in panel~(b) of Supplementary Fig.~\ref{fig:pv_CO_2023}, the overall Keplerian velocity field can also be witnessed in the ALMA C10 $^{29}$SiO v=0 J=8-7 moment-1 map, but that specific angle has changed to $\sim$0--30$^\circ$.
Lastly, as discussed in Supplementary Sect.~\ref{Sec:sensitivity}, the lack of ALMA C10 data can lead to instability in the \texttt{ultranest} retrieval. Their analysis only uses 5 observational constraints to fit 8 free parameters, introducing the risk of overfitting, where the fitting process frequently converges to local maxima or saddle points, rather than the global posterior maximum. An advantage of our methodology is its ability to cover the full prior hypothesis space, in contrast to Ref.~\cite{Montarges2025A&A...699A..22M} who used restrictive priors for the parameters $q$ and $i$. Notably, our retrieved parameters for $q$ and $i$ lie outside of their priors, potentially influencing the other quantities as well.

\clearpage
\section{Supplementary information stellar evolution}

\begin{figure}[!htp]
\centering
\includegraphics[width=.7\textwidth]{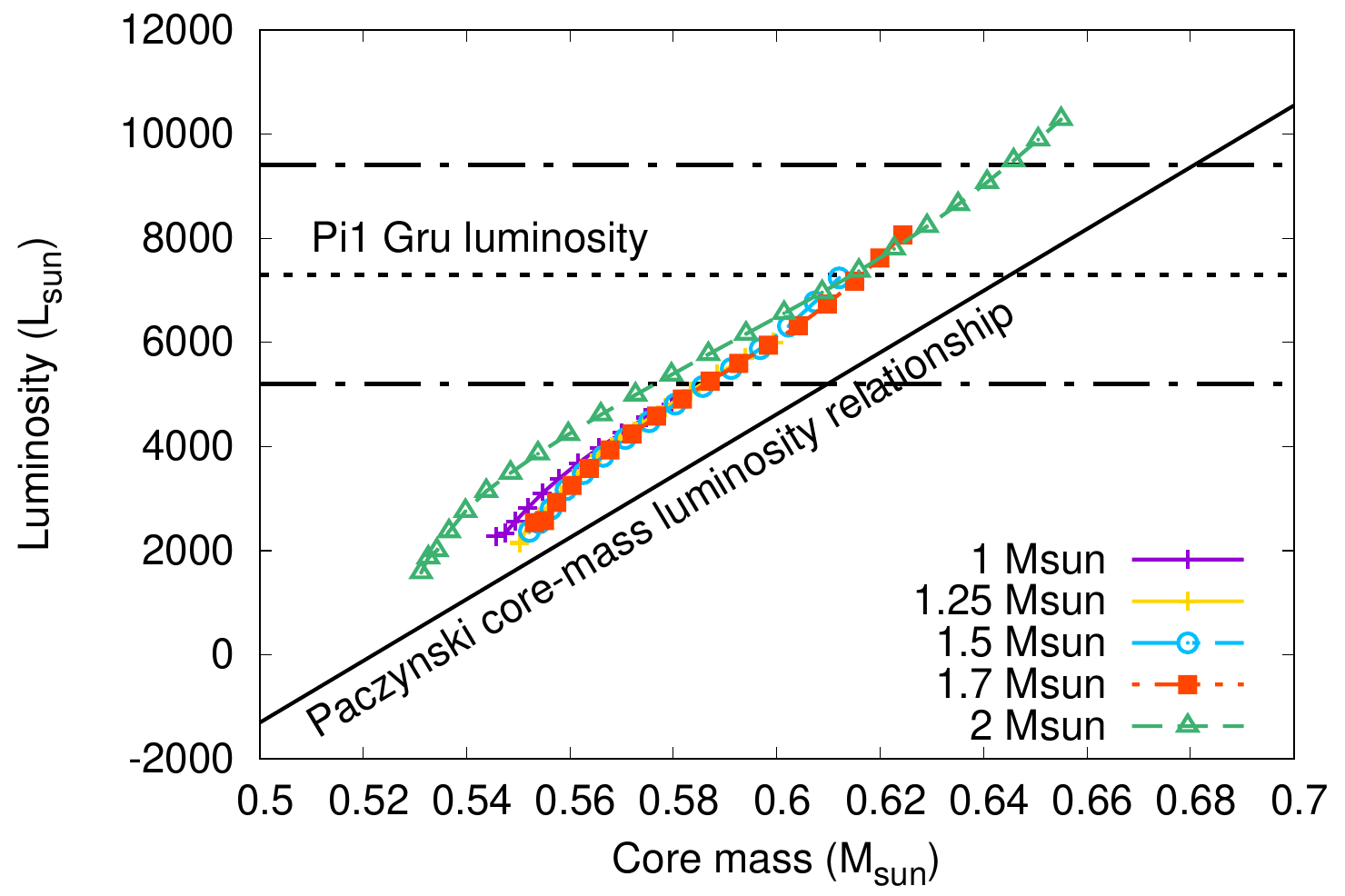}
\caption{\textbf{Core-mass luminosity relation for solar-metallicity AGB models.} Each point shows the predicted maximum pre-thermal pulse luminosity (in solar units, \Lsun) as a function of the H-exhausted core mass (in solar units, \Msun) for AGB models between 1\,\Msun\ and 2\,\Msun\ for solar metallicity ($Z=0.014$). Also included in the plot is the linear relationship derived by Paczy\'{n}ski~\protect\cite{Paczynski1970AcA....20...47P}, and the estimated luminosity of $\pi^1$~Gru~A of 7\,300\,\Lsun\ shown by the horizontal dashed line. The long-dashed horizontal lines show upper and lower bounds to the luminosity of $\pm$ 2\,100\,\Lsun~.}\label{Fig:core-mass-lum}
\end{figure}

\begin{figure}[!htp]
\centering
\includegraphics[width=\textwidth]{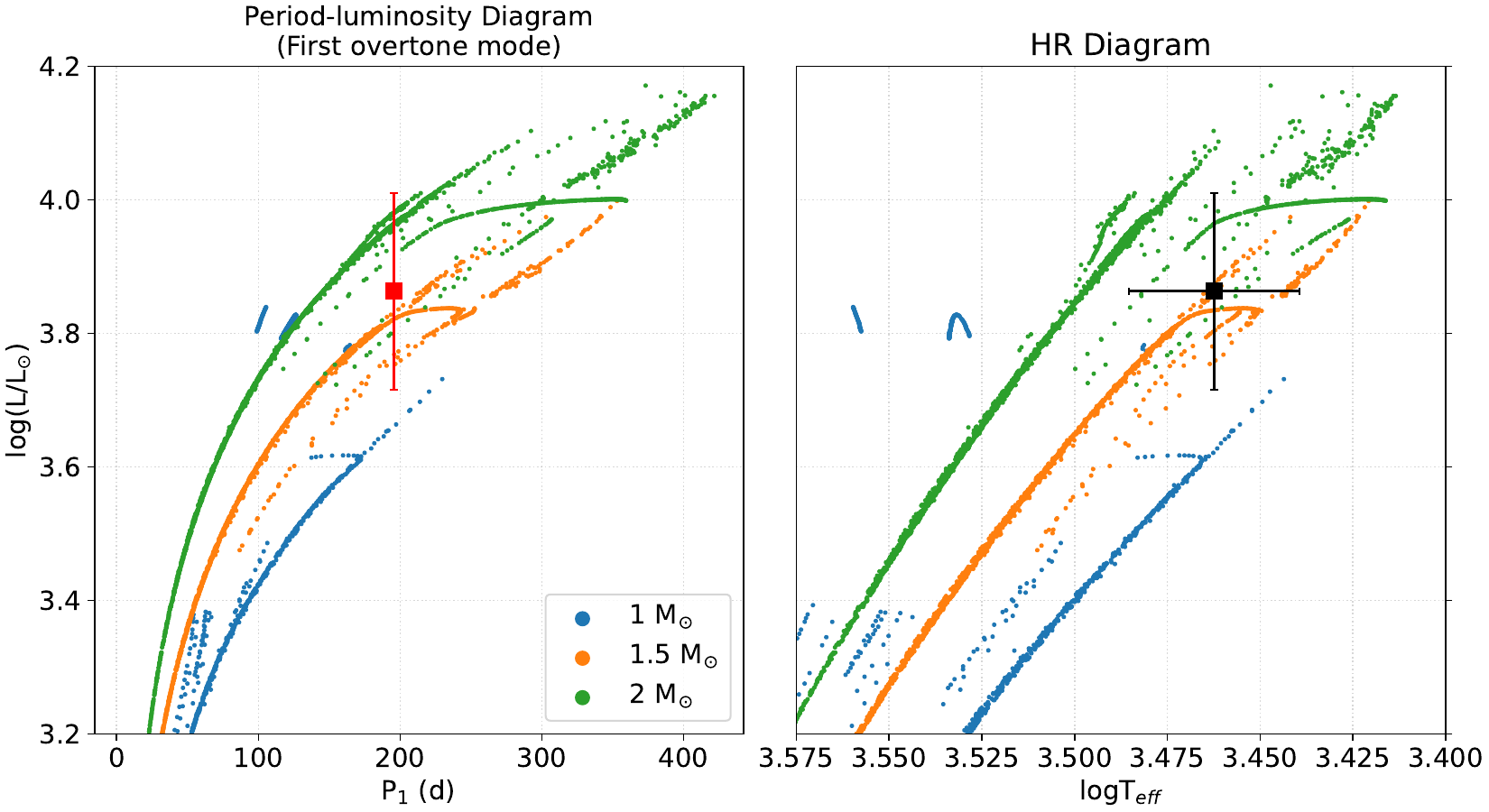}
\caption{\textbf{Theoretical period-luminosity and Hertzsprung-Russell diagrams compared to measurements for $\boldsymbol{\pi^1}$~Gru.} The plot includes detailed model tracks for the first overtone mode period against luminosity in panel (a) and the HR diagram in panel (b), with the measured pulsation period ($195.5$ d) and estimated luminosity ($7300 \pm 2100$ \Lsun) and effective temperature ($2900 \pm 150$ K). Both diagrams only include model points during which the first overtone mode is unstable; this produces the small collections of points to the left of the plots for the $1$\,\Msun\ model, where the first overtone mode becomes briefly active once again during the early part of the interpulse phases late on the TP-AGB. Both the period-luminosity and HR diagrams indicate that $\pi^1$~Gru~A is consistent with an initial mass between 1.5\,--\,2\Msun.}\label{Fig:tpagb-plhr-z014}
\end{figure}

\clearpage
\section{Supplementary information orbital evolution}

    \subsection{Sensitivity analysis of the orbital dynamics}\label{sec:orbevolsensitivity}
    \begin{figure}[!htp]
        \centering
        \includegraphics[width=\linewidth]{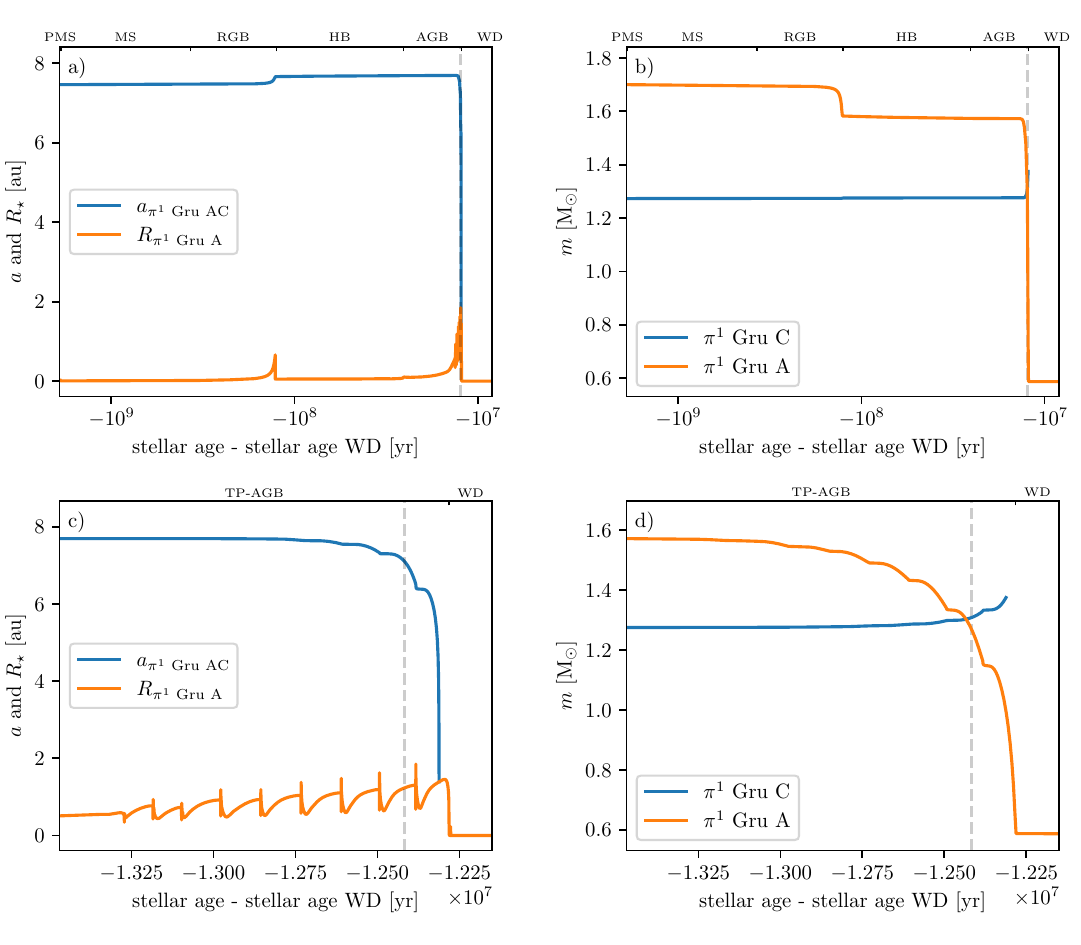}
            \caption{\textbf{Orbital evolution of the $\boldsymbol{\pi^1}$~Gru system during the TP-AGB phase.} Panel~(a) illustrates the variation in orbital separation $a$ over time (blue line), along with the evolution of the radius of $\pi^1$~Gru~A (orange line). Panel~(b) shows the mass evolution of $\pi^1$~Gru~A (orange line) and $\pi^1$~Gru~C (blue line). Panels (c) and (d) show a zoom-in to the TP-AGB phase. These panels show the evolution of the system parameters when the initial mass of $\pi^1$~Gru~A is set to 1.7\,\Msun, and the current mass of $\pi^1$~Gru~C is set to 1.23\,\Msun. The vertical dashed grey line in each panel marks the current age of the $\pi^1$~Gru system.}\label{fig:OrbitalEvolution_prior}
    \end{figure}
    The orbital evolution described above is based on the retrieved mean values of the binary system parameters for the \texttt{circular} model, as listed in Supplementary Table~\ref{table:fit_sensitivity}, with an initial mass of 1.5\ \Msun. However, these parameters come with associated uncertainties, which can affect the precision of the calculated evolution. To assess the impact of these uncertainties, we conduct a sensitivity analysis by recalculating the orbital evolution using a range of parameter values within their standard deviations.
    Specifically, we explore the dependence of the orbital evolution on the semi-major axis, the current mass of the primary star and of the companion star, and the initial mass of the primary star.

     \subsubsection{Dependence on semi-major axis}
            \begin{figure}[!htp]
                \centering
                \includegraphics[width=\linewidth]{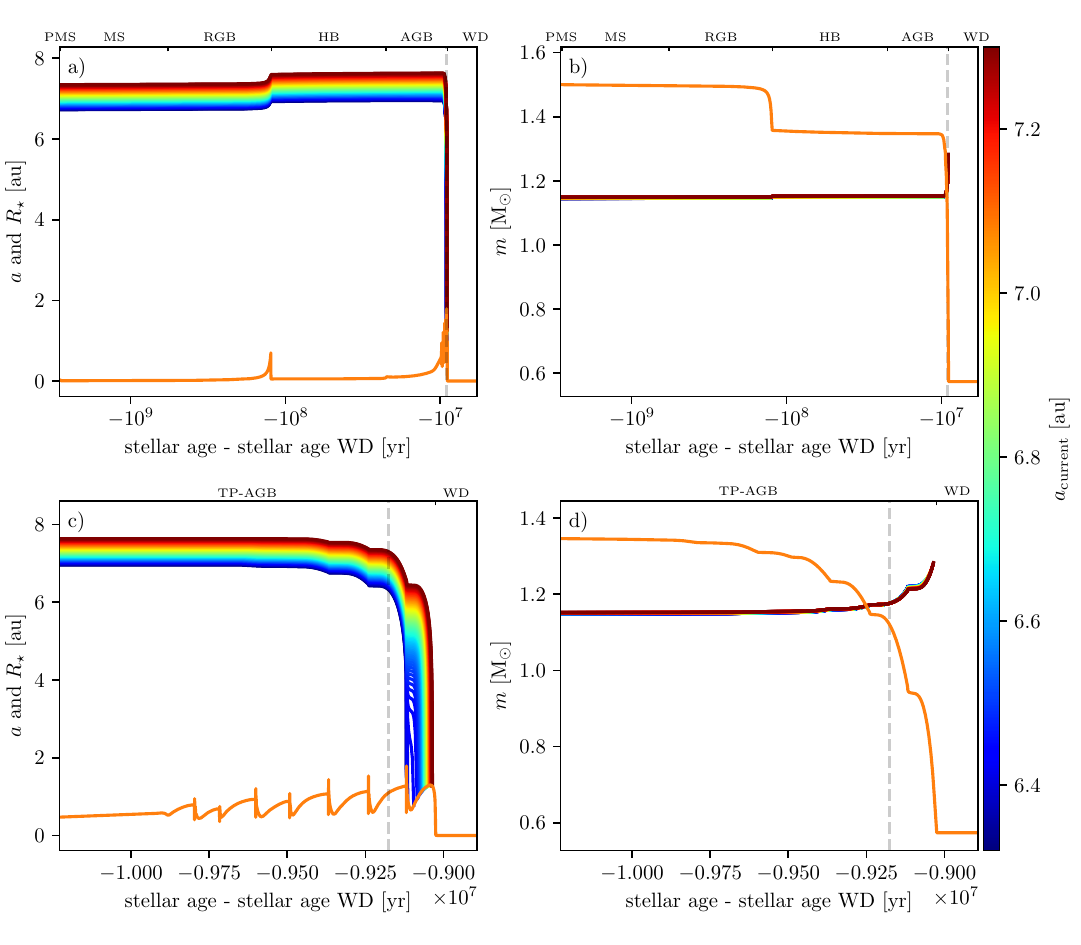}
                \caption{\textbf{Orbital evolution of the $\boldsymbol{\pi^1}$~Gru system with variations in the current semi-major axis.} This figure illustrates how variations in the current semi-major axis (indicated by the colour bar) influence the orbital evolution of the system. Panel~(a) shows the evolution of the orbital separation along with the changing radius of $\pi^1$~Gru~A (orange line). Panel~(b) depicts the mass evolution of $\pi^1$~Gru~A (orange line) and $\pi^1$~Gru~C, with colours corresponding to the semi-major axis values. Panels~(c) and~(d) provide a detailed view of the TP-AGB phase. The vertical dashed grey line in each panel marks the current age of the $\pi^1$~Gru system.}\label{fig:OrbitalEvolutionsma}
            \end{figure}

            The impact of varying the semi-major axis between 6.32 and 7.30~au (reflecting the 1-sigma uncertainty of the circular model, as listed in Supplementary Table~\ref{table:fit_sensitivity}) on the orbital evolution is shown in Supplementary Fig.~\ref{fig:OrbitalEvolutionsma}. An increase in the current orbital separation results in a larger initial orbital separation, and vice versa. The initial orbital separation ranges from 6.7 to 7.4~au, closely aligning with the range of current orbital separations. This alignment occurs because the orbital separation increased during the RGB phase and is currently decreasing during the AGB phase, effectively compensating for these variations. The mass of $\pi^1$~Gru~C is barely affected by the variations in current semi-major axis.

            The future evolution of the system is also affected by changes in the current orbital separation. For smaller current orbital separations, the system's orbit contracts more rapidly, whereas larger current separations result in slower orbital contraction. These variations influence the system's parameters at the onset of the common envelope phase. For the smallest current orbital separation, the common envelope phase begins during the last thermal pulse of the primary star, with a companion mass of 1.22~\Msun. In contrast, for the largest current orbital separation, the common envelope phase initiates at the very end of the TP-AGB phase, with a companion mass of 1.28~\Msun.

        \subsubsection[Dependence on current mass of Pi1~Gru~A]{Dependence on current mass of $\boldsymbol{\pi^1}$~Gru~A}
            \begin{figure}[!htp]
                \centering
                \includegraphics[width=\linewidth]{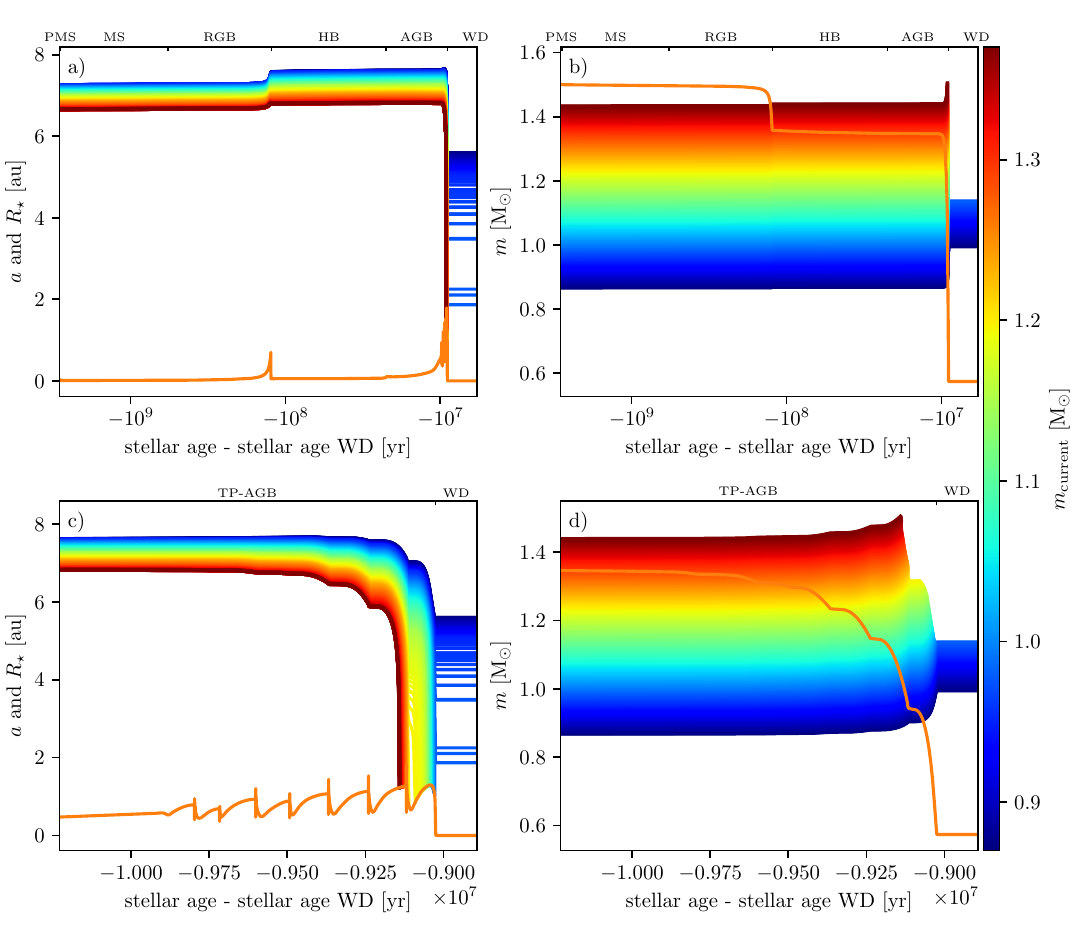}
                \caption{\textbf{Orbital evolution of the $\boldsymbol{\pi^1}$~Gru system with variations in the current mass of $\boldsymbol{\pi^1}$~Gru~A.} This figure illustrates how changes in the current mass of $\pi^1$~Gru~A (represented by the colour bar) affect the orbital evolution of the system over its lifetime. Panel~(a) shows the evolution of the orbital separation for each current value of the mass of $\pi^1$~Gru~A, using colours corresponding to the current mass, alongside the changing radius of $\pi^1$~Gru~A (orange line). Panel~(b) depicts the mass evolution of $\pi^1$~Gru~A (orange line) and $\pi^1$~Gru~C, with colours matching the respective current mass of $\pi^1$~Gru~A. Panel~(c) and (d) show a zoom-in to the TP-AGB phase. The vertical dashed grey line in each panel marks the current age of the $\pi^1$~Gru system.}\label{fig:OrbitalEvolutionM1current}
            \end{figure}

            The impact of varying the current mass of $\pi^1$~Gru~A between 0.87 and 1.37~\Msun\ (reflecting the one-sigma uncertainty of the circular model, as listed in Supplementary Table~\ref{table:fit_sensitivity}) on the orbital evolution is shown in Fig.~\ref{fig:OrbitalEvolutionM1current}. With the initial mass of $\pi^1$~Gru~A held constant (1.5 M$_\odot$), changes in the current mass effectively alter the current age of the system. Simultaneously, the mass ratio $q$ is kept constant (1.05), meaning that the mass of $\pi^1$~Gru~C is adjusted accordingly. A higher current mass corresponds to a younger current age, while a lower current mass indicates an older current age.

            For a younger current age, the semi-major axis has more time left in the AGB phase to evolve under the influence of tidal forces, leading to faster orbital shrinkage and an earlier onset of the common envelope phase. Conversely, for an older current age, the semi-major axis has less time to evolve and is less affected by tidal forces. In the case of the lowest current mass within the uncertainty range, the system avoids entering a common envelope phase altogether, instead ending in a stable configuration.

            However, at both the highest and lowest ends of the current mass range, the assumption that the initial mass of $\pi^1$~Gru~A remains unchanged is no longer valid. Specifically, for the highest current mass within the uncertainty range, $\pi^1$~Gru~A would have already reached this mass during the RGB phase, yet the system is currently observed in the AGB phase. To reconcile this, the initial mass must be increased to ensure $\pi^1$~Gru~A evolves into the AGB phase at the observed current mass. Conversely, for the lowest current mass, $\pi^1$~Gru~A would be at the very end of the (TP-)AGB phase, a stage where a semi-regular variable is unlikely to be observed. In this case, the initial mass must be decreased to yield a system consistent with observations. When these the initial mass is adjusted to account for these effects, the systems will ultimately reach the common envelope phase.

        \subsubsection[Dependence on current mass Pi1~Gru~C]{Dependence on current mass $\boldsymbol{\pi^1}$~Gru~C}
            \begin{figure}[!htp]
                \centering
                \includegraphics[width=\linewidth]{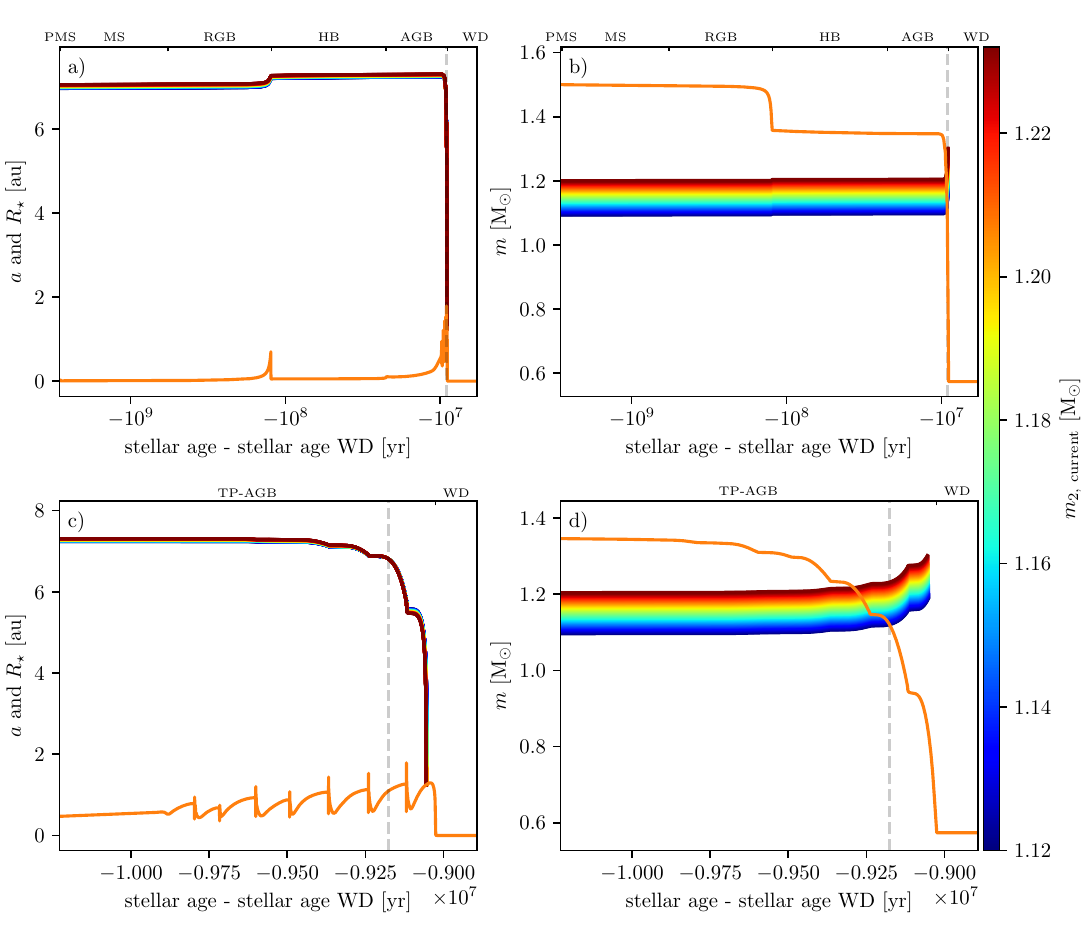}
                \caption{\textbf{Orbital evolution of the $\boldsymbol{\pi^1}$~Gru system with variations in the current mass of $\boldsymbol{\pi^1}$~Gru~C.} This figure illustrates how changes in the current mass of $\pi^1$~Gru~C (represented by the colour bar) affect the orbital evolution of the system over its lifetime. Panel~(a) shows the evolution of the orbital separation for each current value of the mass of $\pi^1$~Gru~C, using colours corresponding to the current mass, alongside the changing radius of $\pi^1$~Gru~A (orange line). Panel~(b) depicts the mass evolution of $\pi^1$~Gru~A (orange line) and $\pi^1$~Gru~C, with colours matching the respective current mass of $\pi^1$~Gru~C. The vertical dashed grey line in each panel indicates the current age of the $\pi^1$~Gru system. Panel~(c) and (d) show a zoom-in to the TP-AGB phase. The vertical dashed grey line in each panel marks the current age of the $\pi^1$~Gru system.}\label{fig:OrbitalEvolutionM2current}
            \end{figure}

            The mass ratio of the system is estimated to be $\sim$1.05. Assuming a current mass of $\pi^1$~Gru~A of 1.12~\Msun, the current mass of the companion is $\sim$1.18~\Msun. Given the spread in the mass ratio $q$ between 1.00 and 1.10, the companion mass is varied between 1.12 and 1.23~\Msun. The impact of these variations on the orbital evolution is shown in Supplementary Fig.~\ref{fig:OrbitalEvolutionM2current}. The results indicate that the current mass of the companion has a minor impact on the orbital evolution.

        \subsubsection[Dependence on initial mass of Pi1~Gru~A]{Dependence on initial mass of $\boldsymbol{\pi^1}$~Gru~A}
            \begin{figure}[!htp]
                \centering
                \includegraphics[width=0.95\linewidth]{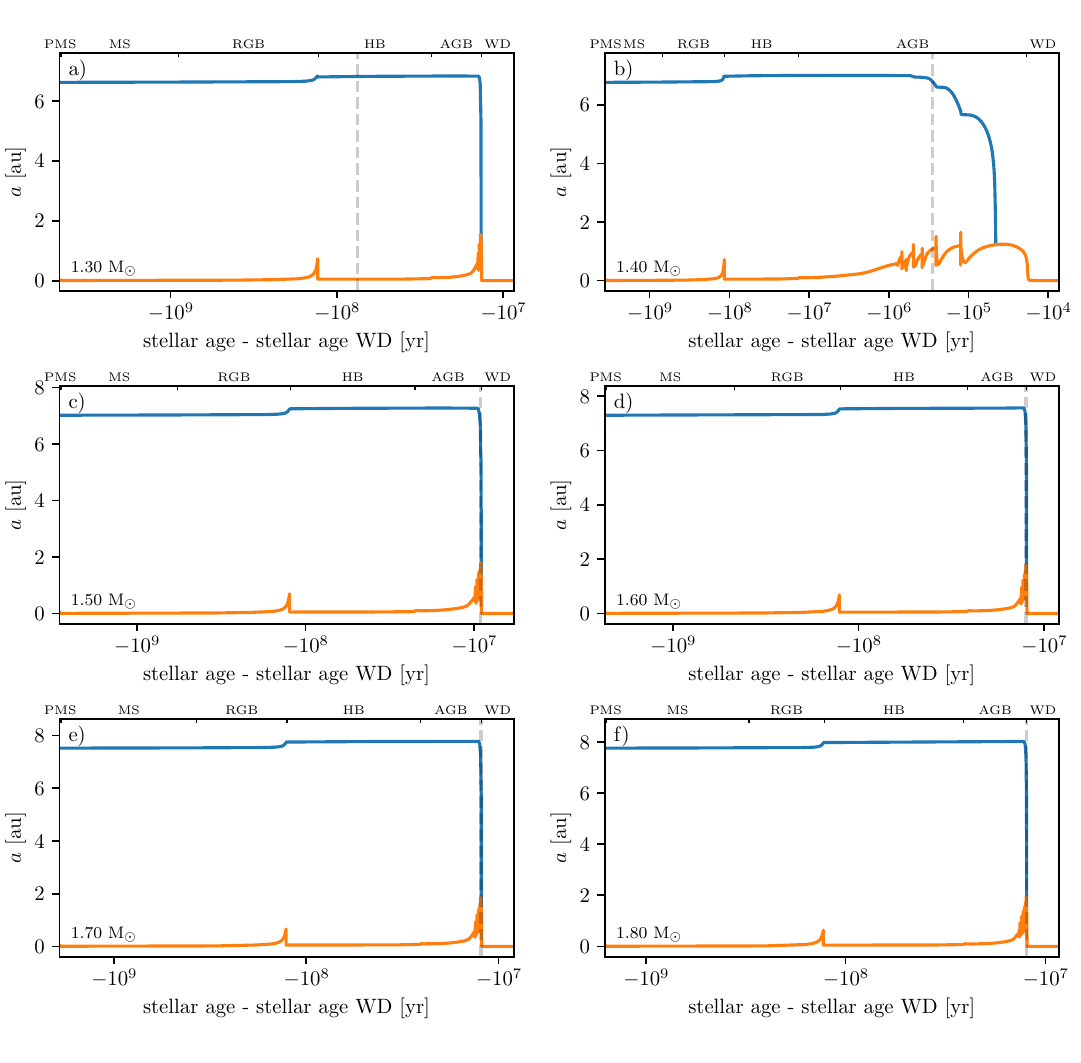}
                \caption{\textbf{Impact of the initial mass of $\boldsymbol{\pi^1}$~Gru~A on the system’s orbital evolution.} Panels (a)--(f) show the evolution of the orbital separation (blue line) and the radius of $\pi^1$~Gru~A (orange line) over the system's lifetime for stellar evolutionary models with initial masses of 1.3\,\Msun\ (a), 1.4\,\Msun\ (b), 1.5\,\Msun\ (c), 1.6\,\Msun\ (d), 1.7\,\Msun\ (e), and 1.8\,\Msun\ (f). The vertical dashed grey line in each panel marks the current age of the $\pi^1$~Gru system.}\label{fig:OrbitalEvolutionM1init}
            \end{figure}
            The initial mass of $\pi^1$~Gru~A is estimated to lie between 1.3 and 1.8\,\Msun\ (see Sect.~\ref{Sec:initial_mass}). To explore this range, \texttt{MESA} stellar evolutionary models have been computed, and the resulting orbital evolution is shown in Supplementary Fig.~\ref{fig:OrbitalEvolutionM1init}. These results indicate that an initial mass of $m_{1,\text{initial}} = 1.3$\,\Msun\ is not a viable option, as it places $\pi^1$~Gru~A in the HB phase rather than the observed AGB phase. In contrast, models with initial masses of 1.4, 1.5, 1.6, 1.7, and 1.8\,\Msun\ are similar in that they place the system in the AGB phase. However, for initial masses of 1.6, 1.7, and 1.8\,\Msun, the system is positioned late in the TP-AGB phase, where a semi-regular variable is not typically expected (see Fig.~\ref{Fig:m1_7_nov2_25}).

            Despite these model-specific limitations, the predicted orbital evolution is similar across the entire range of initial masses. In all cases, the system is expected to enter a common envelope phase in the future. The past orbital evolution also shows notable similarities: higher initial masses result in stronger tidal forces, which drive faster orbital evolution and lead to a larger estimated initial orbital separation for the system.

\clearpage

\afterpage{\clearpage}

\section[The nature of Pi1~Gru~C]{The nature of $\boldsymbol{\pi^1}$~Gru~C}\label{Sec:pi1gruc}

$\pi^1$~Gru~A is an intrinsic S-type star~\cite{VanEck1998A&A...329..971V}, rather than an extrinsic S star, in which case the companion would have already undergone the AGB phase and evolved into a white dwarf.
Given its intrinsic S-type nature, we must consider various evolutionary scenarios for the companion, including the possibility that it is still on the main sequence, a (sub)giant, or has already involved into a white dwarf. In the latter scenario, the white dwarf formed from the initially more massive star in the system, and the two stars were initially well separated and avoided significant mass transfer or tidal interactions during the companion's AGB phase.

The circumstellar envelope of $\pi^1$~Gru~A has been observed to exhibit an elliptical morphology in both CO and dust emission, with an aspect ratio of approximately 2:3~\cite{Homan2020A&A...644A..61H, Chiu2006ApJ...645..605C, Mayer2014A&A...570A.113M}\textsuperscript{,}\cite{Knapp1999A&A...346..175K, Ladjal2010A&A...513A..53L}. Based on the grid of models for systems with parameters similar to the inner $\pi^1$~Gru binary system, as presented by Ref.~\cite{Mastrodemos1999ApJ...523..357M}, this morphology suggests a companion mass of around 1\,\Msun. This result aligns with our retrieval outcomes. A lower-mass companion, with a mass of $\sim$0.5\,\Msun, would not produce the collimated and oblate wind morphology observed. Such a morphology arises from the gravitational influence of the companion on the otherwise spherical wind of the AGB star.

\subsection{Potential evolutionary scenarios}
\begin{figure}[!htp]
    \centering
    \includegraphics[width=\textwidth]{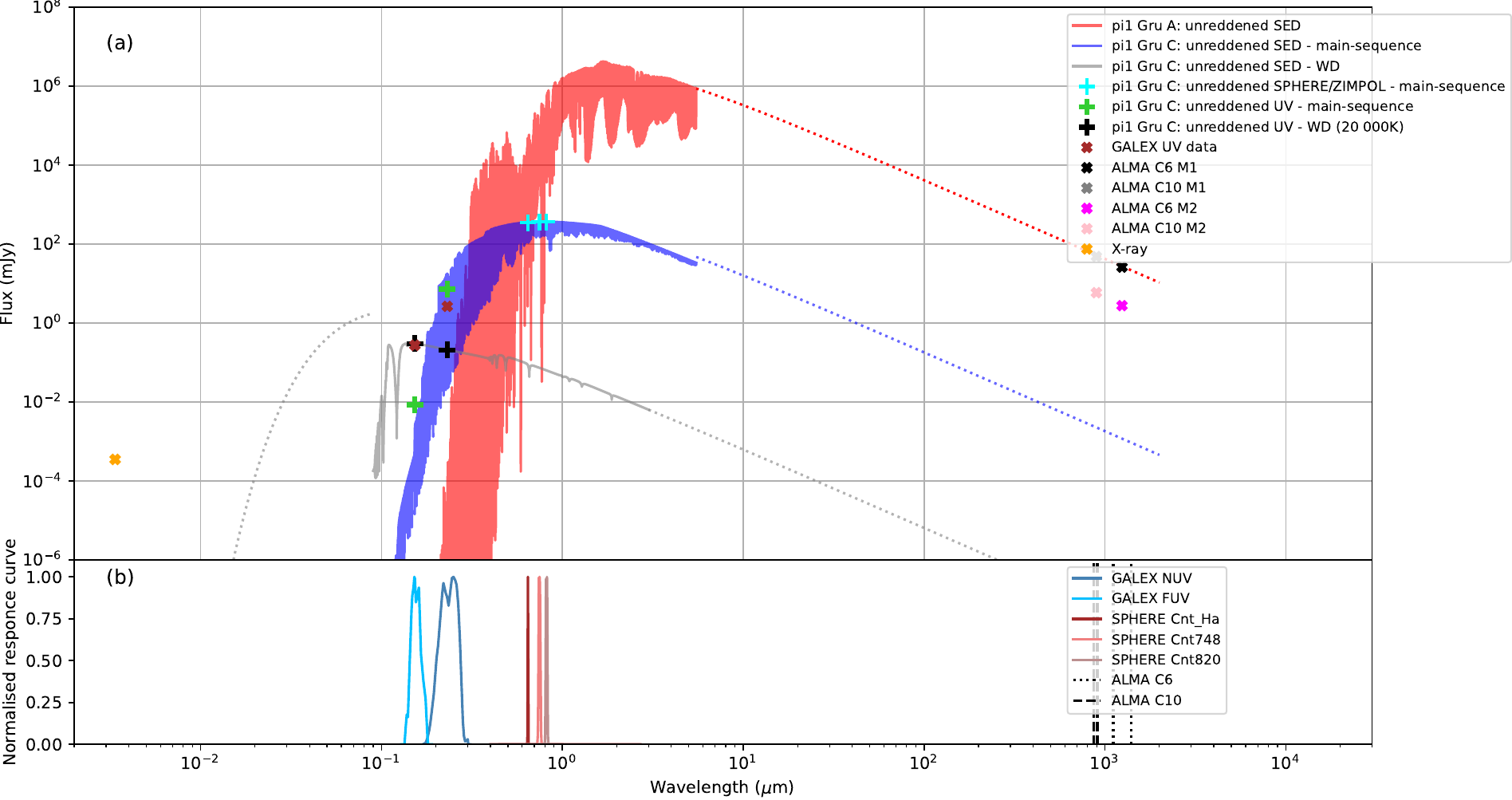}
    \caption{\textbf{Spectral energy distribution of $\boldsymbol{\pi^1}$~Gru~A and C.}
Panel~(a): The red curves represent the unreddened and reddened SEDs of $\pi^1$~Gru~A. The blue and grey curves show the reddened SED of $\pi^1$~Gru~C for the main-sequence and white dwarf scenario, respectively. The cyan crosses indicate the predicted photometric values in the SPHERE/ZIMPOL Cnt$\_$Ha, Cnt748, and Cnt820 filters for the main-sequence scenario. The light green and black crosses correspond to the predicted \textit{GALEX} FUV and NUV photometric values for the main-sequence and white dwarf scenario, respectively. The eROSITA X-ray and \textit{GALEX} UV observations are marked with an orange and brown `x', respectively. The ALMA integrated flux values for $\pi^1$~Gru~A are shown as black and grey crosses for the 2019 ALMA C6 and 2023 ALMA C10 epochs, respectively. The fuchsia and pink crosses represent the ALMA data for $\pi^1$~Gru~C for the same epochs.
Panel~(b): Normalized response curves for the \textit{GALEX} NUV and FUV, and the SPHERE-ZIMPOL Cnt$\_$Ha, Cnt748, and Cnt820 photometric filters. The dotted black lines indicate the minimum and maximum wavelengths of the 2019 ALMA C6 data, while the dashed black lines correspond to the 2023 ALMA C10 data.}\label{Fig:spectra}
\end{figure}

The results of the stellar and orbital evolution calculations indicate that the initial mass of $\pi^1$~Gru~A was $\sim$1.4--1.7\,\Msun, with current mass around 1.1--1.3\,\Msun\ (depending on the prior of $M_1$) and current mass ratio $q\sim1.05$ (Sect.~\ref{Sec:stellar_evolution}--\ref{Sec:orbital_evolution}). The current stellar properties of $\pi^1$~Gru~A have been derived in Sect.~\ref{Sec:Luminosity_A}. The Phoenix~\cite{Husser2013A&A...553A...6H} high-resolution spectrum for a model atmosphere with $T_{\rm eff} =$ 3,000~K, $\log_{10} g = 0.00$, and solar metallicity, scaled to the observed stellar angular diameter, is shown in Supplementary Fig.~\ref{Fig:spectra}. For wavelengths beyond 5.5~$\mu$m, the spectrum has been extrapolated using blackbody emission. Similar to above, we adopt an $R_V$ value of 4.2~\cite{Massey2005ApJ...634.1286M}. Using the Cardelli, Clayton, \& Mathis (CCM) extinction law~\cite{Cardelli1989ApJ...345..245C}, we apply this reddening to the spectral energy distribution (SED) of $\pi^1$~Gru~A (see Supplementary Fig.~\ref{Fig:spectra}).

To explore the nature of $\pi^1$~Gru~C, we thus consider three scenarios. In the first scenario, $\pi^1$~Gru~C is still on the main-sequence with a current mass ratio of $q \sim 1.05$. Based on the main-sequence mass-luminosity and mass-temperature relations~\cite{Eker2018MNRAS.479.5491E}, this implies a current luminosity of $2.25\pm0.10$\,\Lsun, effective temperature of $\sim 6,200\pm120$\,K, and spectral type of F6-F8V. The corresponding stellar radius and surface gravity are estimated to be $1.31\pm0.09$\,\Rsun\ and $\log_{10} g = 4.3\pm0.2$, respectively. By applying the CCM extinction law with the derived parameters and using the corresponding Phoenix~\cite{Husser2013A&A...553A...6H} high-resolution spectrum, we obtain an initial estimate of the reddened SED of $\pi^1$~Gru~C (see Supplementary Fig.~\ref{Fig:spectra}). This estimate does not yet account for additional reddening due to an accretion disk and may vary with the orbital phase of $\pi^1$~Gru~C. However, these contributions are minimal and do not affect the interpretation of the companion's nature.

The estimated main-sequence properties of $\pi^1$~Gru~C align with evolutionary predictions from the \texttt{MESA} stellar evolution code~\cite{Jermyn2023ApJS..265...15J}, assuming a stellar age comparable to that of $\pi^1$~Gru~A. For an initial mass of $m_1$ of 1.5\,\Msun\ and a current mass of $m_1 = 1.12\,\Msun$, \texttt{MESA} predicts a current stellar temperature and radius for $\pi^1$~Gru~C of $\sim$6,200\,K and 1.3\,\Rsun, respectively. When using the current mass estimate derived from the Gaussian prior on $m_1$ of 1.12\,\Msun, the corresponding temperature and radius of $\pi^1$~Gru~C are slightly higher, at $\sim$6,500\,K and 1.4\,\Rsun.

In the second scenario, we investigate whether $\pi^1$~Gru~C could have already evolved off the main sequence. The fact that $\pi^1$~Gru~A is near the tip of the AGB, with a significant difference between its initial and current mass, along with the current mass ratio $q = m_2/m_1$, provides strong constraints. For $\pi^1$~Gru~C to have already evolved into a (sub)giant while $\pi^1$~Gru~A remains in its TP-AGB phase, the difference in initial mass between $m_1$ and $m_2$ would need to be small. However, such a high initial mass for $\pi^1$~Gru~C is incompatible with the current mass ratio of $q \sim 1.05$ given the fact that $\pi^1$~Gru~A already lost a significant portion of its mass. Therefore, the (sub)giant scenario for $\pi^1$~Gru~C can be excluded.

The third scenario is for the case of $\pi^1$~Gru~C being a massive white dwarf (WD). Given the rather narrow mass distribution of WDs around $M\approx 0.6$\,\Msun, the existence of such massive white dwarfs has been proposed to arise either from the merger of two averaged-mass white dwarfs in close binary systems or from the evolution of massive intermediate-mass single stars~\cite{Althaus2010AARv..18..471A,Siess2007A&A...476..893S}. Massive white dwarfs are believed to harbor cores composed of mainly oxygen and neon. The mass-radius relation for a single WD with mean molecular weight of 2 implies a radius of $\sim 4,100$\,km~\cite{Althaus2010AARv..18..471A,Nauenberg1972ApJ...175..417N}, and thus a gravity $\log_{10} g \sim 9$. However, deriving the WD luminosity, and hence effective temperature, is challenging. One route here is using the \textit{Galex} FUV measurement (see Supplementary Sect.~\ref{Sec:SEDs}) as upper limit for the intrinsic luminosity of the WD. Using the FUV filter and accounting for reddening, we derive an upper limit for the effective temperature of $\sim$40,000~K\footnote{The hot temperature of the companion may be due to the accretion of matter from the wind of the giant star.}, corresponding to a luminosity of $\sim$0.07\,\Lsun\footnote{Neglecting reddening, the derived effective temperature is $\sim$20,000~K.}. The corresponding WD spectrum from Ref.~\cite{Koester2010MmSAI..81..921K}~\footnote{available via \url{http://svo2.cab.inta-csic.es/theory/newov2/index.php?models=koester2}} is shown in Supplementary Fig.~\ref{Fig:spectra}. For an atomic mass of 16, this implies a cooling age of $\sim 5 \times 10^7$~yr~\cite{Althaus2010AARv..18..471A}. As the white dwarf (WD) evolves, it continues to cool. After a cooling time of approximately $3 \times 10^9$ years, its luminosity will fall below $10^{-4} \, L_\odot$. Meanwhile, the WD accretes mass from the TP-AGB star. Based on the orbital evolution computations in Sect.~\ref{Sec:orbital_evolution}, the companion has already accreted approximately $0.03\,\Msun$ and is expected to accrete an additional $0.04$ to $0.11\,\Msun$ before entering the common-envelope phase. Assuming a current mass of $1.18\,\Msun$ for $\pi^1$~Gru~C, this accretion is insufficient for the white dwarf to surpass the Chandrasekhar limit. However, the 1-sigma uncertainty on the current mass of $\pi^1$~Gru~A suggests an upper limit of $1.37\,\Msun$, implying a potential companion mass of $1.44\,\Msun$, which exceeds the Chandrasekhar limit. If the companion is a white dwarf, there remains a small possibility that it could reach the Chandrasekhar limit before entering the common-envelope phase. This scenario could result in a Type Ia supernova, potentially allowing the system to avoid a common-envelope fate due to a white dwarf kick.
That being said, the likelihood of $\pi^1$~Gru~C being a massive white dwarf is quite low, not only due to the constraints imposed by its prior evolutionary scenario but also because no evidence of symbiotic activity -- such as accretion signatures, emission lines, or variability typically associated with such systems -- has been reported in the literature.

Hence, based on these statistical considerations tied to the evolutionary scenarios, the most probable evolutionary phase for $\pi^1$~Gru~C is that of a main-sequence star with a spectral type of F8V, making the AGB star the primary in the stellar system. However, prior ALMA data reveal that the circumstellar environment around $\pi^1$~Gru~A exhibits a fast, bipolar outflow with velocities reaching up to $\sim100~\mathrm{km~s^{-1}}$. Various symbiotic stars display similar morphologies, with bipolar outflows or jets formed by interactions between the accretion disk and the compact star~\cite{Ramstedt2018A&A...616A..61R}. If the potential symbiotic nature of the system is confirmed via emission lines characteristic of symbiotic stars -- such as the Balmer series, He~I, and forbidden lines like [O~III] and [Fe~VII] -- $\pi^1$~Gru~A would instead be identified as the secondary star.

A further investigation of the spectral energy distribution (SED; see Supplementary Sect.~\ref{Sec:SEDs}) and of the chemical signatures imprinted in the AGB stellar wind imprint (Supplementary Sect.~\ref{Sec:chemistry}) can shed further light on the proposed scenarios.

\subsection{Distinguishing evolutionary scenarios based on SED}\label{Sec:SEDs}

A first step toward distinguishing between the main-sequence and white-dwarf evolutionary scenarios involves analyzing photometric data spanning from the X-ray to the (sub)millimeter range (see Supplementary Fig.~\ref{Fig:spectra}). This requires the convolution of the spectral energy distribution (SED) with the appropriate response curves for each filter.
For the (sub)millimeter data, we rely on the ALMA Band 6 and Band 7 detections of $\pi^1$~Gru~C, with flux values provided in Sect.~\ref{Sec:observational_input}. Constraints on the optical emission from $\pi^1$~Gru~C are derived from its non-detection in the SPHERE-ZIMPOL observations~\cite{Montarges2023A&A...671A..96M}, which were obtained contemporaneously with the ALMA Band 6 data. During these observations, the minimum projected separation between the companion and the AGB star was 17.295\,mas ($\sim$1.9 times the AGB stellar radius), while the maximum projected separation over the full orbit reaches 37.83\,mas.

 $\pi^1$~Gru was also observed by the \textit{GALEX} satellite, which captured both near-ultraviolet (NUV) and far-ultraviolet (FUV) fluxes, with central wavelengths at 231\,nm and 153\,nm, respectively. The NUV emission is centered on the far-companion $\pi^1$~Gru~B, whereas the FUV emission is centered on $\pi^1$~Gru~A and its close-companion $\pi^1$~Gru~C ($\pi^1$~Gru~A and C are unresolved by the GALEX point-spread function of 4\farcs5 in the FUV)~\cite{Sahai2022Galax..10...62S}. The NUV flux is measured at 2.68038$\pm$0.0258\,mJy, while the FUV flux is 0.277$\pm$0.012\,mJy~\cite{Bianchi2017ApJS..230...24B}. The FUV-to-NUV flux ratio is $\sim$0.1. This places $\pi^1$~Gru in the intermediate domain between \textit{fuvAGBstars} which likely host an interacting companion surrounded by an accretion disk and \textit{nuvAGBstars} where the UV excess is either intrinsic arising from the AGB stellar chromosphere or extrinsic from accretion associated with a close binary companion~\cite{Sahai2022Galax..10...62S}. A significant fraction of \textit{nuvAGBstars}  may also be binarires with active, but weak accretion~\cite{Sahai2022Galax..10...62S}.

 The eROSITA instrument aboard the Spectrum-Roentgen-Gamma mission has detected X-ray emission toward $\pi^1$~Gru~\cite{Schmitt2024A&A...688A...9S} with the matching distance between the eROSITA and \textit{Gaia} position being 1\farcs2. The measured count rate in the 0.2-2.3~keV band is 1.155 counts/s, which is equivalent to a flux of $1.07 \times 10^{-12}$ milliWatt~m$^{-2}$~\cite{Merloni2024A&A...682A..34M} or $3.6 \times 10^{-4}$~mJy. Given the position of $\pi^1$~Gru in the eROSITA count rate versus apparent \textit{Gaia} $G$ magnitude diagram, Ref.~\cite{Schmitt2024A&A...688A...9S} concludes that there is a 98\% probability that the X-ray source is associated with the giant star producing coronal emission, which they classify as `optical contamination'. Nevertheless, given our detection of the close inner companion $\pi^1$~Gru~C, we cannot entirely rule out the possibility that the X-ray emission originates from the close-in companion, which would classify the inner binary system as a D-type symbiotic star. The eROSITA spectral shape, based on data from Ref.~\cite{Merloni2024A&A...682A..34M}, indicates that most X-ray photons originate in the 0.2–0.5\,keV range. According to the X-ray spectral classification scheme for symbiotic stars proposed by Ref.~\cite{Luna2013A&A...559A...6L}, the $\pi^1$~Gru system belongs to the $\alpha$ class, comprising supersoft X-ray sources  of which the likely origin is the collision of the wind from a white dwarf with that from the red giant.

 We do not include infrared data in this SED analysis because, in both scenarios, the infrared emission from the circumstellar envelope surrounding $\pi^1$~Gru~A dominates over that of $\pi^1$~Gru~C and its potential accretion disk. Consequently, these data do not provide additional insights in our search for the nature of $\pi^1$~Gru~C.

For the main-sequence scenario, the predicted optical fluxes in the SPHERE-ZIMPOL Cnt-Ha, Cnt748, and Cnt820 filters are 117\,mJy, 153\,mJy, and 178\,mJy, respectively, and even much lower for the WD scenario (see Supplementary Fig.~\ref{Fig:spectra}). These  low values explain the non-detection of the companion in the SPHERE-ZIMPOL data. Moreover, at the companion's location, the geometrical dilution factor remains approximately 30\%, and the SPHERE-ZIMPOL point spread function (PSF) has a full-width at half-maximum (FWHM) of 26.4\,mas~\cite{Montarges2023A&A...671A..96M}, corresponding to 1.4 times the AGB stellar diameter. Consequently, the circumstellar envelope's contribution to the flux in the SPHERE-ZIMPOL filters is $\sim$50 times stronger than the companion's emission, making the SPHERE-ZIMPOL data insufficient to distinguish between the two scenarios.

The predicted flux values in the X-ray and UV domains differ significantly between the main-sequence and white dwarf scenario. In the main-sequence case, accounting for reddening, the predicted NUV and FUV fluxes are 0.7~mJy and $8 \times 10^{-4}$~mJy, respectively. If reddening is neglected, these fluxes increase by a factor of 10. Given the uncertainties in estimating reddening and recognizing that the observed NUV emission is centered on the distant companion $\pi^1$~Gru~B, the NUV flux aligns reasonably well with the main-sequence scenario. However, the FUV flux would require additional explanation (see below). If the FUV flux is considered as an upper limit for the white dwarf luminosity, the observed NUV flux is approximately 25 times higher than the SED prediction. As mentioned, this discrepancy may be explained by the NUV emission originating from $\pi^1$~Gru~B.

However, neither scenario can explain the observed X-ray and ALMA (sub)millimeter fluxes. While a companion’s coronal activity cannot be excluded as origin for the X-ray emission~\cite{Schmitt2024A&A...688A...9S}, the alternative scenario of symbiotic activity cannot be excluded either. As we discuss in next section, we believe that the (sub)millimeter excess is caused by dust emission from an accretion disk surrounding $\pi^1$~Gru~C. If the observed FUV flux represents the disk accretion luminosity, we can define the efficiency factor $ \eta_{\rm FUV} $ as the fraction of gravitational energy that is converted into FUV radiation. For a mass accretion rate efficiency of $\sim 15\%$ (see Sect.~\ref{Sec:3D_hydro}), $\eta_{\rm FUV} \sim2\%$ for the main-sequence scenario. Note that for the WD scenario, we cannot estimate $\eta_{\rm FUV}$ given our assumption that the FUV flux represents the maximum of the WD luminosity.

In summary, accounting for the uncertainties in reddening, this SED analysis suggests that either $\pi^1$~Gru~C is an F8V main-sequence star, with its FUV emission potentially originating from the companion’s coronal activity or an accretion disk and its interaction with the companion, or that $\pi^1$~Gru~C is a massive white dwarf with an upper temperature limit of 40,000~K, where the X-ray luminosity may indicate symbiotic activity. While R~Aqr, at a distance of 218~pc, is currently the closest known symbiotic star~\cite{Ramstedt2018A&A...616A..61R}, $\pi^1$~Gru would become the closest symbiotic star if its nature were confirmed as such.

We note that H$\beta$, H$\gamma$, and H$\delta$, as well as several Fe lines, were detected in emission in the optical spectrum of $\pi^1$~Gru~\cite{Feast1953MNRAS.113..510F}. Such emission suggests the presence of an ionizing source or a high-density region where collisional excitation and recombination processes are efficient. These regions can be associated with a red giant's chromosphere~\cite{Judge1991ApJS...77...75J}, as well as with accretion disks, jets or outflows in binary systems, or ionized nebulae around hot stars. While chromospheric activity provides a possible explanation, the detection of these lines is also consistent with the potential presence of an accretion disk and our proposed nature of $\pi^1$~Gru~C.

\subsection{Distinguishing evolutionary scenarios based on chemical imprints}\label{Sec:chemistry}
The ultraviolet (UV) radiation from close stellar companions can significantly influence the chemistry of AGB outflows. This effect depends on the type of companion and the density structure of the outflow, which together determine the balance between destructive photoreactions and two-body reactions that increase chemical complexity~\cite{VandeSande2022MNRAS.510.1204V}\textsuperscript{,}\cite{VandeSande2023FaDi..245..586V}. The specific chemical signature is shaped by the radiation field (set by the companion type) and the outflow’s density structure. Photochemically produced molecules in the dense inner wind are a valuable diagnostic tool for detecting and characterizing companions~\cite{Danilovich2024NatAs...8..308D}\textsuperscript{,}\cite{Siebert2022ApJ...941...90S}.

Ref.~\cite{VandeSande2022MNRAS.510.1204V}\textsuperscript{,}\cite{VandeSande2023FaDi..245..586V} explored these effects using a 1D chemical model, considering three companion types: a red dwarf, a solar-like star, and a white dwarf with a surface temperature of 10,000~K. Their results suggest that a red dwarf has negligible influence on the outflow’s composition, whereas a solar-like star or a white dwarf drives the production of specific photochemically produced molecules, such as NS, SiC, and SiN, in both oxygen-rich and carbon-rich outflows. Despite $\pi^1$~Gru~A being an S-type AGB star, we expect similar molecular signatures to appear in the ALMA observations, especially considering their detection in another S-type AGB star, W~Aql~\cite{Danilovich2024NatAs...8..308D}.

These molecules, however, remain undetected in $\pi^1$~Gru~A~\cite{Wallstrom2024A&A...681A..50W}.
This non-detection can be attributed to several factors.
The chemical models assume a companion temperature of 10,000~K, while our estimates place the main-sequence companion of $\pi^1$~Gru~A at $\sim$6,200~K and the white dwarf companion at $\leq 40,000~$K. Additionally, the UV radiation from the accretion disk is not included in current chemical models. A larger UV radiation field might impact the efficiency of photochemical reactions and alter the expected molecular signatures, potentially destroying rather than producing molecules.
Such an apparently molecule-poor outflow is in line with observations, as $\pi^1$~Gru~A is among the most molecule-poor AGB stars observed in the ALMA survey by Ref.~\cite{Wallstrom2024A&A...681A..50W}, alongside U~Del.

Another possible contributing factor is the potential presence of an accretion disk, which can significantly modify the density structure of the outflow. Such a large-scale structure can disrupt the expected chemical processes~\cite{VandeSande2024MNRAS.532..734V}.

To resolve these uncertainties, a chemical model tailored to the specific 3D density structure and UV field of $\pi^1$~Gru~A’s outflow is necessary. Such a model would account for the contributions of the companion, the accretion disk, and the intrinsic properties of the star, providing more accurate predictions of molecular abundance distributions and offering the potential of using the chemical imprints of stellar companions to distinguish evolutionary scenarios.

\section{Accretion disk}\label{Sec:properties_dust_disk}

Neither the main-sequence or white dwarf scenario can explain the ALMA submillimeter Band 6 and Band 7 flux at the position of $\pi^1$~Gru~C (see Supplementary Table~\ref{table:UD}; Supplementary Fig.~\ref{Fig:spectra}). The observed emission has a spectral index, $\alpha^s_{M_2} = 2.3 \pm 0.3$ ($F_\nu \propto \nu^{\alpha^s}$), suggesting that it is dominated by dust. The hydrodynamical simulations presented in Sect.~\ref{Sec:3D_hydro} indicate that the accretion disk around the companion has a mass of {{$\sim 2\times10^{-6}$}}~\Msun. Both the inner and outermost regions of such accretion disks are expected to have very high temperatures due to shocks from the material accreted from the primary star and potential interactions with magnetic field lines\footnote{The surface magnetic field strength for AGB stars has been inferred to be of the order of several Gauss~\cite{Vlemmings2024A&A...686A.274V}, but direct observations of magnetic fields in this type of accretion discs are still lacking. The magnetic field strength for $\pi^1$~Gru~A has not yet been measured, but we note that giant convective cells were recently detected at its surface~\cite{Paladini2018Natur.553..310P}. Giant convective cells also exist on the supergiant Betelgeuse, and a local dynamo causes the detected magnetic field in it~\cite{Auriere2010A&A...516L...2A}. As such, it is plausible that a similar dynamo-driven mechanism could be responsible for the generation of magnetic fields in $\pi^1$~Gru~A. Further observational campaigns focusing on polarimetric measurements or Zeeman splitting could help constrain the magnetic field properties of this system.} In these regions, it is unlikely that dust can survive, as it would either sublimate or be destroyed by energetic collisions. However, in the intermediate disk regions, temperatures can be significantly lower, allowing dust grains to persist. The hydrodynamical simulations indicate that temperatures as high as 5,000\,--\,10,000\,K can be reached in the disk. However, it is important to note that the disk temperatures in these simulations are overestimated due to the fact that only H\,I cooling is accounted for.

Nevertheless, our ALMA observations can be used to estimate the amount of dust present in the disk and, therefore, guide future modeling approaches. The isotropic emission from a spherical dust grain of radius $a_g$ is given by
\begin{equation}
    l_\nu = 4 \pi a_g^2 \pi B_\nu(T_d) Q_\nu
\end{equation}
where $B_\nu(T_d)$ is the Planck function at the dust grain's temperature $T_d$, and $Q_\nu$ is the grain extinction efficiency (i.e., the ratio of the the dust grain's absorption cross-section to geometric cross-section), assumed to follow a power-law dependence on frequency~\cite{Knapp1993ApJS...88..173K}:
\begin{equation}
    Q_\nu = Q_0 {\left( \frac{\nu}{\nu_0} \right)}^\beta \,.
\end{equation}
We start with a simple model assuming that the disc is optcially thin at millimeter wavelengths. The total luminosity of such a spherically symmetric, optically thin dust cloud at a specific frequency $\nu$ can be expressed as
\begin{equation}
    L_\nu = \int_{r_{\rm in}}^{r_{\rm out}} 4 \pi r^2 n_g(r) Q_\nu \pi B_\nu(T_d(r)) 4 \pi a_g^2 \, dr
\end{equation}
where $n_g(r)$ is the grain number density. At low frequencies, the blackbody radiation can be approximated using the Rayleigh-Jeans law, leading to
\begin{equation}
    L_\nu = \frac{32 \pi^3 Q_\nu a_g^2 \nu^2 k_B}{c^2} \int_{r_{\rm in}}^{r_{\rm out}} T_d(r) r^2 n_g(r) \, dr
\end{equation}
where $k_B$ is the Boltzmann constant. Assuming spherical grains of size $a_g$ and specific density $\rho_s$, the total dust mass of the disk is given by
\begin{equation}
    M_d = \frac{16}{3} \pi^2 a_g^3 \rho_s \int_{r_{\rm in}}^{r_{\rm out}} r^2 n_g(r) \, dr \,.
\end{equation}
Hence, for an isothermal dust shell, the flux received at Earth is
\begin{equation}
    F_\nu = \frac{L_\nu}{4 \pi D^2} = \frac{3 Q_\nu \nu^2 T_d k_B M_d}{2 a_g c^2 \rho_s D^2} \,.\label{Eq:Fnu}
\end{equation}
Thus, the flux scales as $F_\nu \propto \nu^{\alpha^s}$, where $\alpha^s = 2 + \beta$. For evolved stars, the dust emissivity index $\beta$ typically ranges between $\sim 0$ and $1.0$~\cite{oGorman2015A&A...573L...1O}. Our observations suggest that, in the accretion disk around the companion, $\beta \sim 0.3 \pm 0.3$. Similar low values have also been found in many post-AGB objects~\cite{Sahai2011ApJ...739L...3S,Sahai2017ApJ...835L..13S}, one post-RGB object~\cite{Sahai2017ApJ...841..110S}, and in a compact dust cloud around the central star in V Hya~\cite{Sahai2022ApJ...929...59S}.

Although Eq.~\eqref{Eq:Fnu} is derived for the case of an optically thin, spherically symmetric cloud, it can easily be shown that the relation is also valid for optically thin disks, where $T_d$ is then representing the average dust temperature in the disk~\cite{Woitke2015EPJWC.10200007W}, i.e.
\begin{equation}
    T_d = <T_d> = \frac{\int T_d(r) \Sigma_d(r) r dr}{\int \Sigma_d(r) r dr }
\end{equation}
where $\Sigma_d(r)$ [g cm$^{-2}$] is the dust column density.

Assuming $Q_0 = 5.65 \times 10^{-4}$ at $\nu_0 = 274.6$~GHz, a dust grain radius of $0.1\,\mu$m, and a specific dust density $\rho_s$ of 3.5~g~cm$^{-3}$~\cite{oGorman2015A&A...573L...1O},
we use the ALMA Band 6 and Band 7 observations to estimate the dust mass in the disk. For a disk temperature of $1,000$\,K, the estimated dust mass is approximately $1.3 \times 10^{-8}$\,\Msun. In comparison, for a disk temperature of $300$\,K, the estimated dust mass increases to approximately $4.4 \times 10^{-8}$\,\Msun. In general, the derived dust mass is inversely dependent on both the assumed dust temperature and emissivity.

We note that our choice of $Q_0$, based on Ref.~\cite{oGorman2015A&A...573L...1O}, implies a grain emissivity of $\kappa_0\hspace{-3pt}=\hspace{-2pt}3 Q_0 / (4 \rho_s a)$ = 12 cm$^2$g$^{-1}$ at 274.6 GHz. This value is relatively high compared to typical literature values, which generally range between 1--3 cm$^2$g$^{-1}$ at $\lambda \sim$1 mm~\cite{Sahai2011ApJ...739L...3S}. Since the disk dust mass is inversely dependent on $Q_0$, reducing $Q_0$ by a factor of $\sim$6 would lead to a corresponding increase in dust disk mass by a similar factor.

In this analysis, the attenuation of the submillimeter $M_2$ emission by the circumstellar envelope (CSE) surrounding $\pi^1$~Gru~A can be neglected. For a gas mass-loss rate of $7.7\hspace{-1pt}\times\hspace{-1pt}10^{-7}$\,\Msun yr$^{-1}$~\cite{Doan2017A&A...605A..28D} and a dust-to-gas mass ratio of 1/100, the total dust optical depth through the CSE at 240\,GHz and 340\,GHz is estimated to be $\tau_\mathrm{CSE} < 0.001$, making its impact on the observations negligible.

We also examine a simple analytical model for a standard spatially thin accretion disk with a vertical structure that is isothermal and where the temperature profile as a function of radius $r$ is approximated by~\cite{Shakura1973A&A....24..337S}:
\begin{equation}
    T_d(r) = {\left(\frac{3 G m_2 \dot{M}_a}{8 \pi \sigma r^3}\right)}^{1/4} \equiv T_0 {\left(\frac{r_0}{r}\right)}^q\label{Eq:T_disk}
\end{equation}
where $G$ is the gravitational constant, $\dot{M}_a$ is the accretion rate, $\sigma$ is the Stefan-Boltzmann constant, $q=3/4$, and $T_0 = {((3 G m_2 \dot{M}_a)/(8 \pi \sigma r_0^3))}^{1/4}$. The hydrodynamical (HD) simulations indicate an accretion efficiency of approximately 15\% for an AGB mass-loss rate of $8 \times 10^{-7}$~\Msun~yr$^{-1}$ (see Sect.~\ref{Sec:3D_hydro}). These simulations also indicate an outer disk radius of $\sim$0.83~au, which is consistent with the fact that the disk remains spatially unresolved in the ALMA images with a spatial resolution of $\sim$19 mas (or $\sim$3 au). For radii between $\sim$0.085\,--1\,au, the disk temperature is expected to range between 1,200\,K and 190\,K, allowing for the existence of dust grains.

For a spatially thin, axisymmetric disk with temperature $T_d(r)$ and surface density $\Sigma_d(r)$ only as functions of the radial distance from $\pi^1$~Gru~C, the flux is given by~\cite{Woitke2015EPJWC.10200007W}
\begin{equation}
    F_\nu = \frac{2 \pi \cos i}{D^2} \int_{R_{\rm in}}^{R_{\rm out}} B_\nu(T(r)) \left(1 - \exp\left( - \frac{\tau_\nu(r)}{\cos i}\right)\right) r \, dr \,,\label{Eq:Fnu_integral}
\end{equation}
where $B_\nu(T(r))$ is the Planck function, $R_{\rm in}$ and $R_{\rm out}$ are the inner and outer disk radii, $D$ is the distance to the source, and $\tau_\nu(r) = \kappa_\nu^{\rm abs} \Sigma_d(r)$ is the dust optical depth across the disk. The flux from a flat, razor-thin disk that is vertically optically thick and inclined at an angle $i$ to the plane of the sky is then given by:
\begin{equation}
    F_\nu = \frac{2 \pi \cos i}{D^2} \int_{R_{\rm in}}^{R_{\rm out}} B_\nu(T(r)) \, r \, dr \,.
\end{equation}
In the Rayleigh-Jeans limit, assuming the temperature profile of the disk given in Eq.~\eqref{Eq:T_disk} and recognizing that $R_{\rm in} \ll R_{\rm out}$, the total flux becomes:
\begin{equation}
    F_\nu = \frac{16 \pi \cos i \, \nu^2 k_B}{5 D^2 c^2} {\left(\frac{3 G m_2 \dot{M}_a}{8 \pi \sigma} \right)}^{1/4} R_{\rm out}^{5/4} \,.
\end{equation}
For $R_{\rm in} = 0.085$\,au and $R_{\rm out} = 1$\,au, the predicted flux at 240~GHz and 340~GHz is 1.14~mJy and 2.19~mJy, respectively, which is approximately a factor of 2.5 lower than the observed values. In this limiting case of an optically thick, razor-thin disk, this discrepancy could suggest that $\dot{M}_a$ is approximately 40 times higher than currently predicted. This difference in mass accretion could potentially be explained by an episodically enhanced mass-loss rate of $3.3 \times 10^{-6}$\,\Msun~yr$^{-1}$, which is 5 times larger than our average mass-loss rate~\cite{Doan2020A&A...633A..13D}. This episodic increase has been proposed as the cause of the fast bipolar outflow observed in the ALMA interferometric images. However, in the mm-cm wavelength range, the disk becomes increasingly transparent, leading to the breakdown of the assumption of optically thick blackbody emission at certain wavelengths. Consequently, the impact of episodic mass loss on the dust emission from the accretion disk should be interpreted with caution.

 As a final refinement, we approximate the integral in Eq.~\eqref{Eq:Fnu_integral} by dividing the disk into two regions: one where $\tau_\nu \geq 1$ and the other where $\tau_\nu < 1$, following the methodology presented in Ref.~\cite{Beckwith1990AJ.....99..924B}. We adopt a power law for the dust surface density:
\begin{equation}
\Sigma_d(r) = \Sigma_{d,0} {\left(\frac{r_{\rm in}}{r}\right)}^p\,.\label{Eq:Sigma_dust}
\end{equation}
Radial integration over Eq.~\eqref{Eq:Sigma_dust} from $R_{\rm in}$ to $R_{\rm out}$ results in the total disk dust mass $M_d$, i.e.
\begin{equation}
    M_d = \frac{ 2 \pi \Sigma_{d,0} R_{\rm in}^p R_{\rm out}^{2-p}}{2-p}\,.
\end{equation}
For realistic disks with $p > 0$, the inner region is optically thick, while the outer region is optically thin, with the transition occurring at $r_1$. For the average disk dust optical depth defined as~\cite{Beckwith1990AJ.....99..924B}:
\begin{equation}
    \bar{\tau}_\nu = \frac{\kappa_\nu M_d}{\cos i \, \pi R_{\rm out}^2}\,,\label{Eq:tau_bar}
\end{equation}
it has been shown that~\cite{Beckwith1990AJ.....99..924B}:
\begin{equation}
    r_1 \approx {\left( \frac{2-p}{2} \right)}^{1/p} \bar{\tau}_\nu^{1/p} R_{\rm out},
\end{equation}
for $p<2$. Defining
\begin{equation}
    \Delta = -\frac{p}{(2-q) \log\left\{ \left[\frac{2-p}{2}\right] \bar{\tau}_\nu\right\}}
\end{equation}
and
\begin{equation}
    A_1 = 16 \pi^2 \frac{k \nu^2}{c^2},
\end{equation}
the flux~\footnote{In contrast to Ref.~\cite{Beckwith1990AJ.....99..924B}, we use luminosity and flux in our analysis rather than luminosity density and flux density, which differ by a factor of $\nu$. Consequently, $A_1$ has a $\nu^2$ dependence instead of a $\nu^3$ dependence.} can be expressed as~\cite{Beckwith1990AJ.....99..924B}:
\begin{equation}
    F_\nu = \frac{A_1\, f_0 \,\cos i \, T(r_1)\, r_1^2}{4 \pi D^2} \frac{{(R_{\rm out}/r_1)}^{(2-p-q)} -1}{2-p-q}\, (1+\Delta),\label{Eq:Fnu_thick_thin}
\end{equation}
where $f_0 \approx 0.8$.

By solving Eqs.~\eqref{Eq:tau_bar}--\eqref{Eq:Fnu_thick_thin} and comparing the predictions at 240~GHz and 340~GHz for an inclination of $39.37^\circ$ (see Table~\ref{table:robustness}) with the observed ALMA band 6 and band 7 data, we derived the average disk dust optical depth $\bar{\tau}_\nu$, and corresponding $r_1$ and disk dust mass. For $p = 1$, we find that for $\bar{\tau}_\nu = 1.9$~\footnote{The listed value for $\bar{\tau}_\nu$ is the average value of $\bar{\tau}_{\rm 240~GHz}$ and $\bar{\tau}_{\rm 340~GHz}$, which only differ by 10\%.}, the predicted flux values at 240~GHz and 340~GHz are 2.94~mJy and 5.65~mJy, respectively. The corresponding value for $r_1$ is 0.82~au, with the disk dust mass being $\sim 8.5 \times 10^{-8}$\,\Msun. As discussed by Ref.~\cite{Beckwith1990AJ.....99..924B}, the derived disk mass depends only weakly on the choice of the density exponent $p$. For $p=0.5$, the derived disk dust mass is $\sim 5.8 \times 10^{-8}$\,\Msun. An outer disk radius of  $\sim 1$~au and a value of $r_1 \sim 0.82$~au implies that the emission is more or less optically-thick everywhere in the disk. Hence, although the disk dust mass derived using the optically thin approach is consistent with the mass derived from the more sophisticated model, which accounts for both optically thin and optically thick regions, this agreement might be coincidental. The latter method should be preferred, as it is more robust and provides more reliable results.

\clearpage
\section{Supplementary information galactic orbit}
To constrain the location where the $\pi^1$~Gru system was formed, we traced back the evolution of the system over its probable lifetime. We used {\tt{gaply}} \cite{2015ApJS..216...29B}, a widely used python library for galactic dynamics. We calculated the current 6D Galactocentric phase-space position of the system based on the celestial coordinates, distance, proper motion and radial velocity of the system's barycentre (see Sect.~\ref{Methods:orbital_results}). We assumed the three-component {\tt{MWPotential2014}} axisymmetric potential of the Milky Way~\cite{2015ApJS..216...29B}, which has been calibrated based on a variety of stellar dynamical data. We fixed the Galactocentric distance of the Sun at $R_0 = 8.23 \pm 0.12~{\text{kpc}}$ \cite{2023MNRAS.519..948L} and the Sun's velocity in the rotation direction at $v_\odot/R_0 = 30.32\pm0.27~{\text{km}}~{\text{s}}^{-1}~{\text{kpc}}^{-1}$ \cite{2019ApJ...885..131R}.

We back-integrated the orbit of the $\pi^1$~Gru system over the past 2.8~Gyr and focused on the part of the orbit corresponding to the time span between 2.8 and 1.8~Gyr ago, corresponding to the boundaries of the estimate of the stellar age of $\pi^1$~Gru~A. We calculated 1\,000 orbits by Monte Carlo sampling the probability distribution of each of the input parameters (we assumed Gaussian probability distributions). A representative sample of 20 orbits is shown in Supplementary Fig.~\ref{pi1Gru-GalacticOrbit.fig}, both projected onto the Galactic plane (left) and in the meridional plane (right). The system oscillates in the radial and vertical directions while orbiting the Galactic centre. The Galactocentric radius of the system varies between 6.8 and 8.6~kpc, while the height oscillates between 15~pc above and below the Galactic plane.

Supplementary Fig.~\ref{pi1Gru-GalacticOrbit-Histograms.fig} shows the relative probability of finding the system at a given Galactocentric radius and height at any moment between 2.8 and 1.8~Gyr ago, based on the 1\,000 calculated orbits. The probability distribution in Galactocentric radius is bimodal with modes near 7.25 and 8.45~kpc, respectively. The formal mean Galactocentric radius is $\langle R \rangle = 7.86\pm0.50$~kpc. The probability distribution in height is also bimodal with modes corresponding to 130~pc above and below the Galactic plane. The formal mean of the distribution is $\langle z \rangle = 0.8\pm90.7~{\text{pc}}$.

Based on this analysis, we conclude that the $\pi^1$~Gru system was formed close to the Galactic plane, either near or slightly within the Solar orbit.

\begin{figure}[!htp]
    \centering
    \begin{tikzpicture}
        \node[anchor=south west,inner sep=0] (image) at (0,7.4) {\includegraphics[width=\textwidth]{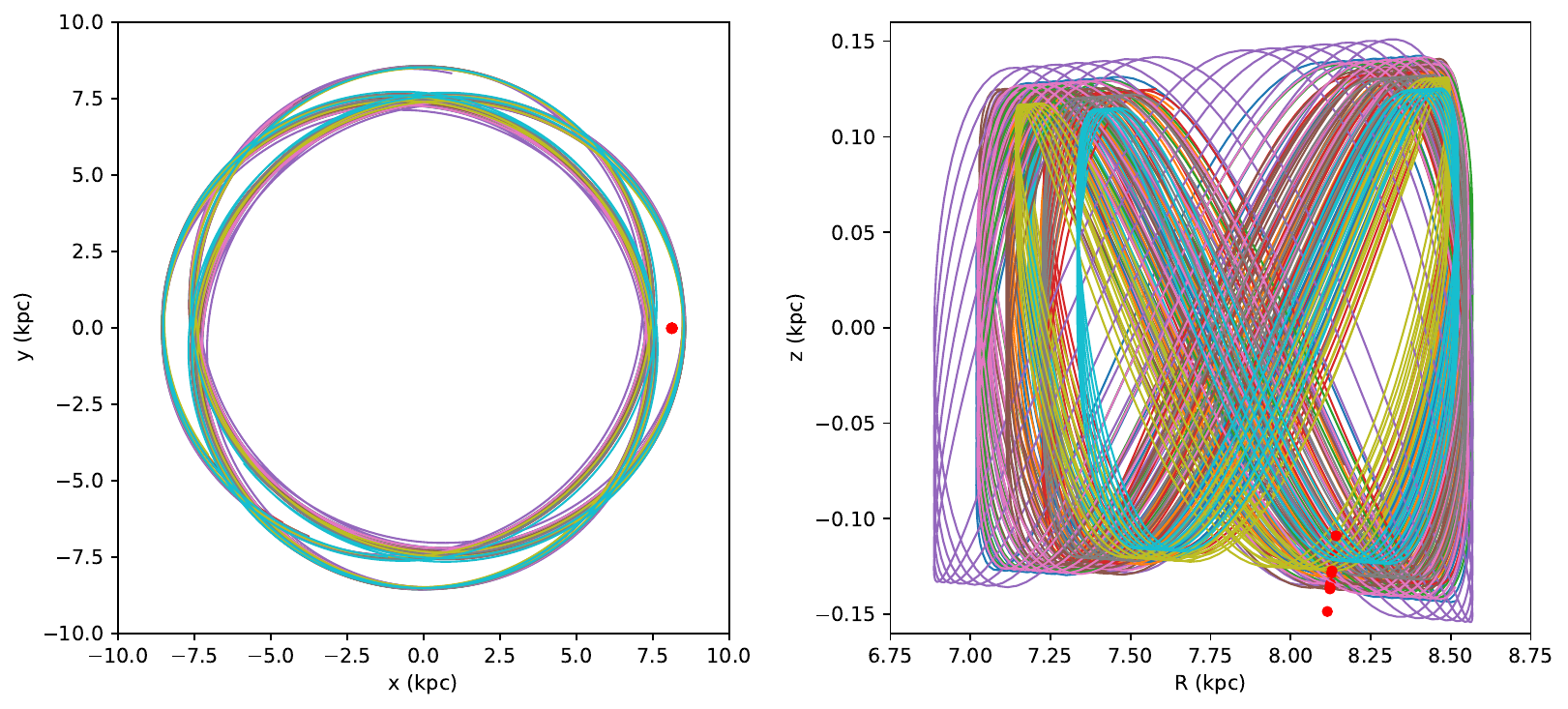}};
        \begin{scope}[x={(image.south east)},y={(image.north west)}]
            \node[anchor=north west] at (0.08,  0.94)  {(a)};
            \node[anchor=north west] at (0.575, 0.697) {(b)};
          \end{scope}
    \end{tikzpicture}
    \caption{\textbf{The Galactic orbit of the $\boldsymbol{\pi^1}$~Gru system.} Predicted orbital motion of the $\pi^1$~Gru system around the Galactic centre where panel~(a) shows the Galactic plane while panel~(b) shows the meridional plane. The different lines correspond to 10 different possible orbits, corresponding to Monte Carlo variations of the initial conditions according to the uncertainties on the observational data. The red dots represent the current location of $\pi^1$~Gru.}\label{pi1Gru-GalacticOrbit.fig}
\end{figure}

\begin{figure}[!htp]
    \centering
    \begin{tikzpicture}
        \node[anchor=south west,inner sep=0] (image) at (0,7.4) {\includegraphics[width=\textwidth]{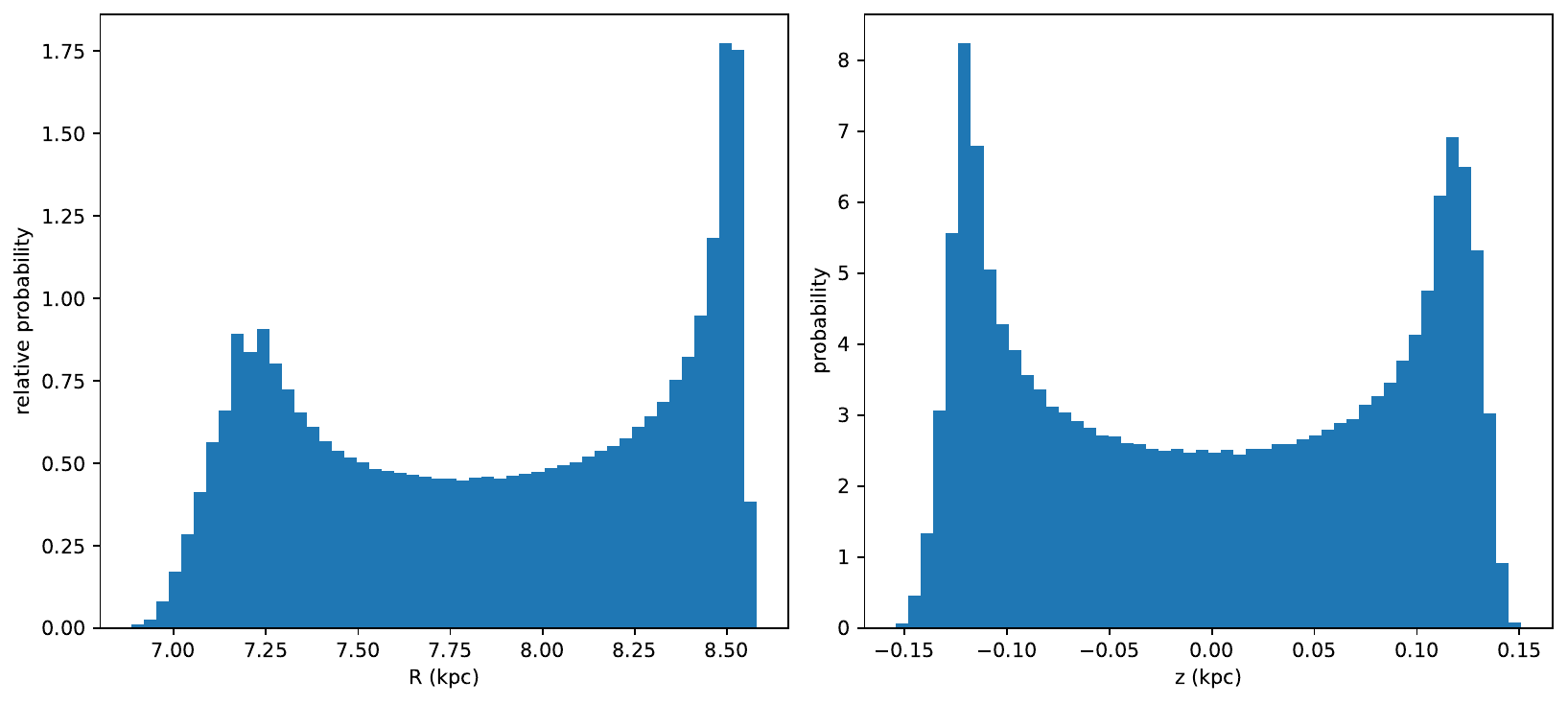}};
        \begin{scope}[x={(image.south east)},y={(image.north west)}]
            \node[anchor=north west] at (0.07,  0.94)  {(a)};
            \node[anchor=north west] at (0.56, 0.697) {(b)};
          \end{scope}
    \end{tikzpicture}
    \caption{\textbf{Formation site of the $\boldsymbol{\pi^1}$~Gru system.} Histograms diplaying the probability distributions of the possible locations of the formation site of the $\pi^1$~Gru system. Panel~(a) shows the probability as a function of Galactocentric radius, and Panel~(b) as a function of height above the Galactic plane.}\label{pi1Gru-GalacticOrbit-Histograms.fig}
\end{figure}

\clearpage
\section{Additional retrieval tables}

\afterpage{
\begin{landscape}
\begin{table*}[htp]
\caption{\textbf{Robustness analysis of the retrieved parameters for the $\boldsymbol{\pi^1}$~Gru system.} Each row presents the results from 10 independent retrieval runs using the \texttt{ultranest} algorithm. The last column shows the corresponding Bayesian evidence (marginalized likelihood), expressed as $\log z$. The first part of the table reports the mean and standard deviation of the retrieved parameters, computed from posterior samples. The mean is derived by weighting the samples according to their likelihood, while the standard deviation reflects the spread of the posterior distribution. The second part provides the best-fit values, corresponding to the maximum posterior probability for each parameter. We refer to the section on `Geometrical Degeneracies' for a discussion on the degeneracy of $i$ and its complement ($360^\circ-i$) and ($\omega$, $\Omega$) with their antipodal values ($\omega+180^\circ$, $\Omega+180^\circ$). Among all the runs, \texttt{run e3} achieved the highest maximum likelihood for its best-fit value. We refer to this outcome as the \texttt{eccentric} model.}\label{table:robustness}
\centering
\small
\setlength{\tabcolsep}{0.8mm}
\begin{tabular} {l|ccccccccccc|l}
\toprule
run & $m_1$ & $q$ & $a$ & $e$ & $T_0$ & $\omega$ & $\Omega$ & $i$ & $D$ & $\mu^G_\alpha$ & $\mu^G_\delta$ & $\log z$\\
 & [$M_\odot$] & & [au] & & [yr] & [$^\circ$] & [$^\circ$] & [$^\circ$] & [pc] & [mas yr$^{-1}$] & [mas yr$^{-1}$] & \\
\midrule
e1 & 1.06$\pm$0.23 & 1.05$\pm$0.05 & 6.7$\pm$0.45 & 0.020$\pm$0.014 & 2021.89$\pm$4.39 & 178$\pm$122.75 & 91$\pm$40 & 12$\pm$7 & 177$\pm$10 & 45.2434$\pm$0.160 & $-$18.791$\pm$0.068 & 75.531\\
e2 & 1.33$\pm$0.25 & 1.05$\pm$0.05 & 7.24$\pm$0.43 & 0.018$\pm$0.013 & 2026.39$\pm$2.35 & 240$\pm$102 & 122$\pm$26 & 12$\pm$7 & 188$\pm$9 & 45.256$\pm$0.152 & $-$18.804$\pm$0.063 & 73.251 \\
\textbf{e3} & \textbf{1.02$\pm$0.20} & \textbf{1.04$\pm$0.05} & \textbf{6.60$\pm$0.41} & \textbf{0.023$\pm$0.017} & \textbf{2026.75$\pm$3.16} & \textbf{101$\pm$98} & \textbf{94$\pm$23} & \textbf{14$\pm$8} & \textbf{174$\pm$9} & \textbf{45.212$\pm$ 0.168} & \textbf{$-$18.773$\pm$0.068} & \textbf{75.192}\\
e4 & 1.03$\pm$0.20 & 1.05$\pm$0.05 & 6.64$\pm$0.40 & 0.019$\pm$0.014 & 2021.22$\pm$3.81 & 144$\pm$124 & 105$\pm$31 & 13$\pm$7 & 176$\pm$9 & 45.222$\pm$0.152 & $-$18.778$\pm$0.066 & 75.256 \\
e5 & 1.11$\pm$0.22 & 1.06$\pm$0.05 & 6.82$\pm$0.42 & 0.015$\pm$0.012 & 2023.41$\pm$4.10 & 174$\pm$96 & 89$\pm$51 & 10$\pm$6 & 180$\pm$9 & 45.231$\pm$0.153 & $-$18.791$\pm$0.065 & 74.560 \\
e6  & 1.13$\pm$0.24 & 1.04$\pm$0.05 & 6.82$\pm$0.46 & 0.021$\pm$0.014 & 2020.60$\pm$4.18 & 93$\pm$105 & 101$\pm$31 & 13$\pm$7 & 178$\pm$10 & 45.227$\pm$0.152 & $-$18.786$\pm$0.065 & 75.050 \\
e7  & 1.00$\pm$0.30 & 1.07$\pm$0.06 & 6.55$\pm$0.60 & 0.023$\pm$0.015 & 2023.43$\pm$4.64 & 201$\pm$132 & 114$\pm$31 & 13$\pm$7 & 174$\pm$11 & 45.244$\pm$0.159 & $-$18.788$\pm$0.068 & 74.075 \\
e8  & 1.09$\pm$0.23 & 1.05$\pm$0.05 & 6.75$\pm$0.46 & 0.021$\pm$0.014 & 2024.27$\pm$4.73 & 87$\pm$84 & 44$\pm$37 & 9$\pm$6 & 178$\pm$9 & 45.289$\pm$0.155 & $-$18.823$\pm$0.064 & 75.781 \\
e9  & 1.06$\pm$0.27 & 1.05$\pm$0.05 & 6.67$\pm$0.55 & 0.022$\pm$0.014 & 2016.58$\pm$2.50 & 85$\pm$96 & 50$\pm$33 & 9$\pm$6 & 176$\pm$11 & 45.292$\pm$0.150 & $-$18.824$\pm$0.060 & 75.755 \\
e10 & 1.23$\pm$0.26 & 1.05$\pm$0.06 & 7.04$\pm$0.47 & 0.022$\pm$0.014 & 2018.86$\pm$4.90 & 224$\pm$130 & 103$\pm$33 & 13$\pm$7 & 183$\pm$10 & 45.248$\pm$0.159 & $-$18.800$\pm$0.066 & 75.182\\
\midrule
e1   & 1.01  & 1.04  & 6.62  & 0.033  & 2017.09  & 14.05  & 109.50  & 19.06  & 173.32  & 45.283  & $-$18.776 & 113.810  \\
e2   & 1.17  & 1.02  & 6.91  & 0.030  & 2029.67  & 37.53  & 109.02  & 20.26  & 177.77  & 45.223  & $-$18.793 & 113.191 \\
\textbf{e3}   & \textbf{0.89}  & \textbf{1.06} & \textbf{6.35}  & \textbf{0.031}  & \textbf{2028.58}  & \textbf{4.09}  & \textbf{107.97}  & \textbf{15.76}  & \textbf{169.38}  & \textbf{45.308}  & \textbf{$-$18.805} &  \textbf{113.937} \\
e4   & 0.97  & 1.04  & 6.52  & 0.027  & 2016.67  & 4.74  & 104.88  & 15.04  & 171.73  & 45.281  & $-$18.818 & 113.915 \\
e5   & 1.09  & 1.05  & 6.80  & 0.018  & 2027.89  & 345.83  & 104.37  & 12.12  & 178.80  & 45.22  & $-$18.787 & 113.348 \\
e6   & 1.03  & 1.02  & 6.63  & 0.032  & 2017.03  & 17.41  & 103.95  & 16.75  & 172.07  & 45.283  & $-$18.793 & 113.823 \\
e7   & 0.93  & 1.05  & 6.44  & 0.029  & 2028.11  & 350.24  & 108.09  & 13.56  & 170.56  & 45.293  & $-$18.823 & 113.771  \\
e8   & 1.08  & 1.05  & 6.77  & 0.021  & 2016.18  & 3.45  & 91.89  & 17.39  & 177.30  & 45.173  & $-$18.773  & 113.363 \\
e9   & 1.01  & 1.05  & 6.60  & 0.033  & 2016.03  & 357.13  & 92.63  & 17.65  & 173.50  & 45.296  & $-$18.813 & 113.695 \\
e10   & 0.97  & 1.08  & 6.54  & 0.034  & 2016.72  & 354.15  & 117.43  & 19.33  & 173.86  & 45.315  & $-$18.817 & 113.722 \\
\bottomrule
\end{tabular}
\end{table*}
\end{landscape}}

\afterpage{
\begin{landscape}
\begin{table*}[htp]

\caption{\textbf{Robustness analysis of the retrieved parameters for a circular orbit describing the $\boldsymbol{\pi^1}$~Gru system.} This table is similar to Supplementary Table~\ref{table:robustness}, but specific to a circular orbit ($e=0$). For a circular orbit, $\omega$ and $T_0$ are undefined; here, $T_0$ denotes the reference time at which the body crosses the ascending node. Among all the runs, \texttt{run c7} achieved the highest maximum likelihood for its best-fit value, referred to here as the \texttt{circular} model.}\label{table:robustness_circular}
\centering
\small
\setlength{\tabcolsep}{0.8mm}
\begin{tabular} {l|ccccccccc|l}
\toprule
run & $m_1$ & $q$ & $a$ & $T_0$ & $\Omega$ & $i$ & $D$ & $\mu^G_\alpha$ & $\mu^G_\delta$ & $\log z$\\
 & [$M_\odot$] & & [au] & [yr] & [$^\circ$] & [$^\circ$] & [pc] & [mas yr$^{-1}$] & [mas yr$^{-1}$] & \\
\midrule
c1 & 1.08$\pm$0.23 & 1.05$\pm$0.05 & 6.73$\pm$0.46 & 2016.52$\pm$1.14 & 105$\pm$35 & 11$\pm$7 & 178$\pm$10 & 45.204$\pm$0.146 & $-$18.775$\pm$0.062 & 78.555\\
c2 & 1.13$\pm$0.26 & 1.04$\pm$0.05 & 6.84$\pm$0.49 & 2016.22$\pm$1.14 & 96$\pm$35  & 11$\pm$7 & 180$\pm$10 & 45.202$\pm$0.143 & $-$18.778$\pm$0.060 & 78.231 \\
c3 & 1.09$\pm$0.26 & 1.04$\pm$0.05 & 6.75$\pm$0.52 & 2016.04$\pm$1.16 & 90$\pm$35  & 10$\pm$6 & 178$\pm$11 & 45.200$\pm$0.142 & $-$18.776$\pm$0.059 & 78.655 \\
c4 & 1.06$\pm$0.25 & 1.04$\pm$0.05 & 6.69$\pm$0.5  & 2016.32$\pm$1.13 & 99$\pm$34  & 10$\pm$6 & 177$\pm$10 & 45.201$\pm$0.143 & $-$18.774$\pm$ 0.061 & 78.354 \\
c5 & 1.18$\pm$0.25 & 1.04$\pm$0.05 & 6.93$\pm$0.47 & 2016.34$\pm$1.09 & 100$\pm$33 & 11$\pm$7 & 182$\pm$10 & 45.204$\pm$0.143 & $-$18.779$\pm$0.059 & 78.333 \\
c6 & 1.11$\pm$0.26 & 1.04$\pm$0.05 & 6.79$\pm$0.51 & 2016.31$\pm$1.13 & 99$\pm$35  & 11$\pm$6 & 179$\pm$10 & 45.201$\pm$0.144 & $-$18.776$\pm$0.060 & 78.265 \\
\textbf{c7} & \textbf{1.12$\pm$0.25} & \textbf{1.05$\pm$0.05} & \textbf{6.81$\pm$0.49} & \textbf{2016.39$\pm$1.18} & \textbf{101$\pm$36} & \textbf{11$\pm$7} & \textbf{180$\pm$10 } & \textbf{45.203$\pm$0.144} & \textbf{$-$18.776$\pm$0.061} & \textbf{78.805} \\
c8 & 1.16$\pm$0.26 & 1.04$\pm$0.05 & 6.90$\pm$0.48 & 2016.34$\pm$1.11 & 100$\pm$34 & 11$\pm$7 & 181$\pm$10 & 45.197$\pm$0.144 & $-$18.775$\pm$0.061 & 78.047\\
c9 & 1.12$\pm$0.25 & 1.05$\pm$0.05 & 6.82$\pm$0.48 & 2016.34$\pm$1.12 & 100$\pm$34 & 11$\pm$7 & 180$\pm$10 & 45.198$\pm$0.143 & $-$18.775$\pm$0.061 & 78.241\\
c10 & 1.05$\pm$0.27 & 1.05$\pm$0.05 & 6.66$\pm$0.55 & 2016.33$\pm$1.14 & 100$\pm$34 & 11$\pm$7 & 177$\pm$11 & 45.199$\pm$0.143 & $-$18.772$\pm$0.062& 78.710\\
\midrule
c1 & 0.96 & 1.04 & 6.52 & 2017.08 & 121.80 & 11.31 & 174.0 & 45.211 & $-$18.765 & 112.697\\
c2 & 1.02 & 1.06 & 6.65 & 2016.82 & 114.37 & 15.01 & 176.94 & 45.170 & $-$18.769 & 112.706 \\
c3 & 0.98 & 1.05 & 6.59 & 2016.77 & 112.90 & 14.56 & 174.92 & 45.143 & $-$18.738 & 112.708\\
c4 & 0.99 & 1.04 & 6.57 & 2016.90 & 116.23 & 13.62 & 174.64 & 45.237 & $-$18.765 & 112.642\\
c5 & 0.90 & 1.07 & 6.40 & 2016.97 & 118.87 & 15.10 & 172.34 & 45.186 & $-$18.761 & 112.689 \\
c6 & 0.89 & 1.05 & 6.36 & 2016.57 & 106.52 & 9.80 & 171.13 & 45.188 & $-$18.774 & 112.731 \\
\textbf{c7} & \textbf{0.93} & \textbf{1.05} & \textbf{6.46} & \textbf{2016.72} & \textbf{111.41} & \textbf{14.04} & \textbf{172.28} & \textbf{45.149} & \textbf{$-$18.740} & \textbf{112.739} \\
c8 & 1.04 & 1.06 & 6.70 & 2016.83 & 114.91 & 13.71 & 178.12 & 45.147 & $-$18.757 & 112.727 \\
c9 & 1.03 & 1.05 & 6.68 & 2017.05 & 121.25 & 13.38 & 176.80 & 45.194 & $-$18.753 & 112.671 \\
c10 & 0.95 & 1.05 & 6.49 & 2016.65 & 109.25 & 11.66 & 174.33 & 45.183 & $-$18.762 & 112.714 \\
\bottomrule
\end{tabular}
\end{table*}
\end{landscape}}

\afterpage{
\begin{landscape}
\begin{table*}[htp]
\caption{\textbf{Robustness analysis of the retrieved parameters for the $\boldsymbol{\pi^1}$~Gru system using the approximate orbital solution.} The columns are defined in the same way as in Supplementary Table~\ref{table:robustness}.}\label{table:approximate}
\centering
\small
\setlength{\tabcolsep}{0.8mm}
\begin{tabular} {l|ccccccccccc|l}
\toprule
run & $m_1$ & $q$ & $a$ & $e$ & $T_0$ & $\omega$ & $\Omega$ & $i$ & $D$ & $\mu^G_\alpha$ & $\mu^G_\delta$ & $\log z$ \\
 & [$M_\odot$] & & [au] & & [yr] & [$^\circ$] & [$^\circ$] & [$^\circ$] & [pc] & [mas yr$^{-1}$] & [mas yr$^{-1}$] & \\
\midrule
1 & 0.68$\pm$0.07 & 1.18$\pm$0.07 & 6.14$\pm$0.19 & 0.031$\pm$0.018 & 2020.5$\pm$3.23 & 79$\pm$92 & 146$\pm$7, & 38$\pm$4 & 155$\pm$5 & $\mu_{HG,\alpha_1}$ & $\mu_{HG,\delta_1}$& 114.394\\
2 & 0.68$\pm$0.07 & 1.18$\pm$0.07 & 6.14$\pm$0.19 & 0.033$\pm$0.017 & 2018.99$\pm$1.14 & 66$\pm$84 & 147$\pm$7 & 38$\pm$4 & 154$\pm$5 & $\mu_{HG,\alpha_1}$ & $\mu_{HG,\delta_1}$&  115.884\\
3 & 0.68$\pm$0.06 & 1.19$\pm$0.07 & 6.13$\pm$0.18 & 0.03$\pm$0.018 & 2019.92$\pm$4.31 & 218$\pm$142 & 145$\pm$7 & 37$\pm$4 & 155$\pm$5 & $\mu_{HG,\alpha_1}$ & $\mu_{HG,\delta_1}$&  114.311\\
4 & 0.68$\pm$0.07 & 1.19$\pm$0.07 & 6.14$\pm$0.18 & 0.032$\pm$0.018 & 2019.0$\pm$1.64 & 80$\pm$101 & 146$\pm$7 & 38$\pm$4 & 154$\pm$5 & $\mu_{HG,\alpha_1}$ & $\mu_{HG,\delta_1}$&  113.901\\
\midrule
1 & 0.60 & 1.20 & 5.91 & 0.031 & 2017.94 & 2.87 & 147.87 & 38.33 & 148.80 & $\mu_{HG,\alpha_1}$ & $\mu_{HG,\delta_1}$& 151.125 \\
2 & 0.60 & 1.23 & 5.94 & 0.034 & 2018.02 & 6.90 & 146.80 & 39.97 & 149.84 & $\mu_{HG,\alpha_1}$ & $\mu_{HG,\delta_1}$& 151.191\\
3 & 0.61 & 1.24 & 5.97 & 0.039 & 2017.71 & 355.76 & 149.20 & 40.73 & 147.78 & $\mu_{HG,\alpha_1}$ & $\mu_{HG,\delta_1}$& 151.037\\
4 & 0.62 & 1.22 & 5.98 & 0.030 & 2018.25 & 11.35 & 149.21 & 39.09 & 150.33 & $\mu_{HG,\alpha_1}$ & $\mu_{HG,\delta_1}$& 151.129\\

\bottomrule
\end{tabular}
\end{table*}
\end{landscape}}

\afterpage{\clearpage}
\newpage

\section{Additional corner plots}

\begin{figure*}[!htp]
\centering
\includegraphics[width=0.98\textwidth]{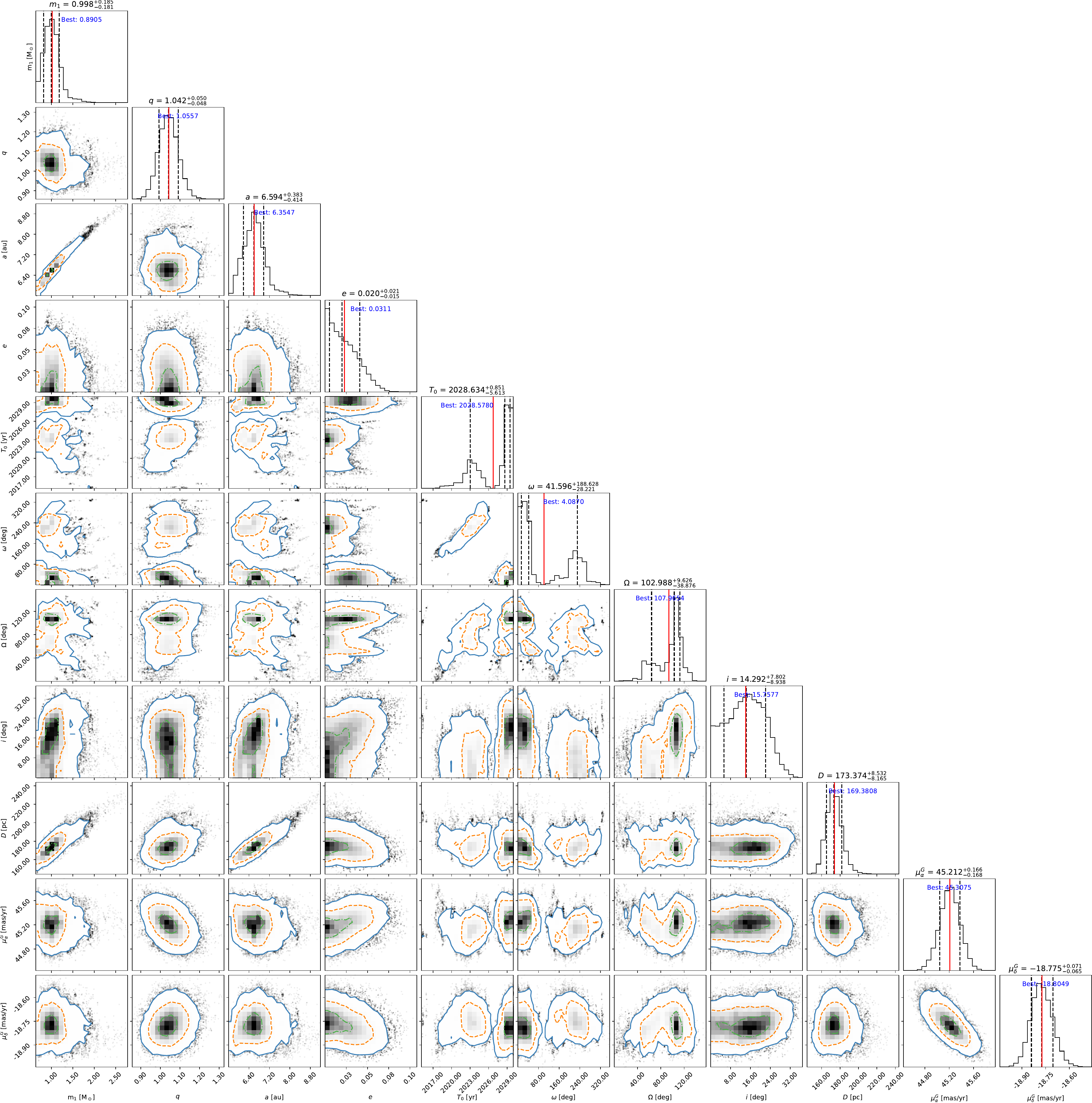}
\caption{\textbf{Corner plot showing the retrieval of the $\boldsymbol{\pi^1}$~Gru orbital parameters using \texttt{ultranest} sampling for an eccentric orbit.} For a detailed explanation, refer to the caption of Extended Data Fig.~\ref{fig:pi1_gru_corner_ultranest}. See also Supplementary Table~\ref{table:fit_sensitivity} and Supplementary Table~\ref{table:best_fit}. The posterior distribution of $T_0$ reveals two maxima: one at $T_0 = 2028.6$ and another at $T_0 - T_{\rm orb}$.}\label{fig:pi1_gru_corner_eccentric}
\end{figure*}

\begin{figure*}[!htp]
\centering
\includegraphics[width=0.98\textwidth]{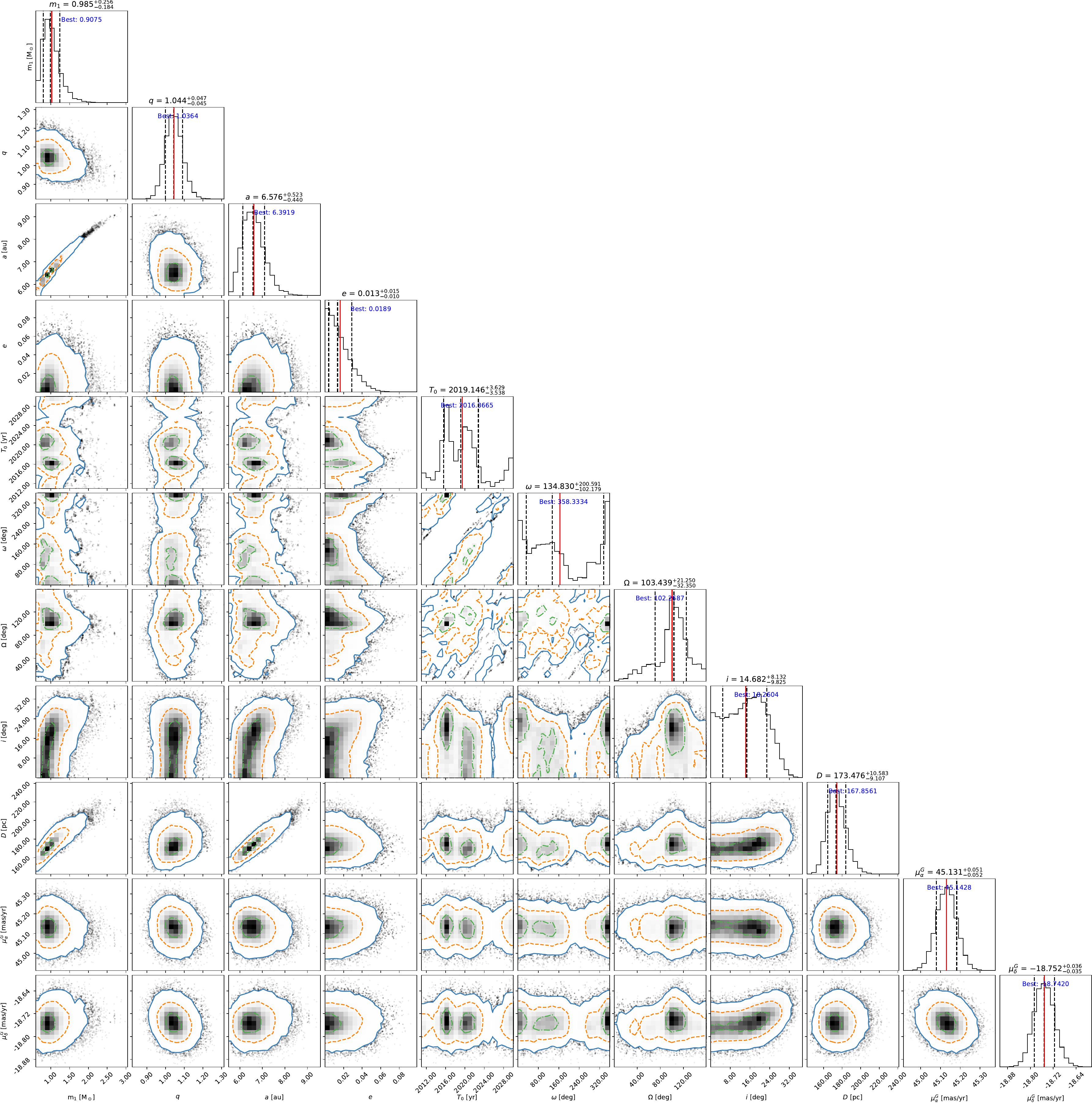}
\caption{\textbf{Corner plot showing the retrieval of the $\boldsymbol{\pi^1}$~Gru orbital parameters using \texttt{ultranest} sampling, not accounting for the \textit{Gaia} and \textit{Hipparcos} covariance matrices.} For a full explanation, we refer to the caption of Extended Data Fig.~\ref{fig:pi1_gru_corner_ultranest}. See also Supplementary Table~\ref{table:fit_sensitivity} and Supplementary Table~\ref{table:best_fit}.}\label{fig:pi1_gru_corner_indep}
\end{figure*}

\begin{figure*}[!htp]
\centering
	\includegraphics[width=0.98\textwidth]{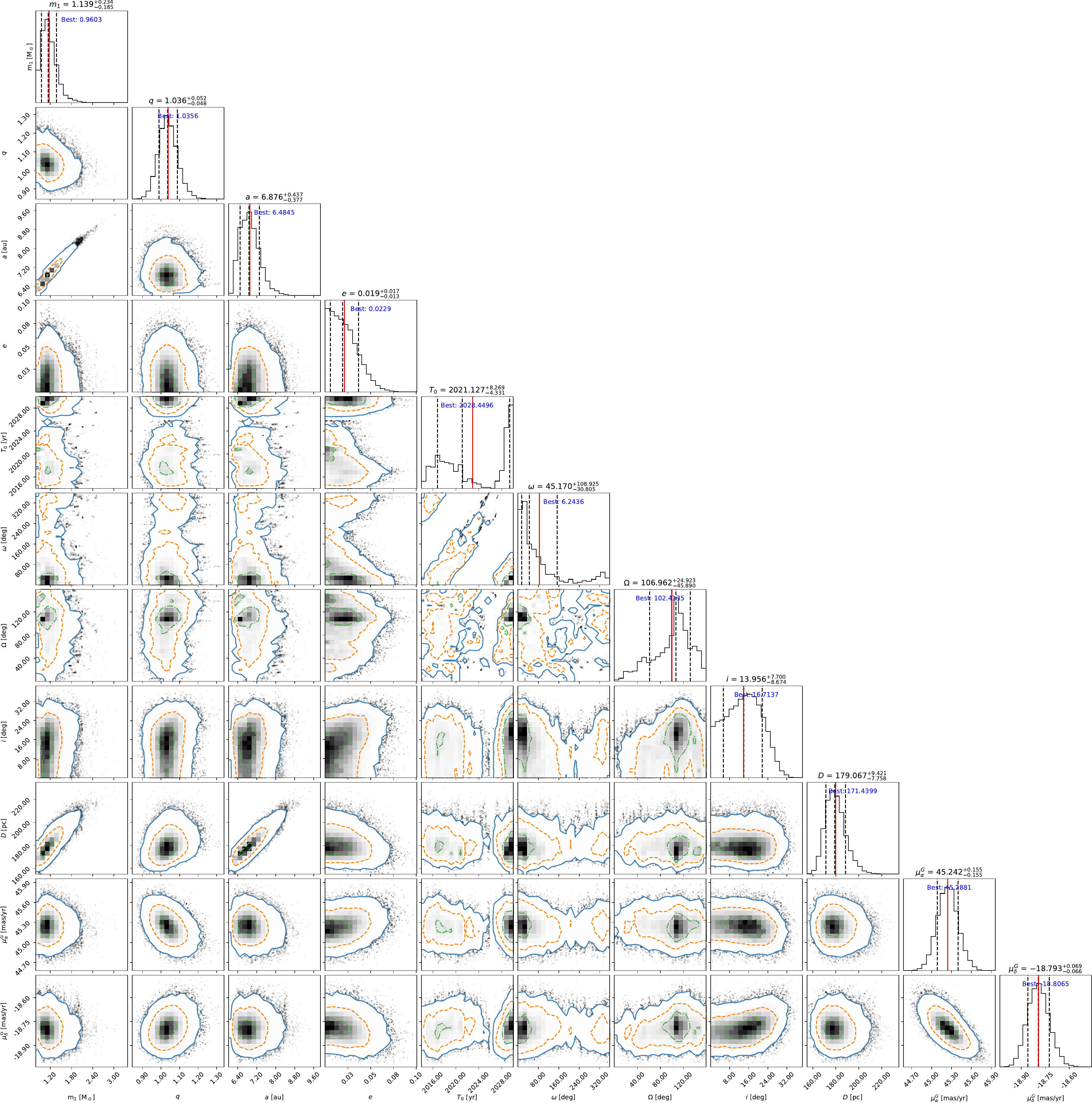}
 \caption{\textbf{Corner plot showing the retrieval of the $\boldsymbol{\pi^1}$~Gru orbital parameters with a Gamma prior for the distance with $L=700$.} For a full explanation, we refer to the caption of Extended Data Fig.~\ref{fig:pi1_gru_corner_ultranest};
 see also Supplementary Table~\ref{table:fit_sensitivity} and Supplementary Table~\ref{table:best_fit}.}\label{fig:pi1_gru_corner_ultranest_prior_D}
\end{figure*}

\begin{figure*}[!htp]
\centering
	\includegraphics[width=0.98\textwidth]{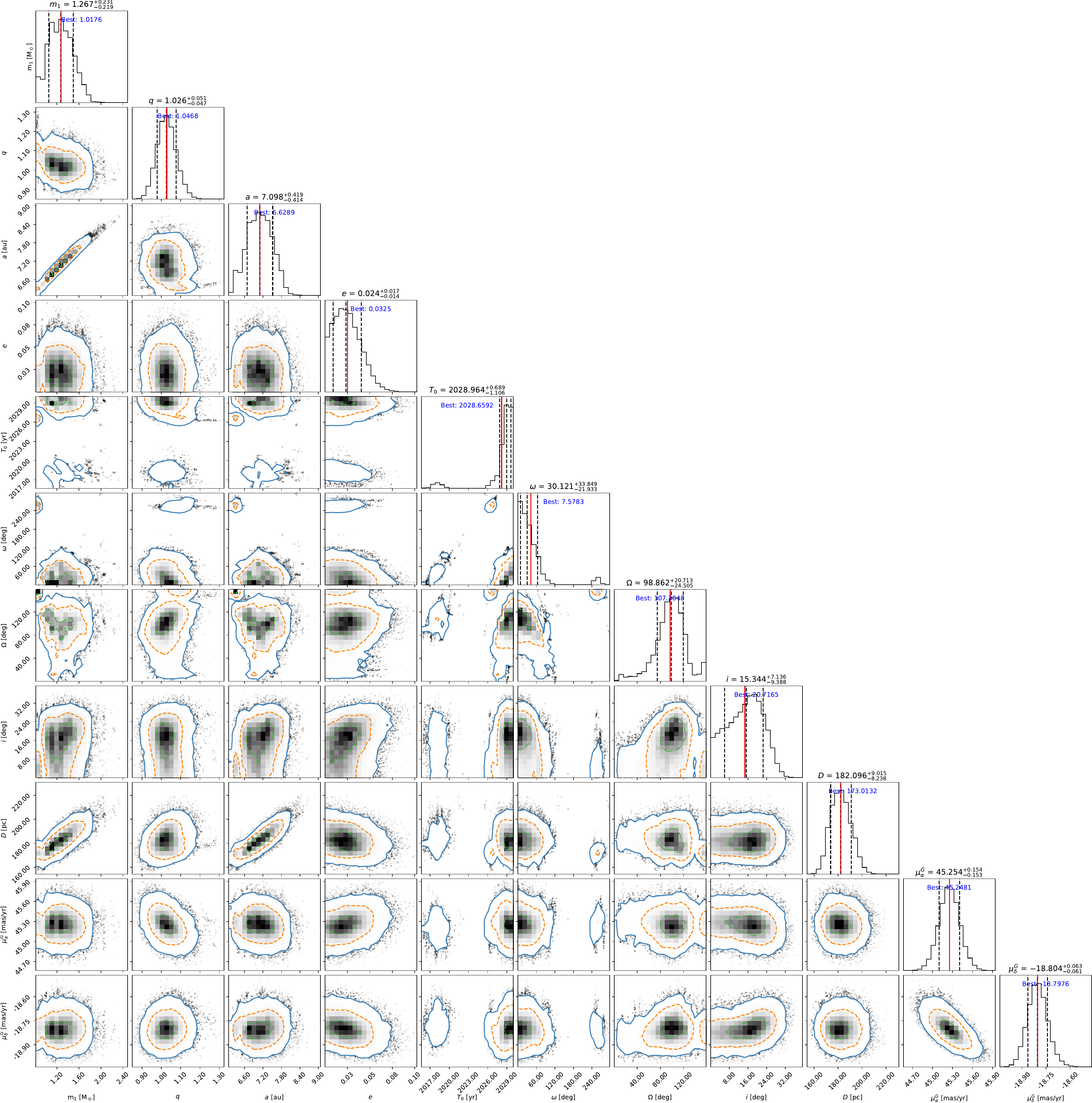}
 \caption{\textbf{Corner plot showing the retrieval of the $\boldsymbol{\pi^1}$~Gru orbital parameters using a Gaussian prior for $m_1$ with mean value of 1.5 \Msun\ and a standard deviation \ of 0.5 \Msun.} For a full explanation, we refer to the caption of Extended Data Fig.~\ref{fig:pi1_gru_corner_ultranest};
 see also Supplementary Table~\ref{table:fit_sensitivity} and Supplementary Table~\ref{table:best_fit}.}\label{fig:pi1_gru_corner_ultranest_prior_M1}
\end{figure*}

\begin{figure*}[!htp]
\centering
\includegraphics[width=0.98\textwidth]{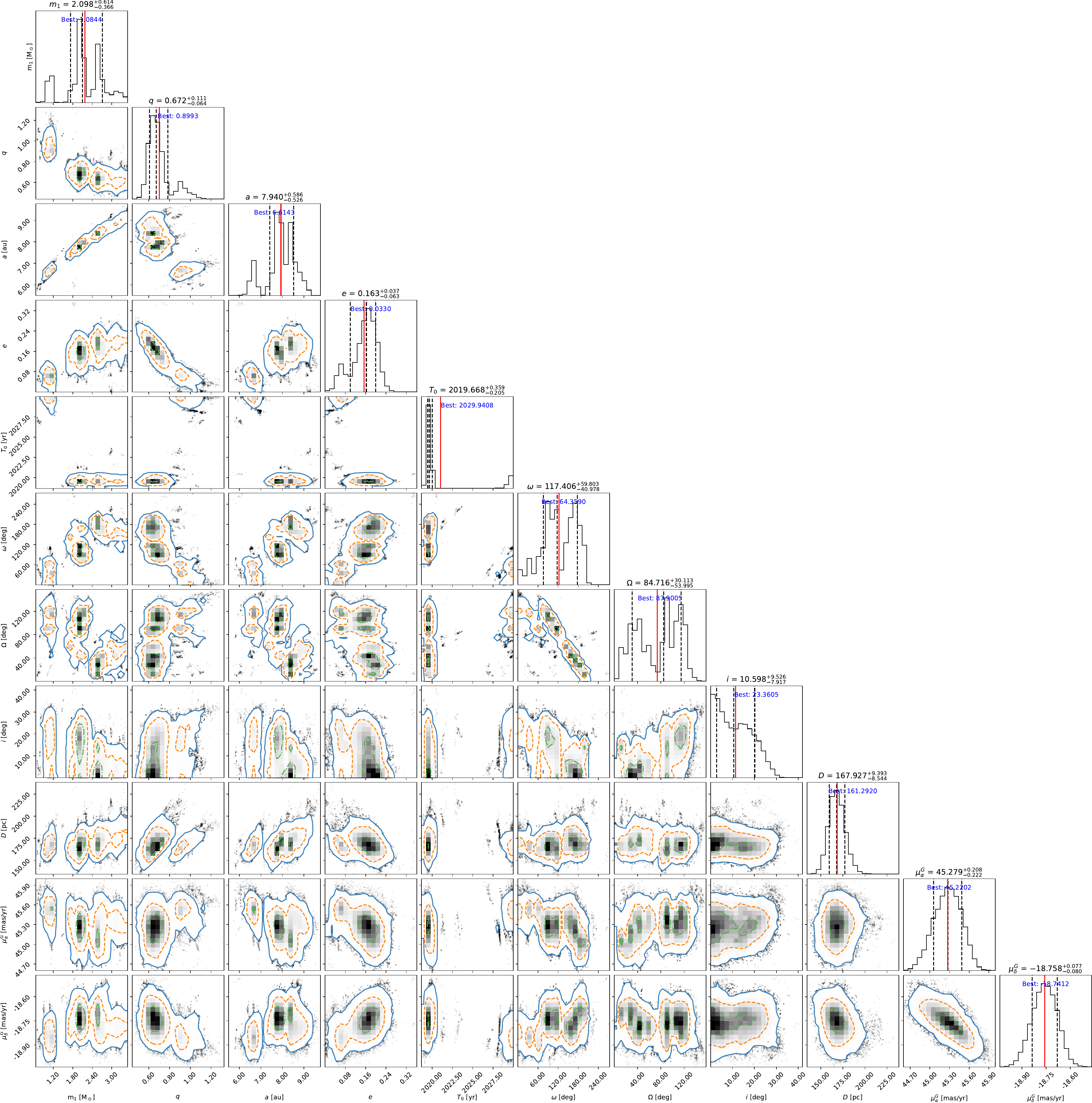}
\caption{\textbf{Corner plot showing the retrieval of the $\boldsymbol{\pi^1}$~Gru orbital parameters using \texttt{ultranest} sampling, excluding the ALMA 2023 C10 data from the fitting.} The corner plot belongs to \texttt{run a1} of the fitting; see text for more information.
For a full explanation, we refer to the caption of Extended Data Fig.~\ref{fig:pi1_gru_corner_ultranest}. See also Supplementary Table~\ref{table:fit_sensitivity} and Supplementary Table~\ref{table:best_fit}.}\label{fig:pi1_gru_corner_no_C10_run1}
\end{figure*}

\begin{figure*}[!htp]
\centering
\includegraphics[width=0.98\textwidth]{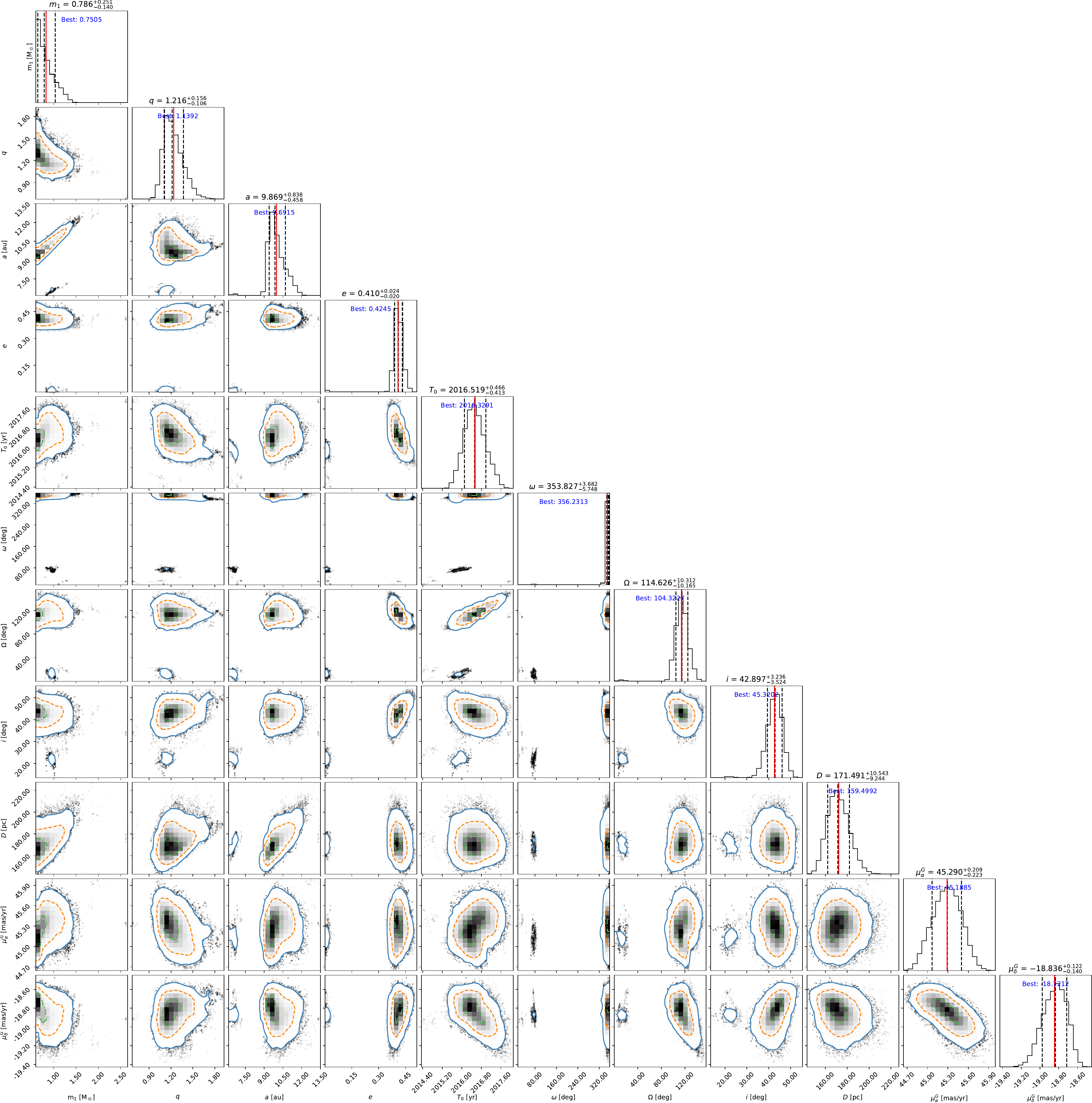}
\caption{\textbf{Corner plot showing the retrieval of the $\boldsymbol{\pi^1}$~Gru orbital parameters using \texttt{ultranest} sampling, excluding the ALMA 2023 C10 data from the fitting.} The corner plot belongs to \texttt{run a2} of the fitting; see text for more information.
For a full explanation, we refer to the caption of Extended Data Fig.~\ref{fig:pi1_gru_corner_ultranest}. See also Supplementary Table~\ref{table:fit_sensitivity} and Supplementary Table~\ref{table:best_fit}.}\label{fig:pi1_gru_corner_no_C10_run2}
\end{figure*}

\begin{figure*}[!htp]
\centering
\includegraphics[width=0.98\textwidth]{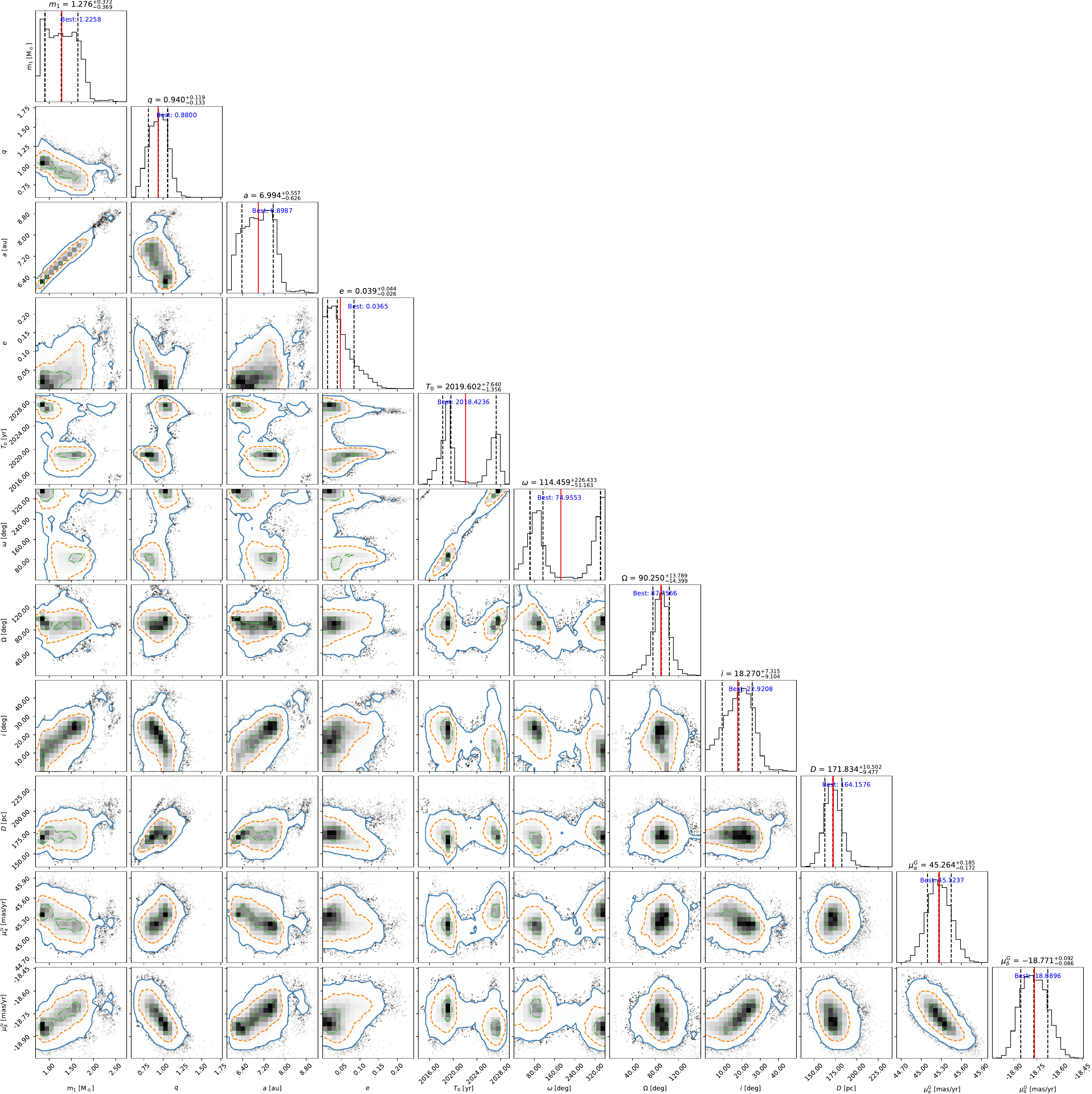}
\caption{\textbf{Corner plot showing the retrieval of the $\boldsymbol{\pi^1}$~Gru orbital parameters using \texttt{ultranest} sampling, excluding the ALMA 2023 C10 data from the fitting.} The corner plot belongs to \texttt{run a3} of the fitting; see text for more information.
For a full explanation, we refer to the caption of Extended Data Fig.~\ref{fig:pi1_gru_corner_ultranest}. See also Supplementary Table~\ref{table:fit_sensitivity} and Supplementary Table~\ref{table:best_fit}.}\label{fig:pi1_gru_corner_no_C10_run3}
\end{figure*}

\begin{figure*}[!htp]
\centering
\includegraphics[width=0.98\textwidth]{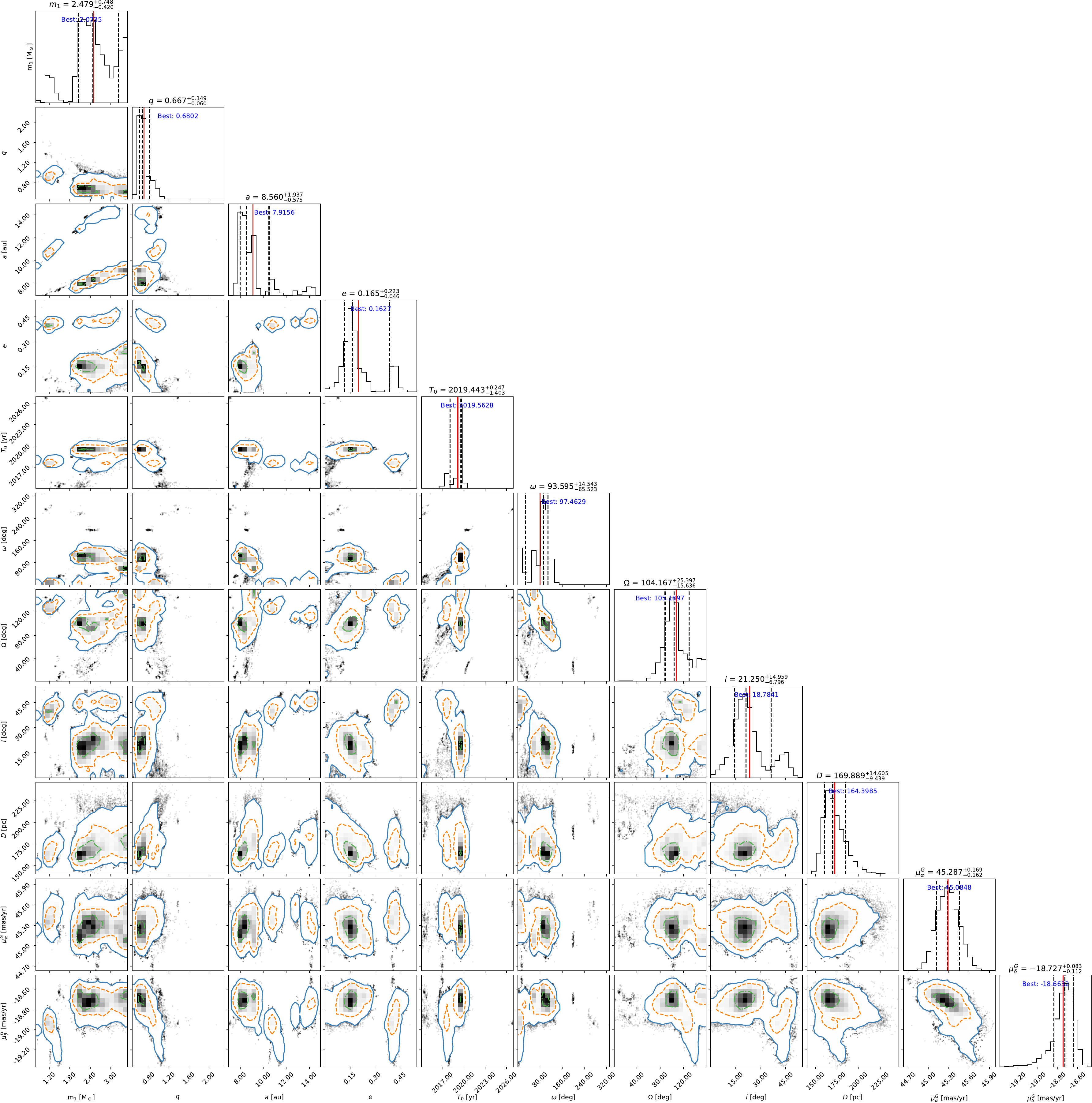}
\caption{\textbf{Corner plot showing the retrieval of the $\boldsymbol{\pi^1}$~Gru orbital parameters using \texttt{ultranest} sampling, excluding the ALMA 2023 C10 data from the fitting.} The corner plot belongs to \texttt{run a4} of the fitting; see text for more information.
For a full explanation, we refer to the caption of Extended Data Fig.~\ref{fig:pi1_gru_corner_ultranest}. See also Supplementary Table~\ref{table:fit_sensitivity} and Supplementary Table~\ref{table:best_fit}.}\label{fig:pi1_gru_corner_no_C10_run4}
\end{figure*}

\begin{figure*}[!htp]
\centering
\includegraphics[width=0.98\textwidth]{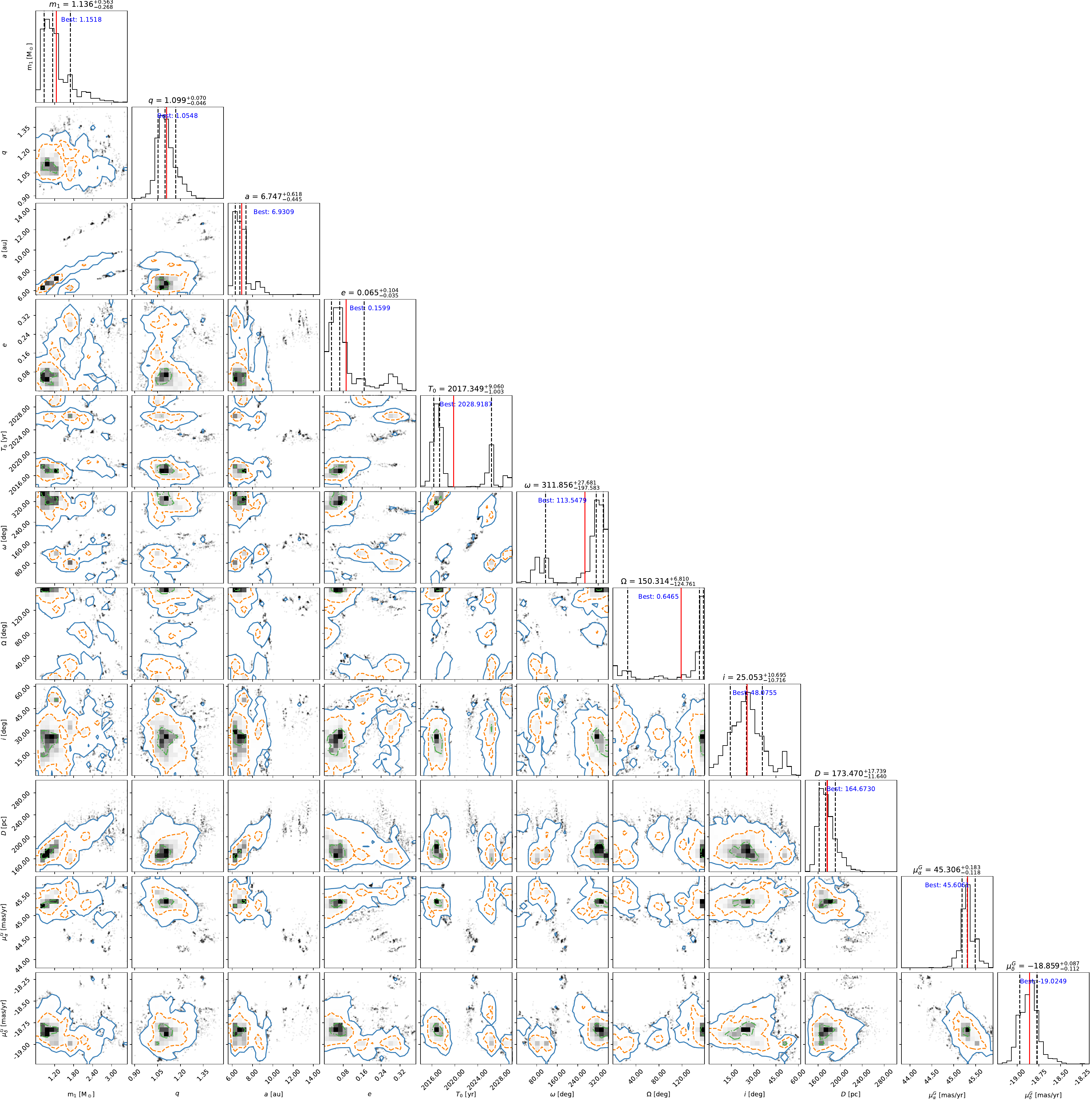}
\caption{\textbf{Corner plot showing the retrieval of the $\boldsymbol{\pi^1}$~Gru orbital parameters using \texttt{ultranest} sampling, excluding the \textit{Gaia} data from the fitting.} For a full explanation, we refer to the caption of Extended Data Fig.~\ref{fig:pi1_gru_corner_ultranest}. See also Supplementary Table~\ref{table:fit_sensitivity} and Supplementary Table~\ref{table:best_fit}.}\label{fig:pi1_gru_corner_no_GAIA}
\end{figure*}

\begin{figure*}[!htp]
\centering
\includegraphics[width=0.98\textwidth]{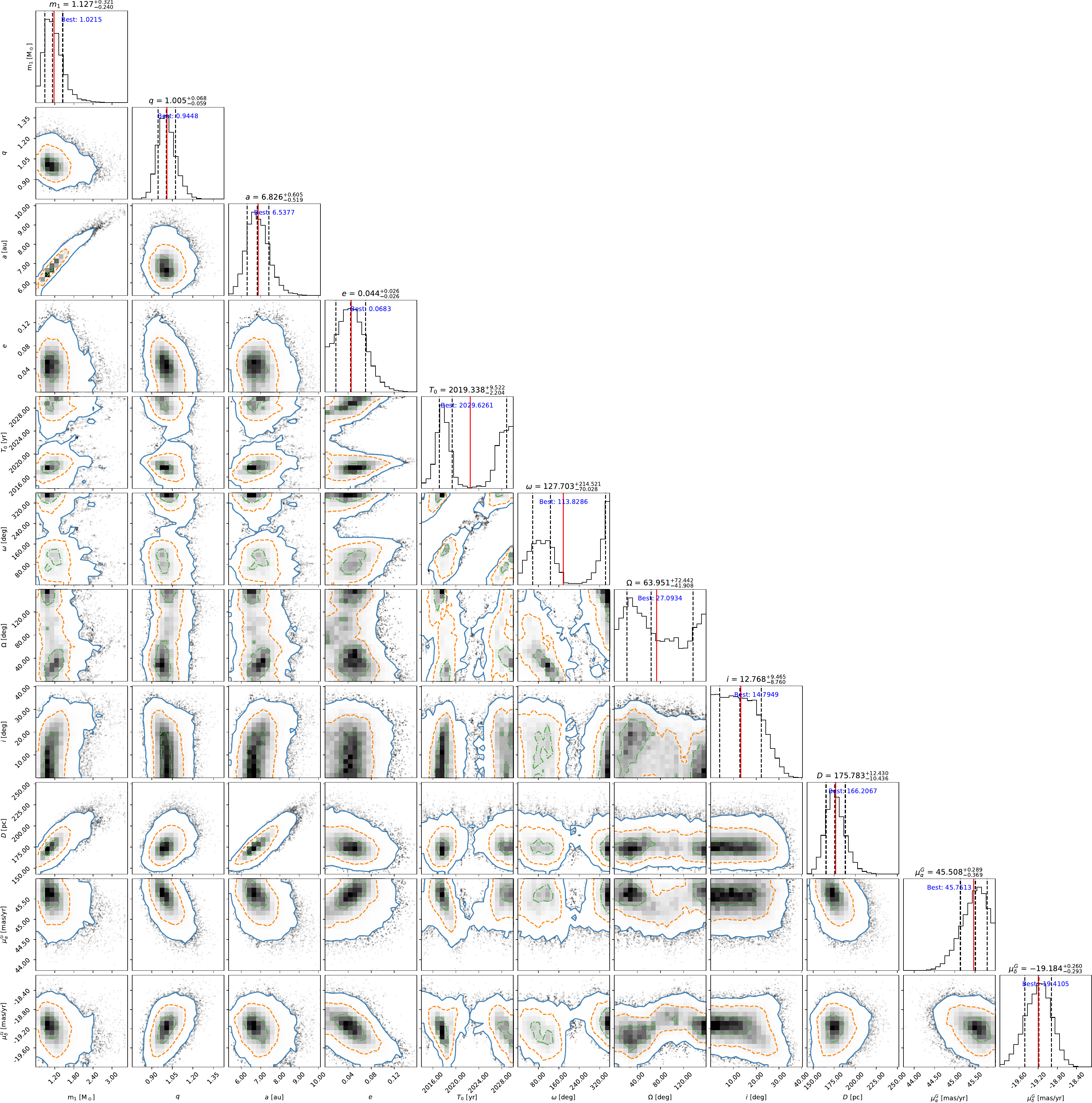}
\caption{\textbf{Corner plot showing the retrieval of the $\boldsymbol{\pi^1}$~Gru orbital parameters using \texttt{ultranest} sampling, excluding the \textit{Hipparcos} data from the fitting.} For a full explanation, we refer to the caption of Extended Data Fig.~\ref{fig:pi1_gru_corner_ultranest}. See also Supplementary Table~\ref{table:fit_sensitivity} and Supplementary Table~\ref{table:best_fit}.}\label{fig:pi1_gru_corner_no_HIP}
\end{figure*}

\begin{figure*}[!htp]
\centering
\includegraphics[width=0.98\textwidth]{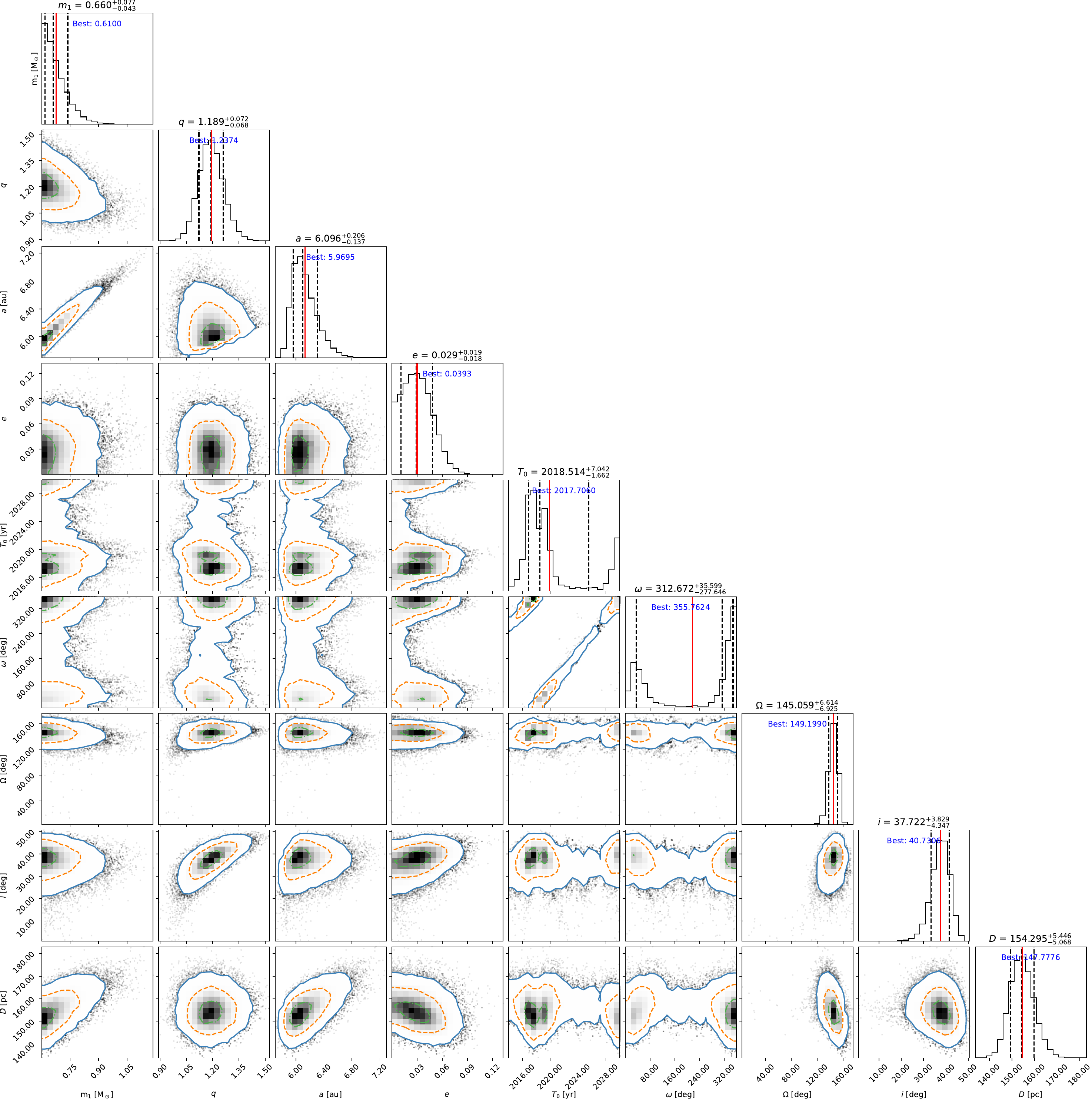}
\caption{\textbf{Corner plot showing the retrieval of the $\boldsymbol{\pi^1}$~Gru orbital parameters using \texttt{ultranest} sampling, where the orbital solution has been approximated in both the astrometric position and the tangential velocity anomaly.} For a full explanation, we refer to the caption of Extended Data Fig.~\ref{fig:pi1_gru_corner_ultranest}. See also Supplementary Table~\ref{table:fit_sensitivity} and Supplementary Table~\ref{table:best_fit}.}\label{fig:pi1_gru_corner_approximate}
\end{figure*}

\afterpage{\clearpage}
\newpage

\section{Additional plots for fit to astrometric positions}
\begin{figure*}[!htp]
\centering
	\includegraphics[width=15cm]{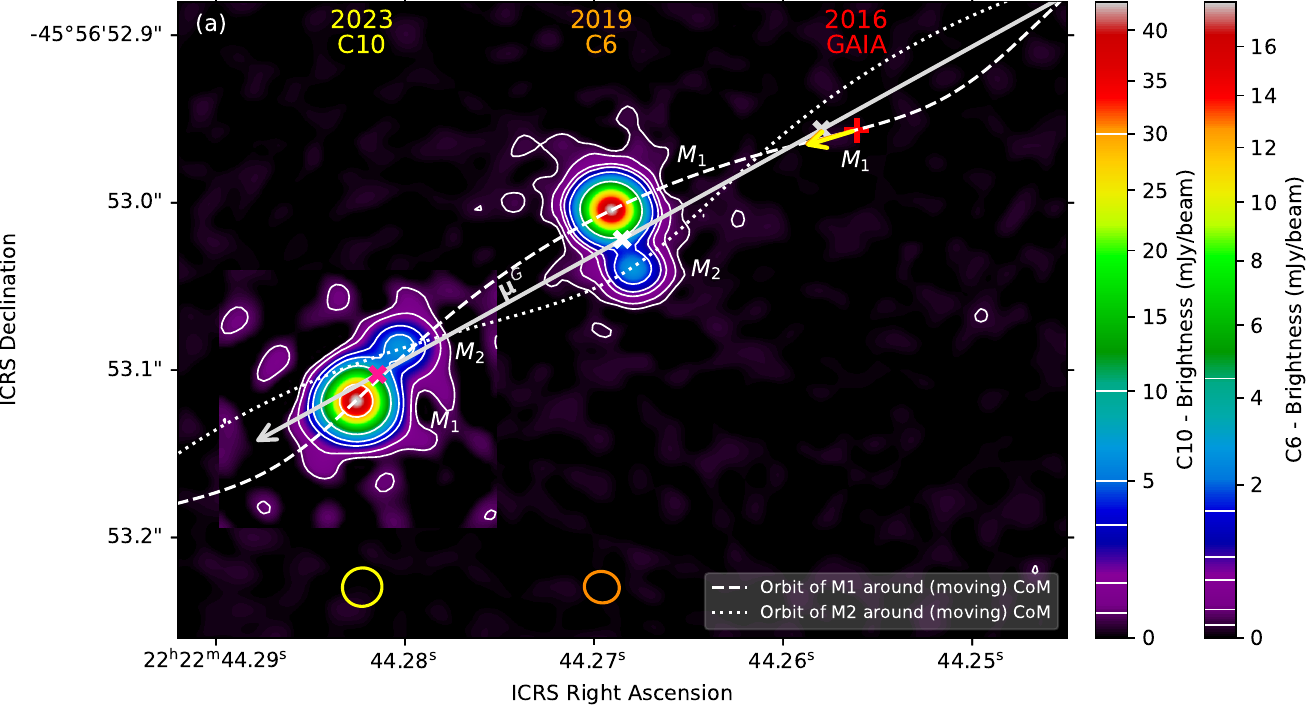}\\
    \includegraphics[width=15cm]{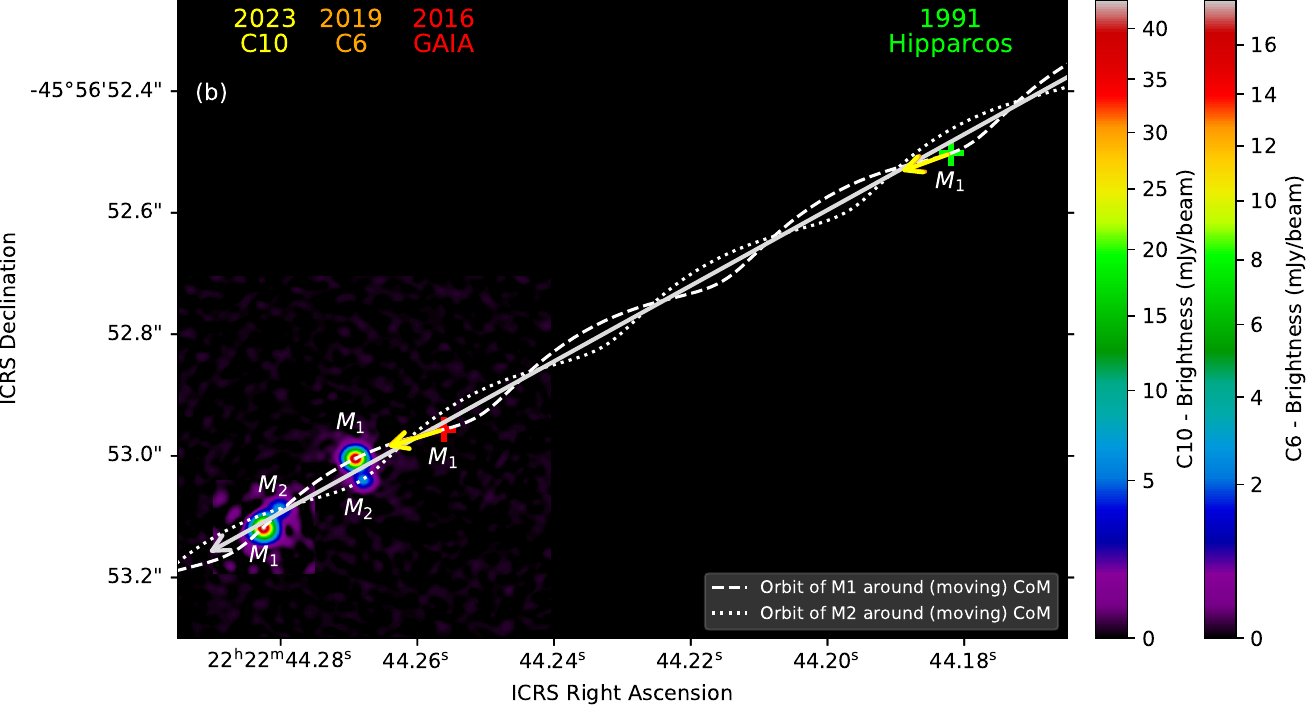}
 \caption{\textbf{Proper motion of the $\boldsymbol{\pi^1}$~Gru system for the \texttt{eccentric} model.} Parameters are listed in Supplementary Table~\ref{table:best_fit}. See the caption of Fig.~\ref{fig:pi1_gru} for a comprehensive explanation. }\label{fig:pi1_gru_motion_eccentric}
\end{figure*}
\afterpage{\clearpage}
\newpage

\begin{figure*}[!htp]
\centering
	\includegraphics[width=15cm]{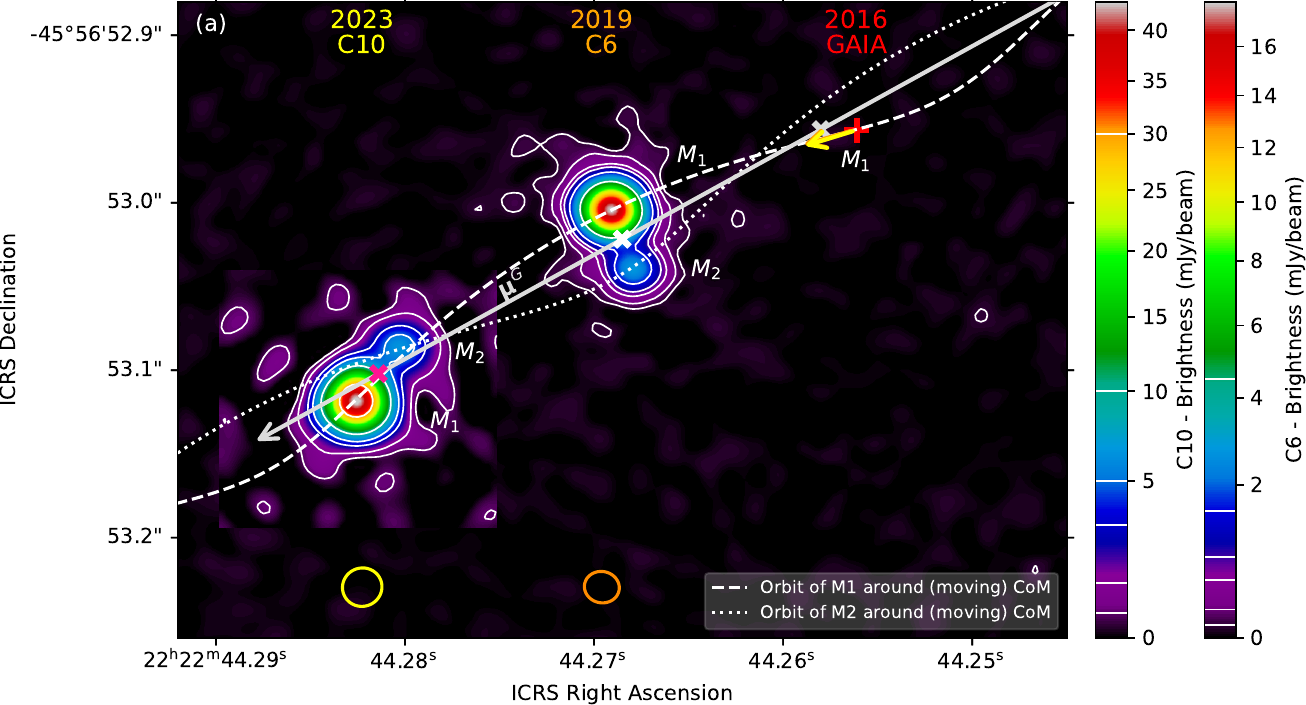}\\
    \includegraphics[width=15cm]{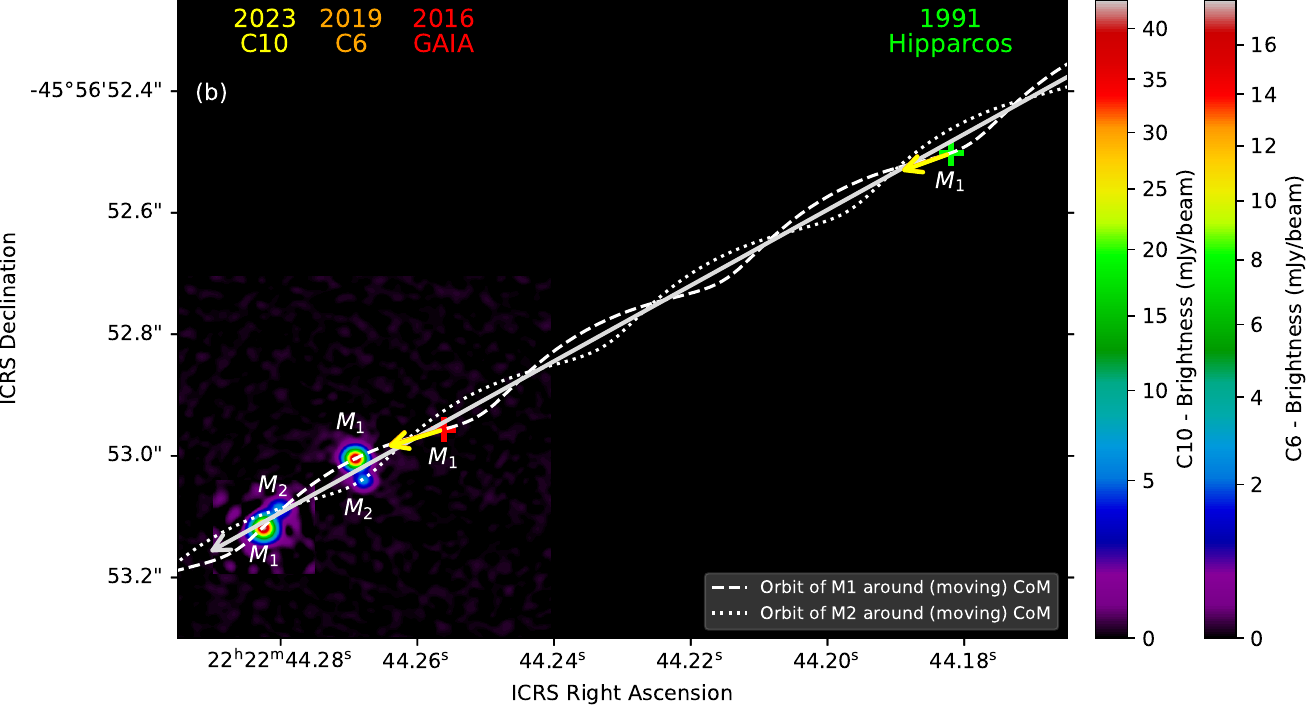}
 \caption{\textbf{Proper motion of the $\boldsymbol{\pi^1}$~Gru system derived when not accounting for the \textit{Gaia} and \textit{Hipparcos} covariance matrices.} Parameters are listed in Supplementary Table~\ref{table:best_fit}. See the caption of Fig.~\ref{fig:pi1_gru} for a comprehensive explanation.}\label{fig:pi1_gru_motion_indep}
\end{figure*}
\afterpage{\clearpage}
\newpage

\begin{figure*}[!htp]
\centering
	\includegraphics[width=15cm]{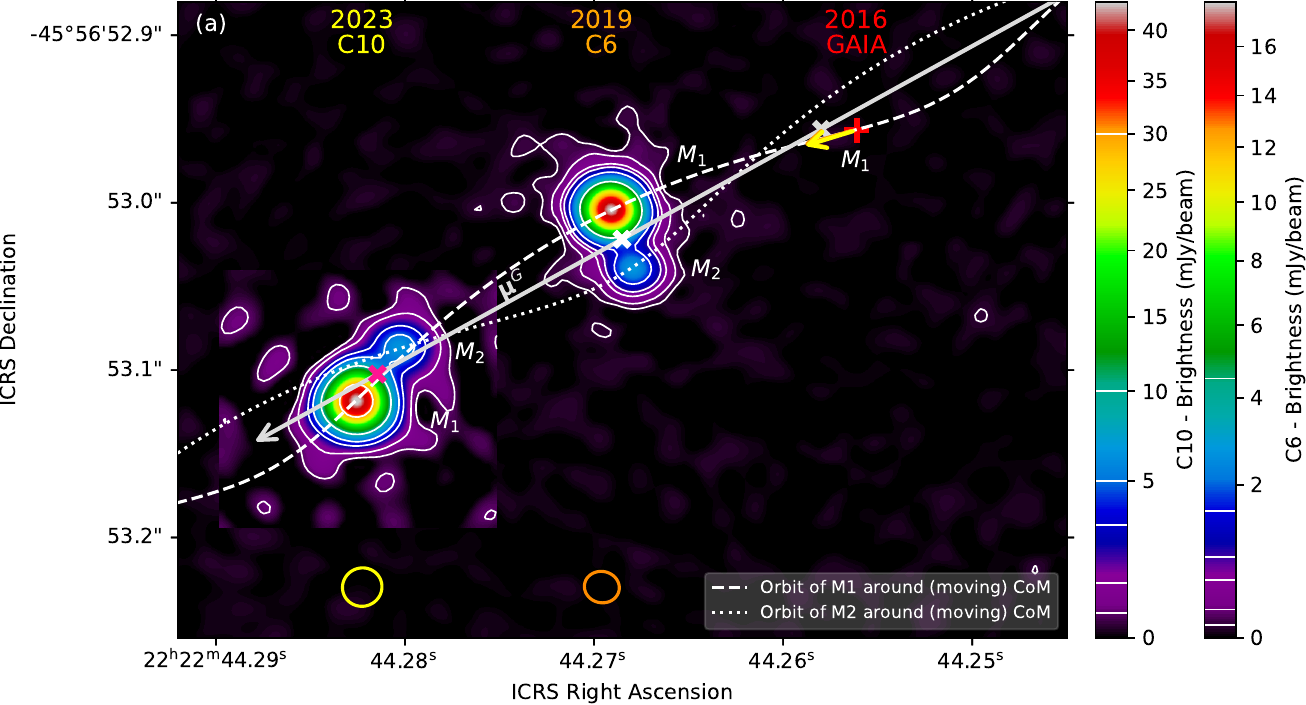}\\
    \includegraphics[width=15cm]{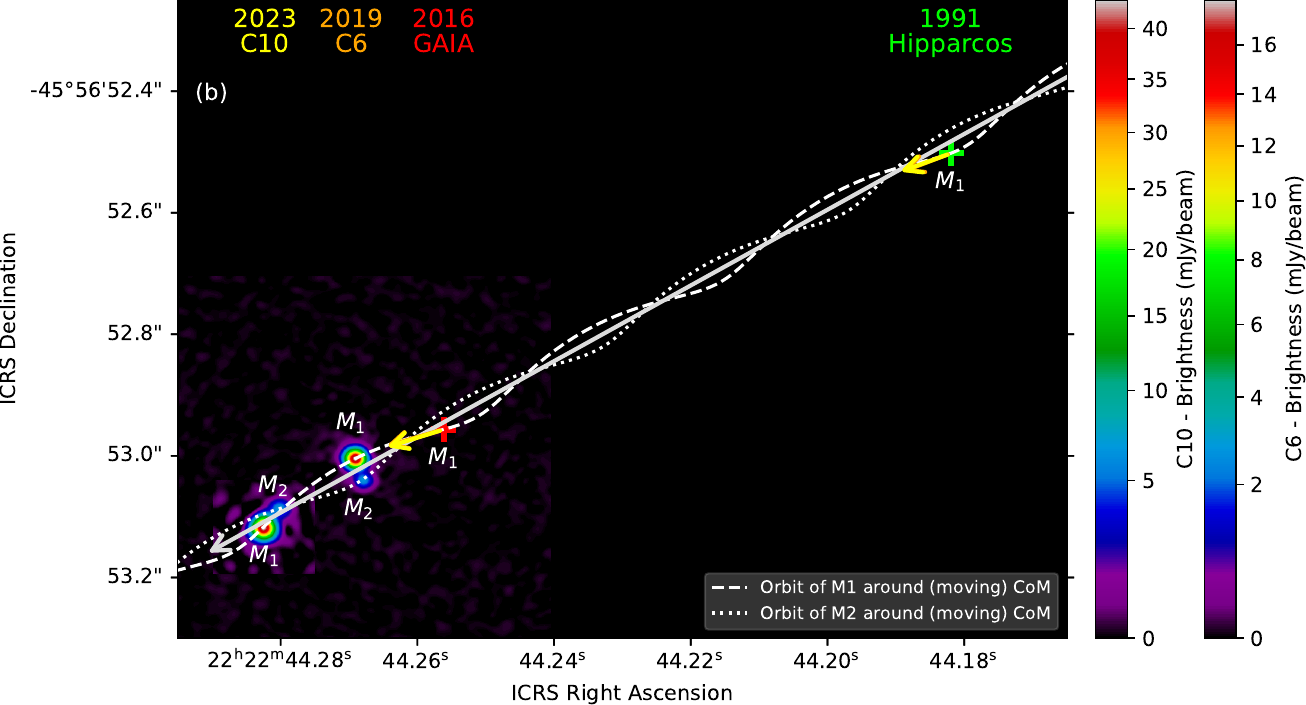}
 \caption{\textbf{Proper motion of the $\boldsymbol{\pi^1}$~Gru system derived using a Gamma prior for the distance with $L=700$.} Parameters are listed in Supplementary Table~\ref{table:best_fit}. See the caption of Fig.~\ref{fig:pi1_gru} for a comprehensive explanation.}\label{fig:pi1_gru_motion_prior_D}
\end{figure*}
\afterpage{\clearpage}
\newpage

\begin{figure*}[!htp]
\centering
	\includegraphics[width=15cm]{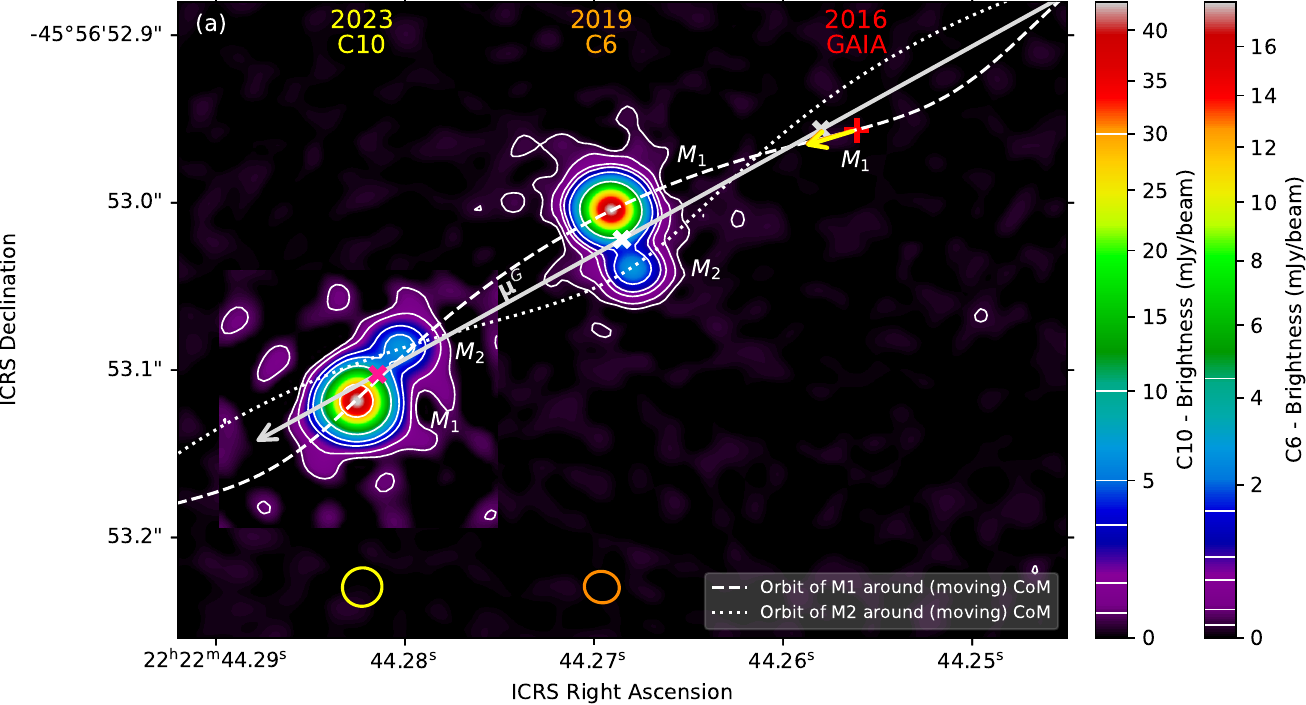}\\
    \includegraphics[width=15cm]{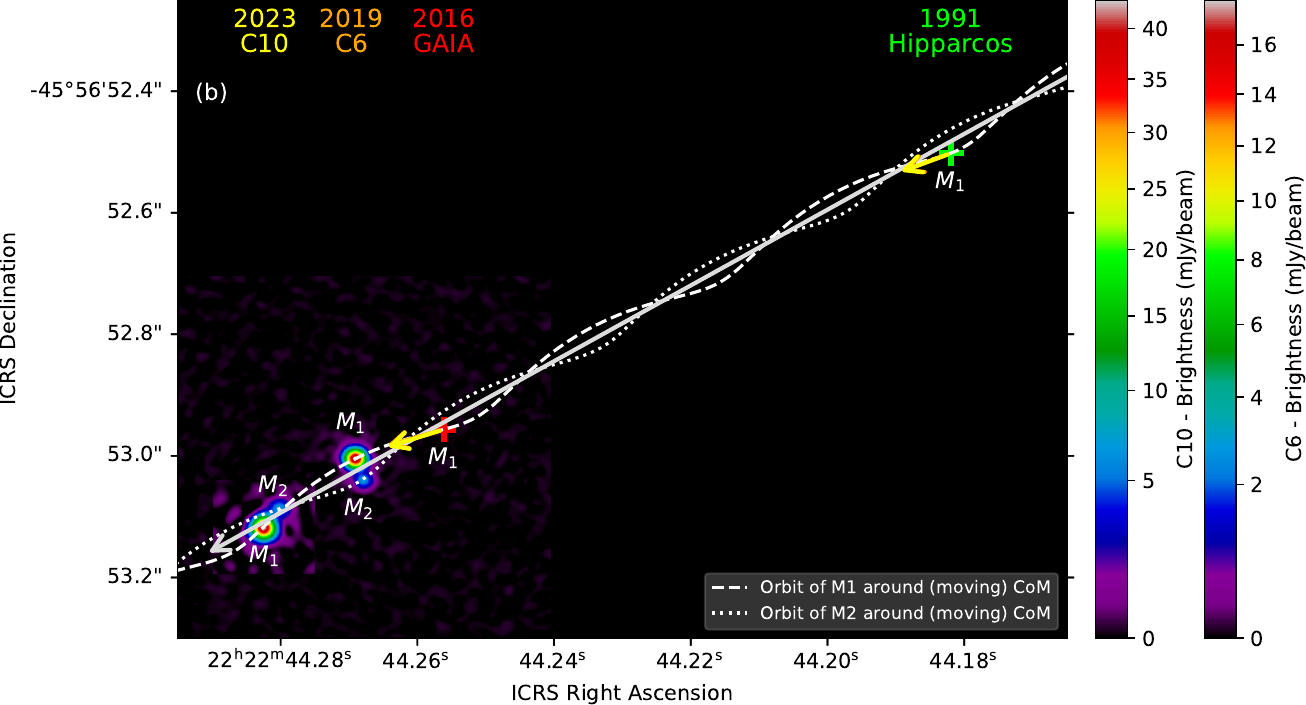}
 \caption{\textbf{Proper motion of the $\boldsymbol{\pi^1}$~Gru system derived using a Gaussian prior for $m_1$ with mean value of 1.5 \Msun\ and a standard deviation \ of 0.5 \Msun.} Parameters are listed in Supplementary Table~\ref{table:best_fit}. See the caption of Fig.~\ref{fig:pi1_gru} for a comprehensive explanation.}\label{fig:pi1_gru_motion_prior_M1}
\end{figure*}
\afterpage{\clearpage}
\newpage

\begin{figure*}[!htp]
\centering
	\includegraphics[width=15cm]{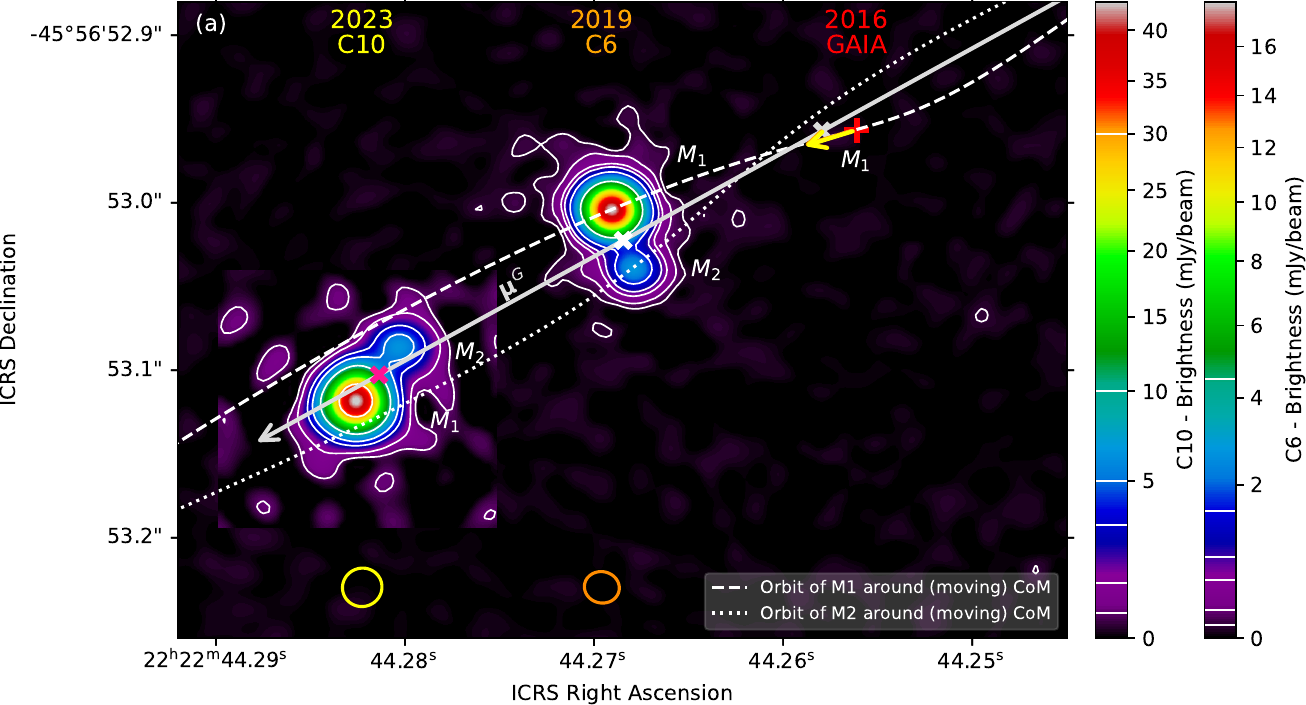}\\
    \includegraphics[width=15cm]{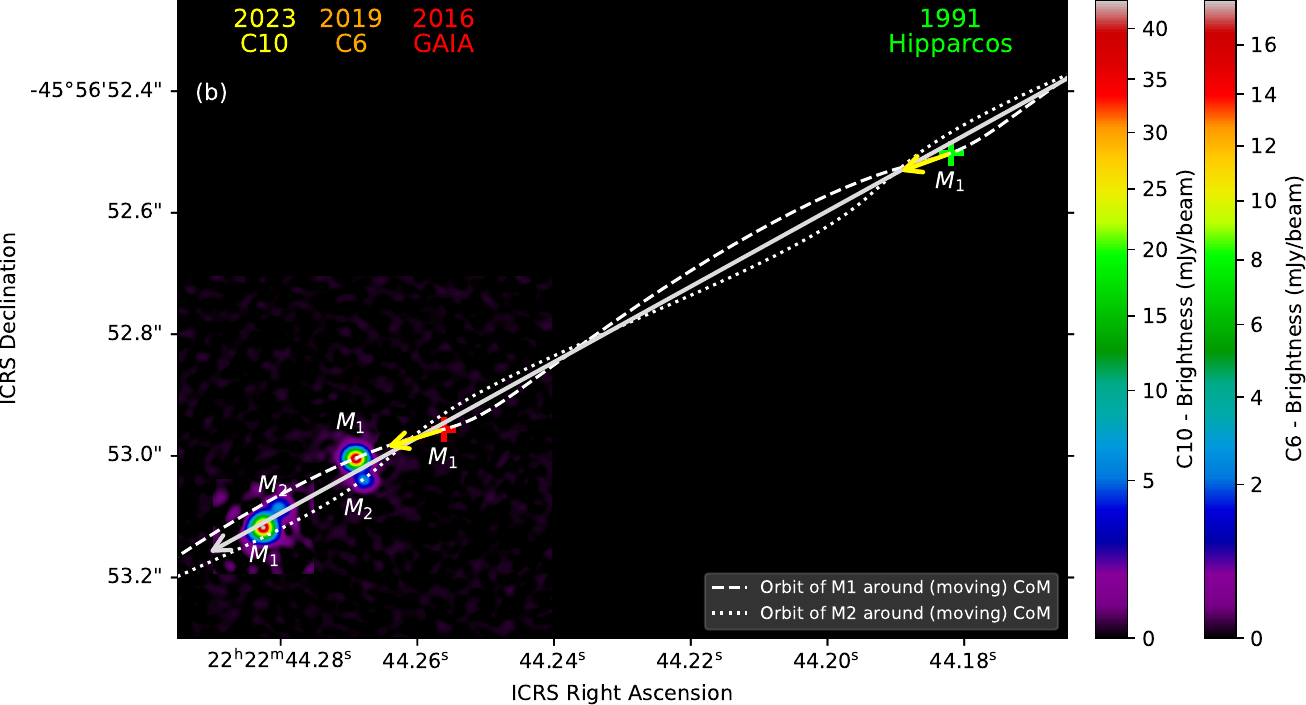}
 \caption{\textbf{Proper motion of the $\boldsymbol{\pi^1}$~Gru system derived when excluding the ALMA 2023 C10 data from the fitting.} Figure belongs to \texttt{run a2} of the fitting with corresponding corner plot and retrieved best-fit parameters shown in Supplementary Fig.~\ref{fig:pi1_gru_corner_no_C10_run2}. See the caption of Fig.~\ref{fig:pi1_gru} for a comprehensive explanation.}\label{fig:pi1_gru_motion_no_C10}
\end{figure*}
\afterpage{\clearpage}
\newpage

\begin{figure*}[!htp]
\centering
	\includegraphics[width=15cm]{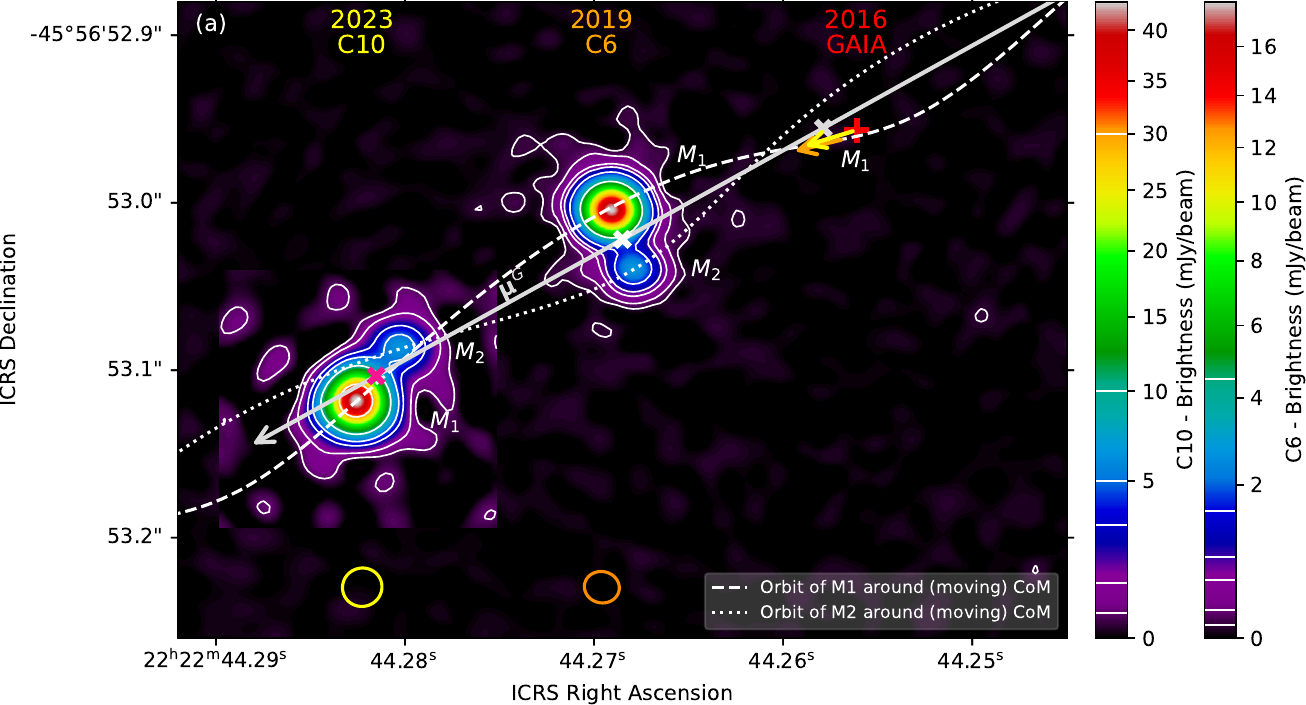}\\
    \includegraphics[width=15cm]{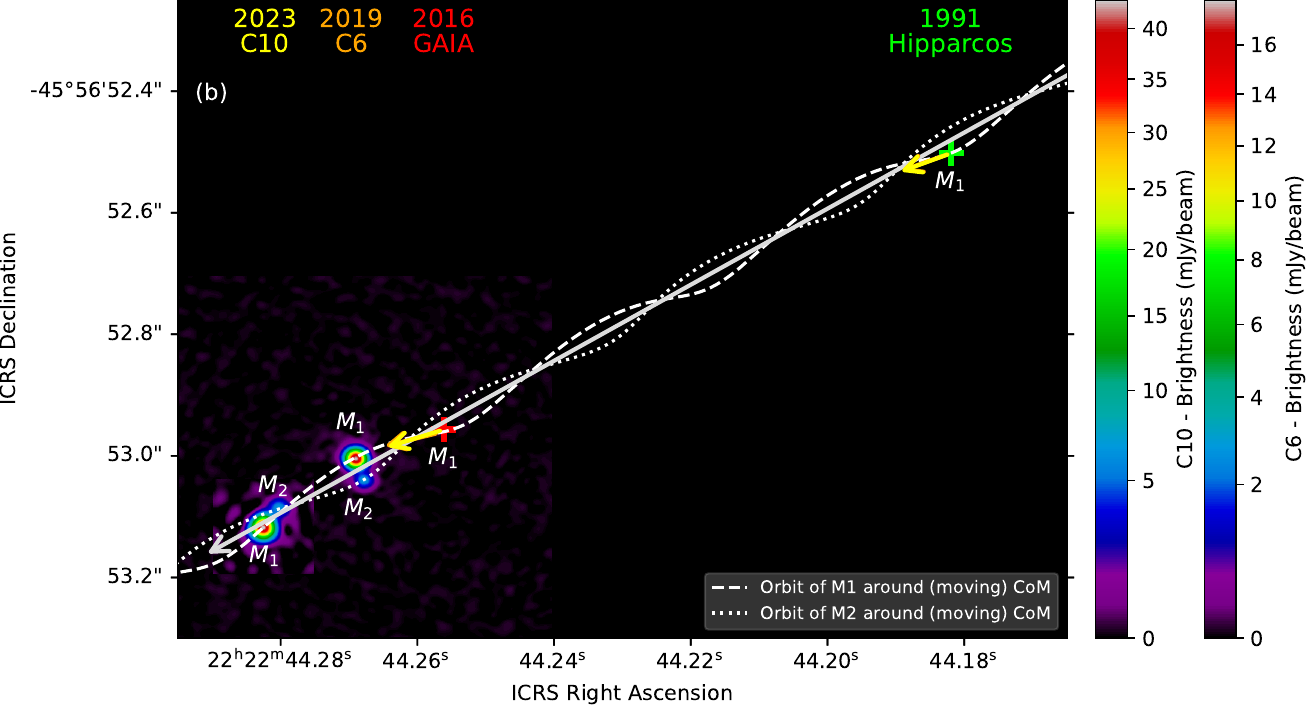}
 \caption{\textbf{Proper motion of the $\boldsymbol{\pi^1}$~Gru system derived when excluding the \textit{Gaia} data from the fitting.} Parameters are listed in Supplementary Table~\ref{table:best_fit}. See the caption of Fig.~\ref{fig:pi1_gru} for a comprehensive explanation.}\label{fig:pi1_gru_motion_no_GAIA}
\end{figure*}
\afterpage{\clearpage}
\newpage

\begin{figure*}[!htp]
\centering
	\includegraphics[width=15cm]{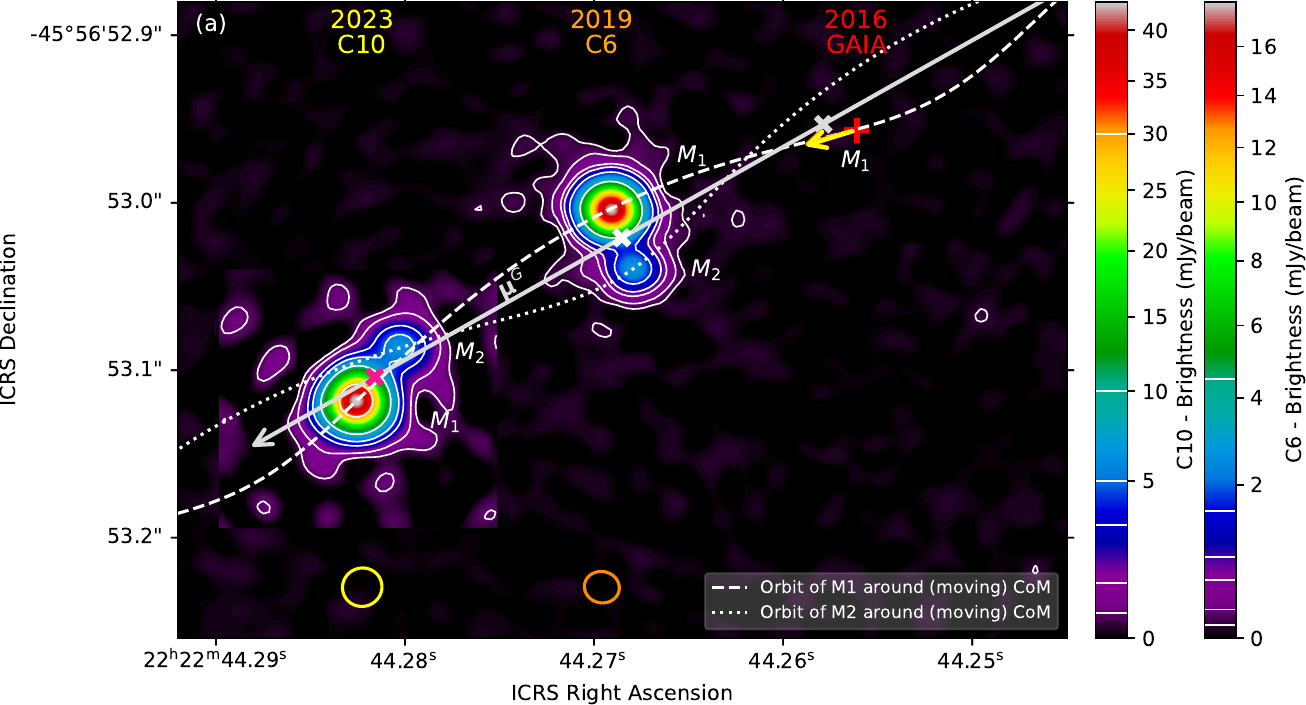}\\
    \includegraphics[width=15cm]{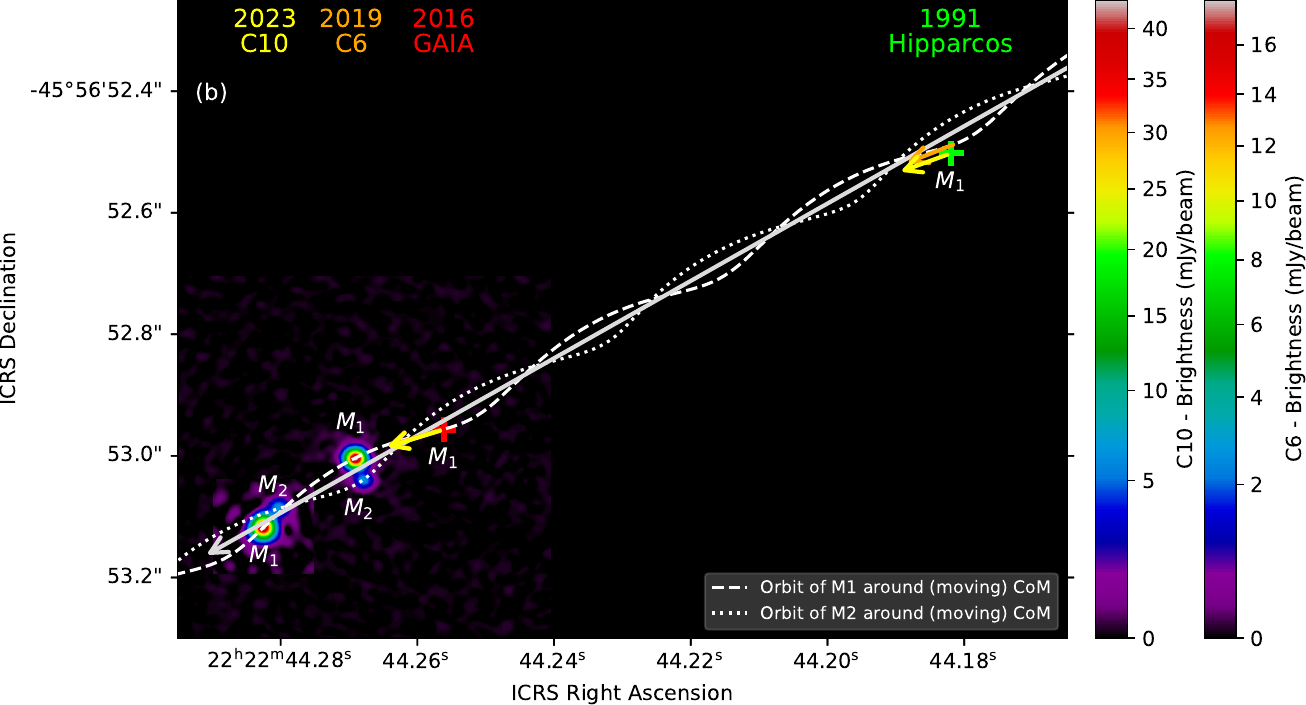}
 \caption{\textbf{Proper motion of the $\boldsymbol{\pi^1}$~Gru system derived when excluding the \textit{Hipparcos} data from the fitting.} Parameters are listed in Supplementary Table~\ref{table:best_fit}. See the caption of Fig.~\ref{fig:pi1_gru} for a comprehensive explanation.}\label{fig:pi1_gru_motion_no_HIP}
\end{figure*}
\afterpage{\clearpage}
\newpage

\begin{figure*}[!htp]
\centering
	\includegraphics[width=15cm]{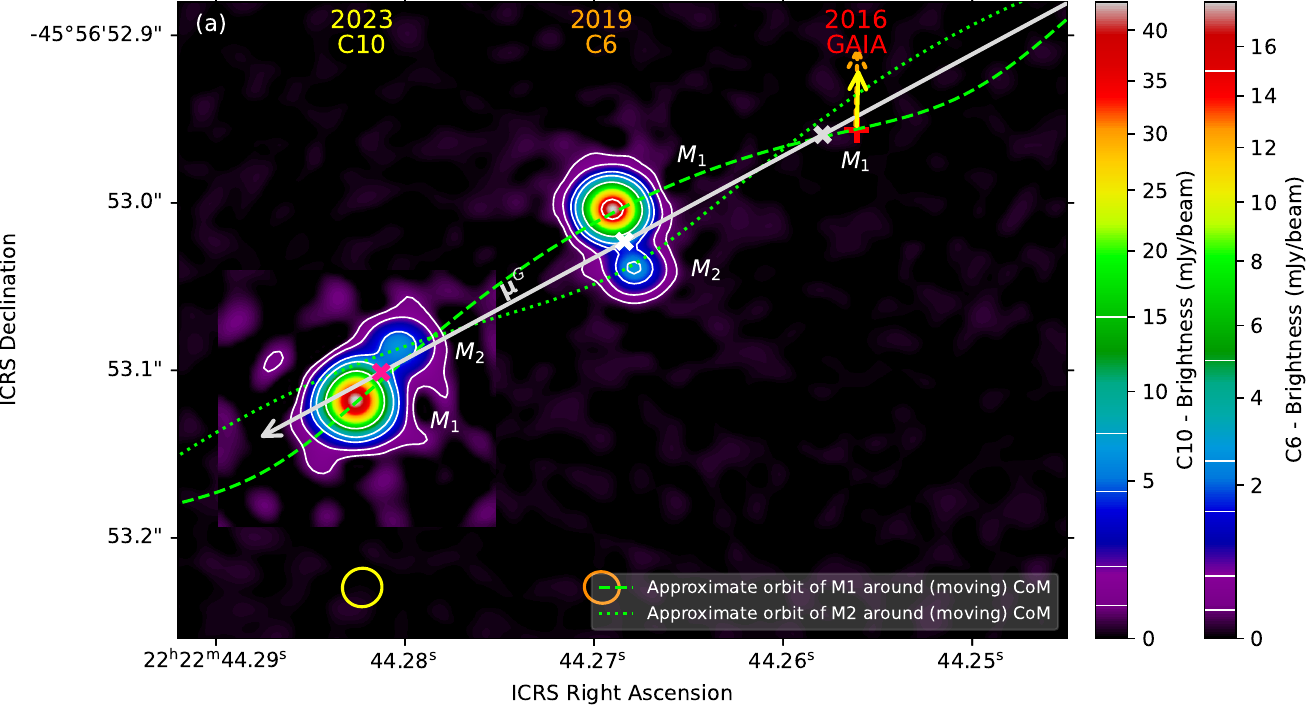}\\
    \includegraphics[width=15cm]{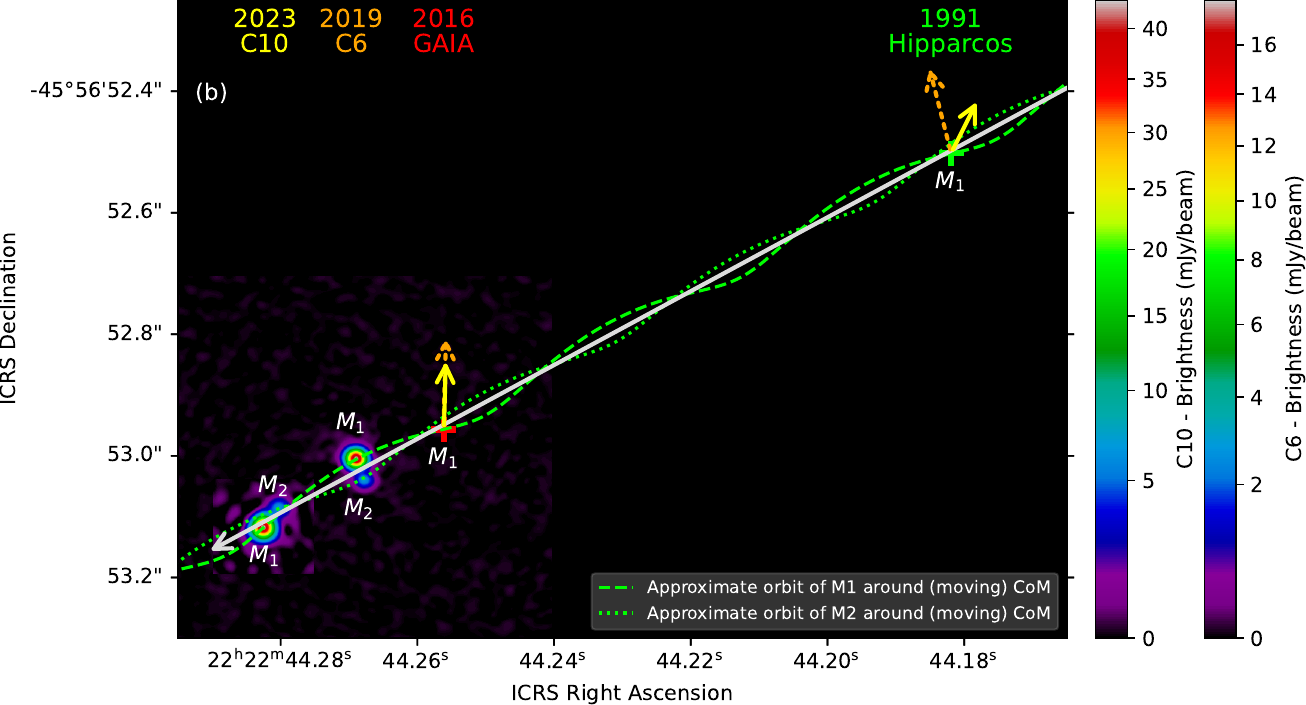}
    \caption{\textbf{Proper motion of the $\boldsymbol{\pi^1}$~Gru system derived using the approximate orbital solution for both the astrometric position and the tangential velocity anomaly.} The corresponding parameters are listed in Supplementary Table~\ref{table:best_fit}. For a comprehensive explanation, refer to the caption of Fig.~\ref{fig:pi1_gru}. Similar to Supplementary Fig.~\ref{Fig:approximate_vtan_xy}, dashed and dotted lime lines represent the proper motion of $M_1$ and $M_2$, respectively, using the approximate solution for the astrometric position. Here, the yellow vectors indicate the observed \textit{Hipparcos} and \textit{Gaia} tangential velocity anomalies, and the dashed orange vectors the predicted anomalies during the \textit{Gaia} and \textit{Hipparcos} observing epochs approximated by using the orbital tangential velocity.}\label{fig:pi1_gru_motion_approximate}
\end{figure*}
\afterpage{\clearpage}
\newpage

\section*{Bibliography}
\longrefs=1
\bibliographystyle{naturemag}
\bibliography{sn-bibliography}


\end{document}